\documentclass[preprint,12pt]{elsart_mod}

\usepackage{amssymb}
\usepackage{cite}
\usepackage{slashed}
\renewcommand{\appendixname}{}

\usepackage{amsmath}

\usepackage[linktocpage=true]{hyperref}

\font\mybb=msbm10 at 11pt
\font\mybbb=msbm10 at 9pt
\newcommand{\bb}[1]{\hbox{\mybb#1}}
\newcommand{\bbb}[1]{\hbox{\mybbb#1}}
\newcommand{\bZ}{\bb{Z}}
\newcommand{\bR}{\bb{R}}
\newcommand{\bC}{\bb{C}}
\newcommand{\bH}{\bb{H}}

\newcommand{\bbR}{\bbb{R}}

\newcommand{\se} {{e}}

\newcommand{\fe} {{\bf e}}

\newcommand{\fiveF}{{F}}

\newcommand{\elF}{{F }}
\newcommand{\elE}{{G }}
\newcommand{\elM}{{\tilde F }}

\newcommand{\iiF}{{F }}

\newcommand{\bbl}{{\bf{\ell}}}

\newcommand{\spc}{\Omega}

\newcommand{\be}{\begin{equation}}
\newcommand{\ee}{\end{equation}}
\newcommand{\bea}{\begin{eqnarray}}
\newcommand{\eea}{\end{eqnarray}}
\newcommand{\la}{\label}
\newcommand{\nn}{\nonumber}

\newcommand{\eps}{\epsilon}
\newcommand{\Ga}{\Gamma}

\newcommand{\al}{\alpha }
\newcommand{\bet}{\beta }

\newcommand{\hga}{\gamma}

\newcommand{\tM}{\text{\tiny $M$}}
\newcommand{\tB}{\text{\tiny $B$}}
\newcommand{\tA}{\text{\tiny $A$}}
\newcommand{\tC}{\text{\tiny $C$}}
\newcommand{\tD}{\text{\tiny $D$}}
\newcommand{\tN}{\text{\tiny $N$}}
\newcommand{\tL}{\text{\tiny $L$}}
\newcommand{\tP}{\text{\tiny $P$}}
\newcommand{\tR}{\text{\tiny $R$}}
\newcommand{\tQ}{\text{\tiny $Q$}}
\newcommand{\tS}{\text{\tiny $S$}}
\newcommand{\tI}{\text{\tiny $I$}}
\newcommand{\tJ}{\text{\tiny $J$}}
\newcommand{\tK}{\text{\tiny $K$}}
\newcommand{\tG}{\text{\tiny $G$}}
\newcommand{\tE}{\text{\tiny $E$}}

\newcommand{\nat} {{\natural}}

\newcommand { \LF} {{LF}}
\newcommand {\LG} {{LG}}
\newcommand { \LP} {{LP}}

\newcommand { \BF} {{BF}}

\newcommand*{\defeq}{\mathrel{\vcenter{\baselineskip0.5ex \lineskiplimit0pt
                     \hbox{\scriptsize.}\hbox{\scriptsize.}}}%
                     =}

\DeclareMathOperator{\AdS}{AdS}
\DeclareMathOperator{\dvol}{dvol}
\DeclareMathOperator{\CW}{CW}
\DeclareMathOperator{\dS}{dS}

\journal{Physics Reports}

\begin{document}

\begin{frontmatter}




\title{Classification, geometry and  applications of supersymmetric  backgrounds}


\author{U.~Gran}
\address{Department of Physics, Division for Theoretical Physics\\
Chalmers University of Technology\\
SE-412 96 G\"oteborg, Sweden}

\author{J.~Gutowski}
\address{Department of Mathematics
\\
University of Surrey \\
Guildford, GU2 7XH, UK}

\author{G.~Papadopoulos}
\address{Department of Theoretical Physics,
CERN\\
1211 Geneva 23, Switzerland
\\
On study leave:  Department of Mathematics, King's College London,
\\
 Strand, London WC2R 2LS, UK}

\begin{abstract}
We review the remarkable progress that has been made the last 15 years towards the classification of supersymmetric solutions with emphasis on the description of the bilinears and spinorial geometry methods. We describe in detail the geometry of  backgrounds of key supergravity theories, which have applications in the context of black holes, string theory, M-theory and the AdS/CFT correspondence  unveiling a plethora of existence and uniqueness theorems. Some other aspects of supersymmetric solutions like the Killing superalgebras and the homogeneity theorem are also presented,  and the non-existence theorem for certain smooth supergravity flux compactifications is outlined. Amongst the applications described is the proof of the emergence of conformal symmetry near black hole horizons and  the classification of warped AdS backgrounds that preserve more than 16 supersymmetries.
\end{abstract}

\end{frontmatter}



\newpage
\tableofcontents

\section{Introduction}

General relativity has brought a  momentous change in  the relationship between physics and geometry as the gravitational  force
at large scales is modelled in terms  of geometry. The relevant geometry involved   is that of  manifolds equipped with a Lorentzian signature metric.  The interplay between general relativity and manifold theory has led to the rapid development of both fields. The investigation of solutions of general relativity has had a profound impact on our understanding of the universe and
the matter it contains. It has led to the introduction of black holes, the discovery of gravitational waves
and to cosmological models which describe the evolution of our universe.

Following the general relativity paradigm,  supersymmetric systems, which include string theory and M-theory,  admit a new class of solutions, the ``supersymmetric solutions'', which  in addition to the field equations also solve the Killing spinor equations (KSEs).  These arise from the vanishing condition  of the  supersymmetry variations of the fermions of these theories.  It was soon realized that such
solutions may saturate certain Bogomol'nyi type bounds, and because of this they are also called
 ``BPS'' solutions.  In  gauge theories supersymmetric solutions include  monopoles and instantons  which play a central role in the
  understanding of strong coupling dynamics and non-perturbative corrections to these theories. In gravity theories supersymmetric solutions include  extreme black hole solutions for Einstein-Maxwell
type  theories, as well as gravitational waves. In string theory and M-theory supersymmetric solutions include compactification vacua, extreme black holes and brane solutions.  The latter are considered as the solitons of these theories and
they have played a central role in unravelling the string dualities and in the microstate counting of black hole entropy, see \cite{pkt-11d, cmhpkt, ascv}.
 These applications have continued in the context of the AdS/CFT correspondence \cite{Maldacena:1997re} as the gravitational backgrounds which correspond to the vacuum states of dual superconformal theories are supersymmetric.   There is a plethora of  supersymmetric solutions and the research is ongoing as they have widespread   applications, for reviews see e.g. \cite{dufflu, ks-11d}.  The first systematic investigation  of supersymmetric solutions was done by Tod who used twistorial techniques to solve the KSE of minimal
${\cal N}=2$ $d=4$ supergravity and classify all such  solutions \cite{Tod:1983}.

The purpose of this review is to summarize the significant progress  that has been made the last 15 years in
classifying the supersymmetric backgrounds. The problem has been solved for a large number of supergravity theories and
 the aim is to present the development and produce a guide to the field. The focus will be to explain the methods that have been
used for this  as well as to describe some of the key results that have been obtained.  These include
 insights into the structure  of all supersymmetric solutions
 in some theories, and the proof of  existence and uniqueness theorems for several classes of solutions. Other aspects of the
  supersymmetric solutions, like their Killing superalgebras, the homogeneity theorem and a non-existence theorem for de-Sitter and Minkowski supergravity flux compactifications, are also included.  The classification of maximal and near maximal supersymmetric backgrounds of some supergravity theories  is also described. Applications of the results will also be considered in the context of black holes, string theory, M-theory and the AdS/CFT correspondence.  Some aspects of the  fascinating connection between supersymmetric backgrounds and   special geometric structures  will be presented. A generalization of classic results like the Lichnerowicz theorem will also be given as part of the proof of the horizon conjecture which explains the emergence of conformal symmetry near supersymmetric Killing horizons.  The classification of warped AdS backgrounds that preserve more than 16 supersymmetries in $d=11$ and $d=10$ type II supergravities is also included.
As part of the review,  many of the proofs  of key statements and examples given in the literature have been extensively reworked. As a result,  their description has become more concise and shorter than that in their original exposition.

 The two methods that have been extensively used to classify supersymmetric backgrounds, and which will be reviewed here, are the bilinears or G-structure method proposed by Gauntlett, Gutowski, Hull, Pakis and Reall \cite{gghpr}
 and the spinorial geometry method proposed by Gillard, Gran and Papadopoulos \cite{jguggp}.  The use of these two methods is sufficient to describe all results that have been obtained in the literature, apart from the maximally supersymmetric solutions which are classified using a technique introduced in \cite{jfofgp1}. Both methods are explained in a simple example, the solution of the  gaugino KSE on $\bR^6$, where their individual features are illustrated.  Moreover, the solution of the  KSE of minimal  ${\cal N}=1$ $d=5$ supergravity is described employing both methods. The spinorial geometry method in particular is used in the classification of backgrounds which preserve a near maximal number of supersymmetries.

Apart from the description of the two methods, the classification of supersymmetric solutions of minimal ${\cal N}=2$ $d=4$, ${\cal{N}}=1,
d=4$,  ${\cal N}=1$ $d=5$ and ${\cal N}=(1,0)$ $d=6$ supergravities will be presented.  In addition, the solution of the KSEs of heterotic supergravity will be described together with  general theorems on the existence of certain classes of solutions.  Aspects of the supersymmetric solutions of $d=11$ and $d=10$ type II
supergravities will be explained.  These include the solutions of the KSEs for one and nearly maximal number of Killing spinors.
The classification of nearly maximal and maximal supersymmetric backgrounds of $d=11$ and $d=10$ type II
supergravities will also be given.  As an application we present the horizon conjecture and its proof in the context of $d=11$ supergravity which amongst other things demonstrates that  $SL(2,\bR)$ generically emerges as a symmetry of near horizon geometries.  The review will conclude with the computation of the Killing superalgebras of warped AdS backgrounds,  and the proof of existence and uniqueness theorems for AdS backgrounds that preserve more than 16 supersymmetries.

\section{Methods for solving KSEs}

\subsection{KSEs and supersymmetry}

The KSEs of supergravity theories are  the vanishing conditions
of the supersymmetry variations of the fields. These are  evaluated
in the sector where all fermions vanish, which in turn implies that the supersymmetry
variations of the bosons are identically zero. The remaining equations are a parallel transport equation for the supercovariant
connection, ${\cal D}$, which is
associated with the supersymmetry variation of the gravitino, $\psi$,
and  some  algebraic equations which are associated with the supersymmetry variations of the
remaining fermions, $\lambda$.
In particular, one has
\bea
\delta\psi_\tM|_{\psi,\lambda=0}={\cal D}_\tM\,\eps=0~,~~~~\delta\lambda|_{\psi,\lambda=0}={\cal A}\,\eps=0~,
\la{kse1}
\eea
where the spinor indices have been suppressed,
\bea
{\cal D}_\tM\defeq \nabla_\tM+\sigma_\tM(e, F)~,
\la{supcon}
\eea
is the supercovariant connection, $\nabla$ is the spin connection of the spacetime acting on the spinors,
\bea
\nabla_\tM\defeq  \partial_\tM+{1\over4} \Omega_{\tM, \tA\tB} \Gamma^{\tA\tB}~,
 \eea
 and $\sigma(e,F)$ is a Clifford algebra element which depends
on the spacetime coframe $e$ and the fluxes $F$.  The expression  of $\sigma(e, F)$ in terms of the fields is theory dependent. The second KSE in (\ref{kse1}) does not involve derivatives on $\eps$, i.e.~it is algebraic, and ${\cal A}$ is a Clifford algebra element that depends on the fields. We use the notation, unless otherwise is explicitly stated,   that capital Latin letters from the middle of the alphabet and onwards denote spacetime indices while capital Latin letters from the beginning
of the alphabet denote coframe indices, i.e.~the relation between the spacetime metric $g$ and the coframe $e$ is
$g_{\tM\tN}= \eta_{\tA\tB} e^\tA_\tM e^\tB_\tN$.
The spinor  $\epsilon$ should be thought of as the
parameter of the supersymmetry transformations and is taken to be commuting, see appendix \ref{app:spinors} for our spinor conventions.

The KSEs (\ref{kse1}) are clearly linear  in $\eps$ and at most first order.
\begin{enumerate}

\item[-] The solutions of the field
equations of supergravity theories
 that admit a non-vanishing $\eps$ which satisfies (\ref{kse1})  are called supersymmetric.

 \item[-]  The
 number, $N$, of supersymmetries preserved by a background is the number of linearly independent
 solutions $\eps$ that the KSEs (\ref{kse1}) admit when they are evaluated on the fields of the background.

\end{enumerate}
Generically, there are always solutions which do not preserve any supersymmetry. Conversely, the maximal number of supersymmetries that a background can preserve is the number of supersymmetry charges of the theory.

\subsection{Holonomy and gauge symmetry}\label{gaugeholonomy}

To understand some of the properties of the KSEs, it is instructive to investigate their
integrability conditions. The first order integrability  conditions can be written schematically as
\bea
{\cal R}_{\tM\tN}\eps\defeq [{\cal D}_\tM, {\cal D}_\tN]\eps=0~,~~~[{\cal D}_\tM, {\cal A}]\eps=0~,~~~[{\cal A}, {\cal A}]\eps=0~,
\label{intconxxx}
\eea
where ${\cal R}$ is the curvature of the supercovariant connection. As we shall describe later
these integrability conditions  are also used in the investigation of the field equations of supersymmetric backgrounds.

It is natural to focus first on the gravitino KSE, which is a parallel transport equation.
For a  d-dimensional spacetime, the (reduced) holonomy group,
$\mathrm {hol}(\nabla)$, of the spin connection $\nabla$ is contained in $Spin(d-1,1)$. However because of the presence of fluxes, and in particular of the sigma term in (\ref{kse}),
the holonomy of the supercovariant connection, $\mathrm {hol}({\cal D})$,   is contained in an SL group rather than a $Spin$ group. In particular, the (reduced) holonomy of the supercovariant connections
 of generic  $d=11$ \cite{hullholonomy, duffholonomy, gptsimpisholonomy} and type II supergravity backgrounds \cite{gptsimpisiib} is contained in $SL(32, \bR)$. A list
of the holonomies of lower dimensional supergravities can be found in \cite{Batrachenko}.

To see this,  note that the Lie algebra of $\mathrm {hol}({\cal D})$  is computed by evaluating the  supercovariant curvature ${\cal R}$ and its covariant derivatives ${\cal D}^k {\cal R}$ on spacetime vector fields, and then look at the span of the resulting expressions. In particular for $d=11$ supergravity ${\cal R}$ is given in (\ref{supelcurv}).  Observe that ${\cal R}(X,Y)$, for any two vector field $X$ and $Y$, is a general Clifford algebra element as it is expanded in all possible skews-symmetric products of gamma matrices apart perhaps from that of the zeroth order. As a consequence of Clifford algebra representation theory, the Lie bracket algebra of all skew-symmetric products of gamma matrices of degree 1 and above is $\mathfrak{sl}(32, \bR)$ in $d=11$.   This in turn leads to the assertion that $\mathrm {hol}({\cal D})$ is $SL(32,\bR)$ as mentioned above.

This property of the holonomy of the  supercovariant connection has important implications in understanding the geometry of supersymmetric backgrounds, see e.g.~section \ref{gloprosol}. An immediate consequence is that standard techniques, like the Berger classification,  which are used to investigate the geometry of manifolds that admit parallel spinors with respect to the Levi-Civita connection do not apply.  As a result a new approach is needed to investigate the solutions of KSEs and determine the geometry of solutions that admit Killing spinors.

Another property of the KSEs, which is essential in understanding the supersymmetric solutions,
is the gauge symmetry. The gauge transformations of the Killing spinor equations are defined as the local transformations
which transform a spacetime coframe
$e$,  fluxes $F$ and spinor $\eps$ but leave the  KSEs (\ref{kse1}) covariant, i.e.
\bea
\ell {\cal D}(e, F) \ell^{-1}={\cal D}(e^\ell, F^\ell)~,~~~\ell {\cal A}(\fe,
 F) \ell^{-1}={\cal A}(e^\ell, F^\ell)~.
\eea
The gauge group $G$ of most supergravity theories is smaller than the $\mathrm {hol}({\cal D})$ of generic backgrounds, and always contains
$Spin(d-1,1)$ as a subgroup.  This will be one of the ingredients of the spinorial geometry method.

\subsection{The spinor bilinears or $G$-structure method}

The bilinears or $G$-structure method was the first one to be used to systematically find all the solutions of minimal ${\cal N}=1$  $d=5$ supergravity in \cite{gghpr}.
It is based on the observation that for  spinors $\eps_1$ and $\eps_2$, one can associate a k-form,
\bea
\tau={1\over k!}\bar\eps_1 \Gamma_{\tM_1\tM_2\dots \tM_k} \eps_2\,dx^{\tM_1}\wedge dx^{\tM_2}\wedge \dots \wedge dx^{\tM_k}~,
\eea
which is clearly bilinear in the spinors $\eps_1$ and $\eps_2$, where any $Spin(d-1,1)$ invariant inner product can be used instead of the Dirac inner product indicated here, see appendix \ref{app:spinors}.   From here on, we shall refer to these forms either as ``k-form bilinears'' or simply ``bilinears''.
The 1-form bilinear is the familiar Dirac current.

The existence of parallel spinors on simply connected Riemannian manifolds
is equivalent to the existence of parallel forms. Indeed,  $\nabla\eps=0$ implies that
$\nabla\tau=0$, where $\tau$ is any form constructed as a bi-linear of the parallel spinors. Conversely, the existence of certain
parallel forms imply that the holonomy of the Levi-Civita connection, $\mathrm {hol}(\nabla)$, reduces to a subgroup of $SO(d)$. Then
the spinor representations decomposed under $\mathrm {hol}(\nabla)$ have singlets which correspond to the parallel spinors.

This way of solving parallel transport equations for spinors can be adapted to the context of supergravity. One of the ingredients is
to   turn the KSEs into equations for the form bilinears $\tau$.
In particular,
the gravitino  and the algebraic  KSEs in (\ref{kse1})  imply that
\bea
\nabla_\tA \tau_{\tB_1\dots \tB_k}- \bar\eps_1\Gamma_{\tB_1\dots \tB_k} \sigma_\tA\eps_2-{\overline{\eps_1 \sigma}}_A \Gamma_{\tB_1\dots \tB_k}\eps_2=0~,
\cr
\bar\eps_1\Gamma_{\tB_1\dots \tB_k}{\cal A} \eps_2=0~,
\la{fkse}
\eea
respectively, for every pair of Killing spinors $\eps_1$ and $\eps_2$. Expanding the $\sigma$ and ${\cal A}$ dependent parts in skew-symmetric
powers of gamma matrices, the above equations can be expressed as equations for the form  bi-linears $\tau$ of $\eps$, their covariant derivatives
$\nabla\tau$, and the fluxes of the supergravity theory.
The resulting equations that typically contain bilinears of different degree are solved
to express some of the fluxes in terms of the form bilinears $\tau$ and their spacetime derivatives. In addition,
one also finds conditions on the spinor bi-linears themselves. These are interpreted as the geometric conditions on the spacetime geometry
required so that it admits a Killing spinor $\eps$.

 Another ingredient that it is used to  understand  the  geometry and topology of spacetime and to solve (\ref{fkse}) are the algebraic relations between the spinor bi-linears $\tau$. These arise as a result  of  Fierz identities.  In particular, these can be used to relate the wedge products of the various form bilinears.  In turn, these provide information about the topological $G$-structure of the underlying manifold. Because of this, this method of solving KSEs  is also referred to as the $G$-structures method.   An illustration of how the method works will be given in section \ref{sec:gaugeex}
to solve the KSE of $d=6$ gauge theory.

\subsection{The spinorial geometry method}

 Spinorial geometry \cite{jguggp} is a method for solving the KSEs working directly with the spinors.  It is based on
 three ingredients.  The first is  the gauge symmetry of the KSEs, the second is a description of spinors in terms of multi-forms
 and third an oscillator basis in the space of Dirac spinors. These three ingredients can be used to solve the KSEs as follows.

As the Killing spinor equations admit a gauge invariance, it is natural to identify
two backgrounds which are related by such transformations.
As a result, the gauge symmetry can be used to set the Killing spinors in a normal or canonical form. This is equivalent to choosing
representatives of the orbits of the gauge group of the theory on the space of spinors.

The description of spinors in terms of multi-forms  is used to
explicitly give the canonical forms
of the Killing spinors up to a gauge transformation and leads to a simplification of the computations.
This realization of spinors in terms of multi-forms is described in appendix \ref{app:spinors}
for both Euclidean and Lorentzian signatures.

 Furthermore, an  oscillator basis
in the space of Dirac spinors, together with the linearity of the Killing spinor
equations, are utilized to express
the Killing spinor equation as a linear system in terms of the fluxes and the geometry.  The latter is
represented by components
of the spin connection of spacetime. This system is solved to express some of the
fluxes in terms of the geometry and to
find the conditions on the geometry required for the existence of Killing spinors.

The solution of the linear system for both the fluxes  and
geometry can always be organized in representations of the isotropy group of the Killing spinors in the gauge group
of the supergravity theory under study. This is  the case even when the linear system is not manifestly expressed in representations
of the isotropy group of the Killing spinors but instead in representations of a subgroup.

In the spinorial geometry approach,
there is also a spacetime coframe, the ``spinorial geometry coframe'', adapted  to the choice of the Killing spinor representatives and to the spinor oscillator basis
that is used to solve the KSEs. The solution of the linear system is initially expressed in this coframe. Typically,
the conditions on the spacetime geometry can be re-expressed as differential relations between the form bilinears
of the Killing spinors. Similarly, the solution for the fluxes can be given in terms of the form bilinears, their derivatives  and the metric.

As spinorial geometry is rather efficient for solving KSEs for a small as well as a large  number $N$  of supersymmetries,
many of the results in this review have been described in this method.  However, the illustrative example below, as well
as the solution of the KSEs of minimal $d=5$ supergravity in section \ref{sec:minimald5}, have been described employing both methods to  provide
a description of both approaches.

\subsection{A gauge theory example}\label{sec:gaugeex}

Before we proceed to describe the solution of the KSEs of supergravity theories, it is instructive to provide a simple example to illustrate  how
the bilinears and spinorial geometry methods work.  For this  consider the gaugino KSE
\bea
F_{\tA\tB} \Gamma^{\tA\tB}\epsilon=0~,
\label{gaugkse}
\eea
 on $\bR^6$ equipped with the standard Euclidean metric, where we have suppressed the gauge and spinor indices and $\epsilon$ is a constant Weyl spinor, $\epsilon\in \Delta^+(\bR^6)$.

 \subsubsection{Solution using bilinears}

To solve the  gaugino KSE (\ref{gaugkse}) in the bilinears method,  consider the  Fierz identity
given by
\begin{equation}
\label{ggfrz2}
\langle \epsilon_1, \epsilon_2 \rangle \langle \epsilon_3, \epsilon_4 \rangle=
{1 \over 4} \langle \epsilon_1, \epsilon_4 \rangle \langle \epsilon_3, \epsilon_2 \rangle
-{1 \over 8} \langle \epsilon_1, \Gamma_{\tA \tB} \epsilon_4 \rangle \langle
\epsilon_3, \Gamma^{\tA \tB} \epsilon_2 \rangle~,
\end{equation}
where $\epsilon_1, \epsilon_2, \epsilon_3, \epsilon_4 \in {}^c\Delta^+
({\mathbb{R}}^6)$ and thus satisfy $\Gamma_7 \epsilon_r=i \epsilon_r$, $r=1,2,3,4$ with respect to chirality operator,   $\Gamma_7\defeq \Gamma_1\cdots \Gamma_6$,    and indices are raised and lowered with the Euclidean metric.
This is also equivalent to
\begin{equation}
\label{ggfrz3}
\langle \epsilon_3, \epsilon_4 \rangle \epsilon_2 = {1 \over 4} \langle \epsilon_3, \epsilon_2 \rangle \epsilon_4 -{1 \over 8} \langle \epsilon_3, \Gamma^{\tA \tB} \epsilon_2 \rangle
\Gamma_{\tA \tB} \epsilon_4
\end{equation}
for $\epsilon_2, \epsilon_3, \epsilon_4 \in {}^c\Delta^+
({\mathbb{R}}^6)$.

To proceed with the analysis, suppose that $\epsilon \in  {}^c\Delta^+
({\mathbb{R}}^6)$. We define the real 2-form $\omega$
\begin{equation}
\omega ={i\over2} \langle \epsilon, \Gamma_{\tM \tN} \epsilon \rangle\, dx^\tM\wedge dx^\tN \ .
\end{equation}
For convenience, we choose the normalization $\parallel \epsilon \parallel=1$. Then,
on setting $\epsilon_1=\epsilon_2=\epsilon_3=\epsilon_4=\epsilon$ in
({\ref{ggfrz2}}),  we find
\begin{equation}
\label{ggid1}
\omega_{\tA \tB} \omega^{\tA \tB}=6~,
\end{equation}
and on setting $\epsilon_2=\epsilon_3=\epsilon_4=\epsilon$ in ({\ref{ggfrz3}}),
one obtains
\begin{equation}
\label{ggid2}
\omega_{\tM \tN} \Gamma^{\tM \tN} \epsilon = 6i \epsilon \ .
\end{equation}
Next, substituting $\epsilon_1=\epsilon_3=\epsilon$ and $\epsilon_2= \Gamma_\tA{}^\tL \epsilon$,
$\epsilon_4= \Gamma_{\tB \tL} \epsilon$ in ({\ref{ggfrz2}}), and making use of ({\ref{ggid1}})
together with the convention
\begin{equation}
\label{ggdual}
\Gamma_{\tL_1 \dots \tL_6} \epsilon=i \epsilon_{\tL_1 \dots \tL_6} \epsilon
\end{equation}
one derives  the condition
\begin{equation}
\omega_{\tA}{}^\tL \omega_{\tB \tL} = \delta_{\tA \tB} \ .
\end{equation}
Therefore, on defining $I$ by $\omega_{\tA \tB} = \delta_{\tA \tC} I^\tC{}_{\tB}$, $I$ is a complex structure on ${\mathbb{R}}^6$.

In order to find the conditions on $F$ for the gaugino KSE  ({\ref{gaugkse}}) to admit a Killing spinor, first note that
this condition implies that
\begin{equation}
F_{\tA \tB} \langle \epsilon, \Gamma^{\tA \tB} \epsilon \rangle =0~,
\end{equation}
and hence
\begin{equation}
\label{ggtr}
F_{\tA \tB}\, \omega^{\tA \tB} =0 \ .
\end{equation}
Furthermore, ({\ref{gaugkse}}) also implies that
\begin{equation}
F_{\tC_1 \tC_2} \langle \epsilon, \Gamma_{\tA \tB} \Gamma^{\tC_1 \tC_2} \epsilon \rangle =0 \ .
\end{equation}
On taking the imaginary part of this identity, we find
\begin{equation}
\label{gg11}
F_{\tC \tD} I^\tC{}_\tA I^{\tD}{}_\tB = F_{\tA \tB} \ .
\end{equation}
The conditions ({\ref{ggtr}}) and ({\ref{gg11}}) imply that $F$ is traceless and $(1,1)$ with respect to $I$, i.e.~$F$ satisfies the Hermitian-Einstein condition on $\bR^6$.

The  conditions (\ref{ggtr}) and (\ref{gg11}) are also sufficient to
ensure that there is a solution $\eps\not=0$ to  ({\ref{gaugkse}})  with no further conditions imposed on $F$.
This can be straightforwardly shown by computing
\begin{equation}
\label{ggaux1}
\parallel F_{\tA \tB} \Gamma^{\tA \tB} \epsilon \parallel^2 = 2 F^2 - \langle \epsilon, F_{\tA \tB} F_{\tC \tD} \Gamma^{\tA \tB \tC \tD} \epsilon \rangle \ .
\end{equation}
The Fierz identities ({\ref{ggid1}}) and ({\ref{ggid2}}) together with ({\ref{ggdual}})
imply that
\begin{equation}
{}^* (\omega \wedge \omega) = 2 \omega~,
\end{equation}
from which it follows that
\begin{equation}
\label{ggaux2}
\langle \epsilon, \Gamma_{\tA \tB \tC \tD} \epsilon \rangle = -{1 \over 2} (\omega \wedge \omega)_{\tA \tB \tC \tD} \ .
\end{equation}
On substituting ({\ref{ggaux2}}) into ({\ref{ggaux1}})
and making use of the conditions ({\ref{ggtr}}) and ({\ref{gg11}}), one can show that
$\parallel F_{\tA \tB}\Gamma^{\tA \tB} \epsilon \parallel =0$, and hence ({\ref{gaugkse}}) holds.

\subsubsection{Solution using spinorial geometry}

As $\mathrm{Spin}(6)=SU(4)$,
 the Weyl representation can be identified with the fundamental representation  of $SU(4)$ on $\bC^4$.  Clearly (\ref{gaugkse}) is covariant
 under rigid $\mathrm{Spin}(6)$ transformations. This can be used to choose $\epsilon$ as follows. Observe  that $\mathrm{Spin}(6)$ has a single type of non-trivial orbit on
 $\bC^4$ which is $S^7$.  As a result $\epsilon$ can be chosen to lie along any direction in $\bC^4$.  Identifying $\bC^4={}^c\Delta^+(\bR^6)=\Lambda^{*\mathrm{ev}}(\bC^3)$, see appendix \ref{app:spinors}, one can choose  $\epsilon$  as $\epsilon=1$.

In such a case, the gaugino KSE (\ref{gaugkse}) can be written in the oscillator basis of appendix \ref{app:spinors} as
\bea
(F_{\bar\alpha\bar\beta} \gamma^{\bar\alpha\bar\beta}+ 2 F_{\alpha\bar\beta} \delta^{\alpha\bar\beta})1=0~,
\eea
where we have used that $\gamma^\alpha1=\sqrt2\, i_{e_\alpha}1=0$.  This implies that
\bea
F_{\bar\alpha\bar\beta}=0~,~~~F_{\alpha\bar\beta} \delta^{\alpha\bar\beta}=0~,
\label{fabbar}
\eea
which can be recognized as the Hermitian-Einstein conditions on $F$ written in complex coordinates on $\bR^6$.  If $F$ is real, then $F_{\alpha\beta}=0$ and so in the
language of complex geometry $F$ is a (1,1)- and Hermitian traceless form.

Clearly the conditions (\ref{fabbar}) on $F$ are written
in representations of the $SU(3)$ isotropy group of the Killing spinors. They can also be written covariantly after using the
2-form bilinear
\bea
\omega={i\over 2}\langle 1, \Gamma_{\tM\tN} 1\rangle\, dx^\tM\wedge dx^\tN=-i\delta_{\alpha\bar\beta}\, dz^\alpha\wedge dz^{\bar\beta}~,
\eea
where $z$ are complex coordinates on $\bR^6$ with respect to the complex structure $I$ defined by  $\omega_{\tA\tB}=\delta_{\tA\tC} I^\tC{}_\tB$.
In particular, the conditions (\ref{fabbar}) can be rewritten as ({\ref{ggtr}}) and ({\ref{gg11}}).
This type of procedure for finding solutions to the linear system presented above
can also be applied to all linear systems that arise in the solutions of the KSEs of supergravity theories.

\section{Minimal ${\cal N}=2$ $d=4$  supergravity}

The bosonic fields of the gravitational multiplet are a metric and an abelian 2-form gauge field strength $F$, $dF=0$. The bosonic action is the Einstein-Maxwell system.
Since it describes the  long range force fields of the cosmos, it has been extensively  investigated and its solutions include black holes
with electric and/or magnetic charges and gravitational waves. It also arises in various limits of higher dimensional theories which include string- and M-theory.
As a result, many of the brane configurations of these theories reduce upon dimensional reduction to solutions of this minimal ${\cal N}=2$ supergravity.

Furthermore, this is the theory for which the KSE was first solved in full generality \cite{Tod:1983} using twistorial  techniques.  Because of this, we shall begin the investigation of  gravitational theories with solving the KSE  of this theory.
Here, we shall present the analysis employing the spinorial geometry method.

 \subsection{Fields  and Killing spinors}

 \subsubsection{KSE and field equations}

 The only fermionic field in the theory is a gravitino whose supersymmetry variation gives the  KSE
 \bea
  {\cal D}_{\tM}\eps=0~,
  \label{4D-KSE}
  \eea
  where the supercovariant connection, ${\cal D}$, is
 \begin{align}
  {\cal D}_{\tM} \defeq \nabla_{\tM} + \tfrac{i}{4} F_{\tA\tB} \Gamma^{\tA\tB} \Gamma_{\tM} \,,
 \label{4dsupercov}
  \end{align}
 $\nabla$ is the spin  connection of the spacetime  and the supersymmetry parameter $\eps$ is a Dirac spinor of $Spin(3,1)$.

The supercovariant  curvature, ${\cal R}$,  is
 \begin{align}
 {\cal R}_{\tM\tN}\defeq [ {\cal D}_{\tM}, {\cal D}_{\tN}] =  &
 \tfrac 14 {R}_{\tM \tN,\tA\tB } \Gamma^{\tA\tB}
 - {1\over2} F_{\tM\tA} F_{\tN\tB} \Gamma^{AB}-{1\over2} {}^*F_{\tM\tA} {}^*F_{\tN\tB} \Gamma^{AB}
  \notag \\
 & -\tfrac 12 i \Gamma_{\tA\tB[\tM}\nabla_{\tN]}F^{\tA\tB}
 - i\nabla_{[\tM}{F_{\tN]}}^{\tA}\Gamma_{\tA}  \,,
 \label{4D-intcond}
 \end{align}
 where $R$ is the Riemann tensor of the spacetime which arises from
the spin connection term   $\nabla$ in $ {\cal D}$ and ${}^*F$ is the Hodge dual of $F$,  ${}^*F_{\tA\tB}=(1/2) F_{\tC\tD} \epsilon^{\tC\tD}{}_{\tA\tB} $, with $\epsilon^{0123}=1$.
To derive (\ref{4D-intcond}), one also uses $dF=0$.

For generic backgrounds the (reduced) holonomy of $ {\cal D}$   is contained in $SL(2,\mathbb{H})$ \cite{Batrachenko}.  The enhancement in holonomy from $Spin(3,1)$ of  $\nabla$ to $SL(2,\mathbb{H})$ of ${\cal D}$ is due to the linear and cubic terms in gamma matrices in the expression for ${\cal R}$ above.  This is a  characteristic property of many  supergravity theories and some of its  consequences will be explained below in section \ref{sec:geomd4n1}, see also section \ref{gloprosol}.

The field equations of the theory can be derived from the supercovariant curvature ${\cal R}$.  In particular one has
\bea
\Gamma^\tN{\cal R}_{\tM\tN}=-{1\over2} E_{\tM\tN}\Gamma^\tN-{i\over2} LF_{\tN}\Gamma^\tN \Gamma_\tM~,
\label{gammatrace4d}
\eea
where
\begin{equation}
  E_{\tM\tN} \defeq  R_{\tM\tN} - 2 F_{\tP\tM} {F^{\tP}}_{\tN} + {1\over2}F^2 g_{\tM\tN}=0\,,~~~LF_{\tM}\defeq\nabla^\tN F_{\tN\tM}=0~,
\end{equation}
are  the Einstein and Maxwell field equations, respectively.   It is significant for the investigation of solutions  that some of the components of ${\cal R}$ are proportional
to the field equations. This will be used to demonstrate that some of the field equations are implied from the KSEs.

\subsubsection{Killing spinors}\label{4dspspgeom}

To solve the KSEs in the spinorial geometry approach, \cite{jguggp}, one has to choose a normal form
for the Killing spinors. As described in appendix \ref{app:spinors}, the space of Dirac spinors can be identified with $\Lambda^*(\bC^2)$.  The Weyl representation of $Spin(3,1)=SL(2,\bC)$ is the fundamental
 representation of $SL(2,\bC)$ on $\bC^2$. The chiral and anti-chiral spinors are identified with the even degree, $\Lambda^{\mathrm{ev}}(\bC^2)$, and odd degree, $\Lambda^{\mathrm{odd}}(\bC^2)$, forms,
 respectively. Observe though that in contrast to the spinors of ${\cal N}=1$ $d=4$ supergravity that will be investigated next, the chiral and anti-chiral  representations are not complex conjugate to each other as the components of a Dirac spinor are independent.  Let us assume without loss of generality that the positive chirality component of $\eps$ does not vanish. As $SL(2,\bC)$ acts transitively on $\bC^2-\{0\}$, the positive chirality component of $\epsilon$ can always be chosen as the spinor $1$.  The isotropy group of the spinor $1$ in $SL(2,\bC)$ is $\bC$ whose Lie algebra is spanned by $\{\gamma^{1-}, \gamma^{\bar 1 -} \}$, see appendix \ref{app:spinors}. The most general anti-chiral component of $\eps$ is $a e_1+b e_2$ for $a,b\in \bC$.  If $b\not=0$, then the $\bC$ isotropy group can be used to set
$a=0$. Thus the first Killing spinor can locally be chosen as
\bea
\text{either}~\eps=1+b\, e_2~,~\text{or}~~\eps=1+ a\, e_1~,
\eea
where $a,b$ become  complex-valued functions on spacetime.  The isotropy groups of $1+b e_2$ and  $1+ a e_1$ are $\{1\}$ and $\bC$ in $SL(2,\bC)$, respectively.
As we shall demonstrate, the two Killing spinors give rise to two distinct types of geometries on spacetime; one is associated with a time-like Killing vector field and the other with a null one. Because of this, they are also referred to as the time-like and null cases, respectively.

Before we proceed with the solution of the gravitino KSE (\ref{4D-KSE}), observe that it  is linear over $\bC$.  This means that if $\eps$ is a Killing spinor, then
$i\,\eps$ will  also  be a Killing spinor.  Furthermore, if $\eps$ is a Killing spinor, then both $r_{\mathrm{B}}\eps=-\Gamma_3*\eps$ and $i\,r_{\mathrm{B}}\eps=-i\Gamma_3*\eps$ will also  be Killing spinors because $r_{\mathrm{B}}$ commutes with the KSE, see appendix \ref{app:spinors} for the definition of $r_{\mathrm{B}}$.  As  these four spinors are linearly independent over $\bR$,  (\ref{4D-KSE}) admits either
four or eight Killing spinors as solutions. Therefore the Einstein-Maxwell system admits supersymmetric solutions  which preserve either half or all of the supersymmetry of the theory.

\subsection{Case 1: $\eps=1+b e_2$}

\subsubsection{Solution of linear system}

To construct the linear system apply the gravitino KSE  (\ref{4D-KSE}) to the spinor $\eps=1+b e_2$ and expand the resulting expression in
 the basis
$\{1, e_{12}, e_1, e_2\}$ in the space of   Dirac spinors. The vanishing of each component in this basis yields
\bea
-{1\over2}  \Omega_{\tM,+-}+{1\over2} \Omega_{\tM, 1\bar1}+{i\over\sqrt2} b\, (F_{-\tM}-i\,{}^*F_{-\tM})&=&0~,
\cr
-\Omega_{\tM,+\bar1}+{i\over\sqrt2} b\,  (F_{\bar1\tM}-i\,{}^*F_{\bar1\tM})&=&0~,
\cr
-b\, \Omega_{\tM, -\bar 1} +{i\over \sqrt 2} (F_{\bar 1 \tM}+i\, {}^*F_{\bar 1\tM})&=&0~,
\cr
\partial_\tM b+{b\over 2} (\Omega_{\tM,+-}+\Omega_{\tM, 1\bar1})+{i\over\sqrt2} (F_{+\tM}+i\, {}^*F_{+\tM})&=&0~,
\label{4dlsysta}
\eea
where ${}^*F$ is the Hodge dual of $F$ and $\epsilon_{-+1\bar 1}=-i$.

This linear system can be solved to give
\begin{eqnarray}
\spc_{+, - +} &=& \partial_+ \log b + \partial_+ \log \bar b \,,
\qquad \spc_{-, - +} = 0\,, \qquad \spc_{1, - +} =
\partial_{1} \log \bar b \,, \nonumber \\
\spc_{+, - \bar 1} &=& \spc_{\bar 1, - \bar 1} = 0\,, \qquad
\spc_{-, - \bar 1} = - (b \bar b)^{-1} \partial_{\bar1} \log b \,, \qquad
\spc_{1, - \bar 1} = \partial_- \log b \,, \nonumber \\
\spc_{+, + \bar 1} &=& - b\,\partial_{\bar1}\bar b\,, \qquad
\spc_{-, + \bar 1} = \spc_{\bar 1, + \bar 1} = 0\,, \qquad
\spc_{1, + \bar 1} = \partial_+ \log \bar b \,, \nonumber \\
\spc_{+, 1 \bar 1} &=& \partial_+ \log \bar b - \partial_+ \log b \,, \qquad
\spc_{-, 1 \bar 1} = 0 \,, \qquad
\spc_{1, 1 \bar 1} = \partial_1 \log \bar b \,,
 \label{4dspinconn}
\end{eqnarray}
with
\begin{align}
  \partial_+ b = |b|^2 \partial_- b \,,
  \label{t-indep}
 \end{align}
and

\bea
F&=&- \tfrac{i}{\sqrt 2}  \partial_- (b- \bar b )\,\fe^-\wedge \fe^++\tfrac{i}{\sqrt 2}  \partial_- (b +\bar b)\, \fe^1\wedge\fe^{\bar 1}
\cr
&&~~~+i \tfrac 1{\sqrt 2} |b|^{-2} (  \partial_{\bar 1} b\,\fe^{\bar 1} -\partial_{1} \bar b\,\fe^ 1)\wedge \fe^-+i \tfrac 1{\sqrt 2} (\partial_{\bar 1} \bar b\, \fe^{\bar 1}-\partial_{ 1}  b\, \fe^{ 1})\wedge \fe^+~.
\label{4dflux}
\eea
%
Therefore all of the components of the spin connection and those of the flux $F$ are determined in terms of the complex function $b$.

\subsubsection{Geometry}\label{d4timegeom}

To identify  the geometry of the spacetime as a consequence of the conditions \eqref{4dspinconn} that arise from the KSE, it is useful to consider the 1-form bilinears of the Killing spinors $\epsilon$ and $\tilde \epsilon = \Gamma_3 * \epsilon$.  These can easily be computed to find that the linearly independent ones can be chosen as
 \begin{align}
  & X={1\over2} D(\epsilon, \Gamma_\tA \epsilon) \fe^\tA ={1\over \sqrt{2}} ( |b|^2 \fe^+ - \fe^-) \,, \nonumber \\
  & W^2={1\over2}D(\epsilon, \Gamma_5 \Gamma_\tA \epsilon) \fe^\tA =  {1\over \sqrt{2}} ( |b|^2 \fe^+ + \fe^-) \,, \nonumber \\
  &W^3+iW^1=-{1\over2}
  D(\tilde \epsilon, \Gamma_5 \Gamma_\tA \epsilon) \fe^\tA =  \sqrt{2}\, i\, b\, \fe^1 \,,
  \label{4D-spinbil}
   \end{align}
which give rise to four real 1-forms $X, W^1, W^2, W^3$ on spacetime. These are orthogonal and $X$ is timelike, $g(X,X)=-|b|^2$, and  the remaining three are spacelike, $g(W^i, W^j)=|b|^2\delta^{ij}$.

The conditions on the geometry can be rewritten in terms of the 1-form bilinears as
\bea
{\cal L}_Xg=0~,~~~dW^i=0~,
\eea
i.e.~$X$ is Killing and $W^i$ are closed.  Furthermore, a consequence of $dF=0$ and (\ref{4dflux}) is that ${\cal L}_X F=0$ and therefore the flux $F$ is invariant as well.
It can be shown that ${\cal L}_X\eps=0$, where ${\cal L}_X$ is the spinorial Lie derivative
\bea
{\cal L}_X=\nabla_X+{1\over8} (dX)_{\tA\tB} \Gamma^{\tA\tB}~,
\label{spinder}
\eea
along the Killing vector field $X$.  The significance of ${\cal L}_X\eps=0$ will become apparent in the description of the Killing superalgebras of supersymmetric backgrounds.

From here on, we denote by $X$ both the 1-form  bilinears of the Killing spinors and their associated vector fields which  leave all  fields of supersymmetric backgrounds invariant. Such an identification is justified because the spacetime metric induces an isomorphism between the cotangent and tangent bundles
of a spacetime.

\subsubsection{Special coordinates}

One can introduce a  set of local coordinates $(t, x^i)$  on the spacetime as $X=\partial_t$ and $W^i=dx^i$.  (\ref{t-indep}) implies that $\partial_t b=0$.
In these coordinates, the metric and flux $F$ are written as
 \bea
 && ds^2 = - |b|^2 (dt + \omega_i\, dx^i)^2 +  |b|^{-2} \delta_{ij} dx^i dx^j \,,
 \cr
 && F = -d(\mathrm{Im}\, b) \wedge (dt + \omega_i dx^i) + \tfrac12 |b|^{-2} *_{{}_3}d(\mathrm{Re}\, b)\,,
 \label{4dlfieldsa}
 \eea
where the Hodge duality operation $*_{{}_3}$ is taken with respect to the Euclidean 3-metric and
\bea
d\omega=-*_{{}_3} Y~,~~~Y_i=i|b|^{-2}  \partial_{ x^i} \log {b\over \bar b}~.
\label{4dlfieldsb}
\eea
The equations (\ref{4dlfieldsa}) and (\ref{4dlfieldsb}) summarize all the conditions on the fields implied by the KSE.

\subsubsection{Solutions}\label{4dintsol}

To find solutions, one has to solve the field equations and the Bianchi identity, $dF=0$, of the theory.  However, the Einstein equation is implied by the KSE,  the Maxwell equation of $F$ and $dF=0$.  To see this, assuming the field equation for $F$ one has from the integrability condition of the KSE  (\ref{gammatrace4d}) that $E_{\tA\tB}\Gamma^B\eps=0$.  Taking the Dirac inner product with the Killing spinor $\eps$, one deduces that
\bea
E_{\tA\tB} X^\tB=0~.
\label{4defielda}
\eea
  So everywhere that $|b|\not=0$, $E_{\tA 0}=0$ as $X= |b| \fe^0$ is along the coframe direction $\fe^0$.
Next acting on $E_{\tA\tB}\Gamma^B\eps=0$ with $E_{\tA\tC}\Gamma^\tC$ and using that $E_{\tA 0}=0$, one finds that
\bea
E_{\tA i} E_{\tA}{}^i=0~,~~~\text{no~summation~over}~A~,
\label{4defieldb}
\eea
as  $\eps\not=0$.   In turn (\ref{4defielda}) and (\ref{4defieldb}) imply that $E_{\tA\tB}=0$ and the Einstein equation is satisfied.  Therefore to find solutions, one has to solve the field equation of $F$ and $dF=0$.

To find electric or magnetic solutions, one has to take $b$ to be imaginary or real, respectively.  In such case, the field equation for $F$, or $dF=0$, imply that
$|b|^{-1}$ is a harmonic function on $\bR^3$. The solutions are static, $d\omega=0$.  For  $|b|^{-1}=1+{Q/|x|}$, one recovers the electric or magnetic extreme Reissner - Nordstr\" om black holes. If $|b|^{-1}$ is chosen
to be a multi-centered harmonic function on $\bR^3$, one finds the  Majumdar and Papapetrou \cite{Majumdar,papapetrou}, multi-black hole solutions.

The theory admits dyonic black holes for  $b=|b| e^{i\alpha}$,  where $\alpha$ is a constant phase and $|b|^{-1}$ is a harmonic function on $\bR^3$.  The
solutions are again static.  It can be  seen from (\ref{4dlfieldsb}) that these are the most general static black hole solutions that preserve some supersymmetry \cite{Gibbons:1982}.
Rotating black hole solutions, $d\omega\not=0$, have been considered by Israel, Wilson and Perjes \cite{Israel,Perjes}. These are the most general stationary supersymmetric  black hole solutions \cite{Tod:1983} of the theory.

\subsection{Case 2: $\eps=1+ a e_1$}

\subsubsection{Solution of linear system}

Evaluating the KSE on $\eps=1+ a e_1$, and expanding the resulting equation in the spinor basis $\{1, e_1, e_2, e_{12}\}$, one finds the linear system
\bea
&&{1\over 2} (\Omega_{\tM, -+}+\Omega_{\tM,1\bar 1})+{i\over\sqrt{2}} a\, (F_{1\tM}-i \,{}^*F_{1\tM})=0~,
\cr
&&
\Omega_{\tM,\bar 1+}+{i\over\sqrt2} a (-F_{+\tM}+i\,{}^*F_{+\tM})=0~,
\cr
&&
\partial_\tM a+{1\over2} a\, (\Omega_{\tM, -+}-\Omega_{\tM,1\bar 1})+{1\over\sqrt2}(F_{\bar 1\tM}+i\,{}^*F_{\bar 1\tM})=0~,
\cr
&&
- a\, \Omega_{\tM,1+}+{i\over\sqrt2} (F_{+\tM}+i\,{}^*F_{+\tM})=0~,
\eea
where ${}^*F$ is the spacetime Hodge dual of $F$ as in (\ref{4dlsysta}).

Suppose first that $a\not=0$.  In such case, the non-vanishing components of the spin connection are $\Omega_{\tM, -1}$, $\Omega_{\tM, -\bar 1}$ and
\bea
\Omega_{-,-+}=-\partial_-\log (a\bar a+1)~,~~~\Omega_{-,1\bar1}={\bar a\partial_-a-a\partial_-\bar a\over a\bar a+1}~,
\label{4dspiconcon}
\eea
with
\bea
\partial_1 a=\partial_{\bar 1} a=\partial_+a=0~,
\eea
and the flux is given by
\bea
F=-{i\over\sqrt 2 (a\bar a+1)} \left(\partial_- \bar a\, \fe^1- \partial_-  a\, \fe^{\bar 1}\right)\wedge \fe^-~.
\eea
On the other hand, if $a=0$, then $F=0$ and the non-vanishing components of the spin connection are $\Omega_{\tM, -1}$ and $\Omega_{\tM, -\bar 1}$.

\subsubsection{Geometry}\label{sec:geomd4n1}

To investigate the geometry of spacetime, one can compute the form ilinears or equivalently explore the restrictions on the coframe that arise from the vanishing conditions
  on the components of the spin connection. In particular one finds that the null 1-form bilinear
\bea
X=(1+a\bar a) \fe^-~,
\eea
is $\nabla$-parallel, $\nabla_\tA X=0$.  Therefore these backgrounds are pp-waves.  The rest of the conditions on the spacetime
can be recovered by asserting that
\bea
\alpha=\beta\, \fe^-\wedge \fe^1~,~~~\partial_-\log\beta=-2(1+a\bar a)^{-1} a \partial_- \bar a~,
\label{4dn2con3x}
\eea
is parallel, $\nabla_\tA \alpha=0$, with $\partial_+\beta=\partial_1\beta=\partial_{\bar 1}\beta=0$.  Therefore the full geometric content of spacetime is to admit a parallel real null  1-form
and a parallel complex null 2-form.  The geometry of backgrounds with $a=0$ can be described in a similar way.

A feature of the geometry of supersymmetric backgrounds in $d=4$ is that the orbit type of the Killing spinor can change under parallel transport.  This is due to the fact that
the holonomy of the supercovariant connection is contained in $SL(2, \bH)$ instead of $Spin(3,1)$. So it is possible to begin with a Killing spinor with isotropy group $\{1\}$
and after parallel transport with the supercovariant connection ${\cal D}$ to end up with a Killing spinor with isotropy group $\bC$. In such a case,
the  1-form bilinear $X$ will  change from timelike to null.  Such a phenomenon occurs in black hole solutions with a Killing horizon for which the stationary Killing vector field coincides
with the vector bilinear. Therefore the description of the geometry here and in section \ref{d4timegeom}
 is local. A more detailed discussion of this, and how it is related to G-structures, will be given
in section \ref{gloprosol}.

\subsubsection{Special coordinates}

As $d\fe^-=0$, introduce a coordinate $v$ and set $\fe^-=dv$.  Furthermore, adapt a coordinate $u$ along $X$, $X=\partial_u$.  As all the fields and form bilinears are invariant under $u$, a coframe can be chosen as
\bea
\fe^-=dv~,~~~\fe^+=(1+a\bar a) (du+Vdv+n_\tI dy^\tI)~,~~~\fe^i=e^i_\tI dy^\tI+ p^i dv~,
\label{coordd4frame}
\eea
where all components  are $u$ independent.
A further simplification is possible as the choice of the coframe $\{\fe^-, \fe^+, \fe^i : i=1, \bar 1\}$, is not unique.
  Indeed the local coframe rotation
\bea
\fe^-\rightarrow \fe^-~,~~~\fe^+\rightarrow \fe^+-q_i  \,\fe^i-{1\over2} q^2 \, \fe^-~,~~~\fe^i\rightarrow   \fe^i+ q^i\, \fe^-~,
\label{4dnullpatching}
\eea
leaves all the geometric data invariant, including the form  bilinears and the fields of the theory.  The parameter $q$ is a local gauge
transformation which takes values in the isotropy group $\bC$ of the Killing spinor.  A more formal treatment of this will be given
in the discussion of the geometry of $d=6$ supergravity backgrounds in section \ref{d6n1geom}.
Thus up to a possible   rotation  (\ref{4dnullpatching}), one can choose a coframe  (\ref{coordd4frame}) with
$p=0$. The remaining description of the geometry and solutions will be conducted  in such a coframe.

 The condition $\nabla_i\alpha=0$ in (\ref{4dn2con3x}) implies
that  the coframe $\{\fe^1, \fe^{\bar 1}\}$ can be chosen to be independent of $y^\tI$.
To summarize, the metric and flux can be chosen as
\bea
ds^2&=&2 (1+a\bar a) dv (du+ V dv+ n_\tI dy^\tI)+ \delta_{ij} e^i_\tI(v) e^j_\tJ(v) dy^\tI dy^\tJ~,
\cr
F&=&-{i\over\sqrt 2 (a\bar a+1)} \left(\partial_- \bar a\, e_\tI^1(v)- \partial_-  a\, e^{\bar 1}_\tI(v)\right) dy^\tI\wedge dv~,
\eea
where $V$ can depend on both $v$ and $y$ coordinates and   $\tilde d n$ depends only on $v$, where $\tilde d n\defeq {1\over2} (dn)_{ij} \, \fe^i\wedge \fe^j$. The latter property arises after computing the
spin connection in this coframe and comparing it with the second condition in (\ref{4dspiconcon}).

\subsubsection{Solutions}

To find solutions, one has to solve the field equations and the Bianchi identity $dF=0$ of the theory. Observe that the field equation for $F$ and its Bianchi identity
are automatically satisfied.
Furthermore, some components of the Einstein equation  are also implied from the KSE.  The argument for this is similar to that presented in section \ref{4dintsol}. In particular, one has from the integrability condition of the KSE  (\ref{gammatrace4d}) that $E_{\tA\tB}\Gamma^B\eps=0$.  Taking the Dirac inner product with the Killing spinor $\eps$, one deduces that $E_{\tA\tB} X^\tB=0$.  So  $E_{\tA +}=0$ as $X= (1+a\bar a) \fe^-$ is along the coframe direction $\fe^-$.
Next, acting on $E_{\tA\tB}\Gamma^B\eps=0$ with $E_{\tA\tC}\Gamma^\tC$ and using that $E_{\tA +}=0$, one finds that $E_{\tA i} E_{\tA}{}^i=0$ as  $\eps\not=0$,  where there is no summation over the index $A$.   Therefore all the field equations are satisfied provided that $E_{--}=0$.

A large class of solutions can be found after assuming in addition that $\partial_v$ leaves all the fields invariant, i.e.~$a$ is constant.  Then $F=0$,  $\tilde d n=0$ and $\fe^i=dy^i$.  The spacetime can locally be viewed
as a fibration having fibre $\bR^2$ with coordinates $(u,v)$ over a base space $B^2=\bR^2$ with coordinates $y^i$.  The Einstein equation $E_{--}=0$ gives that
\bea
\partial^2 V=0~,
\eea
i.e. $V$ is a harmonic function on $\bR^2$.   A large class of solutions can be found for  $V= Q \log |y|^2+A_{ij} y^i y^j$ and  $n=\delta_{ij} y^i dy^j$, where $Q$ is a constant and the constant real matrix $(A_{ij})$ is traceless,
$\delta^{ij} A_{ij}=0$.

\subsection{Maximally supersymmetric solutions}

For maximally supersymmetric solutions the supercovariant curvature ${\cal R}$ in (\ref{4D-intcond})  must vanish. In particular the linear term in gamma matrices gives that
$\nabla_{[\tM} F_{\tN]\tL}=0$ which together with $dF=0$ imply that $\nabla F=0$.
Thus $F$ is parallel. Then the terms quadratic in the gamma matrices in (\ref{4D-intcond}) imply that the spacetime curvature $R$ is parallel as well, $\nabla R=0$.  Therefore the spacetime
is a Lorentzian symmetric space and $F$ is an invariant 2-form. Lorentzian symmetric spaces, up to discrete identifications, are products of   de-Sitter $dS_n$, anti-de-Sitter $AdS_n$, Cahen-Wallach $CW_n$ and Minkowski $\bR^{n-1,1}$ spaces with Euclidean signature symmetric spaces \cite{mcnw-max}.  A description of de-Sitter $dS_n$ and anti-de-Sitter $AdS_n$ spaces can be found in \cite{shge},   and for the Cahen-Wallach spaces see appendix \ref{cwmanifolds}.

After some investigation, the maximal supersymmetric solutions of minimal ${\cal N}=2$ supergravity  are locally isometric to

\begin{enumerate}

 \item[-] $AdS_2\times S^2$ with metric and flux
 \bea
&& ds^2= \ell^2\, d\mathring{s}^2(AdS_2)+  \ell^2\, d\mathring{s}^2(S^2)~,~~~
 \cr
&& F=  \mu\, \mathring{{\rm dvol}} (AdS_2)+\nu\, \mathring{{\rm dvol}} (S^2) ~,~~
 \eea
with $\ell^2=\mu^2+\nu^2$ and $\ell, \mu,\nu\in \bR$, $\ell\not=0$.

 \item[-] the plane wave $CW_4$ with metric and flux
 \bea
&& ds^2=2 dv du+ A_{ij} x^i x^j dv^2+\delta_{ij} dx^i dx^j~,~~~
\cr
&&F = \mu_i  dv \wedge dx^i~,
\label{cwasd4}
 \eea
 with $A=-\mu^2 {\rm
      diag}(1,1)$ and $\mu\not=0$.

 \item[-] and Minkowski spacetime  $\bR^{3,1}$ for which $F=0$.

\end{enumerate}

where $d\mathring{s}^2$ and  $\mathring{{\rm dvol}}$ denote the metrics and volume forms  of the indicated spaces  with radii  normalized to  one, respectively.

\subsection{Classification of   non-minimal ${\cal N}=2$ supergravity solutions}


After the first classification of ${\cal N}=2$ supergravity solutions in \cite{Tod:1983},
further extensions of this work to include dilaton and axion scalar fields were constructed
in \cite{Tod:1995jf}, using the same techniques. The solutions again split into timelike
and null cases; and the timelike solutions have a metric whose form is identical
to that given in ({\ref{4dlfieldsa}}), though the conditions on the terms appearing
in the metric receive modifications if additional scalar fields are present.
The next theory to be considered was the minimal gauged supergravity. It was known for some time that this theory contains dyonic black holes \cite{Kostelecky:1995ei}, so a systematic
understanding of the supersymmetric solutions was of particular interest.
In \cite{Caldarelli:2003} the classification was performed using the bilinears method;
the presence of the negative cosmological constant deforms the transverse 3-manifold,
which appears in the timelike class of solutions as ${\mathbb{R}}^3$, to a more general class of 3-manifold, which admits a Riemannian Weyl structure.
Other interesting solutions whose supersymmetry was investigated
using these techniques are supersymmetric Plebanski-Demianski geometries, and
the C-metric \cite{Klemm:2013eca,Plebanski:1976gy}.

The minimal gauged supergravity analysis was then further extended in \cite{Cacciatori:2004rt},
both in terms of constructing new examples of solutions, and in terms of investigating solutions with extended
supersymmetry. In particular, solutions of the ungauged theory preserve either $N=4$ or
$N=8$ supersymmetry, whereas in the gauged theory $N=2$, $N=4$, $N=6$ and $N=8$ solutions
are in principle allowed. It was shown in \cite{Cacciatori:2004rt} by considering explicitly the integrability conditions of the KSE that all null solutions with $N=6$ solutions must be locally isometric to the unique maximally supersymmetric solution $AdS_4$; this result
was also shown to hold for the timelike class in \cite{Grover}, using spinorial geometry techniques. Further classification, via spinorial geometry,
of the solutions with $N=4$ supersymmetry was done in \cite{Cacciatori:2007},
subject to the assumption, for solutions entirely in the timelike class,
that the spinorial Lie derivative of the additional Killing spinor
with respect to the isometry generated by the first spinor vanishes.

Numerous solutions have also been found for ${\cal N}=2$ supergravity coupled to vector multiplets. For the case of the ungauged theory, a large class of solutions
in the timelike class, including electrically and magnetically charged black holes,
were found in \cite{Behrndt:1997ny}. These solutions, for which the metric takes the
same form as in the minimal theory ({\ref{4dlfieldsa}}), were found by proposing
a particular ansatz. Later, it was shown in the classification of \cite{Meessen:2006tu},
using the bilinears method, that the timelike solutions found in \cite{Behrndt:1997ny}
are in fact the most general possible, and all null solutions were also determined.

Black hole solutions in gauged supergravity coupled to vector multiplets were also constructed in
\cite{Sabra:1999ux}, \cite{Chamseddine:2000bk} and \cite{Chong:2004na}. The systematic classifications of solutions of gauged supergravity coupled to vector multiplets were constructed in
\cite{Cacciatori:2008ek} and \cite{Klemm:2009uw}. In this case, solutions with $N=2$ supersymmetry
were classified using spinorial geometry techniques, and in the timelike class the
general form of the metric is again a $U(1)$ fibration over a 3-dimensional
transverse manifold admitting a Riemannian Weyl structure. Using these results, novel examples of black holes were found \cite{Klemm:2011xw,Colleoni:2012jq};
further examples of supersymmetric asymptotically $AdS_4$ black holes were
considered in \cite{Gnecchi:2013mja} and \cite{Chow:2013gba}.
Solutions in the timelike class with extended $N=4$ supersymmetry were then classified using spinorial geometry techniques in \cite{Klemm:2010mc}, again subject to the assumption that the spinorial Lie derivative of the additional Killing spinor
with respect to the isometry generated by the first spinor vanishes. A further generalization
to include non-abelian vector multiplets, using the bilinears method, was made in \cite{Hubscher:2008yz}.

Additional generalizations have also been made to include both vector and hypermultiplets.
Supersymmetric solutions in the ungauged theory coupled to arbitrary vector and hypermultiplets
were classified in \cite{Huebscher:2006mr}. One novel feature of the results of this work is that
solutions in the timelike class no longer have a metric of the form given in ({\ref{4dlfieldsa}}),
but instead have geometries which are $U(1)$ fibrations over a  3-manifold
whose spin connection is determined by the pull-back of a certain quaternionic $SU(2)$ connection.
The extension of this analysis to gauged supergravity coupled to vector and hypermultiplets
was carried out in \cite{Meessen:2012sr}, and supersymmetric black holes coupled to hypermultiplets were constructed in \cite{Hristov:2010eu}, which also included a (partial) classification of solutions. These classifications all employ the bilinears method.

Higher derivative solutions have also been considered in specific examples.
In \cite{LopesCardoso:2000qm}, an ansatz for stationary solutions in supergravity coupled to
vector and hypermultiplets, including higher derivative terms was considered. The
conditions imposed by supersymmetry were derived, assuming that the Killing spinors satisfied
a certain projection. It would be of interest to construct a systematic classification
of supersymmetric solutions of higher derivative supergravity in four dimensions.

Supersymmetric solutions of various ${\cal{N}}>2$ theories have also been classified in
\cite{Bellorin:2005zc}, further generalizing the earlier analysis of \cite{Tod:1995jf};
as well as that in  \cite{Meessen:2010fh}. Theories with novel signature have been considered as well.
The simplest case is minimal Euclidean supergravity with a single Maxwell field, \cite{Dunajski:2010zp} and
\cite{Dunajski:2010uv}, whose supersymmetric solutions were classified via spinorial geometry.
Solutions of minimal gauged supergravity with (2,2) signature were also classified using the bilinears method in
\cite{Klemm:2015mga}. These include geometries involving Gauduchon-Tod structures, which also appear in various other types of $d=4$ and $d=5$ supergravity theories with non-standard couplings,
such as de-Sitter supergravities and other pseudo-supergravities; we shall not consider such theories here.

\section{ ${\cal N}=1$ $d=4$ supergravity}

Next we shall describe   the solutions to the KSEs  of  ${\cal N}=1$ $d=4$ supergravity
coupled to any number of (non-abelian) vector and scalar multiplets \cite{gggpn1d4}, see also  \cite{to-n1d4} for the treatment of a special case.  For the construction of the theory see \cite{wessbagger} and references therein. This theory is one of the most phenomenologically attractive in the context of supersymmetry.  Furthermore, as we shall demonstrate  the KSEs can be solved exactly for any number of supersymmetries.

 The bosonic fields of the theory, in addition to the
metric, are the vector gauge potentials $A^a$  and the scalars $\phi^i$. The latter are functions on the spacetime with values
in a K\"ahler manifold $S$.  We shall refer to $S$ as  the  ``scalar'' or ``sigma model'' manifold. The K\"ahler geometry
on $S$ arises as a consequence of the invariance of the action under the supersymmetry transformations of the theory. The rest
of  the relevant properties of the  theory, including the couplings, will be described  below along with the KSEs.

\subsection{Fields and spinors}

\subsubsection{Killing spinor equations}\label{killing}


The KSEs  of ${\cal N}=1$ supergravity coupled to any number of (non-abelian) vector
and matter multiplets can easily be read off from the supersymmetry transformations of the fermions of the theory.
These are the gravitino KSE
\bea
\label{graveq}
\nabla_\tM\eps_L+{1\over4} (\partial_i{\mathcal K}\, {\cal D}_\tM \phi^i-\partial_{\bar i}{\mathcal K}\,{\cal D}_\tM \phi^{\bar i})\eps_L+{i\over2} e^{{{\mathcal K}\over2}} W\Ga_\tM\eps_R=0~,
\eea
the gaugino KSE
\bea
\label{gaugeq}
F^a_{\tM\tN}\Ga^{\tM\tN}\eps_L-2i  \mu^a\eps_L=0~,
\eea
and the KSE associated with the scalar multiplets
\bea
\label{mateq}
i \Ga^\tM \eps_R {\cal D}_\tM \phi^i- e^{{{\mathcal K}\over2}} G^{i\bar j} D_{\bar j} \bar W \eps_L=0~,
\eea
where $\nabla$ is the spin connection, $\phi^i$ is a complex scalar  field, ${\mathcal K}={\mathcal K}(\phi^i, \phi^{\bar j})$ is the K\"ahler potential
of the
 scalar manifold $S$, whose metric is $G_{i\bar j}=\partial_i \partial_{\bar j} {\mathcal K}$, $W=W(\phi^i)$ is a (local) holomorphic function on $S$,
\bea
D_iW=\partial_iW+\partial_i{\mathcal K}\, W~,~~~{\cal D}_\tM \phi^{i}=\partial_\tM\phi^i- A_\tM^a \xi^i_a~,
\eea
 $\xi_a$ are  holomorphic Killing  vector fields on $S$, $A^a$ is the gauge connection with field strength $F^a$ and
$\mu^a$ is the moment map, i.e.
\bea
G_{i\bar j} \xi_a^{\bar j}=i\partial_i\mu_a~.
\eea
We mostly follow the metric and spinor conventions of \cite{wessbagger}. In particular, the spacetime metric
has signature mostly plus,
$\eps$ is a Majorana spinor and $\eps_{L,R}={1\over2} (1\pm \Ga_5)\eps$, where $\Ga_5^2=1$.
We have set the gauge coupling to 1.

The gravitino KSE is a parallel transport equation for a connection which, apart
from the Levi-Civita part, contains additional terms that depend on the matter couplings. The gauge group
of the KSEs is $Spin_c(3,1)=Spin(3,1)\times_{\bZ_2} U(1)$. The $Spin(3,1)$ part acts on $\eps$ with the Majorana representation
while $U(1)$ acts on the chiral component $\eps_L$ with the standard 1-dimensional representation and on the anti-chiral $\eps_R$
with its conjugate. The additional $U(1)$ gauge transformation is due to the coupling of the spinor $\eps$ to the
$U(1)$ connection constructed from the K\"ahler potential $K$ associated with the matter couplings. In what follows,
we  use only the $Spin(3,1)$ component of the gauge group to locally choose the representatives of the Killing spinors.

\subsubsection{Spinors}
\label{spinorsn1d4}

We have already described the Dirac spinors of $Spin(3,1)$ in the context of ${\cal N}=2$ theory in section \ref{4dspspgeom}.
Here the difference is that  the supersymmetry parameter $\eps$ is in the Majorana representation of $Spin(3,1)$.  To impose the reality condition required,  let us identify
 the chiral representation with the
even forms, $\Lambda^{{\rm ev}}(\bC^2)$,  and the anti-chiral with odd ones, $\Lambda^{{\rm odd}}(\bC^2)$. The complex conjugation operation
is imposed by the anti-linear map,  $r_A=-\Gamma_{012}*$, $r_A^2=1$, see appendix \ref{app:spinors}. There is one orbit of $Spin(3,1)=SL(2,\bC)$ on $\Lambda^{{\rm ev}}(\bC^2)-\{0\}$, and so the chiral
component of $\eps$ can be locally chosen as $1$.  Applying $r_A$ to the spinor $1$, one finds
that a   Majorana
representative for the orbit is
\bea
\eps=1+\se_1~,~~~~\eps_L=1~,~~~\eps_R=\se_1~.
\la{fks}
\eea
This can be chosen as the first Killing spinor of the theory. The isotropy group of the spinor $1$ in $SL(2,\bC)$ is $\bC$.  This remaining gauge symmetry will be used
later to choose the second Killing spinor.

\subsection{N=1 backgrounds}

\subsubsection{Solution of KSEs}

Evaluating the KSEs on the first Killing spinor $\epsilon=1+\se_1$,
one finds a linear system which relates some components of the spin connection to the fluxes and matter
couplings,  and restricts the geometry of the spacetime. The construction of  the linear system is similar to that already described for  in the minimal  ${\cal N}=2$ $d=4$ theory and therefore we shall not give more details. The solution of this linear system for the
gravitino KSE  gives
\bea
\label{auxgrav1}
&&\Omega_{+,+-} = \Omega_{+,1 \bar{1}} =
\Omega_{+,+1} = \Omega_{-,-+} = \Omega_{1,+ \bar{1}}
= \Omega_{1,+1} = 0
\cr
&&\Omega_{-,+1}+\Omega_{1,+-}=0~,
\eea
and
\bea
\Omega_{-,1 \bar{1}} + {1 \over 2} (\partial_i{\mathcal K}\, {\cal D}_- \phi^i-\partial_{\bar i}{\mathcal K}\,{\cal D}_- \phi^{\bar i})
&=& 0~,
\cr
i\sqrt{2}e^{{{\mathcal K} \over 2}} W+2\Omega_{-,+ \bar{1}}&=&0~,
\cr
\Omega_{-,+1} + \Omega_{1,1 \bar{1}}
+{1 \over 2}  (\partial_i{\mathcal K}\, {\cal D}_1 \phi^i-\partial_{\bar i}{\mathcal K}\,{\cal D}_1 \phi^{\bar i}) &=& 0~.
\la{n1geomb}
\eea
The conditions in (\ref{auxgrav1}) are considered as restrictions on the geometry of spacetime while the
conditions in (\ref{n1geomb}) are thought of as an expression of the fluxes in terms of the geometry.

Similarly,  the solution of the linear system for the  gaugino  ({\ref{gaugeq}})
and the matter multiplet (\ref{mateq}) Killing spinor equations gives
\be
F^a_{+ 1} =F^a_{+-}=0, \qquad   F^a_{1 \bar{1}} -i\mu^{a}=0~,
\la{n1gsol}
\ee
and
\be
{\cal D}_+ \phi^i =0, \qquad \sqrt{2} i {\cal{D}}_1 \phi^i = e^{{\mathcal K} \over 2} G^{i \bar{j}} D_{\bar{j}} {\bar{W}}~,
\la{n1msol}
\ee
respectively.
In what follows, we explore the consequences of the above conditions on the geometry of spacetime.

\subsubsection{Geometry}
To proceed, the metric in  the spinorial geometry coframe is
\bea
ds^2= 2 {\bf{e}}^- {\bf{e}}^++2 \fe^1 {\fe}^{\bar 1}~.
\la{lcmetr}
\eea
The form bilinears associated with the Killing spinor $\epsilon$ are
\bea
X={\bf{e}}^-~,~~~\tau={\bf{e}}^-\wedge ({\bf{e}}^1+
{\bf{e}}^{\bar 1})~,
\label{n1formb}
\eea
 and their
spacetime duals.  Observe that $X$ is invariant under the $U(1)$ transformations generated by the K\"ahler potential, while $\tau$ is not,
and thus it is a section of a $U(1)$ bundle.
The conditions on the geometry
 ({\ref{auxgrav1}}) can now be re-written as
\be
{\cal L}_X g =0,  \qquad {\bf{e}}^{-} \wedge {d\bf{e}}^{-} =0,\qquad  {\fe}^{-}\wedge {\fe}^{\bar 1 } \wedge d {\fe}^1 =0 \ .
\la{geomcon4}
\ee
Observe also that ${\bf{e}}^{-}\wedge {\bf{e}}^1 \wedge d {\bf{e}}^1 =0$ and ${\cal L}_X\eps=0$, where ${\cal L}_X$ is
the spinorial Lie derivative (\ref{spinder}).

The first condition in (\ref{geomcon4}) implies that the metric admits a null Killing vector field.
While the second implies that the distribution defined by $X$ is integrable.  Therefore, there is locally a function $h$ such that
$\fe^-=h\, dv$ for some coordinate $v$.
Adapting also a coordinate along $X$, $X=\partial_u$, and after a coframe rotation as in (\ref{4dnullpatching}),    the metric can be written as in (\ref{lcmetr}) with
\bea
{\bf{e}}^-=h\, dv~,~~~{\bf{e}}^+ = du + V dv + w_\tI dx^\tI~,~~~\fe^1=\beta_{\tI} dx^{\tI}~,
\la{frame}
\eea
where $u,v,x^{\tI}$, $\tI=1,2$, are real coordinates,  and  $h,V,w_{\tI}$ are real and $\beta_1, \beta_2$ are complex
spacetime functions, respectively.  As the metric and the form bilinears are invariant under the action of $X$, the coframe above  can  be chosen to be $u$ independent.




The  conditions that relate the fluxes to the geometry in (\ref{n1geomb}) can be rewritten in terms of the form bilinears $X$ and $\tau$  as
\bea
 {1 \over 2} (\partial_i{\mathcal K}\, {\cal D}_- \phi^i-\partial_{\bar i}{\mathcal K}\,{\cal D}_- \phi^{\bar i})+\nabla_- \tau_{-1}
&=& 0~,
\cr
\sqrt{2} e^{{{\mathcal K}\over2}} W \fe^--\star (\fe^1\wedge d\fe^-)&=&0~,
\cr
\star d(\fe^-\wedge \fe^{\bar 1})-{1\over \sqrt{2}} e^{{{\mathcal K}\over2}} \bar W \fe^--
{i\over 2} (\partial_i{\mathcal K}\, {\cal D}_1 \phi^i-\partial_{\bar i}{\mathcal K}\,{\cal D}_1 \phi^{\bar i})\fe^- &=& 0~,
\eea
where the orientation of the spacetime is chosen as $\epsilon_{-+1\bar 1}=-i$.

To solve (\ref{n1gsol}), one can locally  always choose the gauge $A_+^a=0$. The first two conditions in (\ref{n1gsol}) will then imply that
the remaining components of $A$ are independent of $u$. There is no general procedure to give an explicit solution
for the last condition (\ref{n1gsol}).

Next we turn to the conditions (\ref{n1msol}) that arise from the Killing spinor equations of the matter multiplet.
In the gauge $A_+^a=0$, the first condition in (\ref{n1msol}) implies that the scalar fields can be taken
to be independent of $u$, $\partial_u\phi=0$. The last condition in (\ref{n1msol}) can be interpreted as a
holomorphic flow equation. The construction of explicit solutions will depend on the form of the K\"ahler potential
and $W$, i.e.~on the details of the model. This concludes the solution of the KSEs for one Killing spinor of  ${\cal N}=1$ $d=4$ supergravity coupled to any number of vector and scalar multiplets.

\subsection{N=2 backgrounds}

\subsubsection{Killing spinors}

The first Killing spinor is the same as that of the $N=1$ case investigated above. So we set $\eps_1=\eps$, where $\eps$ is given in (\ref{fks}).
To choose the second Killing spinor, consider the most general Majorana spinor
\bea
\eps_2= a 1+be_{12}+ r_A (a 1+be_{12})~,~~~a,b\in \bC~.
\eea
The isotropy group of  $\eps_1$  in $Spin(3,1)$ is  $\bC$. This  can be used to simplify the
expression for $\eps_2$.
There are two cases to consider.
First if $b=0$, the $\bC$ isotropy transformations leave $\eps_2$ invariant.  Therefore, one can set
\bea
\eps_2=a 1+ \bar a \se_1~.
\la{sksa}
\eea
Linear independence of $\eps_1$ and $\eps_2$ requires that ${\rm Im}\, a\not=0$.

Next suppose that  $b\not=0$.  Acting on $\epsilon_2$ with the isotropy group  $\bC$ of the first Killing spinor  with parameter $\lambda$,  one has
\bea
\eps_2'= (a+\lambda b) 1+ b \se_{12}+r_A[(a+\lambda b) 1+ b \se_{12}]~.
\eea
Setting  $\lambda=-{a\over b}$, one can choose a normal form for $\eps_2$ as
\bea
\eps_2=b \se_{12}-\bar b \se_2~.
\la{sksb}
\eea
So the second Killing spinor $\epsilon_2$ can be locally chosen either as in (\ref{sksa}) or as in (\ref{sksb}) with $a,b$ promoted
to complex spacetime functions.


\subsubsection{Solution of KSEs for $\eps_2=a 1+ \bar a \se_1$}
Consider first the case for which $\eps_2=a 1+ \bar a \se_1$. The linear system is easy to read off from that of the $N=1$ case.
In particular, the supercovariant connection along the $-$ light-cone direction gives
\bea
2a \Omega_{-,+\bar 1}+i \sqrt{ 2} \bar a e^{{{\mathcal K}\over2}} W=0~.
\eea
Comparing this condition with those of the $N=1$ case in (\ref{n1geomb}), one concludes that either $W=0$ on the field configurations $\phi$ of the solution
or $a=\bar a$. If the latter is the
case, then it turns out that $a$ is also constant and so $\eps_2$ is not linearly independent of $\eps_1$.

Therefore, for these $N=2$ solutions, we have to choose $W=0$. After some further investigation of the gravitino and scalar multiplets KSEs, we find that the conditions for $N=2$ supersymmetry are
\bea
\label{auxgrav2}
\Omega_{\tA,+\tB} = \Omega_{+,1 \bar{1}} =0~,~~~\partial_\tA a=0~,
\eea
and
\bea
\Omega_{-,1 \bar{1}} + {1 \over 2} (\partial_i{\mathcal K}\, {\cal D}_- \phi^i-\partial_{\bar i}{\mathcal K}\,{\cal D}_- \phi^{\bar i})= 0~,~~~\Omega_{1,1 \bar{1}}
-{1 \over 2} \partial_{\bar i}{\mathcal K}\,{\cal D}_1 \phi^{\bar i} = 0~,
\cr
W=\partial_jW=0~,~~~{\cal D}_1\phi^i={\cal D}_+\phi^i=0~.
\la{n2pp}
\eea
Therefore $a$ is constant.
There are no additional conditions that arise from the gaugino Killing spinor equation apart from those that we have found
in the $N=1$ case (\ref{n1gsol}).

\subsubsection{Solution of KSEs for $\eps_2=b \se_{12}-\bar b \se_2$}

A direct substitution of $\eps_2=b \se_{12}-\bar b \se_2$ into the gravitino KSE  reveals that
\bea
\partial_+b=0~,~~~~ b \Omega_{+,-1}+ \bar b \Omega_{-,+\bar 1}=0~,
\cr
\partial_-b-\Omega_{-,1\bar1} b=0~,~~~\Omega_{-,-1}=0~,
\cr
\partial_1 b-b(\Omega_{1,-+}+\Omega_{+,-1}+\Omega_{1,1\bar1})=0~,~~~\Omega_{1,-1}=0~,
\cr
\partial_{\bar 1}b-b\Omega_{\bar1,1\bar1}=0~,~~~\Omega_{\bar1, -1}=0~,
\la{n2dsol}
\eea
where we have used the $N=1$ relations in (\ref{auxgrav1}) and (\ref{n1geomb})   to simplify the expressions.
Similarly, the gaugino KSE gives
\bea
F^a_{-1}=0~,~~~~F^a_{1\bar 1}+i \mu^a=0~.
\la{n2gsol}
\eea
Furthermore, the KSEs associated with the matter multiplets evaluated on $\epsilon_2$ reveal that
\bea
{\cal D}_-\phi^i=0~,~~~i\sqrt{2}\,\, \bar b \,\,{\cal D}_{\bar 1} \phi^i+b e^{{{\mathcal K}\over2}} G^{i\bar j}D_{\bar j}\bar W=0~.
\la{n2msol}
\eea
One can easily combine the above conditions with those described in   (\ref{auxgrav1}), (\ref{n1geomb}), (\ref{n1gsol}) and (\ref{n1msol})
which arise  from demanding that  $\eps_1$ is a Killing spinor. Below we shall describe the consequences that all these conditions have on the spacetime geometry and
the restrictions they impose on the fields.

\subsubsection{Geometry of $N=2$ backgrounds }

Case 1: {$\eps_2=a 1+ \bar a \se_1$}

The geometric conditions  (\ref{auxgrav2}) imply that
\bea
\nabla_A X_B=0~,~~~{\fe}^{-}\wedge {\fe}^{\bar 1 } \wedge d {\fe}^1 =0~,
\eea
where $X={\bf{e}}^-$.  Thus the spacetime admits a null parallel   vector field $X$.  Note also that as
a consequence of the above conditions $ {\cal L}_X \epsilon_1={\cal L}_X\epsilon_2=0$,
where ${\cal L}_X$ is the spinorial Lie derivative defined in (\ref{spinder}).

Apart from $X$, the spacetime admits a 2-form bilinear $\rho=\fe^-\wedge \fe^1$.  Observe that  $\tau$ in (\ref{n1formb}) is the
real part of $\rho$.  The investigation of the  conditions in (\ref{n2pp}) is similar to
those in (\ref{n1geomb}) and (\ref{n1msol}) in the $N=1$ case.    In particular, we have
\bea
 {1 \over 2} (\partial_i{\mathcal K}\, {\cal D}_- \phi^i-\partial_{\bar i}{\mathcal K}\,{\cal D}_- \phi^{\bar i})+\nabla_- \tau_{-1}
&=& 0~,
\cr
\star d(\fe^-\wedge \fe^{\bar 1})+
{i\over 2} \partial_{\bar i}{\mathcal K}\,{\cal D}_1 \phi^{\bar i} \fe^- &=& 0~,
\cr
W=\partial_jW=0~,~~~{\cal D}_1\phi^i={\cal D}_+\phi^i&=&0~.
\eea
It is apparent from this that the scalar fields must lie on both the vanishing locus and critical points of $W$.
Furthermore, they obey a light-cone holomorphicity condition as a consequence of the last two conditions in the above equation.  Observe that the distribution spanned
by $(\partial_+, \partial_1)$ is integrable as $d\fe^-=0$, and $d\fe^{\bar 1}(\partial_+, \partial_1)=0$.  A coframe can be chosen as in (\ref{frame}) but now with $h=1$.  Further simplifications are possible in special gauges.  For example
one can choose $A_+^a=A_1^a=0$ as $F_{+1}=0$. For more details on the geometry of these solutions see \cite{gggpn1d4}.

The physical interpretation of spacetimes with a null parallel Killing vector field is that of a pp-wave.
However this class also includes
 the cosmic string solutions \cite{cosmicstrings} and their generalizations \cite{jggp-n1d4,gdrkavp-n1d4}.

Case 2: {$\eps_2=b \se_{12}-\bar b \se_2$}

To analyze the conditions (\ref{n2dsol}) which arise from the KSEs,
it is convenient to define the 1-forms
\be
X= {\bf{e}}^-, \qquad Y= |b|^2 {\bf{e}}^+, \qquad Z = {\bar{b}} {\bf{e}}^1 + b {\bf{e}}^{\bar{1}},
\qquad V =  i {\bar{b}} {\bf{e}}^1 -i b {\bf{e}}^{\bar{1}} \ .
\la{frem}
\ee
Observe that $Z$ is orthogonal to $X, Y, V$, and $V$ is orthogonal to $X, Y, Z$.
Then it is straightforward to show that the Killing spinor equations imply that
$X$, $Y$ and $Z$ are all Killing vectors. Furthermore, $V$ is closed, $d V =0$.
In addition, one finds the following  commutators
\be
[V, X] = [V, Y] = [V, Z] =0~,
\ee
and
\be
[X,Y] = c Z, \qquad [X, Z]=-2c K, \qquad [Y, Z]=2cY~,
\ee
where $c=b(\Omega_{-,+1}-\Omega_{+,-1})$.

Consider the commutator $[X,Y] = c Z$. Since $V$ commutes with the other three vector field, the Jacobi identity
implies that $Vc=0$. Similarly, the Jacobi identity for $Z, X$ and $Y$ together with the linear independence
of these three vector field imply that $Xc=Yc=Zc=0$. So $c$ can be taken to be a constant.

Next, since $Z$ and $V$ commute one can choose coordinates $x, y$ such that $Z=\partial_x$ and $V=\partial_y$. Moreover,
the rest of the commutators imply that there are additional coordinates  $u$, $v$ such that
\be
X= e^{2cx} \partial_u , \qquad Y= e^{-2cx}\bigg(c^2 u^2 \partial_u
+cu \partial_x + \partial_v  \bigg) \ ,
\la{vf}
\ee
Using (\ref{frem}),  one can compute the coframe
in terms of the coordinates $x,y,v,u$ to find
\bea
\fe^-&=&e^{2cx} |b|^2 dv~,~~~\fe^+= e^{-2cx} (du-c^2 u^2 dv)~,~~~
\cr
\fe^{1}&=&b [(dx-idy)- cu dv]~,~~
\fe^{\bar 1}=\bar b [(dx+idy)- cu dv]~.
\eea

Hence the spacetime metric can be written as
\bea
ds^2=2|b|^2 [ds^2(M_3)+dy^2]~,
\la{metrb}
\eea
where
\be
ds^2(M_3) = dv (du-c^2 u^2 dv)+(dx-cu dv)^2 \ .
\ee
Thus $M_3$ is either $\bR^{2,1}$ if $c=0$, or $AdS_3$ if $c \neq 0$.

The function $b$ depends only on $y$. After some computation, one finds that
\be
{d b \over dy} = \sqrt{2} |b|^2 e^{{{\mathcal K} \over 2}} W
+{1 \over \sqrt{2}} e^{{{\mathcal K} \over 2}} b \big( b \partial_i {\mathcal K}\, G^{i \bar{j}}
D_{\bar{j}} \bar{W} - \bar{b} \partial_{\bar{i}} {\mathcal K}\, G^{\bar{i} j} D_j W \big) \ .
\la{f1}
\ee
which relates the only unknown component of the metric to the scalar fields.

Combining the conditions that arise from gaugino KSE on both $\epsilon_1$ and $\epsilon_2$ spinors, one finds   that
\be
F^a =0, \qquad \mu^a =0~.
\ee
So the gauge connection is flat and can locally be set to zero. The vanishing of the moment map restricts the
scalars to lie on a K\"ahler quotient $S//H$ of $S$, where $H$ is the gauge group.

Setting  $A=0$ locally,
the conditions on ${\cal{D}} \phi^i$
imply that $\partial_u\phi^i=\partial_x\phi^i=\partial_v\phi^i=0$.
Moreover, the remaining Killing spinor equations of the scalar multiplet (\ref{n2msol})  gives
\be
{d \phi^i \over dy} = -\sqrt{2}\, b e^{{{\mathcal K} \over 2}} G^{i \bar{j}} D_{\bar{j}} {\bar{W}}~.
\la{f2}
\ee
Observe that this expression depends on $b$. This is again a flow equation driven by the holomorphic potential $W$.
One can change parameterization to simplify the flow equations (\ref{f1}) and (\ref{f2}). The construction of
explicit solutions depends on the details of the models.

Clearly, the spacetime is of cohomogeneity one with a homogenous section either $AdS_3$ or $\bR^{2,1}$. So
this class of $N=2$ solutions can be thought of as domain wall spacetimes. For a review of the domains walls
in supergravity theories as well as their applications see \cite{cveticwalls}.

\subsection{N=3 and N=4 backgrounds}

\subsubsection{Killing spinors for $N=3$ backgrounds}

Let us first consider the $N=3$ backgrounds. It is clear that after choosing the first  two Killing spinors using the
$Spin(3,1)$ covariance of the theory, there is little or no more gauge symmetry left to restrict the choice of the third
Killing spinor. This could potentially lead to difficulties with solving the KSEs. Because of this, we use instead  a technique which was originally applied to classify the
near maximally supersymmetric backgrounds of IIB supergravity in \cite{ugjggpdr-nmax}. As at each point in spacetime, the three Killing spinors
span a hyperplane in the space of Majorana spinors, we use the $Spin(3,1)$ gauge symmetry of the theory to restrict the form
of the normal $\nu$ to the hyper-plane of the Killing spinors.  As $Spin(3,1)$ has a single non-trivial orbit on the space of Majorana
spinors, we can always chose $\nu=i(\se_2+\se_{12})$.  The orthogonal directions to $\nu$ with respect to Majorana inner product, $A(\zeta, \eta)=\langle\Gamma_{12}\zeta^*, \eta\rangle$, see appendix \ref{app:spinors}, are
 $\{\eta_r\}=\{1+\se_1, \se_2-\se_{12}, i(\se_2+\se_{12})\}$. So
the three Killing spinors can be chosen as
\bea
\eps_r=\sum_s f_{rs} \eta_s~,~~~r,s=1,2,3~,
\la{n3ks}
\eea
where $f_{rs}$ is a real $3\times 3$ invertible matrix of spacetime functions. Schematically we write $\eps=f\eta$.

 In $N=4$ backgrounds, the Killing spinors can again be written as a linear combination of the basis $\{1+\se_1, i(1-\se_1), \se_2-\se_{12}, i(\se_2+\se_{12})\}$
of Majorana spinors with real spacetime functions as coefficients. Next we shall solve the Killing spinor equations for both cases.

\subsubsection{Local $N=3$ supersymmetry implies $N=4$}

Let us begin with the $N=3$ case. We shall first solve the Killing spinor equations locally.
To proceed, observe that (\ref{n3ks}) implies that schematically $\eps_L=f \eta_L$ and $\eps_R=f \eta_R$. Substituting this
into the  gaugino (\ref{gaugeq}) and scalar multiplet  (\ref{mateq}) KSEs, one finds
that the
dependence on $f$ can be eliminated, because $f$ is invertible.
Moreover the conditions that one obtains  are those of (\ref{n1gsol}) and (\ref{n1msol}), or
(\ref{n2gsol}) and (\ref{n2msol}) for $b=1$ or $b=i$, respectively. These  imply that
\bea
F^a_{\tM\tN}={\cal D}_\tM\phi^i=D_i W=\mu^a=0~.
\la{n3sol}
\eea
Since the gauge connection is flat, we can locally set the gauge potential to vanish, $A^a_\tM=0$. As a result the second equation implies that
$\phi^i$ are constant. Substituting these data into the gravitino Killing spinor equation, and computing its integrability condition,
we obtain
\bea
R_{\tM\tN, \tR\tS} \Ga^{\tR\tS}\eta_L+2 e^{{\mathcal K}} W\bar W \Ga_{\tM\tN}\eta_L=0~.
\la{n3int1}
\eea
Clearly the integrability condition takes values in $\mathfrak{spin}(3,1)$. Since the isotropy group of three linearly independent spinors
in $Spin(3,1)$ is the identity, (\ref{n3int}) implies that
\bea
R_{\tM\tN, \tR\tS}=- e^{{\mathcal K}} W\bar W (g_{\tM\tR} g_{\tN\tS}-g_{\tM\tS} g_{\tN\tR})~.
\la{n3int}
\eea
It is easy to see that (\ref{n3sol}) and (\ref{n3int}) are precisely the conditions that one finds  for backgrounds that admit maximal $N=4$ supersymmetry. So one concludes that $N=3$ backgrounds locally admit an additional
Killing spinor  and are  therefore locally maximally supersymmetric. Furthermore, (\ref{n3int}) implies that the spacetime
is either $\bR^{3,1}$ or $AdS_4$. In the former case,  $e^{{\mathcal K}}|W|^2=0$  and in the latter $e^{{\mathcal K}}|W|^2\not=0$ when
 evaluated at the constant maps $\phi$, respectively.

The moment map condition in (\ref{n3sol}), $\mu^a=0$, together with the remaining constant gauge transformations imply
that the constant maps $\phi$ take values in a K\"ahler quotient $S//H$ of the scalar manifold $S$. It remains to investigate
$D_i W=0$. Suppose that we have chosen some constant maps $\phi=\phi_0$. If $W(\phi_0)=0$, then  $D_i W=0$ implies that
$\partial_i W(\phi_0)=0$. So $W$ and its first derivative vanish at $\phi=\phi_0$.
On the other hand if $W(\phi_0)\not=0$, $D_i W=0$ relates the value of the first derivative of $W$ to that
of the K\"ahler potential at $\phi=\phi_0$.

The physical interpretation of the $N=4$ backgrounds  is that they are the supersymmetric vacua of the supergravity theory.
The spacetime geometry is either Minkowski or $AdS_4$.

Although the existence of local geometries which preserve strictly $N=3$ supersymmetry has been ruled out, the possibility still
remains that such backgrounds can be constructed as discrete  quotients of maximally supersymmetric ones.  We shall not give the
details here.  This question has been raised in \cite{fofpreons11}
in the context of ${\cal{N}}=2$ supergravity theory. One can show that a background that preserves strictly $N=3$ supersymmetries can be
constructed as discrete  quotient of a maximally supersymmetric $AdS_4$ solution \cite{fofgppreons4}.

\subsection{A reflection on the results}

One of the conclusions from the results presented is that there is a systematic way to find the solutions
of the KSEs of  ${\cal N}=1$  $d=4$ supergravity coupled to any number of vector and scalar multiplets.
Although this is not sufficient to find all  supersymmetric solutions of the theory as for those the
field equations have to be solved as well, a  narrative emerges regarding the geometry of the solutions.
This is especially apparent for those  that preserve $N>1$ supersymmetry as such  backgrounds are sufficiently
constrained. As a result, their physical interpretation is more apparent and can be extracted
from their geometric properties.

The observations made here regarding the geometry of supersymmetric backgrounds extend to other supersymmetric theories.
A more involved example is the solution of the KSEs of heterotic supergravity.  As for the  ${\cal N}=1$ $d=4$
supergravity, the geometry of all supersymmetric heterotic backgrounds can be identified.  This leads to the categorization
of all the supersymmetric solutions of the theory that have been found as well as to new directions
 that remain to be investigated.

\section{Minimal  ${\cal N}=1$ $d=5$   supergravity}\label{sec:minimald5}

The bosonic fields of the theory are a metric and a $U(1)$ gauge field $A$ with field strength $F=dA$.
The action of the theory is the Einstein-Maxwell system with the addition of a Chern-Simons term for the $U(1)$
field.  This theory has found widespread applications in the microscopic
counting of black hole entropy within string theory \cite{ascv}.  This is because some brane configurations of 10- and 11-dimensional supergravities
dimensionally reduce to black hole solutions of this theory and brane techniques in string theory can be used to do the counting.

Minimal ${\cal N}=1$ $d=5$  supergravity is  the first theory
whose KSE was systematically solved using the bilinears method \cite{gghpr}.  Here we shall present the solution
of the KSE employing both the bilinears and spinorial geometry methods to provide a comprehensive
example for both methods used to solve KSEs.

\subsection{KSE and field equations}

The only fermion of the theory is the gravitino whose supersymmetry variation leads to the  KSE
\bea
{\cal D}_\tA\eps=0~,
\label{kspinv}
\eea
where the supercovariant connection is
\bea
\label{kspinvb}
{\cal D}_\tA\defeq \nabla_\tA -\frac{i}{4\sqrt{3}}
   \left( \Gamma_\tA{}^{\tB\tC} - 4
\delta^{\tB}_{\tA}\Gamma^\tC
   \right) \fiveF_{\tB\tC}~,
\eea
and $\eps$ is  in the Dirac representation of $Spin(4,1)$. The supercovariant curvature can be written as
\bea
 {\cal R}_{\tA\tB}&=&{1\over4} \hat R_{\tA\tB, \tC\tD}\Gamma^{\tC\tD}+{i\over \sqrt 3}\left( \hat\nabla_\tA F_{\tB\tC}- \hat\nabla_\tB F_{\tA\tC}\right) \Gamma^\tC
 \cr && +{ 2i\over 3} H_{\tA\tB}{}^\tD
 F_{\tD\tC} \Gamma^\tC-{2\over3} F_{\tA\tC} F_{\tB\tD} \Gamma^{\tC\tD}~,
 \label{5dsupercov}
\eea
where $\hat R$ is the curvature of the connection $\hat \nabla_\tA Y^\tB\defeq \nabla_\tA Y^\tB+ (1/\sqrt3) H^\tB{}_{\tA\tC}  Y^\tC$ and $H=*F$ with $\epsilon^{01234}=1$.

Furthermore, a straightforward computation using $dF=0$ reveals that
\bea
 \Gamma^{\tB}{\cal R}_{\tA\tB}=-{1\over2} E_{\tA\tB} \Gamma^\tB-{1\over12\sqrt 3} LF_{\tA\tB_1\tB_2\tB_3} \Gamma^{\tB_1\tB_2\tB_3}+{i\over\sqrt 3} {}^*LF_\tA~,
 \label{5dintfeqs}
 \eea
 where
 \bea
\label{5dfeqns}
E_{\tA\tB}&\defeq &R_{\tA\tB}-2(\fiveF_{\tA\tC} \fiveF_\tB{}^\tC
-\frac{1}{6}g_{\tA\tB}{}\fiveF^2)=0~,
\nn \\
LF&\defeq &d{}^*{}F - \frac{2}{\sqrt{3}} F \wedge F=0~,
\eea
 are the field equations of the theory~.

It turns out that as in the minimal  ${\cal N}=2$ $d=4$ supergravity, the KSE (\ref{kspinv}) admits either four or eight Killing spinors.  To see this, first observe
that (\ref{kspinv}) is linear over the complex numbers so if $\eps$ is a Killing spinor so is $i\,\eps$.    Moreover, if $\epsilon$
is a Killing spinor so is $r_{{\mathrm A}}\eps$, where
 $r_{{\mathrm A}}=\Gamma_{12}*$, see appendix \ref{app:spinors} for the definition and properties of these Clifford algebra operations. As $\eps$, $i\, \eps$, $r_{{\mathrm A}}\eps$ and
 $i\, r_{{\mathrm A}}\eps$ are linearly independent, the solutions of (\ref{kspinv}) come as multiples of four.

\subsection{Solution of the KSE using the bilinears method}

To begin the analysis of the KSE using the bilinears method \cite{gghpr}, we use Fierz identities
 to obtain algebraic conditions on various form  bilinears.
Then further conditions on the geometry and flux are determined
by an analysis of the KSE. To start, we define a real scalar $f$
and a real form  bilinear $X$ by

\begin{eqnarray}
\label{sp5bilin1}
if^2\defeq D(\epsilon,\epsilon)~, \quad X\defeq D(\epsilon,\Gamma_\tA \epsilon)\, \fe^\tA~.
\end{eqnarray}
Note that the  bilinear $X$ cannot vanish identically, as by definition
$X_0 = \parallel \Gamma_0 \epsilon \parallel^2 = \parallel \epsilon \parallel^2\not=0$.
We have also chosen a convention for which $D(\epsilon,\epsilon)=if^2$.
If a spinor ${\hat{\epsilon}}$ satisfies
$D({\hat\epsilon},{\hat \epsilon})=-if^2$, then there exists a
spin transformation, which lies in a disconnected component of the
spin group, relating ${\hat{\epsilon}}$ to a spinor $\epsilon$ satisfying
$D(\epsilon,\epsilon)=if^2$. Hence, without loss of generality, we take
 $D(\epsilon,\epsilon)=if^2$.

Further 2-form bilinears $\omega_1$ and $\xi$ are given by
\begin{eqnarray}
\label{sp5bilin2}
\omega_1\defeq {1 \over 2}
D(\epsilon,\Gamma_{\tA \tB} \epsilon) \fe^\tA \wedge \fe^\tB \ ,
\quad \xi \defeq {1 \over 2} D(\epsilon, \Gamma_{\tA \tB} r_{\mathrm A} \epsilon )\fe^\tA \wedge \fe^\tB\,,
\end{eqnarray}
where $\omega_1$ is real and $\xi$ is complex. There are no other non-vanishing scalar or 1-form bilinears.

The bilinears satisfy a number of algebraic
conditions due to the Fierz identities ({\ref{fierz5a}}) and ({\ref{fierz5b}}), which are derived in detail in Appendix {\ref{fierzfive}}.
On defining $\xi = \omega_2+i\omega_3$, for real 2-forms $\omega_2, \omega_3$, these algebraic conditions on the bilinears can be rewritten as
\begin{equation}
\label{ac51}
i_X \omega_r =0\,, \quad i_X {}^* \omega_r = -f^2 \omega_r\,, \quad \omega_r \wedge \omega_s = -2f^2 \delta_{rs}\, {}^* X \, ,
\end{equation}
\begin{equation}
\label{ac53}
(\omega_r)_{\tC \tA}(\omega_s)^\tC{}_\tB = \delta_{rs} (f^4 g_{\tA \tB} + X_\tA X_\tB)+\epsilon_{rs}{}^p f^2 (\omega_p)_{\tA \tB} \ ,
\end{equation}
and
\begin{equation}
\label{ac54}
X^2=-f^4\,, \qquad r, s, p=1,2,3\ .
\end{equation}
The spinor $\epsilon$ also satisfies several conditions as a consequence
of the Fierz identities, which are
\begin{eqnarray}
\label{proj5}
X_\tA \Gamma^\tA \epsilon=if^2 \epsilon, \quad (\omega_1)_{\tA \tB} \Gamma^{\tA \tB} \epsilon =-4if^2 \epsilon\,,
\nonumber \\
(\omega_2)_{\tA \tB} \Gamma^{\tA \tB} \epsilon =-4if^2 r_{\mathrm A} \epsilon\,, \quad (\omega_3)_{\tA \tB} \Gamma^{\tA \tB} \epsilon =-4f^2 r_{\mathrm A} \epsilon \ .
\end{eqnarray}

Having obtained the algebraic conditions ({\ref{ac51}})-({\ref{ac54}}) and
({\ref{proj5}}), the conditions obtained from the Killing spinor equation can be determined. These are
\begin{equation}
\label{cc51}
df^2 = {2 \over \sqrt{3}} i_XF \ ,\quad {\cal L}_X g=0 \ ,
\end{equation}
\begin{equation}
\label{cc53}
dX = {4 \over \sqrt{3}}f^2 F+ {2 \over \sqrt{3}} {}^* (F \wedge X) \ ,
\end{equation}
and
\begin{eqnarray}
\label{cc54}
\nabla_\tA (\omega_r)_{\tB \tC} &=& -{2 \over \sqrt{3}} F_\tA{}^\tD  ({}^* \omega_r)_{\tD \tB \tC} +{2 \over \sqrt{3}}
F_{[\tB}{}^\tD ({}^* \omega_r)_{\tC] \tA \tD}
\nonumber \\
&-&{1 \over \sqrt{3}} g_{\tA[ \tB}({}^* \omega_r)_{\tC]}{}^{\tD_1 \tD_2}F_{\tD_1 \tD_2} \ .
\end{eqnarray}
In particular, the first condition in ({\ref{cc51}}) implies that ${\cal L}_X F=0$, and also
${\cal L}_X f^2=0$. Also, ({\ref{cc54}}) implies that
\begin{equation}
\label{cc55}
d\omega_r=0
\end{equation}
and this together with ({\ref{ac51}}) imply that ${\cal L}_X \omega_r=0$.  Therefore the fields $g$ and $F$, as well as the bilinears, are invariant under the action of $X$.

The analysis of these conditions splits into two cases, according as to whether
the vector field $X$ is timelike or null, corresponding to the cases
$f \neq 0$ and $f=0$ respectively.

\subsubsection{Geometry and supersymmetry of timelike solutions} \label{fivetime}

In the timelike class of solutions for which $f \neq 0$, it is convenient to
introduce a local spacetime coframe $\{ {\bf{e}}^0, {\bf{e}}^i ; i=1,2,3,4 \}$ such that
$X\defeq f^2 {\bf{e}}^0$.

Adapting a coordinate $t$ along $X$,  $X=\partial_t$, ${\bf{e}}^0$ can be written as  ${\bf{e}}^0 = f^2 (dt+\alpha)$,
where $\alpha=\alpha_i {\bf{e}}^i$. It is also useful to define
${\bf{e}}^i \defeq f^{-1} {\mathring{\bf{e}}^i}$ in which case the metric  becomes
\begin{equation}
ds^2 = -f^4(dt+\alpha)^2 + f^{-2}  {d\mathring{s}^2}~,
\end{equation}
where ${d\mathring{s}^2} = \delta_{ij} {\mathring{\bf{e}}^i}{\mathring{\bf{e}}^j}$. The metric ${d\mathring{s}^2}$,
as well as $f$, $\alpha$, $\omega_r$ and $F$ are all $t$-independent because as has been mentioned they are invariant under the action of $X$.

Locally the spacetime can be viewed as a fibration over a 4-dimensional base manifold $B$ with fibres the orbits of $X$.
 The volume form ${\rm dvol}_B$ on the 4-dimensional base manifold $B$, equipped with metric
${d\mathring{s}^2}$, is related to the 5-dimensional volume form by
${\rm dvol}_5 = f^{-4} {\bf{e}}^0 \wedge {\rm d \mathring{vol}}_B$.
The conditions ({\ref{cc51}}) and ({\ref{cc53}}) then determine the Maxwell field strength via
\begin{equation}
F= {\sqrt{3} \over 2} d {\bf{e}}^0 -{1 \over \sqrt{3}} f^2 (d \alpha)_{{\rm asd}}~,
\end{equation}
where $(d \alpha)_{{\rm asd}}$ denotes the anti-self dual part of $d \alpha$ on $B$.

The base space $B$ admits a hyper-K\"ahler structure, associated with the three 2-form bilinears $\omega_r$, $r=1,2,3$.  To see this, note that
$\omega_r$, $r=1,2,3$ descend on $B$ as ${\cal L}_X\omega_r=0$ and $i_X\omega_r=0$.  The 2-forms  $\omega_r$ are self-dual on $B$, as a consequence of ({\ref{ac51}}).  Moreover the associated complex structures, $I_r$, $\omega_r(Y,Z)=\mathring g(Y, I_rZ)$,  satisfy the algebra of the imaginary quaternions, $I_1^2=I_2^2=-{\bf 1}$, $I_3=I_1 I_2$,  $I_1I_2=-I_2I_1$ on $B$, as a consequence of ({\ref{ac53}}). The remaining
content of the condition (\ref{cc54}) is
\begin{equation}
{\mathring{\nabla}} \omega_r =0
\end{equation}
where ${\mathring{\nabla}}$ denotes the Levi-Civita connection on $B$.
Hence the $\omega_r$ define a hyper-K\"ahler structure on $B$.

This analysis exhausts the content of the algebraic and differential conditions which we have obtained on the  bilinears. It remains to determine the remaining conditions imposed by the Killing spinor equations, given the conditions on the geometry and flux obtained so far. This can be done by noting that the spinor $\epsilon$ must satisfy the condition
\begin{equation}
\label{zproj}
\Gamma_0 \epsilon = -i \epsilon~.
\end{equation}
This follows from the first identity in ({\ref{proj5}}). In turn, this implies that
$\Lambda_{ij} \Gamma^{ij} \epsilon=0$
for any anti-self-dual 2-form $\Lambda$ on $B$. Then the $M=0$ component of ({\ref{kspinv}}) is equivalent to
\begin{equation}
\label{tcomp5}
\partial_t \epsilon=0~,
\end{equation}
and the remaining components of ({\ref{kspinv}}) are equivalent to
\begin{equation}
\label{bcomp5}
{\mathring{\nabla}} (f^{-1} \epsilon)=0 \ .
\end{equation}
It follows that the spinor $\epsilon$ is given by $\epsilon=f \eta$, where
$\eta$ is a covariantly constant $t$-independent spinor on the hyper-K\"ahler base $B$, $
\Gamma_0 \eta = -i \eta$.
The conditions appearing in ({\ref{proj5}}) which involve $\omega_r$ are not involved
in the evaluation of ({\ref{tcomp5}}). Hence any covariantly constant spinor $\eta$ on $B$ satisfying $
\Gamma_0 \eta = -i \eta$ gives rise to a solution $\epsilon = f \eta$ of the KSE of the $d=5$ theory.

This analysis exhausts the content of the KSE. It remains to impose the Bianchi identity and field equations; the resulting conditions are common to both the bilinears method and the spinorial geometry approach to solving the KSE, and will be presented after discussing the spinorial geometry analysis.

\subsubsection{Geometry and supersymmetry of null solutions} \label{fivenull}

In the null class of solutions, for which $f=0$, and $X$ is null, it is convenient to
introduce a local spacetime coframe $\{{\bf{e}}^-, {\bf{e}}^+, {\bf{e}}^i: i=1,2,3 \}$ such that
$X\defeq  {\bf{e}}^-$.
The algebraic identities ({\ref{ac51}}) then imply that
\begin{equation}
\omega_r = {\bf{e}}^- \wedge \tau_r~,
\end{equation}
where without loss of generality we take $(\tau_r)_-=0$. Furthermore,
({\ref{ac53}}) simplifies to
\begin{equation}
(\omega_r)_{\tC \tA} (\omega_s)^\tC{}_{\tB} = \delta_{rs} X_\tA X_\tB \ .
\end{equation}
Setting $s=r$, $A=-$, $B=+$ then implies that $(\tau_r)_+=0$, and setting
$A=B=-$ further implies that
\begin{equation}
(\tau_r)_\tC (\tau_s)^\tC=\delta_{rs}
\end{equation}
Hence we can choose a basis for which $\tau_r = \delta_{ri} {\bf{e}}^i$.
The condition ({\ref{cc51}}) implies that $ i_X  F=0$, so the Maxwell field strength decomposes as
\begin{equation}
F=F_{-i} \fe^-\wedge \fe^i+{1\over2} F_{ij} \fe^i\wedge \fe^j \ .
\end{equation}
In order to introduce co-ordinates, we shall introduce a
local co-ordinate $u$ such that $X=\partial_u$. Also, the algebraic conditions ({\ref{cc51}}) and ({\ref{cc53}}) imply that
\begin{equation}
X \wedge dX=0~,
\end{equation}
and it follows that a further local co-ordinate $v$ can be found such that
\begin{equation}
{\bf{e}}^-= h^{-1} dv~,
\end{equation}
 for some function $h$. Next, consider the closure condition ({\ref{cc55}}); this implies that
\begin{equation}
dv \wedge d(h^{-1} {\bf{e}}^i)=0~,
\end{equation}
and hence there exist co-ordinates $x^I$, $I=1,2,3$ and functions $q^I$, $I=1,2,3$
such that
\begin{equation}
{\bf{e}}^i = \delta^i_I \big(h dx^\tI + p^\tI dv \big) \ .
\end{equation}
Using a change of basis as in (\ref{4dnullpatching}), which leaves ${\bf{e}}^-$ invariant, one can  set  without loss of generality
$p^\tI=0$, so ${\bf{e}}^i = h \delta^i_I  dx^\tI$.
The condition ${\cal{L}}_{\partial_u} X=0$ implies that $h$ is $u$-independent;
and the condition that $X$ is Killing then implies that ${\cal{L}}_{\partial_u} {\bf{e}}^+=0$. The basis can therefore be written as follows
\begin{equation}
{\bf{e}}^+ = du+ V dv + n_\tI dx^\tI, \qquad {\bf{e}}^- = h^{-1} dv, \qquad
{\bf{e}}^i = h\, \delta^i_I\,  dx^\tI~,
\end{equation}
where $V, h, n_\tI$ are $u$-independent. It remains to determine the components
$F_{-i}$ and $F_{ij}$ of the flux. To do this, we first write
5-dimensional volume form as
${\rm dvol}_5 = -h^3\, {\bf{e}}^+ \wedge {\bf{e}}^- \wedge {\mathring \epsilon}$,
where $\mathring \epsilon$ is the volume form of the flat 3-metric. The components $F_{ij}$ are determined by the condition ({\ref{cc53}}) as
\begin{equation}
{1 \over 2} F_{ij}\, {\bf{e}}^i \wedge {\bf{e}}^j = -{\sqrt3\over4} \mathring\epsilon_{\tI\tJ}{}^\tK \partial_\tK h \,dx^\tI\wedge dx^\tJ \ .
\end{equation}
The remaining components $F_{-i}$ are obtained from the condition ({\ref{cc54}})
on setting $A=M=-$, $N=j$ to find
\begin{equation}
F_{-k}\, {\bf{e}}^- \wedge {\bf{e}}^k = - {1\over 4\sqrt3} \mathring\epsilon_\tI{}^{\tJ\tK} h^{-2} dn_{\tJ\tK} dv\wedge dx^\tI~,
\end{equation}
and hence
\begin{equation}
F = - {1\over 4\sqrt3} \mathring\epsilon_\tI{}^{\tJ\tK} h^{-2} dn_{\tJ\tK} dv\wedge dx^\tI
-{\sqrt3\over4} \mathring\epsilon_{\tI\tJ}{}^\tK \partial_\tK h \,dx^\tI\wedge dx^\tJ \ .
\end{equation}
On substituting these conditions back into the Killing spinor equation ({\ref{kspinv}}), we find that the spinor is constant,
$\partial_\tM \epsilon =0$,
and  as a consequence of the first identity in ({\ref{proj5}}) satisfies $\Gamma_+ \epsilon =0$.

This analysis exhausts the content of the KSE. It remains to impose the Bianchi identity and field equations. Again,  the resulting conditions are common to both the bilinears and the spinorial geometry approaches to solving the KSE, and will be
presented after the spinorial geometry analysis.

 \subsection{Solution of the KSE using the spinorial geometry method}

 One way to describe the Dirac representation of $Spin(4,1)$ in terms of forms is to begin from that of $Spin(4)$ and identify $\Gamma^0=i\Gamma^{1234}$.
 Therefore the Dirac spinors of $Spin(4,1)$ are identified with $\Lambda^*(\bC^2)$, where the action of the gamma matrices of $Spin(4)$ on $\Lambda^*(\bC^2)$ is described in appendix \ref{app:spinors}.  Furthermore $Spin(4)=SU(2)\times SU(2)$ and acts on $\Lambda^{\mathrm{ev}}(\bC^2)$ and $\Lambda^{\mathrm{odd}}(\bC^2)$ with the $({\bf{2}}, {\bf 1})$ and
 $({\bf{1}}, {\bf 2})$ representations, respectively. As the  orbits of $SU(2)$ on $\bC^2-\{0\}$ are 3-spheres, a representative of the first Killing spinor can be chosen up to a $Spin(4)$
 transformation as $\eps=f_1 1+ f_2 e_1$, where $f_1$ and $f_2$ are real constants.
To further simplify this spinor, consider the $SO(1,1)$ transformation generated
by $\Gamma_{03}$. There are three possibilities to consider, according as
$|f_1|>|f_2|$, $|f_1|<|f_2|$ and $|f_1|=|f_2|$. If $|f_1|>|f_2|$, then this transformation can be used to
set $f_2=0$, whereas if $|f_2|>|f_1|$, then this transformation can be used to set $f_1=0$.
These two cases, for which either $f_1=0$ or $f_2=0$, are further related  as  $\Gamma_{03}e_1=-1$, where $\Gamma_{03}$  lies in a disconnected component of the spin group.
In the remaining case, $|f_1|=|f_2|$, the $SO(1,1)$ gauge transformations generated by
$\Gamma_{03}$ and $\Gamma_{13}$ can be used to set  $f_1=f_2=1$.  Therefore, the first  Killing spinor  can be chosen without loss of generality as
 \bea
 \text{either}~~~\eps= f \,1~,~~~\text{or}~~~\eps=1+e_1~,
 \eea
  where $f$ is promoted to a real spacetime function.  The isotropy group of $f 1$ and $1+e_1$  in $Spin(4,1)$ is $SU(2)$ and $\bR^3$, respectively.

\subsection{Case 1: $\eps= f 1$}

\subsubsection{Solution of the linear system}
The KSE (\ref{kspinv}) can be easily evaluated on $f 1$ and after expanding in the basis of Dirac spinors as described in appendix \ref{app:spinors}, one finds for $A=0$, $A=\alpha$ and $A=\bar\alpha$  the following linear system

\bea
\label{time5block1}
&& \partial_0 f +  {1 \over 2} f \Omega_{0,\beta}{}^\beta -{1 \over 2 \sqrt{3}} f \fiveF_\beta{}^\beta
= 0~,~~~
\fiveF_{0 \bar{\beta}} -{\sqrt{3} \over 2} \Omega_{0,0 \bar{\beta}} =0~,
\cr
&&\fiveF_{\alpha \beta} -\sqrt{3} \Omega_{0,\alpha \beta} = 0 \  ,~~~\partial_\alpha f + {1 \over 2} f \Omega_{\alpha, \beta}{}^\beta + {\sqrt{3} \over 2}f \fiveF_{0 \alpha} =0~,
\cr
&&- \Omega_{\alpha, 0 \bar{\beta}} -{1 \over \sqrt{3}} \fiveF_\gamma{}^\gamma \delta_{\alpha \bar{\beta}}
+ \sqrt{3} \fiveF_{\alpha \bar{\beta}} =0~,~~~
 \Omega_{\alpha, \bar{\beta} \bar{\gamma}}
+{2 \over \sqrt{3}} \delta_{\alpha[\bar\beta}\fiveF_{ \bar{\gamma}]0}  = 0 \ ,
\cr
&&\partial_{\bar{\alpha}} f +{1 \over 2} f \Omega_{\bar{\alpha}, \gamma}{}^\gamma +{1 \over 2 \sqrt{3}}f
\fiveF_{0 \bar{\alpha}} =0~,~~~
- \Omega_{\bar{\alpha},0 \bar{\beta}} + {1 \over \sqrt{3}} \fiveF_{\bar{\alpha} \bar{\beta}} = 0~,~~~
\cr
&&
\Omega_{\bar{\alpha}, \bar{\beta} \bar{\gamma}} =0 \ .
\eea
This system can be easily solved  to express the flux $F$ in terms of the geometry as
\bea
F&=&\sqrt3\, d\log f\wedge \fe^0+{\sqrt 3\over 2} \Omega_{\alpha, 0\beta}\, \fe^\alpha\wedge \fe^\beta+{\sqrt 3\over 2} \Omega_{\bar\alpha, 0\bar\beta}\, \fe^{\bar\alpha}\wedge \fe^{\bar\beta}
\cr
&&~~~
+{1\over\sqrt3} (\Omega_{\alpha, 0\bar\beta}+\delta_{\alpha\bar\beta} \Omega_{\gamma,0}{}^\gamma)\, \fe^\alpha\wedge \fe^{\bar\beta}~,
\eea
and to find the conditions
\bea
&&\partial_0f=0~,~~~\Omega_{\alpha,0}{}^\alpha-\Omega_{0,\alpha}{}^\alpha=0~,~~~\Omega_{0,0\alpha}=-2\partial_\alpha\log f~,
\cr
&&\Omega_{\alpha, \beta}{}^\beta=\partial_\alpha\log f~,~~~
\Omega_{\alpha, 0\beta}=\Omega_{0,\alpha\beta}~,~~~\Omega_{\alpha,0\bar\beta}+\Omega_{\bar\beta,0\alpha}=0~,~~~
\cr
&&\Omega_{\alpha, \bar\beta\bar\gamma}=-2 \delta_{\alpha[\bar\beta} \partial_{\bar\gamma]}\log f~,~~~
 \Omega_{\alpha,\beta\gamma}=0~,
\label{5dgeomcon}
\eea
on the spacetime geometry.

\subsubsection{Geometry}

To investigate the geometry of spacetime, let us note that form bilinears are generated by
\bea
&&X=D( f1, \Gamma_\tA f1)\,\fe^\tA= f^2 \fe^0~,~~~
\cr
&&\omega_1={1\over2}D(f1, \Gamma_{\tA\tB} f1)\, \fe^\tA\wedge \fe^\tB=-i f^2 \delta_{\alpha\bar\beta} \fe^\alpha\wedge \fe^{\bar\beta}~,
\cr
&&\omega_2+i\omega_3={1\over2} D(f1, \Gamma_{\tA\tB}\, i\, r_{\mathrm A}f1)\, \fe^\tA\wedge \fe^\tB={1\over2}  f^2 \epsilon_{\alpha\beta}\fe^\alpha\wedge \fe^\beta~,
\label{5dformbil}
\eea
with $\epsilon_{12}=1$.  All the geometric conditions in (\ref{5dgeomcon}) that involve a $\fe^0$ coframe direction can be expressed as
\bea
{\cal L}_X g=0~,~~~{\cal L}_X \omega_r=0~,~~~r=1,2,3~.
\eea
Therefore $X$ is Killing and leaves the other three 2-form bilinears invariant. In addition, the flux $F$ is also invariant under $X$,  ${\cal L}_XF=0$.

The conditions on the geometry imposed by the remaining three conditions in (\ref{5dgeomcon}) are
\bea
d  \omega_r=0~.
\eea
Therefore the spacetime admits three closed 2-forms.

Locally the spacetime can be viewed as a fibration with fibres given by the integral curves of $X$.  As $i_X \omega_r=0$ and ${\cal L}_X \omega_r=0$, these forms descend
to 2-forms on the base space $B$ of the fibration. As $ \omega_r$  are closed and Hermitian with respect to the metric $d\mathring{s}^2=f^2 \delta_{ij} \fe^i \fe^j$ and the associated
complex structures $I_r$, $\omega_r(Y,Z)=\mathring g(Y, I_rZ)$,  satisfy the algebra of imaginary unit quaternions, $I_1^2=I_2^2=-{\bf 1}$, $I_3=I_1 I_2$,  $I_1I_2=-I_2I_1$ on $B$,    $B$ is
a hyper-K\"ahler manifold.  Therefore the spacetime $M$ is a local fibration over a 4-dimensional  hyper-K\"ahler manifold.  Notice that $B$ admits a (weak) hyper-K\"ahler structure with torsion (HKT) \cite{hgphkt}
with respect to $d\tilde s^2=\delta_{ij} \fe^i \fe^j$ and $f^{-2}\omega_r$. As we shall see such structures arise frequently in the investigation of geometries
of supersymmetric backgrounds.

These conditions correspond to those obtained via the bilinears method for solving the KSE. Just as in that analysis, special co-ordinates can be found
in which the solution can be written in a particularly simple form. In particular,
one can adapt a coordinate $t$ such that $X=\partial_t$. As all fields and form bilinears are invariant under $X$, a coframe can be chosen on the spacetime which does not depend explicitly on $t$. The remaining decomposition of the metric and Maxwell field strength is identical to the calculation presented in Section
({\ref{fivetime}}).

\subsubsection{Solutions}

The field equations of the theory are implied as a consequence of the KSE, the Bianchi identity of $F$, $dF=0$ and the vanishing of the electric component of the field equation of $F$, ${}^*LF_0=0$.  This follows from the integrability condition of  the KSE in (\ref{5dintfeqs}) which can be rewritten as
\bea
-{1\over2} E_{\tA\tB} \Gamma^\tB \eps+{i\over2\sqrt 3} {}^*LF_{\tB} \Gamma^{\tB}{}_\tA \eps+{i\over\sqrt 3} {}^*LF_\tA\eps=0~.
\eea
As $\eps=f 1$, taking the Dirac inner product again with $\eps$, one finds that
\bea
-{1\over2} E_{\tA\tB} X^A+{i\over2\sqrt 3} {}^*LF_{\tB} (\omega_1)^{\tB}{}_\tA -{1\over\sqrt 3} f^2\, {}^*LF_\tA=0~.
\eea
where the bilinear $\omega_1$ has been defined in (\ref{5dformbil}). For $A=0$, one finds that $E_{00}=0$.  On the other hand for $A=i$,
one gets that $E_{i0}=0$ and ${}^*LF_i=0$.  Therefore, if ${}^*LF_0=0$, then the field equation of $F$ will be satisfied.  The vanishing of the rest of the components
of the Einstein equation follows from an argument similar to that presented in section \ref{4dintsol} for the minimal  ${\cal N}=2$ $d=4$ supergravity.

 Therefore to find solutions, one has to solve the Bianchi and the electric component of the field equation for $F$, which in turn gives
 \bea
d\left( f^2(d\alpha)_{\mathrm{asd}}\right)=0~,~~~
{\mathring{\nabla}}^2 f^{-2} = {2 \over 9}\, f^4\, \mathring g^{ij} \mathring g^{mn}\left(d \alpha_{\mathrm{asd}}\right)_{im} (d \alpha_{\mathrm{asd}})_{ jn}~,
\label{gaugeF}
 \eea
 respectively, where $\mathring\nabla$ is the Levi-Civita connection of the hyper-K\"ahler metric on $B$.

A large class of solutions can be found  provided that $d \alpha_{\mathrm{asd}}=0$.  In such a case $f^{-2}$ is a harmonic function on a hyper-K\"ahler
manifold $B$.  For $B=\bR^4$ and $f^{-2}=1+\sum_a Q_a/|y-y_a|^2$, the solutions are   rotating multi-black holes.  The rotation is associated with the self-dual
part of $d\alpha$ \cite{bmpv}.

Many solutions also arise in the case for which the base space is a Gibbons-Hawking manifold, which admits a tri-holomorphic isometry \cite{Gibbons:1979zt}. If this tri-holomorphic isometry is a symmetry of the full solution, then the complete solution
is determined by a choice of four harmonic functions on $\mathbb{R}^3$. To illustrate this construction, we take the base space metric to be
\bea
{d\mathring{s}^2} = H^{-1} (dz+\chi)^2 + H \delta_{rs} dx^r dx^s \ , \qquad r,s=1,2,3~,
\eea
where $H$ is a harmonic function on ${\mathbb{R}}^3$ and $\chi=\chi_r dx^r$ is a 1-form on ${\mathbb{R}}^3$ satisfying
\bea
\star_3 d \chi = dH \ .
\eea
The Hodge dual $\star_3$ is taken on ${\mathbb{R}}^3$, and the volume form on the base and the volume form on ${\mathbb{R}}^3$ are related by ${\rm d \mathring{vol}}_B = H {\rm dvol}_3 \wedge dz$.

These conditions imply that the base metric is hyper-K\"ahler with tri-holomorphic isometry
${\partial \over \partial z}$. With this base space convention, the hyper-K\"ahler structure is given by
\bea
\omega_r = \delta_{rp}(dz + \chi) \wedge dx^p -{1 \over 2} H \epsilon_{rpq} dx^p \wedge dx^q, \qquad r, p, q=1,2,3 \ .
\eea
To construct the solution for which the tri-holomorphic isometry ${\partial \over \partial z}$ is
a symmetry of the full solution, decompose $\alpha$ as
\bea
\alpha = \Psi (dz+\chi)+\sigma~,
\eea
where $\Psi$ is a function on ${\mathbb{R}}^3$ and $\sigma$ is a 1-form on ${\mathbb{R}}^3$. The anti-self-dual part of $d \alpha$ is then
\bea
d\alpha_{\mathrm{asd}}&=&{1 \over 2} (dz + \chi)\wedge \big(-d \Psi + H^{-1} \Psi dH + H^{-1} \star_3 d \sigma \big)
\nonumber \\
&+&{1 \over 2} \big(d \sigma+ \Psi \star_3 dH -H \star_3 d \Psi\big) \ .
\eea
The Bianchi identity from ({\ref{gaugeF}}) implies that
\bea
d \bigg(f^2 \big(d \Psi - H^{-1} \Psi dH - H^{-1} \star_3 d \sigma \big) \bigg) =0~,
\eea
and hence there locally exists a function $\rho$ on ${\mathbb{R}}^3$ such that
\bea
\label{ghaux1}
f^2 \big(d \Psi - H^{-1} \Psi dH - H^{-1} \star_3 d \sigma \big) = d \rho \ .
\eea
The remaining content of the Bianchi identity can then be written as
\bea
\Box_3 (H \rho)=0~,
\eea
where $\Box_3$ denotes the Laplacian on ${\mathbb{R}}^3$. It follows that there exists a harmonic function $K$ on  ${\mathbb{R}}^3$ such that
\bea
\rho=3KH^{-1} \ .
\eea
The gauge field equation given in $({\ref{gaugeF}})$ can then be rewritten as
\bea
\Box_3 f^{-2}= \Box_3 \big(K^2 H^{-1} \big)~,
\eea
so there exists a further harmonic function $L$ on ${\mathbb{R}}^3$ such that
\bea
f^{-2} = K^2 H^{-1}+L \ .
\eea
Having determined $f$ in terms of these harmonic functions, we determine $\Psi$ by
making use of ({\ref{ghaux1}}), which can be rewritten as
\bea
\label{ghaux2}
H d \Psi - \Psi d H - \star_3 d \sigma = 3(K^2+LH) d(KH^{-1}) \ .
\eea
Taking the divergence of this condition gives
\bea
\Box_3 \Psi = \Box_3 \big(H^{-2} K^3 +{3 \over 2} H^{-1} KL \big)~,
\eea
which implies that there exists a harmonic function $M$ on ${\mathbb{R}}^3$ such that
\bea
\Psi = H^{-2} K^3 +{3 \over 2} H^{-1} KL+M \ .
\eea
The 1-form $\sigma$ is then fixed by substituting this expression into ({\ref{ghaux2}})
to give
\bea
\star_3 d \sigma= H dM - M dH +{3 \over 2}(K dL -L dK) \ .
\eea
This procedure therefore determines the complete solution entirely in
terms of the harmonic functions $\{H,K,L,M\}$; although there is some freedom to redefine these harmonic functions. In particular, the solutions generated by $\{H,K,L,M\}$
and $\{H,K',L',M'\}$ are identical provided that
\bea
K&=&K'+\mu H~,\quad
L=L'-2\mu K' - \mu^2 H~,
\cr
M&=&M'+{1 \over 2}\mu^3 H -{3 \over 2} \mu L' +{3 \over 2} \mu^2 K'~,
\eea
for constant $\mu$. Also, the harmonic function $M$ is only defined up to an additive constant $\nu$ with
\bea
\label{hredef2}
M={\hat{M}}+\nu, \qquad \sigma={\hat{\sigma}}-\nu \chi~,
\eea
and the harmonic functions $H, K, L$ are unchanged.
Furthermore, it is also possible for the same solution to be described by two different Gibbons-Hawking base spaces. For example, the maximally supersymmetric $AdS_2 \times S^3$ solution can be constructed from both a flat base space,
as well as a singular Eguchi-Hanson base.

An example which describes
a large family of solutions preserving $N=4$ supersymmetry is given by taking $H={1 \over r}$, so that the
base space is ${\mathbb{R}}^4$ together with
\bea
K&=&-{1 \over 2} \sum_{i=1}^P q_i h_i~, \quad
L=1+{1 \over 4} \sum_{i=1}^P (Q_i-q_i^2)h_i~,
\cr
M&=& {3 \over 4} \sum_{i=1}^P q_i \bigg(1-|{\bf{y}}_i| h_i \bigg)~,
\eea
where
$h_i= {1 \over |{\bf{x}}-{\bf{y}}_i|}$
and  $Q_i, q_i, {\bf{y}}_i$ are constant.  In the case of a single pole, $P=1$, there are two possibilities. If ${\bf{y}}_1 = {\bf{0}}$ then the solution will
describe a single rotating BMPV black hole, which  is static provided that $3Q_1=q_1^2$.
The generic multi-BMPV black hole solution does not however lie within this family
of solutions, because although the base space is ${\mathbb{R}}^4$, the tri-holomorphic
isometry is not a symmetry of the full solution.
On the other hand if ${\bf{y}}_1 \neq {\bf{0}}$,  then the solution is the supersymmetric black ring.
Further generalization can be made by taking multiple poles. Such solutions include
configurations of concentric black rings as well as Black Saturn type of solutions found
in \cite{Gauntlett:2004wh}. In addition, all of the maximally supersymmetric solutions can be written as solutions
in the timelike class with a Gibbons-Hawking base space for which the tri-holomorphic isometry is a symmetry of the solution.

\subsection{Case 2: $\eps=1+e_1$}

\subsubsection{Solution of the linear system}

To find the linear system  that arises after evaluating the KSE (\ref{kspinv}) on the Killing spinor $1+e_1$ observe that $(-\Gamma_0+\Gamma_3)(1+e_1)=0$ in the conventions of appendix \ref{app:spinors} for $Spin(4)$ with $\Gamma^0=i\Gamma^{1234}$. Because of this it is convenient to change basis  to  $(\Gamma^+, \Gamma^-, \Gamma^1, \Gamma^2, \Gamma^{\bar 2})$   with  $\sqrt 2\,\Gamma^\pm=\pm\Gamma_0+\Gamma_3$, $\Gamma^1=e_1\wedge +i_{e_1 }$, $\Gamma^2=\sqrt2\, i_{e_2}$ and $\Gamma^{\bar 2}=\sqrt2\, e_2\wedge$.
The calculation is similar to the ones we have already presented  and thus we shall not elaborate.  The solution of the linear system can be written as
\bea
F={1\over 2 \sqrt 3} \epsilon_i{}^{jk} \Omega_{-,jk}\fe^-\wedge \fe^i+{\sqrt3\over 2} \epsilon_{ij}{}^k \Omega_{-,+k} \fe^i\wedge \fe^j~,
\eea
where $\epsilon_{12\bar 2}=-i$.

The conditions on the geometry are
\bea
&&\Omega_{\tA,+\tB}+\Omega_{\tB,+\tA}=0~,~~~\Omega_{+,ij}=0~,~~~\Omega_{i,+j}=0~,~~~\Omega_{2,12}=\Omega_{1,2\bar2}=0~,
\cr
&&2 \Omega_{-,+2}+\Omega_{1,12}=0~,~~~2\Omega_{2,+-}+\Omega_{2,2\bar2}=0~,~~~2\Omega_{1,+-}+\Omega_{2,1\bar2}=0~.
\label{5dnullgeom}
\eea
This is a full content of the KSE.

\subsubsection{Geometry}

A basis of the form  bilinears constructed from the Killing spinors $\eps$ and $ r_{\mathrm{A}}\eps$ is
\bea
X=\fe^-~,~~~\omega_r=\delta_{ri}\,\fe^-\wedge \fe^i~,
\eea
where $\{\fe^-, \fe^+, \fe^i : i=1,2,3\}$ is the  spinorial geometry coframe.
The conditions on the geometry (\ref{5dnullgeom}) can be expressed as
\bea
{\cal L}_X g=0~,~~~X\wedge dX=0~,~~~d\omega_r=0~.
\eea
Therefore, $X$ is a Killing vector field which in addition leaves $F$ invariant, ${\cal L}_XF=0$.

Again, these conditions correspond to those obtained via the bilinears method for solving the KSE, where also special co-ordinates can be
adopted. These co-ordinates $u, v$ and $x^\tI$, as well as the
decomposition of the metric and Maxwell field strength, are obtained in exactly the same
was as described in Section ({\ref{fivenull}}).

\subsubsection{Solutions}

An investigation of the integrability conditions (\ref{5dintfeqs}) reveals that all field equations are satisfied provided that the $E_{--}$ and ${}^*LF_-$ components of the field equations vanish together with $dF=0$.  The latter implies that
\bea
\delta^{\tI\tJ}\partial_\tI\partial_\tJ h=0~,~~~\partial_v\partial_\tI h=-{1\over3} \delta^{\tJ\tK} \partial_\tJ( dn_{\tK\tI} h^{-2})~.
\eea
The Einstein equation $E_{--}=0$  gives
\bea
&&h^{-3} \delta^{\tI\tJ}  \partial_\tI (-\partial_\tJ V h+  \partial_v n_\tJ)- 3h \partial_v^2 h-3 (\partial_v h)^2+{3\over2} \delta^{\tI\tJ} (\partial_\tI V \partial_\tJ h
\cr
&&~~~-\partial_v n_\tI h^{-2}
\partial_\tJ h)+{1\over6} \delta^{\tI\tJ} \delta^{\tK\tL} dn_{\tI\tK}dn_{\tJ\tL}=0~,
\eea
and ${}^*LF_-=0$ is  satisfied with no further conditions.

These equations can be solved and solutions include a magnetic multi-string solution for $V=n=0$ and $h=1+\sum_a Q_a/|x-x_a|$ and a multi pp-wave solution for
$h=1$, $n=0$ and $V=\sum_a Q_a/|x-x_a|$.

\subsection{Maximally supersymmetric backgrounds}

The  supercovariant curvature (\ref{5dsupercov}) of maximally supersymmetric backgrounds vanishes.  It is a consequence of the homogeneity theorem \cite{fofjh1,fofjh2}, which will be   demonstrated in section \ref{homtheorem}, that the maximally supersymmetric solutions must be homogeneous spaces.   Upon using  the Bianchi identity $dF=0$, one can establish from the term  linear  in  gamma matrices in the supercovariant connection  that
\bea
\check\nabla_\tA F_{\tB\tC}=0~,~~~H^\tD{}_{[\tA\tB} F_{\tC]\tD}=0~,
\eea
where $\check\nabla=\nabla-(1/\sqrt3)H$.
The latter relation also follows from  $H={}^*F$.   The quadratic term  in gamma matrices in the supercovariant connection gives that
\bea
\check R_{\tA\tB, \tC\tD}={4\over3} F_{\tA\tB}  F_{\tC\tD}~.
\eea
Therefore, the Riemann curvature $R$ of the spacetime, as well as the flux $F$, are parallel with respect to the connection $\check\nabla$. These data are compatible with
a Lorentzian homogeneous structure on the spacetime with canonical connection $\check \nabla$, which has  torsion $(-2/\sqrt3)H$ and  curvature $(-2/\sqrt3)F$, see appendix \ref{homospaces}.  Therefore, we shall take that the spacetime locally admits a Lorentzian transitive 6-dimensional group with Lie algebra $\mathfrak{g}=\mathfrak{h}\oplus \mathfrak{m}$ which has   self-dual
structure constants.   The commutation relations are
\bea
[m_\tA, m_\tB]= {2\over\sqrt3}( F_{\tA \tB} h + H_{\tA\tB}{}^\tC m_\tC )~,~~~[h, m_\tA]= {2\over\sqrt3} F_\tA{}^\tB m_\tB~,
\label{5comrel}
\eea
where $h$ is the generator of an abelian subalgebra $\mathfrak{h}$ and $\{m_\tA\}$ is a basis in  $\mathfrak{m}$.
Note that the indices have been raised with respect to the spacetime metric.

There are three Lorentzian 6-dimensional Lie algebras with self-dual structure constants which are isomorphic to
\bea
\mathfrak{sl}(2,\bR)\oplus \mathfrak{su}(2)~,~~~\mathfrak{cw}_6~,~~~\bR^{5,1}~,
\eea
where $\mathfrak{cw}_6$ is the Lie algebra of the $CW_6$ group manifold described in appendix \ref{cwmanifolds}.
As the structure constants of $\bR^{5,1}$ are zero, all the  maximally supersymmetric solutions associated to it  are locally isometric to the Minkowski space $\bR^{4,1}$ with $F=0$.

Consider the commutation relations
\bea
[t_2, t_\pm]=\pm2 t_\pm~,~~~[t_+,t_-]=t_2~~~\text{and}~~~[w_a, w_b]=2\epsilon_{ab}{}^c w_c~,
\eea
of  $\mathfrak{sl}(2,\bR)$ and $\mathfrak{su}(2)$ Lie algebras, respectively,  where
$a,b,c=3,4,5$.  The most general choice of the generator $h$ which gives rise to a reductive Lorentzian 5-dimensional homogeneous space, see appendix \ref{homospaces},  is
$h=\alpha t_2+\beta w_5$, where $\alpha, \beta\in \bR$.  Then $\mathfrak{m}$ is spanned by $\{ t_\pm, p, u_3, u_4\}$, where $p= \alpha t_2-\beta u_3$.
Let $\ell=\ell^+ t_++\ell^- t_-+ \ell^2 p+\ell^3 u_3+\ell^4 u_4$ be the left-invariant coframe on the homogeneous space.  One can verify that for $\alpha,\beta\not=0$ this homogeneous space admits a three parameter family of
Lorentzian invariant metrics and a two-parameter family of invariant 2-forms.  However, imposing the relation between the structure constants as indicated in (\ref{5comrel}), and that $H$ is dual to $F$, one finds that the fields are
\bea
ds^2&=& {1\over \alpha^2} \ell^+\ell^-+{1\over 2 \beta^2} \left((\ell^3)^2+ (\ell^4)^2\right)+ {4\alpha^6\over\beta^6} (\ell^2)^2~,
\cr
F&=&{\sqrt3\over4 \alpha} \ell^+\wedge \ell^-+{\sqrt3\over4 \beta}\ell^3\wedge \ell^4~.
\eea
The homogeneous space is locally isometric to $(SL(2,\bR)\times SU(2))/\bR_{\alpha,\beta}$ where the inclusion of $\bR$ in $SL(2,\bR)\times SU(2)$ is $(\text{diag}( e^{\alpha t}, e^{-\alpha t}), \text{diag} (e^{i\beta t}, e^{-i\beta t}))$. If either $\alpha=0$ or $\beta=0$, then the spacetime is locally isometric to $AdS_3\times S^2$ or  $AdS_2\times S^3$ with magnetic or electric
flux, respectively.  These are the near horizon geometries of the magnetic string and extreme Reissner-Nordstr\"om black hole, respectively. While for $\alpha \beta\not=0$, the maximally supersymmetric background is the near horizon geometry of the BMPV black hole.

The non-vanishing commutation relations of  $\mathfrak{cw}_6$ are given in (\ref{cwncomrel}).  As $\beta$ is skew-symmetric and non-degenerate it can be brought, up to an $O(4)$ transformation, to a block-diagonal form. Therefore,
 it can be determined by up to two real constants $\lambda_1$ and $\lambda_2$. Requiring that the structure constants are self-dual, one finds that $\lambda\defeq \lambda_1=\lambda_2$ and thus $\beta= \lambda dx^1\wedge dx^2+\lambda dx^3\wedge dx^4$. Two choices of a subalgebra  of a Lie algebra  related by a conjugation are considered as equivalent.  Therefore the generator $h=h^i t_i+h^- t_-+h^+ t_+$ of the abelian subalgebra $\mathfrak{h}$ in $\mathfrak{cw}_6$ can be chosen   up to a conjugation.  For this observe that the adjoint action with $w=w^it_i+w^- t_-+w^+t_+$ gives
   \bea
   &&h^i\rightarrow h^i-w^- \beta^i{}_j h^j+ h^- \beta^i{}_j w^j~,
 ~~h^-\rightarrow h^-~,~~h^+\rightarrow h^+-\beta_{ij} w^i h^j~.
   \eea
Therefore all $h=h^+ t_+$ elements represent independent conjugacy classes  in $\mathfrak{cw}_6$.  On the other hand, if $h=h^i t_i+h^+ t_+$, then the adjoint action acts on $h$ with translations on $h^+$ and rotations on $h^i$ generated by $\beta$.  Thus the independent conjugacy
 classes can be represented by $h=h^i t_i$ up to  identifications of $h^i$ with rotations generated by $\beta$. Finally, if $h^-\not=0$,
 then the adjoint action of $\mathfrak{cw}_6$ acts with translations on both $h^i$ and $h^+$.  The former  can be used to set $h^i=0$.  In such case, the  $h^+$ remains invariant as the translations acting on it are $h^i$ dependent. The conjugacy classes can be represented
 by $h=h^- t_-+ h^+ t_+$. Thus for what follows it suffices to choose the generator of $\mathfrak{h}$ as either  $h=h^i t_i$ or  $h=h^- t_-+ h^+ t_+$.

First consider $h=h^i t_i$.  Then  without loss of generality one can choose $h= t_4$.  This is because
the $\mathfrak{cw}_6$ algebra has an underlying $SU(2)\times SO(2)$ symmetry generated by the anti-self-dual rotations acting on the
generators $t_i$ which leave $\beta$ invariant. The $SO(2)$ symmetry is generated by $\beta$. These $SU(2)$ transformations can be used to
set $h$ to the form above. If $\mathfrak{h}$ is spanned by $t_4$, then a basis in $\mathfrak{m}$ is  $\{ t_-, t_+,  t_1, t_2, t_3\}$. A local section $s$ of the coset can be chosen by setting $x^3=x^4$, see appendix \ref{homospaces}.  A left-invariant coframe on the coset space, see (\ref{cwcoord}),
is
\bea
&&\bbl^-=dv~,~~~\bbl^a=dx^a+\beta^a{}_b x^b dv~,~~~\bbl^3=dx^3+\lambda x^3 dv~,~~~
\cr
&&\bbl^+=du+{1\over2} \beta_{ab} x^a dx^b-{1\over2} \lambda^2 \delta_{ab} x^a x^b dv-\lambda^2 (x^3)^2~,
\eea
and the canonical connection is $\Psi= dx^3-\beta x^3 dv$, where $a,b=1,2$.  There are several invariant metrics that can be
put on the coset space.  However requiring that $H$ is skew-symmetric, one find that the most general one up to an overall scale is
\bea
ds^2=2 \bbl^- \bbl^++\delta_{ab} \bbl^a \bbl^b+ (\bbl^3)^2~.
\eea
After a change of coordinates $y^a=(e^{{1\over2}v\beta})^a{}_b x^b$, $y^3=x^3$ and $u'=u+{1\over2} \lambda (x^3)^2$, one finds that
\bea
ds^2&=&2 dv (du-{1\over2} \lambda^2 (y^3)^2 dv-{1\over8} \lambda^2 \delta_{ab} y^a y^b dv)+ \delta_{ab} dy^a dy^b+(dy^3)^3~,
\cr
F&=&-{\sqrt3\over 2} \lambda dv\wedge dy^3~,
\eea
where we have reset $u=u'$ and $F$ can be read off after comparing the commutation relations of the homogeneous space $\mathfrak{h}\oplus \mathfrak{m}$
with those in (\ref{5comrel}).  All the field equations and KSEs are satisfied.  This is the maximally supersymmetric plane wave solution
of $d=5$ minimal supergravity.

Next consider the conjugacy classes represented by $h=h^- t_-+h^+ t_+$.  If $\mathfrak{h}$ is spanned by  $h=h^- t_-+h^+ t_+$, then a basis
in $\mathfrak{m}$ is $\{w, t_i\}$, where $w=h^- t_--h^+ t_+$.  The non-vanishing commutation relations  (\ref{cwncomrel})  in  the  basis adapted to the homogeneous space are
\bea
&&[h, t_i]=h^- \beta^j{}_i t_j~,~~~[w, t_i]=\alpha \beta^j{}_i t_j~,
\cr
&&[t_i, t_j]=-{1\over 2 h^+} \beta_{ij} h+{1\over 2 h^+} w~.
\label{godelalg}
\eea
Consider the local left-invariant coframe $\bbl=\bbl^0  w+ \bbl^i t_i$, where  a local section $s$ of the coset space has been chosen by setting $v=0$. The most general left-invariant metric on the homogeneous space is
\bea
ds^2=\alpha (\bbl^0)^2+ \gamma_{ij} \bbl^i \bbl^j~,
\eea
where  $\alpha\in \bR-\{0\}$ is a constant and $\gamma$ is a constant non-degenerate symmetric matrix.  Comparing (\ref{godelalg}) with (\ref{5comrel}) and requiring that $H$ is skew-symmetric, one finds that $\gamma_{ij}=\delta_{ij}$
and $\alpha=-2 h^-h^+$.  A similar comparison of the terms containing  $F$ gives $h^- h^+={1\over2}$.  Therefore the solution is
\bea
ds^2=-(du+\beta_{ij} x^i dx^j)^2+ \delta_{ij} dx^i dx^j~,~~~F=-{\sqrt3\over 2\sqrt2} \beta_{ij} dx^i\wedge dx^j~,
\eea
where we used the coordinates obtained in (\ref{cwcoord}) and restricted the $CW_6$ left-invariant coframe to $v=0$.  Furthermore, we have set $h^+=1$.  This is the maximally supersymmetric G\"odel universe solution of $d=5$ supergravity.

\subsection{Solutions of other  d=5 supergravity KSEs and applications}
Following the work of \cite{gghpr} on the minimal ungauged ${\cal N}=1$ $d=5$  supergravity, numerous extensions of
the programme has been made to other $d=5$ supergravities.
First of these has been the classification of the supersymmetric solutions
of minimal gauged supergravity \cite{Gauntlett:2003fk} using the bilinears method. This has been used in \cite{Gutowski:2004ez}  to find the first example of
a supersymmetric asymptotically $AdS_5$ black hole with a regular event
horizon. $AdS_5$ is the unique maximally supersymmetric solution of this theory.
In the near-horizon limit, the black hole preserves 4 supersymmetries, however
in the bulk it preserves only 2 supersymmetries.
Further examples of supersymmetric black holes in minimal gauged supergravity have
also been constructed \cite{Chong:2005hr,Blazquez-Salcedo:2017ghg}.
 A key difference between the gauged and ungauged
theories in five dimensions is that the number of  supersymmetries preserved
in the gauged theory can be 2, 4, 6, or 8 while, as we have seen, in the ungauged theory they are either 4 or 8. All solutions with 6 supersymmetries are locally isometric to $AdS_5$ \cite{Grover:2006ps}, however it is known that there exist discrete quotients
of $AdS_5$ which admit exactly 6 globally well-defined Killing spinors
\cite{fofgppreons4}.

 In terms of the description of the geometry,
the $N=2$ solutions in the minimal gauged theory once again split into a
timelike and a null class. As a consequence of the gravitino KSE,  the spacetime in the timelike class, which is of particular
relevance for black holes, is a local  fibration over a 4-dimensional K\"ahler
base space.
The weakening of the geometric condition on the base space, when compared
to the hyper-K\"ahler condition which arises in the ungauged theory, is
consistent with the reduction in the amount of supersymmetry preserved.
For the ungauged theory, solutions such as the BMPV black hole, and also the supersymmetric black ring
found in \cite{Elvang:2004rt}, have base space $\mathbb{R}^4$.
However, more general black hole geometries, such as those found in
\cite{Horowitz:2017fyg}, as well as the black lens solutions of
\cite{Kunduri:2014kja} and \cite{Tomizawa:2016kjh},
are known to exhibit a Gibbons-Hawking base space.
The Gibbons-Hawking metric
is particularly useful as it allows for a complete integration of all
of the supersymmetry conditions and field equations in terms of
harmonic functions on $\mathbb{R}^3$, provided that the
tri-holomorphic isometry extends to a symmetry of the full theory.
There are no known examples of black hole solutions in a closed
form for which the base space does not possess a tri-holomorphic isometry,
though solutions of this type have been considered numerically \cite{Ghezelbash:2008jt}.
In contrast, the K\"ahler base
space for black holes in the gauged theory cannot be chosen to be hyper-K\"ahler.
This is because the Ricci scalar of the base space cannot vanish as it must diverge
at the location of the Killing horizon. Although the base space is
singular, the 5-dimensional geometry remains regular at the horizon.

In addition, the solution of the KSEs  of
 the ungauged ${\cal N}=1$ theory coupled to vector multiplets   \cite{Gauntlett:2004qy}
has been used to find a large class of supersymmetric ``black Saturn" solutions.  These  consist
of a black hole with $S^3$ spherical horizon topology surrounded by arbitrarily many
concentric black rings. The analogous result for the gauged ${\cal N}=1$ theory coupled to vector multiplets, \cite{Gutowski:2004yv} and \cite{Gutowski:2005id}, has been utilized  to
generalize further the black hole solution of \cite{Gutowski:2004ez}.
Spinorial geometry techniques were also employed to
classify all $N=4$ solutions in gauged supergravity coupled to vector multiplets in \cite{Gutowski:2007ai} and \cite{Grover:2008ih}.

Black hole and black ring \cite{Emparan:2001wn, Elvang:2004rt} type solutions in the theories
coupled to vector multiplets also lie within the
timelike class of solutions. The presence of the vector multiplets does
not affect the restriction on  geometry on the 4-dimensional base space of the spacetime as derived for the minimal theories.
In the ungauged theory and for  a Gibbons-Hawking  base space,
the full solution can again be constructed explicitly.  All explicitly
known examples of black holes, rings and black Saturns have a Gibbons-Hawking base space.
Large classes of
candidate black hole microstate geometries, which are smooth and horizonless,
have also been constructed using these techniques \cite{Bena:2004de,Berglund:2005vb,Bena:2005va,Bena:2006kb,Bena:2007qc}.
Closely related methods have also been used to construct examples of black rings with
varying charge density \cite{Bena:2004td}, though it has been shown that such
solutions typically do not possess smooth horizons \cite{Horowitz:2004je}.
One notable property of the  Gibbons-Hawking manifolds that has been utilized  to describe microstate geometries is
that they are ambipolar, i.e.~the signature flips from
$+4$ to $-4$ across certain surfaces, though the five-dimensional
solution retains a standard signature, and is
regular. The geometry of such ambipolar base spaces has been considered in \cite{Tyukov:2018ypq}.

More recent work has been done to find the solutions of the KSEs in generic
$d=5$ supergravities.
In \cite{Bellorin:2006yr}, all such  solutions
of the ungauged theory coupled to vector multiplets and hypermultiplets
have been presented. In the timelike case, the presence of the hypermultiplets implies that the 4-dimensional base space used to describe such solutions is no longer
hyper-K\"ahler, but rather admits a  weaker set of conditions. Further extension of this work has been made in \cite{Bellorin:2008we}, where
all solutions of the KSEs of the (non-abelian) gauged
theory coupled to arbitrary many vector, tensor and hypermultiplets have been described, and new examples of solutions have been found.
Further work on the solution of the KSEs  of  ${\cal{N}}=4$ supergravity  has been done in \cite{Liu:2006is}.

Higher derivative supergravity solutions have also been investigated in some detail.
In \cite{Castro:2008ne,Castro:2007hc,Castro:2007ci}, an off-shell formalism together with the bilinears method have been employed to consider solutions of the theory comprising of the Weyl multiplet, coupled to arbitrary many (ungauged) vector multiplets, and one hypermultiplet.
Spinorial geometry techniques have also been
used to classify  the solutions \cite{Bonetti:2018lfb}. This theory lies within a large class of
higher-derivative supergravity theories constructed in
\cite{Nishino:2000cz, Fujita:2001kv, Hanaki:2006pj}.
In this case, the structure of the gravitino KSE is identical to that  given in ({\ref{kspinvb}}), but with $F$ replaced by an auxiliary 2-form field which lies within the Weyl multiplet.
As a consequence, some of the geometric conditions are common to those of the minimal 2-derivative theory; for example in the timelike case the
base space is again hyper-K\"ahler. However, the remaining conditions
on the geometry are modified in a highly nontrivial fashion due to
the curvature couplings. Other types of higher-derivative 5-dimensional supergravities are known to exist, such as
\cite{Bergshoeff:2011xn} and \cite{Sloane:2014xya}, and it would be of interest to further extend the classification programme to these theories.

\section{Minimal  ${\cal N}=(1,0)$ $d=6$ supergravity}

Theories in 6 dimensions  have played a significant role in the investigation of string dualities and brane dynamics, see e.g.~\cite{cmhpkt, duffd6dual, wittendualities}.  For example, it has been argued that IIA string theory
 on $\bR^{5,1}\times K_3$ is dual to the heterotic string theory on $\bR^{5,1}\times T^4$.   There is also evidence for a $d=6$ self-dual   string theory  \cite{wittend6}.  Moreover, it is  expected from the AdS/CFT correspondence that a  $d=6$ superconformal theory  models the dynamics of coincident   multiple M5-branes.

 Here we shall focus on the solution of the KSE of minimal   ${\cal N}=(1,0)$ $d=6$ supergravity theory \cite{marcus}, i.e.~8 real supercharges.  This solution has been worked out by the authors of \cite{jgdmhr} using the bilinears method.  The maximal supersymmetric solutions
   have also been found in \cite{acfofws}.  The solution of the KSEs of the  ${\cal  N}=(1,0)$ gauged  supergravity  theory  coupled to any number of tensor, vector and scalar multiplets \cite{nishino, ferrarad6, riccioni}
   have been given by the authors   of  \cite{magpd6} using the spinorial geometry method.

\subsection{Fields and solution of the KSE}

\subsubsection{Fields and KSE}

The bosonic fields of minimal  ${\cal N}=(1,0)$ $d=6$ supergravity  are a metric and an anti-self-dual 3-form field strength $H$ which is closed, $dH=0$.  The KSE of the theory is the vanishing condition of the gravitino supersymmetry variation evaluated at the locus where
the gravitino vanishes.  This reads as
\bea
\hat\nabla_\tA\epsilon=0~,
\la{6kse}
\eea
where $\hat \nabla$ is a connection with skew-symmetric torsion $H$
\bea
\hat\nabla_\tA Y^\tB\defeq \nabla_\tA Y^\tB+{1\over2} H^\tB{}_{\tA\tC} Y^\tC~,
\eea
and $\epsilon$ is a symplectic Majorana-Weyl spinor.  Clearly (\ref{6kse}) is a parallel transport equation with respect to a
connection with holonomy contained in $Spin(5,1)$. A similar gravitino KSE also arises in heterotic supergravity which
will be investigated in section \ref{sec:heterotic}. The symplectic-Majorana condition on $\epsilon$ will be explained later in the description of the
spinors of the theory.

The integrability condition of the gravitino KSE (\ref{6kse}) is
\bea
\hat R_{\tA\tB, \tC\tD} \Gamma^{\tC\tD}\epsilon=0~,
\label{6int}
\eea
where $\hat R$ is the curvature of $\hat\nabla$.
Using that $dH=0$ and that $H$ is anti-self-dual, one  also finds that
\bea
\Gamma^\tB \hat R_{\tA\tB, \tC\tD} \Gamma^{\tC\tD}\epsilon= -2 E_{\tA\tB}\Gamma^\tB\epsilon=0~,
\label{6intb}
\eea
where $E_{\tA\tB}=0$ is the Einstein field equation of the theory.

\subsubsection{Spinors}
\label{laforms}

Given  a pair of Weyl $Spin(5,1)$ spinors $\eps^a$, the symplectic-Majorana reality condition is
$\epsilon^a= \epsilon^{ab} C *\epsilon^T_b$,
where $C$ is the charge conjugation matrix and $\epsilon^{ab}$ is the symplectic invariant
form of $Sp(1)$. It arises for supersymmetric theories in $d=6$ because $Spin(5,1)$ does not admit a real spin representation which is needed
for the supersymmetry transformations to preserve the reality  of the bosonic fields of the theory. Clearly $\epsilon^1$ and $\epsilon^2$
are not linearly independent.

The most effective way to understand   the symplectic-Majorana $Spin(5,1)$ spinors  is to identify them with the $SU(2)$ invariant  Majorana-Weyl spinors of  $Spin(9,1)$ \cite{sghet1, magpd6}.
 To do this explicitly,   the Dirac spinors
of $Spin(5,1)$ and  $Spin(9,1)$ are identified with $\Lambda^*(\bC^3)$ and $\Lambda^*(\bC^5)$, respectively.  Positive chirality spinors are the even degree forms while the  negative chirality
spinors are the odd degree forms. A realization of the gamma matrices of both Clifford algebras
${\rm Cliff }(\bR^{5,1})$ and ${\rm Cliff }(\bR^{9,1})$ is given in appendix \ref{app:spinors}.

Given a Hermitian basis $\{e_1, \dots, e_5\}$ in $\bC^5$, ${\rm Cliff }(\bR^{5,1})$ can be included in ${\rm Cliff }(\bR^{9,1})$ by choosing
the subspace $\bC^3$ in $\bC^5$ as $\bC^3=\bC\langle e_1,e_2, e_5\rangle$.  Therefore the positive chirality Weyl spinors of $Spin(5,1)=SL(2,\bH)$ are given by
$\Lambda^{\rm ev}(\bC\langle e_1, e_2, e_5\rangle)=\bH^2$.
The symplectic Majorana-Weyl condition of $Spin(5,1)$ is  the Majorana-Weyl condition
of $Spin(9,1)$ spinors, i.e.
\bea
\epsilon^*=\Gamma_{6789} \epsilon~,
\eea
where $\epsilon \in \Lambda^{\rm ev} \bC\langle e_1, e_2, e_5\rangle\otimes \Lambda^*\bC\langle e_{34}\rangle$. In particular, a basis for the
symplectic Majorana-Weyl spinors is
\bea
&&1+e_{1234}~,~~~i(1-e_{1234})~,~~~e_{12}- e_{34}~,~~~i(e_{12}+ e_{34})~,~~~
\cr
&&e_{15}+e_{2534}~,~~~i(e_{15}-e_{2534})~,~~~e_{25}-e_{1534}~,~~~i(e_{25}+e_{1534})~.~~~
\label{smw}
\eea
Observe that the above basis selects the diagonal of two copies of the Weyl representation of $Spin(5,1)$, where
the first copy is $\Lambda^{\rm ev}(\bC\langle e_1, e_2, e_5\rangle)$ while the second copy is $\Lambda^{\rm ev}(\bC\langle e_1, e_2, e_5\rangle) \otimes \bC\langle e_{34}\rangle$. The $SU(2)=Sp(1)$ group, whose Lie algebra generators are
\bea
\rho^{1} =  \frac{1}{2}(\Gamma_{38}+\Gamma_{49})~,~~~\rho^{2}= \frac{1}{2}(\Gamma_{89}-\Gamma_{34})~,~~~
\rho^{3}  =  \frac{1}{2}(\Gamma_{39}-\Gamma_{48})~,
\label{spgen}
\eea
 acts on the auxiliary directions $e_3$ and $e_4$ and leaves the basis (\ref{smw}) invariant.

\subsubsection{Solution of the gravitino  KSE}

In the absence of matter multiplets,  supersymmetric solutions of the minimal ${\cal N}=(1,0)$ theory  preserve either $N=4$ or $N=8$ supersymmetries, i.e.
the solutions are either half supersymmetric or maximally supersymmetric.

To see this, observe that the KSE (\ref{6kse}) is covariant under the action of $Spin(5,1)$.  This can be used to choose a representative for the first Killing spinor.
As $Spin(5,1)=SL(2,\bH)$ and the space of spinors is essentially identified with $\bH^2$, one can always choose
the first Killing spinor as $\epsilon=1+e_{1234}$.

Next, it is straightforward to see  that the covariant derivative $\hat\nabla$ in the gravitino KSE (\ref{6kse})
commutes with the generators of $Sp(1)$ in (\ref{spgen}).  So if $\epsilon=1+e_{1234}$ is a Killing spinor, then all of the first four spinors
in (\ref{smw})
\bea
1+e_{1234}~,~~~i(1-e_{1234})~,~~~e_{12}- e_{34}~,~~~i(e_{12}+ e_{34})~,
\label{6killingsp}
\eea
will also be Killing.  Therefore the backgrounds must preserve four supersymmetries.  The isotropy group of all four Killing spinors is
 $Sp(1)\times\bH$ in $Spin(5,1)$.  Therefore $\mathrm{hol}(\hat\nabla)\subseteq Sp(1)\times\bH$.

Furthermore, if a background admits an additional Killing spinor to the first four already chosen, then it will preserve all supersymmetries
and will thus be maximally supersymmetric. This can be easily seen upon inspection of the basis (\ref{smw}) and after  the application of the
argument above with the generators of $Sp(1)$ in (\ref{spgen}).

\subsection{$N=4$ solutions}

\subsubsection{Geometry}\label{d6n1geom}

As the only KSE of minimal ${\cal N}=(1,0)$ supergravity theory is the gravitino KSE, the geometry of the background is characterized by the assertion that
the holonomy of supercovariant connection $\hat\nabla$ is contained in the isotropy group of the Killing spinors in $Spin(5,1)$, $\mathrm{hol}(\hat\nabla)\subseteq Sp(1)\ltimes\bH$.
To investigate the consequences of this on  the geometry of spacetime, one can to compute the form  bi-linears. Given two spinors $\epsilon_1$ and $\epsilon_2$, these are given by
\bea
\tau={1\over k!} B(\epsilon_1, \Gamma_{\tA_1\dots \tA_k} \epsilon_2)\,\, \fe^{\tA_1}\wedge\dots \wedge \fe^{\tA_k}~,
\eea
where $B$ is the Majorana inner product of $Spin(9,1)$, see appendix \ref{app:spinors}, and $\fe^\tA$ is a spacetime coframe. Assuming that $\epsilon_1$ and $\epsilon_2$ satisfy the
gravitino KSE, it is easy to see that
\bea
\hat\nabla_\tA \tau=0~.
\eea
The forms $\tau$ are covariantly constant with respect to $\hat\nabla$.

Applying this to the $N=4$ backgrounds under investigation, the linearly independent bi-linears of the Killing spinors (\ref{6killingsp})
in  the spinorial geometry coframe, see (\ref{Lor-Her})  in appendix \ref{app:spinors},   are
\bea
X\defeq \fe^-~,~~~~\tau_r\defeq \fe^-\wedge \omega_r~,~~~r=1,2,3~,
\label{6funforms}
\eea
where
\bea
&&\omega_1=-i\delta_{\alpha\bar\beta} \fe^\alpha\wedge \fe^{\bar\beta}~,~~~\omega_2=-\fe^1\wedge \fe^2-\fe^{\bar 1}\wedge \fe^{\bar 2}~,~~~
\cr
&&\omega_3=i(\fe^1\wedge \fe^2-\fe^{\bar 1}\wedge \fe^{\bar 2})~,~~~\alpha, \beta=1,2~.
\label{6dforms2}
\eea
Therefore $X$ is a null one-form and $\omega_r$ are the fundamental forms of $Sp(1)$.  Note that in the same coframe the metric is written as  $ds^2=2 \fe^-\fe^++2 \delta_{\alpha\bar\beta} \fe^\alpha \fe^{\bar\beta}$.

It what follows it is convenient to choose the real coframe basis $\{\fe^i\}=\{\fe^\alpha, \fe^{\bar\alpha}\}$ and write the metric and 3-form $H$ as
\bea
ds^2&=& 2 \fe^- \fe^++\delta_{ij} \fe^i \fe^j~,
\cr
H&=&H_{-+i}\, \fe^-\wedge \fe^+\wedge \fe^i+{1\over2} H_{+ij}\, \fe^+\wedge \fe^i\wedge \fe^j
\cr &&~~~+{1\over2} H_{-ij}\, \fe^-\wedge \fe^i\wedge \fe^j
+\tilde H~,
\eea
where $\tilde H={1\over3!} H_{ijk} \fe^i\wedge \fe^j\wedge \fe^k$.  The anti-self duality of $H$ implies that $H_{+ij}$ is anti-self-dual
and $H_{-ij}$ is self-dual as 2-forms in the directions transverse to the lightcone, and $H_{ijk}=\epsilon_{ijk}{}^l H_{-+l}$.

The choice of the local coframe $\{\fe^-, \fe^+, \fe^i: i=1,\dots,4\}$ is not unique. The isotropy  group of the Killing spinors, $Sp(1)\ltimes \bR^8$, acts on the coframe with local transformations as
\bea
\fe^-\rightarrow \fe^-~,~~~\fe^+\rightarrow \fe^+-q_i\,O^i{}_{j}  \,\fe^j-{1\over2} q^2 \, \fe^-~,~~~\fe^i\rightarrow  O^i{}_j\, \fe^j+ q^i\, \fe^-~,
\label{nullpatching}
\eea
where $O$ takes values in $Sp(1)$ and $q$ in $\bH$.   So there is no natural way to choose the transverse directions to the lightcone.
This is a special case of a more general phenomenon that occurs in  all supersymmetric backgrounds that admit Killing spinors which have isotropy group $K\ltimes \bR^m$ in a spin group, where $K$ is a compact group, see also section \ref{sec:heterotic}. The
associated coframe transforms as in (\ref{nullpatching}) with $O\in K$ and $q\in\bR^m$.

 Nevertheless one can proceed as follows.
 The existence of a parallel null 1-form $\fe^-$ defines a trivial sub-bundle $I$ in the cotangent bundle $T^*M$ of spacetime.
Consider the orthogonal sub-bundle  of $I$, $I^\perp$,  in $T^*M$.  As the fibres of $I$ are along the null direction $\fe^-$, the fibres of $I^\perp$ are spanned by $\{\fe^-, \fe^i: i=1, \dots, 4\}$ so $I\subset I^\perp$.  The ``transverse bundle'' (or screening space) to the lightcone, ${\cal T}$,
is defined as ${\cal T}=I^\perp/I$.  This description
of ${\cal T}$ generalizes to all backgrounds with Killing spinors that have isotropy group $K\ltimes \bR^m$.

 The 2-forms $\omega_r$ in (\ref{6dforms2}) are sections of $\Lambda^2({\cal T})$, where we have identified ${\cal T}$ and its dual with the transverse metric $\tilde g=\delta_{ij} \fe^i\fe^j$.
 Furthermore one can define (almost) complex structures $I_r$ on ${\cal T}$, $\omega_r(\tilde Y,\tilde Z)=\tilde g(\tilde Y, I_r\tilde Z)$, where $\tilde Y=\tilde Y^i \fe_i$ and similarly for $\tilde Z$. One can verify that
   $I_1^2=I_2^2=-{\bf 1}$, $I_1I_2=-I_2I_1$ and  $I_3=I_1 I_2$.

The conditions that the gravitino KSE imposes on the spacetime geometry can be written as
\bea
\hat\nabla_\tA X=0~,~~~\hat\nabla_\tA \tau_r=0~.
\la{6par1}
\eea
The first condition in (\ref{6par1}) implies that
\bea
i_XH=dX~,~~~{\cal L}_X g=0~.
\label{6cona}
\eea
  Therefore the  $i_XH$ component of $H$ is determined in terms
of the geometry and $X$ is a Killing vector field on the spacetime $M$.  As $H$ is closed, the first condition in (\ref{6cona}) also implies that $H$ is invariant
under the action of the vector field $X$, ${\cal L}_X H=0$.

It remains to solve the second condition in (\ref{6par1}).  This can be decomposed as
\bea
\nabla_+(\omega_r)_{ij}&=&  H_+{}^k{}_{[i}(\omega_r)_{j]k}=0~,
\cr
\nabla_-(\omega_r)_{ij}&=& H_-{}^k{}_{[i}(\omega_r)_{j]k}~,
\cr
\hat\nabla_i(\omega_r)_{jk}&=&\nabla_i(\omega_r)_{jk}+H^m{}_{i[j} (\omega_r)_{k]m}= 0~,~~~~
\la{6lalala}
\eea
where we have used that in the coframe $\{\fe^-, \fe^+, \fe^i: i=1,\dots, 4\}$ the condition $\mathrm{hol}(\hat\nabla)\subseteq Sp(1)\times\bH$ implies that $\hat\Omega_{\tA,+\tB}=0$.
The right-hand-side of the first condition in (\ref{6lalala}) vanishes because $H_{+ij}$ is anti-self-dual while $\omega_r$ are self-dual forms. Thus $\nabla_X\omega_r=0$
and as a result
\bea
{\cal L}_X(\fe^-\wedge \omega_r)=0~.
\eea
On the other hand
the second condition in (\ref{6lalala}), together with the fact that $H_{-ij}$ is self-dual, implies that $H_{-ij}$ is entirely determined in terms of the geometry.
The last condition in (\ref{6lalala}) is that which one expects for a  manifold with metric $d\tilde s^2=\delta_{ij} \fe^i \fe^j$,  torsion $\tilde H$ and cotangent bundle
${\cal T}$ to admit a HKT structure  \cite{hkt}.  As ${\cal T}$ has rank four, the HKT condition implies that the complex structures $I_r$ are integrable.  After solving all the conditions that arise from the KSE, the fields can be written as
\bea
ds^2&=&2\fe^- \fe^++\delta_{ij} \fe^i \fe^j~,
\cr
H&=&\fe^+\wedge d\fe^-- \left({1\over16}(\omega_r)_{kl} \nabla_-(\omega_s)^{ kl} \epsilon^{rst}
\right)\,(\omega_t)_{ ij}\,\, \fe^-\wedge \fe^i\wedge \fe^j
\cr
&&~~~~~~~~~~~~~-{1\over 3!} (d\fe^-)_{-\ell}\,\,\epsilon^\ell{}_{ijk} \,\, \fe^i\wedge \fe^j\wedge \fe^k~.
\la{6sumn1}
\eea
It is clear that $H$ is determined in terms of the geometry.

Before we proceed with the investigation of the geometry of these backgrounds in more detail, and describe the explicit solutions, let us comment on the solution of the KSEs of gauged $(1,0)$ supergravity coupled to any number of tensor, vector and scalar multiplets which has been carried out in \cite{magpd6}.
 First, the holonomy of the supercovariant connection of such a theory
is in $Spin(5,1)\cdot Sp(1)$ instead of $Spin(5,1)$, where the additional $Sp(1)$ subgroup is due to the inclusion of scalar multiplets. There are also a priori  backgrounds that
preserve $N=1$, $N=2$, $N=3$, $N=4$ and $N=8$ supersymmetries. Furthermore there are two kinds of bilinears.  Those that are forms on the spacetime, like $\fe^-$,  and those that are forms which are twisted
with an $Sp(1)$ bundle like $\fe^-\wedge\omega_r$. The additional $Sp(1)$ twist changes the geometry of the spacetime. For example the geometry of the $N=1$ backgrounds is now based on quaternionic K\"ahler geometries
with torsion \cite{qkt} instead of HKT geometries. Of course there are many more additional conditions that arise from the KSEs of the matter multiplets. For more details, see the original publication.

\subsubsection{Special coordinates}\label{6spcoor}

To give a local description of the geometry of $Sp(1)\ltimes \bH$ backgrounds, one can proceed  as follows.
First let us adapt a coordinate $u$ along the $\hat\nabla$ parallel vector field $X=\partial_u$.  Then a coframe can be chosen as
\bea
&&\fe^-=W^{-1} (dv+m_\tI dy^\tI)~,~~~\fe^+=du+V dv+ n_\tI dy^\tI~,~~~
\cr
&&\fe^i= \tilde e^i_\tI\, dy^\tI+p^i dv~,
\label{6stframe}
\eea
where $v, y^\tI$ are the rest of the spacetime coordinates and $W, V, \tilde e$ and $p$ can depend on all spacetime coordinates.
After performing a coframe rotation as in (\ref{nullpatching}) with $O={\bf 1}$, one can set $p=0$ after  a possible redefinition of $V, n$ and $\tilde e$.  Note that this is equivalent to choosing a splitting $I^\perp=I\oplus{\cal T}$.

As all the geometric data of the theory, including the metric, $H$ and the fundamental forms in (\ref{6funforms}), are invariant  under $X$, a coframe  can also be
chosen such that it is independent of $u$.  In such a case, the expression for the 3-form flux in (\ref{6sumn1}) can be simplified to
\bea
H=d(\fe^-\wedge \fe^+)-{1\over 3!} (d\fe^-)_{-\ell}\,\,\epsilon^\ell{}_{ijk} \,\, \fe^i\wedge \fe^j\wedge \fe^k~,
\label{6fluxh}
\eea
where $\fe^i=\tilde e^i_\tI dy^\tI$.  As a result $dH=0$ leads to a harmonic-like condition on $W$ which will be explored to find solutions.

\subsubsection{Solutions}

Before we proceed to give some examples of solutions, the KSE implies that some of the field equations of the theory are automatically satisfied.  To see this observe that the integrability condition (\ref{6intb})
upon taking an appropriate Majorana inner product with $\epsilon$ implies that $E_{A-}=0$.  Using this and acting on (\ref{6intb}) with $E_{\tA\tC} \Gamma^\tC$, one finds
that $E_{\tA i} E_{\tA}{}^i=0$ as $\epsilon\not=0$, where there is no summation over $A$.  Thus all field equations are implied provided that one imposes $E_{--}=0$
and the Bianchi identity $dH=0$.

A large class of solutions can be found after assuming in addition that $\partial_v$ leaves all the fields invariant.  Then the spacetime can be viewed
locally as a fibration having fibre $\bR^2$ with coordinates $(u,v)$ over a base space $B^4$ which is a 4-dimensional HKT manifold. The co-tangent space $T^*B^4$ of the HKT manifold
is identified with ${\cal T}$. It follows from (\ref{6stframe}) and (\ref{6fluxh}) that  $\tilde H\sim\star_{4} (W^{-1} dW)$. It is known that    4-dimensional HKT manifolds with a co-exact torsion $\tilde H$ are conformal  to  hyper-K\"ahler.
Collecting these data together, the metric and 3-form field strength can be written as
\bea
ds^2&=& 2 \fe^-\fe^++ W ds^2_{\mathrm{hk}}~,~~~
H=d(\fe^-\wedge \fe^+)+ \star_{\mathrm{hk}} dW~,
\eea
where $\{\fe^-, \fe^+\}$ are given in (\ref{6stframe}) but now all components are independent of both the coordinates $u$ and $v$,  $ds^2_{\mathrm{hk}}$ is a hyper-K\"ahler metric and the Hodge
duality operation has been taken with respect to  $ds^2_{\mathrm{hk}}$ as indicated.

These backgrounds solve the KSEs and the field equations provided that
\bea
&&\nabla^2_{\mathrm{hk}}W=\nabla^2_{\mathrm{hk}} V=0~,~~~ \star_{\mathrm{hk}} dm=-dm~,~~~
\cr
&&\star_{\mathrm{hk}}(dn-W dV\wedge m)=(dn-W dV\wedge m)~,
\label{6allsusycon}
\eea
i.e. $W$ and $V$ are harmonic functions on the hyper-K\"ahler manifold $B^4$.  The remaining two conditions in (\ref{6allsusycon}) are implied
by the requirement that $H_{+ij}$ and $H_{-ij}$ are anti-self-dual and self-dual, respectively.

For $\fe^-\wedge d\fe^-=0$, i.e.~$m=0$, the solutions have  the interpretation of  rotating dyonic strings   with a pp-waves propagating on them.  The space transverse to the strings is the hyper-K\"ahler
manifold $B^4$.  For $B^4=\bR^4$, one obtains  planar dyonic strings with $W=1+(Q_s/|y|^2)$ and $V= Q_w/|y|^2$, where $Q_s$ is the charge of the string and $Q_w$ is the momentum along the wave. For $V=0$, one recovers the dyonic string of \cite{dufflu}. A solution for $n$ can also be found, see e.g.~\cite{rotatinggp}. Many more solutions
can be constructed for different choices of the hyper-K\"ahler metric, such as those used to describe certain black hole
microstate geometries \cite{Ford:2006yb, Bena:2011dd},
which are described by a Gibbons-Hawking base space.
  An extensive investigation of the solutions can be found in \cite{vandorend6, Cariglia:2004kk, ortind6}.

\subsection{Maximally supersymmetric backgrounds}

For the maximally supersymmetric backgrounds $\hat R=0$.  Therefore, these are parallelizable  6-dimensional Lorentzian manifolds with respect to
a connection with skew-symmetric torsion. These are Lorentzian signature group manifolds equipped with the bi-invariant metric, and up to a local isometry they can be identified with
\bea
\bR^{5,1}~,~~~~SL(2,\bR)\times SU(2)~,~~~CW_6~,
\eea
where $H$ is given in terms of their structure constants. The anti-self-duality condition on $H$ implies that the radii of $SL(2, \bR)$ and $SU(2)$ must be equal and that the structure constants $\beta$ of the Cahen-Wallach
group manifold $CW_6$ are self-dual, see appendix \ref{cwmanifolds}.

\section{Geometry of heterotic supergravity backgrounds}\label{sec:heterotic}

The effective theory of the heterotic string \cite{dgjhemrr1-het, dgjhemrr2-het} can be described by a type I theory, i.e.~a theory with 16 supercharges in 10 dimensions, which includes
 higher curvature corrections. These can be  organized  in an expansion in terms of the string length square parameter,  $\al'$,  and coupling constant, $g_s$.
 The $\al'$ corrections can be computed by a sigma model loop calculation. The light bosonic fields of heterotic strings  are
the spacetime metric $g$, the NS-NS 3-form field strength $H$, the dilaton $\Phi$ and the 2-form gauge field $F$ with gauge groups
either $E_8\times E_8$ or $SO(32)/\bZ_2$.
These  theories exhibit several attractive features as they are chiral and
upon a compactification on 6-dimensional  Calabi-Yau manifolds  give rise to  ${\cal N}=1$ theories in
4 dimensions.  As a result they have been  extensively explored in   string phenomenology.

The feature that will be described  here is the solution  the KSEs of heterotic supergravity. This has been done in all cases \cite{sghet1, sghet2, het3} and as a result the  geometry
of all backgrounds that preserve some of the supersymmetry of the underlying theory can be systematically  described.  Progress towards the solution of the KSEs of the related common sector in type II theories has also been made, see e.g.~\cite{ugplgp-typeII}.

\subsection{Fields, KSEs,  integrability conditions and spinors}

\subsubsection{KSEs  and field equations}

 The world-sheet theory
 of the heterotic string is chiral and it exhibits an anomaly which is canceled by an anomaly cancelation mechanism \cite{mbgjhs-het}. This
  modifies the
Bianchi identity of the three-form field strength $H$ as
\bea
dH=- \tfrac14 \al' \big( {\rm tr} \tilde R^2- {\rm tr} F^2\big)+{\cal O}(\al'^2)~,
\la{anb}
\eea
where $\tilde R$ is the curvature of a spacetime connection $\tilde \nabla$. The choice of $\tilde R$ is renormalization scheme dependent
and we shall leave $\tilde R$ arbitrary at the moment.

Up to and including  two-loops in the sigma model perturbation theory \cite{cmhpkt-het},
 the KSEs of the effective theory \cite{strominger-het, cmh-het, ebmderoo-het} can be written as
\bea
 {\cal D}_\tM \eps &\defeq& \hat\nabla_\tM\eps+{\cal O}(\al'^2)=0~,~~~
 \cr
 {\cal A} \eps &\defeq& (\Gamma^\tM\partial_\tM\Phi-{1\over12} H_{\tM\tN\tL}\Gamma^{\tM\tN\tL})\eps+{\cal O}(\al'^2)=0~,~~~
 \cr
  {\cal F} \eps  &\defeq& F_{\tM\tN}\Gamma^{\tM\tN} \eps+{\cal O}(\al'^2)=0~,
\la{kse}
\eea
where
\be
\hat\nabla_\tN Y^\tM\defeq\nabla_\tN Y^\tM+{1\over2} H^\tM{}_{\tN\tR} Y^\tR~.
\ee
The first equation is the gravitino KSE for a metric connection $\hat\nabla$
with torsion given by the 3-form field strength $H$. The second equation is the dilatino KSE and the last is the gaugino KSE. The KSEs retain their one loop form as they do not receive an explicit 2-loop contribution.

Furthermore, the field equations of the theory after including the two-loop sigma model correction are
\bea
E_{\tM\tN}&\defeq&R_{\tM\tN}+{1\over4} H^\tR{}_{\tM\tL} H^\tL{}_{\tN\tR}+2\nabla_\tM\partial_\tN\Phi~~~~~~~~~~~~~~~~~~~~~~~~~~~~~
\cr~~~~&&+{\al'\over4} [\tilde R_{\tM\tP,\tQ\tR} \tilde R_\tN{}^{\tP,\tQ\tR}-F_{\tM\tP ab} F_\tN{}^{\tP ab}]+{\cal O}(\al'^2)=0~,
\cr
LH_{\tP\tR}&\defeq&\nabla_\tM[e^{-2\Phi} H^\tM{}_{\tP\tR}]+{\cal O}(\al'^2)=0~,
\cr
L\Phi&\defeq&\nabla^2\Phi-2\nabla_\tM\Phi \nabla^\tM\Phi-{1\over 12} H_{\tM\tN\tP} H^{\tM\tN\tP}
\cr&&
+{\al'\over12}\left(\tilde R_{\tN_1\tN_2\tN_3\tN_4} \tilde R^{\tN_1\tN_2\tN_3\tN_4}- F_{\tM\tN ab} F^{\tM\tN ab}\right)+{\cal O}(\al'^2)=0~,
\cr
LF_\tM&\defeq&\hat\nabla^\tM[e^{-2\Phi} F_{\tM\tN}]+{\cal O}(\al'^2)=0~.
\la{hetfeqn}
\eea
The Einstein and dilaton field equations receive a 2-loop contribution while the field equations for $H$ and $F$ retain
their one-loop form.

In the investigation of solutions of the theory two    distinct cases have been considered distinguished by whether the anomaly contribution to $dH$ and the two-loop
contribution to the field equations vanish or not. These can vanish provided an appropriate choice is made for $F$ and $\tilde R$, i.e.~$F=\tilde R$.  Of course in such a case $dH=0$.

For the solution of the KSEs of heterotic supergravity that  follows, we shall initially
assume that both the anomaly and the two-loop contributions to the field equations vanish. Later in section \ref{hordercor},  we shall explain how our analysis can be modified
to describe the geometry of supersymmetric solutions with non-vanishing anomaly and two-loop contributions.

\subsubsection{Integrability conditions of KSEs}

It is well-known that the KSEs imply  some of the field equations of  supersymmetric backgrounds. To investigate this, as well as to identify some additional consistency conditions required for the compatibility between the Bianchi identity, the  KSEs and the field equations of the theory, see section \ref{hordercor}, let us consider the integrability conditions of the KSEs (\ref{kse}) of the heterotic supergravity.  These are, see also  \cite{dewit-het},
\bea
{\cal R}_{\tM\tN}\eps\defeq [{\cal D}_\tM, {\cal D}_\tN]\epsilon={1\over4} \hat R_{\tM\tN, \tA\tB} \Gamma^{\tA\tB}\epsilon&=&{\cal O}(\al'^2)~,
\cr
[{\cal D}_\tM, {\cal F}]\epsilon=[\hat\nabla_\tM, F_{\tN\tL} \Gamma^{RS}]\epsilon&=&{\cal O}(\al'^2)~,
\cr
[{\cal D}_\tM, {\cal A}]\epsilon=[\hat\nabla_\tM, \partial_\tN\Phi\Gamma^\tN-{1\over12} H_{\tN\tP\tQ} \Gamma^{\tN\tP\tQ}]\epsilon&=&{\cal O}(\al'^2)~,
\cr
[{\cal F},{\cal A} ]\epsilon=[F_{\tR\tS} \Gamma^{\tR\tS},\partial_\tN\Phi\Gamma^\tN-{1\over12} H_{\tN\tP\tQ} \Gamma^{\tN\tP\tQ}]\epsilon&=&{\cal O}(\al'^2)~.
\eea
 Multiplying the first expression above with $\Gamma^\tN$ and using appropriately the remaining integrability
 conditions,
one finds that
\bea
-2 E^{(0)}_{\tM\tN} \Gamma^\tN \epsilon- e^{2\Phi} LH_{\tM\tN} \Gamma^\tN \epsilon-{1\over6} BH_{\tM\tA\tB\tC} \Gamma^{\tA\tB\tC}\epsilon&=&{\cal O}(\al'^2)~,
\cr
L\Phi^{(0)}\epsilon -{1\over4} e^{2\Phi} LH_{\tM\tN}\Gamma^{\tM\tN}\epsilon-{1\over48} BH_{\tM\tN\tP\tQ}\Gamma^{\tM\tN\tP\tQ}\epsilon&=&{\cal O}(\al'^2)~,
\cr
{1\over3} BF_{\tM\tN\tP}\Gamma^{\tM\tN\tP}\epsilon+ 2 e^{2\Phi} LF_\tN\Gamma^\tN\epsilon&=&{\cal O}(\al'^2)~,
\la{intcon}
\eea
where $BH_{\tM\tN\tP\tQ}\defeq dH_{\tM\tN\tP\tQ}$,  $BF_{\tM\tN\tR}\defeq 3\nabla_{[\tM} F_{\tN\tR]}$ and we have expanded
the Einstein equation as $E=E^{(0)}+\al' E^{(1)}+{\cal O}(\al'^2)$ and similarly for $L\Phi$ the field equation for the dilaton. We have also used that
$LH=LH^{(0)}+{\cal O}(\al'^2)$,  $LF=LF^{(0)}+{\cal O}(\al'^2)$ and $BF=BF^{(0)}+{\cal O}(\al'^2)$.  Of course $BH=\al' BH^{(1)}+{\cal O}(\al'^2)$.
These integrability conditions assume a rather simple form whenever $BH=dH=0$ and $BF=0$ which we shall explore later.

\subsubsection{Spinors}

The general description of spinors in terms of forms is given in appendix \ref{app:spinors}. In particular, the Dirac spinors
of $Spin(9,1)$ are identified with $\Lambda^*(\bC^5)$, and the positive chirality spinors are in $\Lambda^{\mathrm{ev}}(\bC^5)$.
A basis in $\Lambda^{\mathrm{ev}}(\bC^5)$ is
\bea
e_{i_1\dots i_k}~,~~~k=0,2~;~~~e_{i_1\dots i_k5}~,~~~k=1,3~,
\la{comspin91b}
\eea
where $i_1,\dots, i_k=1,\dots,4$. This is a complex basis. To describe the Killing spinors of heterotic supergravity,  one has to impose
a reality condition. This is done using the reality map $r_{\mathrm{B}}=-\Gamma_0 b*=\Gamma_{6789}*$. As a result, a real basis is
\bea
&&e_{i_1\dots i_k}+(-1)^{[k/2]} \star e_{i_1\dots i_k}~,~~~i(e_{i_1\dots i_k}-(-1)^{[k/2]} \star e_{i_1\dots i_k})~,~~~k=0,2,
\cr
&& e_{i5}- \star e_{i5}~,~~~i(e_{i5}+ \star e_{i5})~,
\la{reaspin91b}
\eea
where $i_1<\dots<i_k$ and star is the Hodge operation in $\Lambda^*(\bC^4)$. The Killing spinors of Heterotic supergravity
have real components in the basis (\ref{reaspin91b}).

\subsection{Solution of the Killing spinor equations for $dH=0$}

To begin, let us assume that the anomaly contribution to the Bianchi identity vanishes so $dH=0$.
It is convenient to solve the KSEs in the order gravitino, gaugino and dilatino. One of the issues that arises
is whether all the spinors that solve the gravitino KSE also solve the other two.  We shall mainly focus on the
description of the geometry of those backgrounds for which all solutions of the gravitino KSE also solve the other two.
Then we shall describe some of the properties of the descendants, i.e.~the backgrounds for which only some of the
solutions of the gravitino KSE solve also the gaugino and dilatino KSEs.

\subsubsection{Solving the Gravitino KSE}

The gravitino Killing spinor equation is a parallel transport equation for a metric connection with skew-symmetric torsion, $\hat\nabla$.
The (reduced) holonomy of $\hat\nabla$ for generic backgrounds  is in $Spin(9,1)$, i.e.~in the same group as the gauge symmetry of the theory.
The integrability condition of the gravitino KSE
\bea
\hat R{\tM\tN, \tA\tB} \Gamma^{\tA\tB}\eps=0~,
\label{hetgravint}
\eea
implies that either the Killing spinors have a non-trivial isotropy group in $Spin(9,1)$ or
the spacetime is parallelizable, $\hat R=0$.

In the latter case, the spacetime is a Lorentzian group  manifold. In particular, it decomposes, up to discrete identifications,   to a suitable product of $SL(2,\bR)$, $\bR^{n,1}$, $\bR^n$, $SU(2)$, $SU(3)$ and
$\CW_{n}(A)$ $n\geq 4$ groups. The latter groups have been defined in appendix \ref{cwmanifolds}.  The metric on the semi-simple group manifolds is required to be  bi-invariant  as
the group manifold structure constants are identified with $H$ which is a 3-form.

Next suppose that the parallel spinors $\eps_1, \dots, \eps_\tL$ have a non-trivial isotropy, or stability,  group,
 ${\mathrm Stab}(\eps_1, \dots, \eps_\tL)\subset Spin(9,1)$. The  spinors $\eps_1, \dots, \eps_\tL$ are solutions of the gravitino KSE  provided that the holonomy of $\hat\nabla$, $\mathrm{hol}(\hat\nabla)$, is contained in ${\mathrm Stab}(\eps_1, \dots, \eps_\tL)$,
\bea
\mathrm{hol}(\hat\nabla)\subseteq \mathrm {Stab}(\eps_1, \dots, \eps_\tL)~.
\la{holcon}
\eea
The isotropy groups  as well
 as representatives of the parallel spinors have been tabulated in table \ref{tableholsigma}.
 There are two types of isotropy groups distinguished by whether their topology is compact or not.  As in the previous theories we have described,  the geometric properties of spacetime depend on the topology of the isotropy group.

The requirement (\ref{holcon}) completely describes the solutions of the gravitino KSE. A consequence of (\ref{holcon}) is that
and the k-form bi-linears $\tau$ of the $\hat\nabla$-parallel spinors defined in appendix  \ref{app:spinors} are also parallel, i.e.
\bea
\hat\nabla_\tM\tau_{\tN_1\dots\tN_k}=0~.
\la{parform}
\eea
The forms  $\tau$ are the fundamental forms of the group  ${Stab}(\eps_1, \dots, \eps_\ell)$ viewed as the structure group of the
spacetime. Therefore  (\ref{parform}) can be interpreted as the conditions required for a manifold with structure group ${Stab}(\eps_1, \dots, \eps_\ell)$ to admit a compatible connection with
with skew-symmetric torsion.

\begin{table}
\centering
\begin{tabular}{ccc}
\hline

$L$ & ${\mathrm Stab}(e_1,\dots,e_L)$ & ${\rm parallel}\,\,\epsilon$
 \\
 \hline\hline
$1$ & $Spin(7)\ltimes{ \bR^8}$& $1+e_{1234}$ \\
\hline
$2$ &  $SU(4)\ltimes{\bR^8}$&$ 1$
\\
\hline
$3$ & $Sp(2)\ltimes{\bR^8}$&$1,\,\, i(e_{12}+e_{34})$
\\
\hline
$4$ & $\times^2 SU(2)\ltimes{\bR^8}$&$1,\,\, e_{12}$
\\ \\
\hline
$5$ &{ $SU(2)\ltimes\bR^8$}&$1,\,\, e_{12},\,\, e_{13}+e_{24}$
\\
\hline
$6$ &{ $U(1)\ltimes\bR^8$}&$1,\,\, e_{12},\,\, e_{13}$
\\
\hline
$8$ & { $\bR^8$}&$ 1,\,\, e_{12},\,\, e_{13},\,\, e_{14}$
\\
\hline\hline
$2$ & $G_2$&$1+e_{1234},\,\, e_{15}+e_{2345}$
\\
\hline
$4$ & $SU(3)$&$1,\,\, e_{15}$
\\
\hline
$8$ & $SU(2)$&$1,\,\, e_{12}, \,\,e_{15},\,\, e_{25}$
\\
\hline
$16$ & $\{1\}$&$\Delta^+_{{\bf 16}}$
\\
\hline
\hline
\end{tabular}
\vskip 0.2cm
\caption{\la{tableholsigma}  {\small
In the  columns are listed the number of invariant and therefore $\hat\nabla$-parallel spinors,   their isotropy groups in $Spin(9,1)$ and a basis in the space of
invariant spinors, respectively.
The basis of parallel spinors is  always real. So if a complex spinor is given as a basis spinor  it is understood that one should
always take the real and imaginary parts.} }
\end{table}

\subsubsection{Solving the Gaugino KSE}

The gaugino KSE in (\ref{kse}) can be viewed as the invariance condition of the spinor $\eps$ under  infinitesimal  $Spin(9,1)$ rotations generated by the 2-form field strength $F$.  As such it is similar to the integrability
condition of the gravitino KSE (\ref{hetgravint}).  Therefore either the solutions $\eps$ of the gaugino KSE have a non-trivial
isotropy group in $Spin(9,1)$ or $F=0$. In the former case, the restriction on $F$ is to lie in the
Lie algebra of the isotropy group of spinors.  These are the Lie algebras of the groups already tabulated in table  \ref{tableholsigma}.

Suppose now that we have already solved the gravitino KSE and $\hat\nabla$ has holonomy $\mathrm{hol}(\hat\nabla)=G$, where $G$ is one of the isotropy groups in table \ref{tableholsigma}.  The solutions of the gaugino KSE  span a subspace of the solutions of the gravitino one.
It turns out after some investigation \cite{uggpdr-het} that this subspace can always be identified with the Lie algebra of one of the isotropy groups $K$ in table \ref{tableholsigma} that is included in the Lie algebra of $G$, $\mathfrak{Lie}\, K\subseteq \mathfrak{Lie}\, G$.

To proceed  one can either solve the gravitino
KSE on the spinors with isotropy group  $G$ and then separately solve the gaugino KSE by requiring $F$ to lie in the subalgebras of $\mathfrak{Lie}\, G$ contained in table \ref{tableholsigma},
or we can solve both gravitino and gaugino KSEs for the spinors invariant under $K$, i.e.~the group that leaves invariant the solutions of the  gaugino
KSE. The geometry of spacetime with $\mathrm{hol}(\hat\nabla)=K$ is less constrained from that with $\mathrm{hol}(\hat\nabla)=G$. As
a result the solutions to both gravitino and gaugino KSEs with $\mathrm{hol}(\hat\nabla)=K$ include all those with the conditions $\mathrm{hol}(\hat\nabla)=G$ and $F$ restricted to lie in $\mathfrak{Lie}\,K$.  So without loss of generality, one can consider only
the backgrounds for which all solutions of the gravitino KSE are also solutions of the gaugino KSE.  This is the strategy that we shall
adopt from here on.

\subsubsection{Solving the Dilatino KSE}

To solve the dilatino KSE,  let us assume we have solved  both the gravitino and gaugino ones and  denote  the space of their solutions with ${\cal K}_{\tG}$.  ${\cal K}_{\tG}$ is identified with the span of parallel spinors given in table \ref{tableholsigma}.
The space of solutions ${\cal K}_\tD$ of the dilatino KSE is a subspace in ${\cal K}_{\tG}$, ${\cal K}_{\tD}\subseteq {\cal K}_{\tG}$.
To find all solutions of the KSEs of heterotic supergravity, one has to identify all subspaces of ${\cal K}_{\tG}$ which
after solving the dilatino KSE give rise to a distinct spacetime geometry.

The method that has been introduced in \cite{sghet2} to solve this problem is to consider the subgroups $\Sigma({\cal K}_{\tG})\subset Spin(9,1)$ which act almost effectively
on ${\cal K}_{\tG}$.  Clearly such transformation preserve the solutions of the gravitino and gaugino KSEs  and can be used to choose
${\cal K}_\tD$ in ${\cal K}_{\tG}$. As all KSEs are covariant under $Spin(9,1)$ gauge transformations, acting with elements of $\Sigma({\cal K}_{\tG})$ does not change the geometry of spacetime.

A detailed investigation of the action of $\Sigma({\cal K}_{\tG})$ on ${\cal K}_{\tG}$ reveals the following. If the isotropy
group of the solutions of both the gravitino and gaugino KSEs is non-compact, then up to a  $\Sigma({\cal K}_{\tG})$ transformation
${\cal K}_{\tD}$ can be identified with one of spaces of the invariant spinors in table \ref{tableholsigma} contained in  ${\cal K}_{\tG}$.  Using the same argument as for the solution of the dilatino KSE, we can conclude that
it suffices to consider only the cases for which all $\hat\nabla$-parallel spinors also solve the gaugino and dilatino KSEs.
This is because these cases contain all other supersymmetric solutions of the theory for which only some of $\hat\nabla$-parallel spinors
solve the other two KSEs.

It remains now to consider the case for which the isotropy group of the solutions of the gaugino and dilatino KSEs is compact.  Again
$\Sigma({\cal K}_{\tG})$ can be employed to choose ${\cal K}_\tD$ in ${\cal K}_{\tG}$.  All the possibilities have been described
in \cite{sghet2, uggpdr-het, gp-het}.  A further simplification can be made provided that $dH=0$.  In such a case, the Bianchi identity of $\hat R$,
\bea
\hat R_{\tA[\tB,\tC\tD]}=-{1\over3} \hat\nabla_\tA H_{\tB\tC\tD}~,
\label{biadh}
\eea
 and the field equations imply that
\bea
\hat R_{\tM\tN,\tP\tQ} \Gamma^\tN \Gamma^{\tP\tQ}\eps={1\over36} \hat\nabla_\tM \left(\Gamma^\tN\partial_\tN\Phi-{1\over12} H_{\tN\tP\tQ} \Gamma^{\tN\tP\tQ}\right)\epsilon~.
\label{intdil}
\eea
Thus if $\epsilon$ solves the gravitino KSE but not the dilatino one, the gaugino KSE is covariantly constant with respect to $\hat\nabla$.
This signals that the holonomy of $\hat\nabla$ reduces to a subgroup of the isotropy group of the $\hat\nabla$-parallel spinors.
It can be shown that if there is a holonomy reduction,  the pattern of reduction is
\bea
G_2\supset SU(3) \supset SU(2)\supset \{1\}
\eea
Thus it suffices to investigate the descendants  for the cases in which the holonomy of $\hat\nabla$ is strictly $G_2$, $SU(3)$, $SU(2)$ and $\{1\}$. Using this together with the classification of Lorentzian Lie algebras up to dimension six, one establishes the results described
in table \ref{dh0cases}.  It turns out that the $G_2~ (N=1)$ and $SU(3)~(N=1)$ solutions are included in the $Spin(7)\times \bR^8~(N=1)$
backgrounds, and the $SU(3)~(N=2)$ solutions are included in either the $G_2~(N=2)$  or in the $SU(4)\ltimes \bR^8~(N=2)$ solutions.

\begin{table}[ht]
 \begin{center}
\begin{tabular}{|c|c|}\hline
   ${\rm hol}(\hat\nabla)$ &$N$
 \\ \hline \hline
  $Spin(7)\ltimes\bR^8$& 1 \\
\hline
$SU(4)\ltimes\bR^8$&$\nearrow$, 2
\\ \hline
$Sp(2)\ltimes\bR^8$&$\nearrow$, $\nearrow$, 3
\\ \hline
$\times^2SU(2)\ltimes\bR^8 $&$\nearrow$, $\nearrow$, $\nearrow$, 4
\\ \hline
$SU(2)\ltimes\bR^8$&$\nearrow$, $\nearrow$, $\nearrow$, $\nearrow$, 5
\\ \hline
$U(1)\ltimes\bR^8$&$\nearrow$, $\nearrow$, $\nearrow$, $\nearrow$, $\nearrow$, 6
\\ \hline
$\bR^8$&$\nearrow$, $\nearrow$, $\nearrow$, $\nearrow$, $\nearrow$, $\nearrow$, $-$, 8
\\ \hline \hline
$G_2$& 1, 2
\\ \hline
$SU(3)$&1, 2, $-$ , 4
\\ \hline
$SU(2)$&$-$, 2, $-$, 4, $-$, 6, $-$, 8
\\ \hline
$\{1\}$& 8, 10, 12, 14, 16
\\ \hline
\end{tabular}
\end{center}
\caption{\label{dh0cases}\small
In the  columns are listed the holonomy  groups that arise in the solution of the gravitino KSE
and the number $N$  of supersymmetries, respectively. $\nearrow$ and $-$ denote the entries
 in table 2 of \cite{het3} that
are special cases of backgrounds for which all parallel spinors are Killing and those that do not occur, respectively.  }
\end{table}

\subsection{Geometry of backgrounds with non-compact holonomy}

From the results collected in table \ref{dh0cases} regarding the description of all independent solutions of the KSEs whose parallel spinors
have a non-compact isotropy group, it suffices to describe the geometry of the backgrounds for which
all parallel spinors are Killing, i.e.~the parallel spinors solve all KSEs. So there are seven cases to consider all of which share common
geometric properties.  Because of this, to begin with we shall describe all of them together and only at the end provide formulae
particular to each case.

\subsubsection{Solving the Gravitino KSE} \la{hetnoncomp}

To begin let us denote the isotropy group of the Killing spinors as $K\ltimes \bR^8$, where $K$ is the compact subgroup,
see table \ref{tableholsigma}.   In the spinorial geometry  coframe  $\{\fe^-, \fe^+, \fe^i :i=1,\dots,8\}$, the form bilinears of the Killing spinors can be written as
\bea
X= \fe^-~,~~~\tau= \fe^-\wedge \phi~,
\eea
where $\fe^-$ is a null one-form and  $\phi$ denotes collectively all  the fundamental forms of $K$. As the Killing spinors $\epsilon$
satisfy $\hat\nabla\epsilon=0$, the form bilinears also satisfy
\bea
\hat\nabla X=0~,~~~\hat\nabla\tau=0~.
\label{parbihet}
\eea
As for the $N=4$ backgrounds of minimal (1,0) $d=6$ supergravity, the coframe $\{\fe^-, \fe^+, \fe^i :i=1,\dots,8\}$ transforms under the isotropy group $K\ltimes \bR^8$ as in (\ref{nullpatching}),  where now $O$ takes values in $K$ and $q$ in $\bR^8$. There is no natural definition
of the $\fe^+$ light-cone direction or that of the $\fe^i$ directions ``transverse''  to the lightcone. Nevertheless the
directions transverse to the lightcone can be identified as the orthogonal directions to $\fe^-$,  which are spanned by $\{\fe^-, \fe^i :i=1,\dots,8\}$, up to identifications along $\fe^-$.  So again the bundle transverse to the lightcone is  ${\cal T}=I^\perp/I$, where $I$ is the trivial sub-bundle of $T^*M$ with fibre spanned by $\fe^-$ and
$I^\perp$ is its orthogonal complement in $T^*M$.

Let us now consider the condition (\ref{parbihet}).  First, one finds  that
\bea
\hat\nabla_A X = 0 &\Longleftrightarrow& dX=i_X H~,~~~~{\cal L}_Xg=0~.
\la{un1}
\eea
Thus the vector field $X$ is Killing. As $dH=0$, it leaves $H$ invariant as well. The same applies for the
dilaton $\Phi$ as well.

To continue observe that in the coframe we have chosen, the restriction of $\hat\nabla$ to have holonomy contained in $K\ltimes\bR^8$ implies that $\hat\Omega_{\tA,+\tB}=0$.  Using this, the second condition in (\ref{parbihet}) can be rewritten as
\bea
\nabla_+\phi_{j_1\dots j_k}&=&(-1)^k {k\over2}H_+{}^i{}_{[j_1}\phi_{j_2\dots j_k]i}~,
\cr
\nabla_-\phi_{j_1\dots j_k}&=&(-1)^{k}{k\over2} H_-{}^i{}_{[j_1}\phi_{j_2\dots j_k]i}~,
\cr
\hat\nabla_i\phi_{j_1\dots j_k}&=&0~.~~~~
\la{lalala}
\eea
 To investigate the geometric significance of (\ref{lalala}) observe that  $K\subset Spin(8)$ and as the Lie algebra  of $Spin(8)$ is $\mathfrak{spin}(8)=\Lambda^2(\bR^8)$, the Lie algebra of $K$, $\mathfrak{k}$, is a subspace of $\Lambda^2(\bR^8)$, $\mathfrak{k}\subset \Lambda^2(\bR^8)$.  Therefore we can write $\Lambda^2(\bR^8)=\mathfrak{k}\oplus \mathfrak{k}^\perp$, where $\mathfrak{k}^\perp$
 is the orthogonal complement of $\mathfrak{k}$ in $\Lambda^2(\bR^8)$.  This decomposition of vector spaces leads to a decomposition
 of $\Lambda^2({\cal T})$ as  $\Lambda^2({\cal T})=\Lambda^2_{\mathfrak{k}}\oplus \Lambda^2_{\mathfrak{k}^\perp}$, where $\Lambda^2_{\mathfrak{k}}$ and $\Lambda^2_{\mathfrak{k}^\perp}$  have typical fibre $\mathfrak{k}$ and $\mathfrak{k}^\perp$, respectively.

 Next let us focus on the first equation in (\ref{lalala}).
It is clear that this condition does not depend on $i_XH|_{\Lambda^2_{\mathfrak{k}}}$ and expresses $i_XH|_{\Lambda^2_{\mathfrak{k}^\perp}}$ in terms
of the covariant derivative of $\phi$ along  $X$.
However, $i_XH|_{\Lambda^2_{\mathfrak{k}^\perp}}$ is also expressed in terms of the $\Lambda^2_{\mathfrak{k}^\perp}$ component
of  $d\fe^-$ in (\ref{un1}). Therefore consistency requires that schematically
\bea
(d\fe^-)|_{\Lambda^2_{\mathfrak{k}^\perp}}=(\nabla_X\phi)|_{\Lambda^2_{\mathfrak{k}^\perp}}~,
\la{geomcon1}
\eea
which  is interpreted as a condition on the geometry.

Similarly, one can see from the second condition in (\ref{lalala}) that $(H_-)|_{\Lambda^2_{\mathfrak{k}}}$ is not restricted
by the gravitino  KSE while $(H_-)|_{\Lambda^2_{\mathfrak{k}^\perp}}$
is expressed in terms of the $\nabla_-\phi$.

It remains to investigate the last condition in (\ref{lalala}). This  can be analyzed as though
it is examined on an   $8$-dimensional manifold with tangent space ${\cal T}$ admitting a $K$-structure compatible with a connection with skew-symmetric torsion.
The end result depends on
the $K$ structure at hand and it may or may not give additional conditions on the geometry. In all cases, the component of $H$ along $\Lambda^3({\cal T})$ is entirely
determined in terms of the geometry.

Furthermore, notice that a consequence of (\ref{parbihet}), (\ref{un1}) and (\ref{lalala}) is that the only not trivial component
of ${\cal L}_X\tau$ is
\bea
 {\cal L}_X\tau_{-i_1\dots i_k}=k (i_X H)^j{}_{[i_1}\tau_{i_2\dots i_k]j-}~.~~~
 \label{ltauhet}
\eea
Thus the form bilinears $\tau$ are invariant under the vector field $X$,  ${\cal L}_X\tau=0$, if $i_X H|_{\Lambda^2_{\mathfrak{k}^\perp}}=0$. This turns out to be useful
in the investigation of geometry of some backgrounds.

\subsubsection{Solving the Gaugino KSE}

The gaugino Killing spinor equation implies that
\bea
F\in \Lambda^2_{{\mathfrak{k}\oplus\bR^8}}\otimes \mathfrak{h}~,
\eea
where $\mathfrak{h}$ is the Lie algebra of the gauge group.
This means that there is a 1-form $P$ and a 2-form $Q$ along the transverse directions with values in the Lie algebra of the gauge group such that
\bea
F=\fe^-\wedge P+ Q~,
\eea
where $Q\in \Lambda^2_{{\mathfrak{k}}}\otimes \mathfrak{h}$. The condition on $Q$ is a standard instanton like condition which arises on manifolds
with structure group $K$. These will be described later for the individual cases that occur.

\subsubsection{Solving the Dilatino KSE}

It remains to investigate the conditions that arise on the fields form the dilatino KSE. In all cases, the dilatino KSE implies that
\bea
i_Xd\Phi=0~,~~~dX|_{\Lambda^2_{\mathfrak{k}^\perp}}=0~, ~~~ 2\partial_i\Phi-(\tilde\theta_{\phi})_i-H_{-+i}=0~.
\la{un2}
\eea
So the dilaton is invariant under the action of $X$.
The second condition  is equivalent to requiring that $d\fe^-\in \Lambda^2_{\mathfrak{k}\oplus\bR^8}$ and it is exactly the same
condition as that which arises from the gaugino KSE on the field strength $F$.
Thus (\ref{un1}), (\ref{un2})  and (\ref{ltauhet}) imply that
\bea
{\cal L}_X\tau=0~,
\la{un3}
\eea
and that the geometric condition (\ref{geomcon1}) is automatically satisfied.
The third condition in (\ref{un2}) is a geometric condition which  relates the Lee form  $\tilde\theta_{\phi}$ of $\phi$ to the dilaton $\Phi$
and a component of $H$.
The expression of the Lee form depends on $\phi$ and it will be given later during the investigation
of individual cases.

The dilatino Killing spinor equation implies additional conditions to those given in (\ref{un2}).    However, these depend
on the choice of holonomy group  $K\ltimes \bR^8$.

The conditions (\ref{un1}), (\ref{un2}) and (\ref{un3}) are common for all cases and we shall refer them as universal. It remains
to explain the non-universal conditions on the spacetime geometry  which depend on the case under consideration. These arise from
  both the  $\hat\nabla\tau=0$ condition restricted  along the directions transverse to the light-cone and from  the dilatino KSE.

\subsubsection{Fields}

After solving the KSEs, one can demonstrate  that in all cases the metric and 3-form field strength can be written as
\bea
ds^2&=& 2\fe^- \fe^++ \delta_{ij} \fe^i \fe^j~,
\cr
H&=&d\fe^-\wedge \fe^++ \fe^-\wedge (\tilde h+\tilde k) + \tilde H~,
\label{fhetn}
\eea
where $\tilde h\in \Lambda^2_{\mathfrak{k}}$, $\tilde k\in \Lambda^2_{\mathfrak{k}^\perp}$ and  $\tilde H=H\vert_{\cal T}={1\over 3!} H_{ijk}\fe^i\wedge \fe^j\wedge \fe^k$.
The  $\tilde h$ component of $H$ is not restricted by the KSEs and it remains arbitrary. However the restriction  $\tilde h\in \Lambda^2_{\mathfrak{k}}$ imposes an instanton like condition on $\tilde h$ associated to each $K$. The $\tilde k$ and  $\tilde H$ components of $H$ are always expressed in terms of the geometry of spacetime but
 the precise relation depends on the choice of the holonomy group.  In what follows, we shall not express $\tilde k$ in terms of the geometry as
 it does not appear in local calculations after an appropriate choice of a coframe.

\subsubsection{$Spin(7)\ltimes \bR^8$, $N=1$}

The form bilinears are $\fe^-$ and  $\tau=\fe^-\wedge \phi$, where $\phi$
 is the self-dual fundamental 4-form of $Spin(7)$, see \cite{sghet1} for an expression of $\phi$ in the spinorial geometry coframe.
The instanton condition on the $\tilde h$ component of $H$, the $Q$ component of $F$ and the $d\fe^-\vert_{\Lambda^2_{{\cal T}}}$ component of $d\fe^-$ to be in $\Lambda^2_{\mathfrak{spin}(7)}$
is
\bea
\tilde h_{ij}={1\over2} \phi_{ij}{}^{kl} \tilde h_{kl}~,
\eea
and similarly for the other two fields.

The remaining conditions of the gravitino KSE  give
\bea
\tilde H=-\star_{{}_8} \tilde d\phi+\star_{{}_8} (\tilde\theta_\phi\wedge \phi)~,
\eea
where the Lee form of $\phi$ is
\bea
\tilde \theta_\phi=-{1\over6}\star_{{}_8}(\star_{{}_8} \tilde d\phi\wedge \phi)~,
\label{leespin7}
\eea
the Hodge duality operation is that in  ${\cal T}$,  and $\tilde d$ denotes the exterior derivative like operation defined on 1-forms restricted to ${\cal T}$ as
$\tilde d \tau=(\partial_i\tau_j-\Omega_{i,}{}^k{}_j \tau_k) \fe^i\wedge \fe^j$  and similarly extended to k-forms restricted to ${\cal T}$.

The dilatino KSE does not give any additional conditions to those already presented  in (\ref{un2}).  The Lee form
 that appears in the last equation of (\ref{un2}) is  (\ref{leespin7}).

It is clear that $\tilde H$ can be expressed in terms of the fundamental $Spin(7)$
form $\phi$.  The expression is similar to that for a $d=8$ Euclidean signature manifold with a $Spin(7)$ structure and compatible
connection with skew-symmetric torsion \cite{sispin7-het}. There are no further geometric conditions implied by the gravitino  KSE as any 8-dimensional manifold with a $Spin(7)$ structure admits a unique such connection.

\subsubsection{$SU(4)\ltimes \bR^8$, $N=2$}

The $\hat\nabla$-parallel forms are $\fe^-$,  $\fe^-\wedge \omega_I$ and $\fe^-\wedge \chi$, where
 $\omega_I$ and $\chi$ are the Hermitian 2-form associated with an almost complex structure $I$ on ${\cal T}$ and the (4,0) fundamental forms of $SU(4)$, respectively.  The normalization of the fundamental  forms chosen  is ${1\over4!}\wedge^4\omega_I=2^{-4} \chi\wedge \bar\chi=d\mathrm{vol}_{{\cal T}}$, where $d\mathrm{vol}_{{\cal T}}$ is the volume form of the metric on ${\cal T}$.

The condition that $h$, $Q$ and $d\fe^-\vert_{{\cal T}}$ must be in $\Lambda^2_{\mathfrak{su}(4)}$ reads
\bea
\tilde h_{kl}  I^k{}_i I^l{}_j=\tilde h_{ij}~,~~~\tilde h_{ij} \omega_I^{ij}=0~,
\eea
and similarly for the other two fields.

The  gravitino
KSE together with the second condition in (\ref{solsu4}) below gives
\bea
\tilde H=-i_{\tilde I}\tilde d\omega_I=\star_{{}_8}( \tilde d\omega_I\wedge \omega_I)-{1\over2}\star_{{}_8} (\tilde\theta_{\omega_I}\wedge \omega_I\wedge \omega_I)~.
\la{hsu4}
\eea
The expression for $\tilde H$ is as that
   for complex  $d=8$  manifolds with either a $U(4)$  or an $SU(4)$ structure and compatible connection with skew-symmetric torsion.   Manifolds with $SU(n)$ structures   have  extensively been explored in the literature, see e.g.~\cite{strominger-het, cmh-het, sigp-het, gpat-het, dlgz-het, scss-het, Gauntlett:2002sc, Gauntlett:2003cy}.

The additional geometric conditions implied by the KSEs are
\bea
\tilde \theta_{\omega_I}=\tilde \theta_{{\rm Re}\,\chi}~,~~~\tilde {\cal N}(I)=0~,
\la{solsu4}
\eea
where
\bea
\tilde \theta_{\omega_I}=-\star_{{}_8}(\star_{{}_8} \tilde d\omega_I\wedge \omega_I)~,~~~
\tilde\theta_{{\rm Re}\,\chi}=-{1\over4}\star_{{}_8}(\star_{{}_8} \tilde d{\rm Re}\,\chi\wedge {\rm Re}\,\chi)~,
\la{hlee}
\eea
are the Lee forms of ${\omega_I}$ and ${{\rm Re}\,\chi}$, respectively, and
 $\tilde{\cal N}$ is the Nijenhuis tensor of $I$ restricted along the
transverse directions. The vanishing condition of the Nijenhuis tensor is equivalent to requiring
that $\tilde H$ is a (2,1)- and (1,2)-form with respect to $I$.

The vanishing of the Nijenhuis tensor is implied by the dilatino KSE. A consequence of this is that the complexified tangent bundle
$TM\otimes\bC$ of the spacetime admits a Lorentzian complex structure, i.e.~it is a Robinson manifold \cite{pnat-het}. To see this observe that  the 1-forms
  $(\fe^-, \fe^\alpha)$, where $I(\fe^\alpha)=i\fe^\alpha$, induce an integrable distribution on $TM\otimes\bC$.

  On the other hand, the equality of the two Lee forms
is required for the existence of a connection with skew-symmetric torsion compatible with an $SU(4)$ structure.
This geometric condition arises because  $\tilde H$ is uniquely determined in (\ref{hsu4}) in terms of $I$ and $\omega_I$.  However,
it is also required that $\chi$ must be covariantly constant with respect to $\hat\nabla$ along  ${\cal T}$. This also
expresses some of the components of $\tilde H$ in terms of the metric and $\chi$.  So the compatibility of $\hat\nabla\omega_I\vert_{{\cal T}}=0$
and $\hat\nabla\chi\vert_{{\cal T}}=0$ gives the equality of the two Lee forms in (\ref{hlee}).

\subsubsection{The geometry of solutions with $3\leq N\leq 6$}

The holonomy of $\hat\nabla$ for all these cases is
$Sp(2)\ltimes \bR^8$, $N=3$; $\times^2 SU(2)\ltimes \bR^8$, $N=4$;  $SU(2)\ltimes \bR^8$, $N=5$; and  $U(1)\times \bR^8$,
$N=6$.  The geometry \cite{sghet2} of all these solutions can be summarized as follows. The $\hat\nabla$-parallel forms which are relevant for  the description of the spacetime geometry are $\fe^-$ and
$\fe^-\wedge \omega_r$,
where
$\omega_r=\tilde\omega_r$, $r=1,\dots N-1$, are Hermitian forms on ${\cal T}$. The associated (almost)  complex structures $I_r$ satisfy the relations
of the standard basis of the Clifford algebra $\mathrm{Cliff}(\bR^{N-1})$ equipped with a negative definite
inner product.  For example in the $Sp(2)\ltimes \bR^8$ case, the associated Clifford algebra is  $\mathrm{Cliff}(\bR^2)$.
Therefore there are two (almost) complex structures $I_1$ and $I_2$ that satisfy the relations $I_1^2=I_2^2=-{\bf 1}$ and $I_1I_2+I_2I_1=0$.
The third (almost) complex structure that arises in the description of manifolds with holonomy $Sp(2)$ is given by $I_3=I_1I_2$.

Turning now to the description of the fields and geometry for these solutions, the condition that $\tilde h$, $F$ and $d\fe^-\vert_{\cal T}$
must lie in $\Lambda^2_{\mathfrak{k}}$ can be written as
\bea
\tilde h_{kl} (I_r)^k{}_i (I_r)^l{}_j=\tilde h_{ij}~,~~~\mathrm{no~ summation ~over}~r~,
\eea
and similarly for the other two fields.

Now $\tilde H$ can be given as in (\ref{hsu4}) with respect to any of the (almost) complex structures $I_r$.  Similarly, the last condition
in (\ref{un2}) is valid with respect to    the Lee form $\theta_r$ of each  of the Hermitian forms  $\omega_r$. Therefore one finds the geometric conditions
\bea
i_{ I_r}\tilde d\omega_r&=&i_{ I_s}\tilde d\omega_s~,~~~r\not=s~,~~~\mathrm{no~ summation ~over}~r~\mathrm{or}~s~,
\cr
\tilde \theta_r&=&\tilde \theta_s~,~~~~~~~r\not=s~.
\eea
The expression for the Lee form $\tilde \theta_r$ of $\omega_r$ is the same as that for $\omega_I$ in  (\ref{hlee}).  The only additional geometric condition
that arises is
\bea
\tilde {\cal N}(I_r)=0~,
\eea
which is the integrability condition of $I_r$.    The above  conditions we have described  on the geometry of the spacetime can be derived from those we have stated for the   $SU(4)\ltimes\bR^8$ backgrounds
 but now require that they are valid for
each of the $I_r$  complex structures.

\subsubsection{$\bR^8$, $N=8$} \la{r8}

It remains to give the conditions imposed by the KSEs on the fields of backgrounds with holonomy  $\bR^8$ which preserve $N=8$ supersymmetries. These are
\bea
\fe^-\wedge d\fe^-=0~,~~~\tilde H=0~,~~~2\partial_i\Phi-H_{-+i}=0~,~~~Q=0~.
\la{conr8}
\eea
 Observe in particular that $\tilde h=0$ as well. Furthermore $F=\fe^-\wedge P$ and hence it is null.

\subsubsection{Field equations from KSEs}\label{noncompfeqs}

Some of the field equations are implied by the Killing spinor equations. It turns out that
in all the $K\ltimes\bR^8$ cases, the field equations of the theory are implied by the KSEs and the   Bianchi identities, $BH\equiv dH=0$ and $BF=0$,  provided that in addition the following components of the field equations   (\ref{hetfeqn})
\bea
E_{--}=0~,~~~ LH_{-A}=0~,~~~LF_-=0~,
\la{noncompfeqn}
\eea
  are also imposed.

To see this consider the first integrability condition in (\ref{intcon}).  Taking the (Dirac) inner product with $\epsilon$, one finds
\bea
2E_{\tA+}+e^{2\Phi} LH_{A+}=0~.
\label{hetintcon1}
\eea
On the other hand, acting on the first integrability condition with $(2E_{\tA\tC}+e^{2\Phi} LH_{\tA\tC})\Gamma^{\tC}$, one finds
that
\bea
(2E_{\tA\tB}+e^{2\Phi} LH_{\tA\tB}) (2E_{\tA}{}^\tB+e^{2\Phi} LH_{\tA}{}^\tB)=0,~~\mathrm{no~ summation ~over}~A~,
\label{hetintcon2}
\eea
as $\epsilon\not=0$.  Combining (\ref{hetintcon1}) and (\ref{hetintcon2}), one can derive the first two conditions in (\ref{noncompfeqn}).
A similar argument using the remaining integrability conditions in (\ref{intcon}) establishes the last condition in (\ref{noncompfeqn})
as well.  It should be noted that the set of field equations that have to be imposed in addition to the KSEs to find solutions can be further refined compared to those in (\ref{noncompfeqn}) for backgrounds that preserve $N>1$ supersymmetries.

\subsubsection{Special local coordinates}

Local coordinates to describe the geometry of $K\ltimes\bR^8$ backgrounds can be chosen in a way similar to that we have described for the $Sp(1)\ltimes\bH$ solutions
 of ${\cal N}=(1,0)$ $d=6$ supergravity in section \ref{6spcoor}.   After adapting coordinates to the Killing vector field $X$  as $X=\partial_u$ and using a coframe rotation as
 in (\ref{nullpatching}), a coframe can be chosen locally as
\bea
&&\fe^-=W(dv+m_\tI dy^\tI)~,~~~\fe^+=du+ Vdv+n_\tI dy^\tI~,~~~\fe^i=e^i_\tI dy^\tI~.~~
\eea
Furthermore, as the metric, $H$ and the fundamental forms are invariant under the action of $X$,  all component of the coframe can be chosen to be independent of $u$ though they can depend on the $y$ and $v$ coordinates.   In such a case, the 3-form flux $H$ can be rewritten as
\bea
H= d(\fe^-\wedge \fe^+)+ \fe^-\wedge \tilde w+\tilde H~,
\eea
for some $\tilde w\in \Lambda_{\mathfrak{k}}$, generally $\tilde w\not=\tilde h$,  which is not specified by the KSEs.  Note that $\tilde w$ satisfies the same instanton like conditions as $\tilde h$. The above
expression of $H$ is more helpful as it is more straightforward to impose the Bianchi identity.

 As we have seen in the $\bR^8$, $N=8$ case, $\fe^-\wedge d\fe^-=0$. Therefore there is a function $h=h(v,y)$ such that
$\fe^-=h\, dv$. This is useful for the classification of  half supersymmetric solutions \cite{gp2-het} which will be described below.

\subsection{Geometry of backgrounds with compact holonomy}

\subsubsection{Gravitino KSE}
\la{hetcomp}

The $\hat\nabla$-parallel forms on the spacetime $M$  for the compact holonomy groups $K$ in table \ref{tableholsigma}, ${\rm hol}(\hat\nabla)\subseteq K$, are
\bea
\lambda^a=\fe^a~,~~~~\phi={1\over k!} \phi_{i_1\dots i_k} \fe^{i_1}\wedge \cdots \wedge \fe^{i_k}~,
\eea
where $\{\fe^a, \fe^i\}$ is a spinorial geometry coframe, $\lambda^a$ are 1-forms, and $\phi$ represents collectively  the fundamental forms of $K$. In particular, there is always one time-like $\hat\nabla$-parallel
1-form $\lambda^a$, and $2$, $3$ and $5$ space-like 1-forms for $K=G_2, SU(3)$ and $SU(2)$ in table \ref{tableholsigma}, respectively.  Furthermore
$i_a\phi=0$, where $i_a$ denotes inner derivation with respect to (the associated vector field of) $\lambda^a$. The change of notation from $X$ to $\lambda$ will become apparent below.

Moreover, one has  that
\bea
ds^2=\eta_{ab} \lambda^a \lambda^b+ d{\tilde s}^2~,~~~d{\tilde s}^2=\delta_{ij} \fe^i \fe^j~,
\label{compmetrhet}
\eea
where $\eta$ is a constant Lorentzian signature metric as the inner product  of $\lambda^a$ is  constant because they are parallel
with respect to the metric connection $\hat\nabla$.

The condition $\hat\nabla \lambda^a=0$ implies that
\bea
 d\lambda^a=\eta^{ab} i_b H~,~~~~{\cal L}_ag=0~.
\la{unk1}
\eea
Therefore the $i_a H$ components of $H$ are determined in terms of $d\lambda^a$ and $\lambda^a$
are Killing vector fields.

Suppose that $\phi$ is an additional $\hat\nabla$-parallel form, then the condition $\hat\nabla\phi=0$  evaluated along the
coframe $(\lambda^a, e^i)$ gives
\bea
 \nabla_a\phi_{j_1\dots j_k}
&=&
{k\over2}(-1)^k H_a{}^i{}_{[j_1} \phi_{j_2\dots j_k]i}~,~~~
\cr
\hat\nabla_i\phi_{j_1\dots j_k}&=&0~,
\la{parcon}
\eea
where  we have  used that
$i_a\phi=0$.

To continue, first observe that $TM=\Xi\oplus {\cal T}$, where $\Xi$ is spanned by the parallel, and thus nowhere vanishing,
vector fields $\lambda^a$, and ${\cal T}$ is the
orthogonal complement of $\Xi$ with respect to the metric, where we again denote the 1-forms $\lambda^a$ and the associated vector fields with the same symbol. As in the null case, we   refer to ${\cal T}$ as the ``transverse space''.  As the structure group of $M$ has reduced to $K$,  $\Lambda^2({\cal T})$ decomposes as
 $\Lambda^2({\cal T})=\Lambda^2_{\mathfrak{k}}\oplus \Lambda^2_{\mathfrak{k}^\perp}$, $\mathfrak{k}$ is the Lie algebra of $K$. The argument for this decomposition has
 already been presented in the null case.  Next observe that  $i_a H\vert_{\Lambda^2_{\mathfrak{k}}}$ is determined in terms of
 the geometry as a consequence of the first equation in (\ref{parcon}). However the first equation in (\ref{unk1}) also gives  $i_a H\vert_{\Lambda^2_{\mathfrak{k}}}$  in terms of $d\lambda^a\vert_{\Lambda^2_{\mathfrak{k}}}$. As a result,
consistency requires that we have a restriction on the geometry which schematically can be written as
\bea
(d\lambda^a)\vert_{\Lambda^2_{\mathfrak{k}}}=\eta^{ab}(\nabla_b\phi)\vert_{\Lambda^2_{\mathfrak{k}}}~.
\la{geomcon}
\eea

It remains to investigate the last condition in (\ref{parcon}) $(\hat\nabla\phi)\vert_{{\cal T}}=0$. This condition can be solved
as that which arises for manifolds with a $K$-structure compatible  with a connection with skew-symmetric torsion and tangent space ${\cal T}$.  In all cases, $\tilde H=H\vert_{{\cal T}}$ is entirely
determined in terms of the geometry. We shall not give further details here but we describe the final result separately for
 each case.

Using $\hat\nabla \lambda^a=\hat\nabla\phi=0$, one can also compute the  Lie derivative
of $\lambda^a$ and $\phi$   along $\lambda^a$ to find
\bea
[\lambda^a, \lambda^b]&=&-H^{ab}{}_c \lambda_c-H^{abi}\fe_i
\cr
 {\cal L}_a\phi_{i_1i_2\dots i_k}&=&k(-1)^k H_a{}^j{}_{[i_1} \phi_{i_2\dots i_k]j}~,~~~~
 \cr
{\cal L}_a\phi_{bi_1\dots i_{k-1}}&=&(-1)^k H_a{}^j{}_{b} \phi_{i_1\dots i_{k-1}j}~.
\label{lieconhet}
\eea
To analyze these conditions observe that if the span of $\lambda_a$ closes under Lie brackets, i.e.~symbolically  $[\Xi, \Xi]\subseteq \Xi$,
then  $H_{abi}=0$. Also $\phi$ is invariant under the action of the vector fields $\lambda^a$, i.e.~${\cal L}_a\phi=0$,  provided that
 $i_aH\vert_{\Lambda^2_{\mathfrak{k}}}$ vanishes and $[\Xi, \Xi]\subseteq \Xi$.
Moreover observe from (\ref{unk1}) that if $dH=0$, then ${\cal L}_aH=0$.

In what follows, we shall assume that the  algebra of  vector field bilinears $\lambda^a$ closes. One reason for this is
the results of \cite{fofehjgm-super} where it has been demonstrated that the Killing superalgebras of supersymmetric backgrounds close
 on the vector generators
 constructed as Killing spinor  bilinears.  Another reason is that if the commutator
of two such vector fields does not close, it is nevertheless $\hat\nabla$-parallel and so the holonomy of $\hat\nabla$ reduces
further yielding more parallel spinors. So if we insist that the number of parallel spinors is fixed, we are required to take $H_{abi}=0$

Let $\mathfrak{g}$ be the  Lie algebra of  the Killing vector fields $\lambda^a$. The structure constants of $\mathfrak{g}$ are given by $H_{abc}$ and as this is skew-symmetric the metric $\eta$ must be bi-invariant. The Lorentzian Lie algebras $\mathfrak{g}$ up to dimension 6 that are relevant here have been tabulated in table
\ref{tablelie}. There are many ways to utilize the above data to write the spacetime metric and $H$. The most economical way is to assume that the infinitesimal action  generated by the vector fields $\lambda^a$ can be integrated to a free action by a group $G$ with Lie algebra $\mathfrak{g}$. Then
 the spacetime is a principal bundle, $M=P(G,B, \pi)$, equipped with a  principal bundle
connection, $\lambda^a$.
In this case, one finds
\bea
ds^2&=&\eta_{ab} \lambda^a \lambda^b+ \delta_{ij} \fe^i \fe^j~,
\cr
H&=&{1\over3} \eta_{ab} \lambda^a\wedge  d\lambda^b+{2\over3} \eta_{ab} \lambda^a \wedge {\cal F}^b+\tilde H~,~~~\tilde H=H|_{{\cal T}}~,
\la{compgH}
\eea
where
\bea
{\cal F}^a\defeq{1\over2} H^a{}_{ij} \fe^i\wedge \fe^j=d\lambda^a-{1\over2} H^a{}_{bc} \lambda^b\wedge \lambda^c~,
\eea
is the curvature of the principal bundle.
 As ${\cal L}_a \tilde H=0$ and $i_a\tilde H=0$, $\tilde H$ is the pull back  of a 3-form on the base space $B$. Therefore, $H$ is the sum of the Chern-Simons form of the
principal
bundle connection $\lambda$ and the pull-back of a 3-form  on $B$ which we again denote with $\tilde H$. As a consequence,
\bea
dH=\eta_{ab}{\cal F}^a\wedge {\cal F}^b+d\tilde H~.
\label{dhtwist}
\eea
As $dH=0$, the right-hand-side of the equation above must vanish. This condition resembles the anomalous Bianchi identity of $H$ in (\ref{anb}), where $H$ is replaced by $\tilde H$, and the curvature of the spacetime $\tilde R$ and that  of the gauge connection $F$ are replaced by the curvature of the principal fibration ${\cal F}$.

\begin{table}
\begin{center}
\begin{tabular}{|c|c|c|}
\hline
${\mathrm Stab}(\eps_1,\dots,\eps_L)$ & $1-{\rm forms}$&$\mathfrak{Lie}\,G$
 \\
 \hline
$G_2$&$3$&$\bR^{1,2}~,~\mathfrak{sl}(2,\bR)$
\\
\hline
$SU(3)$ & $4$&$\bR^{1,3}~,~\mathfrak{sl}(2,\bR)\oplus \bR~,~ \mathfrak{su}(2)\oplus \bR~,~\mathfrak{cw}_4$
\\
\hline
$SU(2)$ & $6$&$ \bR^{1,5}~,~\mathfrak{sl}(2,\bR)\oplus\mathfrak{su}(2)~,~\mathfrak{cw}_6$
\\
\hline
\end{tabular}
\end{center}
\vskip 0.2cm
\caption{\la{tablelie}
In the first column, the compact isotropy groups of spinors are stated. In the second column, the number of
1-form  bilinear is given. In the third column, the associated Lorentzian Lie algebras are exhibited.
The structure constants of the 6-dimensional Lorentzian Lie algebras of the $SU(2)$ case
are self-dual.}
\end{table}

\subsubsection{Gaugino KSE}

 The gaugino Killing spinor equation implies that
\bea
F\in \Lambda^2_{\mathfrak{k}}\otimes\mathfrak{h}~,
\la{bps}
\eea
where $\mathfrak{h}$ is the Lie algebra of the gauge group.
This is an instanton like condition associated with the holonomy  group $K$.  As this depends on $K$, it will be stated
in each case separately.

\subsubsection{Dilatino KSE}

To simplify the description of the solutions of the dilatino KSE, let us assume that all solutions of the gravitino and gaugino KSEs also solve   dilatino one.  There are descendants for compact holonomy groups and these have been investigated in detail    in \cite{gp-het}.
Under this assumption, the dilatino KSE implies the universal conditions
\bea
i_ad\Phi=0~,~~~~2\partial_i\Phi-(\tilde \theta_\phi)_i=0~,
\la{unk2}
\eea
where $\tilde\theta_\phi$ is the Lee form of $\phi$ which will be given for each case separately.
Therefore the dilaton $\Phi$ is invariant under the action of vector field bilinears and therefore a function of the base space $B$
of the spacetime fibration. In all cases that the Lee form of a $\phi$ satisfies (\ref{unk2}), $\phi$ and also its Lee form $\tilde \theta_\phi$, are pull-backs of a form and a Lee form on $B$, respectively. So the second condition in (\ref{unk2}) implies that $B$ is ``conformally balanced'' with respect to $\phi$. There are additional conditions that are implied
by the dilatino KSE but they depend on the holonomy group $K$.

 \subsubsection{Field equations from KSEs}\label{ccompfeqs}

 All the field equations
of backgrounds with compact holonomy group are implied by the KSEs after imposing the Bianchi identities of the theory.
The proof of this is similar to that described  in section \ref{noncompfeqs} for solutions with non-compact holonomy group.
The main difference is now that there is a time-like $\hat\nabla$-parallel vector field bilinear while in the non-compact cases the bilinear is null. Thus to find solutions, one has to impose  $dH=0$ and $BF=0$, where $dH$ is given in (\ref{dhtwist}).

\subsubsection{$G_2$, $N=2$}

The $\hat\nabla$-parallel forms are $\lambda^a$, $a=0,1,2$ which span $\Xi$, and $\varphi\in \Lambda^3({\cal T})$, where  $\varphi$ is the fundamental $G_2$ form, see \cite{sghet1} for an expression of $\varphi$ in the spinorial geometry coframe.
In addition to the conditions in (\ref{unk2}), the dilatino KSE implies that
\bea
{\cal F}\in \Lambda^2_{\mathfrak{g}_2}\otimes\mathfrak{g}~,~~~\epsilon^{abc} H_{abc}+H_{ijk}\varphi^{ijk}=0~,
\la{g2con}
\eea
where $\mathfrak{g}$ is the Lie algebra of $\lambda^a$ vector fields.
As a consequence of the  first condition above,   $\varphi$  is invariant under the action of $\lambda^a$, see (\ref{lieconhet}). Using the principal bundle language to describe the geometry of $M$ and since $i_a\varphi=0$ as well, $\varphi$  is the pull-back
of a 3-form  on base space $B^7$ which again is denoted with $\varphi$.  The fibre is either $\bR^{1,2}$ or $SL(2,\bR)$ up to discrete identifications.  Moreover as a consequence of the last condition in (\ref{parcon}),  $B^7$ has a $G_2$ structure compatible with a metric connection, $\hat{\tilde \nabla}$, with skew-symmetric torsion.
The metric, torsion and $G_2$ fundamental form  on $B^7$ are given by $d\tilde s^2$ in (\ref{compmetrhet}),  $\tilde H$
and  $\varphi$, respectively.
This in particular this implies that
\bea
&&\tilde H=-{1\over6} (\tilde d\varphi, \star_{{}_7} \varphi)\, \varphi+ \star_{{}_7} \tilde d\varphi-\star_{{}_7} (\tilde\theta_\varphi\wedge\varphi)~,
\cr
&&\tilde d\star_{{}_7}\varphi=\tilde\theta_\varphi\wedge \star_{{}_7}\varphi~,
\la{g2h}
\eea
where
\bea
\tilde\theta_\varphi=-{1\over3} \star_{{}_7}(\star_{{}_7} \tilde d\varphi\wedge \varphi)~,
\eea
is the Lee form of $\varphi$.  The first condition is the expression of the torsion of $B^7$, and so of $\tilde H$, in terms of
the geometry while the second condition is required for a 7-dimensional manifold with a $G_2$-structure to admit a compatible
connection with skew-symmetric torsion \cite{tfsig2-het, tfsig22-het}, see also \cite{Gauntlett:2002sc, Gauntlett:2003cy}. The second universal condition in (\ref{unk2}) is $2d\Phi-\tilde \theta_\varphi=0$
and so $B^7$ is conformally balanced.

The expression for $\tilde H$ in (\ref{g2h}) depends on whether $G$ is abelian or not, see table \ref{tablelie}. As can be seen from (\ref{g2con}), if $G$ is abelian,
then the first term in (\ref{g2h}) for
 $\tilde H$ will vanish.   The requirement that $(\tilde d\varphi, \star_{{}_7} \varphi)=0$  becomes a condition on the geometry. On the other hand if $G=SL(2,\bR)$, then the same term becomes proportional
 to the volume form of $SL(2,\bR)$.

 The curvature of the fibration ${\cal F}$ is a $G_2$ instanton,
\bea
{\cal F}_{ij}={1\over2} \star_{{}_7}\varphi_{ij}{}^{kl} {\cal F}_{kl}~,
\eea
 with gauge group either
$\bR^{1,2}$ or $SL(2,\bR)$. A similar condition is satisfied by  gauge field strength $F$ as a consequence
of the gaugino KSE (\ref{bps}).

\subsubsection{$SU(3)$, $N=4$}
The $\hat\nabla$-parallel forms are the 1-forms $\lambda^a$, $a=0,1,2,3$ which span $\Xi$,  a Hermitian form $\omega_I\in \Lambda^2({\cal T})$  associated
with an almost complex structure $I$ and
(3,0)-form $\chi\in \Lambda^3({\cal T})$. Both $\omega_I$ and $\chi$ are the fundamental forms of $SU(3)$ and satisfy the normalization
conditions ${1\over 3!} \wedge^3\omega_I=-i \, 2^{-3}\, \chi\wedge \bar\chi= d\mathrm{vol}_{{\cal T}}$.

The dilatino KSE implies the additional conditions
\bea
&&{1\over3!} \epsilon^{abcd} H_{bcd}-{1\over2}{\cal F}^a_{ij}\omega_I^{ij}=0~,~~~~{\cal F}^a_{kl} I^k{}_i I^l{}_j={\cal F}^a_{ij}~,
\cr
&&\tilde{\cal N}(I)=0~.
\la{consu3}
\eea
From the second condition,  ${\cal F}^a$ is a (2,0) and (2,0) form  with respect to $I$. In turn this gives that
${\cal L}_a\omega_I=0$.  As $i_a\omega_I=0$, the Hermitian form $\omega_I$ is the pull-back of a Hermitian form
on the base space $B^6$  which
again we denote by $\omega_I$,.  Furthermore the last condition in (\ref{consu3}) implies that $I$ is integrable.  As a result
$B^6$ is a K\"ahler manifold with torsion and so
\bea
\tilde H=-i_{ I}\tilde d \omega_I=\star_{{}_6}\tilde d\omega_I-\star_{{}_6}(\tilde\theta_{\omega_I}\wedge \omega_I)~.
\label{su3torhet}
\eea
The last condition in (\ref{parcon}) that arises from the gravitino KSE on $M$ implies an additional geometric condition  as some components of $\tilde H$ can be expressed
in terms on both $\omega_I$ and $\chi$.  This is
\bea
\tilde\theta_\omega=\tilde\theta_{{\rm Re}\chi}~,
\la{consu3x}
\eea
where
\bea
\tilde\theta_\omega=-\star_{{}_6}(\star_{{}_6} d\omega_I\wedge \omega_I)~,~~~
\tilde\theta_{{\rm Re}\chi}=-{1\over2} \star_{{}_6}(\star_{{}_6} d{\rm Re}\chi\wedge {\rm Re}\chi)~,
\eea
are the Lee forms of $\omega_I$ and $\chi$ on $M$, respectively.

To make further progress on the geometry of $B^6$, suppose that   $G$ is abelian. The first two conditions in (\ref{consu3})
imply  that ${\cal F}\in \Lambda^2_{\mathfrak{su}(3)}\otimes \mathfrak{g}$ and   $\chi$ is invariant under the action of $G$ as a consequence of
 (\ref{lieconhet}).  As $i_a\chi=0$, $\chi$ is the pull-back of a (3,0)-form on $B^6$. The base space, $B^6$, is a manifold with an $SU(3)$ structure compatible with a metric connection, $\hat{\tilde \nabla}$. The  metric, skew-symmetric torsion $\tilde H$, and fundamental forms are given by $d{\tilde s}^2$ in (\ref{compmetrhet}), $\tilde H$ in (\ref{su3torhet}), and $\omega_I$ and $\chi$, respectively.
   The condition (\ref{consu3x})  is that required for $\mathrm{hol}(\hat\tilde \nabla)\subseteq SU(3)$.  The second universal
   condition in (\ref{unk2}) implies that $2d\Phi-\tilde \theta_\omega=0$ and thus $B^6$ is conformally balanced with respect to $\omega_I$.

Next suppose that $G$ is non-abelian and therefore is  either $\bR\times SU(2)$ or $SL(2,\bR)\times U(1)$ up to discrete identifications, see table \ref{tablelie}.
It is clear from the first condition in (\ref{consu3}) that $ {\cal  F}\in \Lambda^2_{\mathfrak{su}(3)\oplus \bR}\otimes\mathfrak{g}$ and so
 $\chi$ is not invariant under the $\bR$ and $U(1)$ group actions, respectively.  As a result, the
canonical bundle of $B^6$ is twisted and therefore $B^6$ does not have   an $SU(3)$ structure but rather a $U(3)$ one.  Hence, $B^6$ is a
manifold with $U(3)$ structure compatible with a connection with skew-symmetric torsion $\hat{\tilde \nabla}$, i.e.~$B^6$ is a K\"ahler manifold with torsion.  The metric, torsion and fundamental form are given by $d\tilde s^2$ in (\ref{compmetrhet}), $\tilde H$ in (\ref{su3torhet}) and $\omega_I$, respectively.
Moreover $B^6$ is conformally balanced with respect to $\omega_I$ as a consequence of the universal conditions in (\ref{unk2}).

The gaugino KSE implies that the curvature of the gauge connection $F$ satisfies $F\in \Lambda^2_{\mathfrak{su}(3)}\otimes \mathfrak{h}$.  So $F$ as a 2-form in $\Lambda^2({\cal T})$
is (1,1) with respect to the complex structure $I$ and $\omega_I$-traceless, $\omega_I^{ij} F_{ij}=0$.  This is the standard instanton
condition on complex manifolds with an $SU(3)$ structure.

\subsubsection{$SU(2)$, $N=8$} \la{su2}

The $\hat\nabla$-parallel forms are the 1-forms  $\lambda^a$, $a=0,\dots,5$, which span $\Xi$ and the Hermitian forms $\omega_r\in \Lambda^2({\cal T})$, $r=1,2,3$,
associated to the (almost) complex structures, $I_r$, on ${\cal T}$ such $I_3=I_1 I_2$ and $I_1I_2=-I_2I_1$.

In addition to the universal conditions (\ref{unk2}), the dilatino KSE implies
\bea
&&H_{a_1a_2a_3}+{1\over3!} \epsilon_{a_1a_2a_3}{}^{b_1b_2b_3} H_{b_1b_2b_3}=0~,
\cr
&&
{\cal N}(I_r)=0~,~~~{\cal F}^a\in \Lambda^2_{\mathfrak{su}(2)}\otimes \mathfrak{g}~.
\la{consu2}
\eea
As $i_a\omega_r=0$, the last condition above implies that $\omega_r$ are the pull-backs of
2-forms on the base space $B^4$  which again we  denote with $\omega_r$. The metric $d\tilde s^2$ is also the pull-back
 of a metric on $B^4$.  This together with $\omega_r$ imply that $B^4$ admits three almost complex structures again denoted by $I_r$.
 These are integrable as a consequence of the second condition  in (\ref{consu2}).  In fact $B^4$ has an $SU(2)$ structure compatible
 with a connection, $\hat\tilde \nabla$, with skew-symmetric torsion and thus $B^4$ is an HKT  manifold. The metric, torsion and fundamental forms are
 $d\tilde s^2$, $\tilde H$ and $\omega_r$, respectively.  The last equation in  (\ref{parcon}) on $\omega_r$  gives
\bea
\tilde H=-i_{I_r}\tilde d\omega_r~,~~~\mathrm{no~summation~over}~r~.
\eea
The second universal condition in (\ref{unk2}) implies that $2d\Phi-\tilde \theta_{\omega_r}=0$, where $\tilde \theta_{\omega_r}$ is
 the Lee form of $\omega_r$. So $B^4$ is
a conformally balanced HKT manifold.  All conformally balanced 4-dimensional HKT manifolds are conformal to  hyper-K\"ahler ones.
So all the above conditions on $B^4$ can be solved to find that
\bea
d\tilde s^2= e^{2\Phi} \, d\tilde s_{\rm hk}^2~,~~~
\tilde H=-\star_{\rm hk} \tilde de^{2\Phi}~,
\eea
where  $d\tilde s_{\rm hk}^2$ is a  hyper-K\"ahler metric on $B^4$.

The first condition in (\ref{consu2}) implies that the structure constants
of the Lie algebra of 1-form bilinears are anti-self-dual. These are given in table \ref{tablelie}.  Therefore
up to discrete identifications the fibre Lie groups are
\bea
\bR^{5,1}~,~~~SL(2,\bR)\times SU(2)~,~~~CW_6~,
\label{su2fibre}
\eea
where the radii of $SL(2,\bR)$ and $SU(2)$ are equal and the structure constants $\beta$ of $CW_6$ obey a self-duality condition, see
appendix \ref{cwmanifolds}.

 The last condition in (\ref{consu2}) implies that the curvature of  the principal bundle connection $\lambda^a$ is an anti-self-dual instanton on $B^4$ with gauge
 group one of those in (\ref{su2fibre}). The gaugino KSE also implies that $F\in \Lambda^2({\cal T})\otimes\mathfrak{h}$ is an anti-self-dual
 instanton as well.  This completes the description of geometry for these backgrounds

\subsection{Including $\alpha'$ corrections}\label{hordercor}

Before we proceed to investigate some of the solutions of the theory, let us make some remarks regarding the geometry
of solutions after taking into account the $\alpha'$ corrections up to and including two loops in sigma model perturbation theory.

To begin suppose that the Bianchi identity of $H$ is modified by the anomaly as in (\ref{anb}),  and the field equations
and KSEs are appropriately modified as (\ref{hetfeqn}) and (\ref{kse}), respectively.  First let us clarify the role of the spacetime curvature $\tilde R$ which enters in the expression for the anomaly and the field equations. It turns out that $\tilde R$
must satisfy the gaugino KSE in order for the anomalous Bianchi identity of $H$ and the KSEs to be compatible with the field equations.
To see this, in the presence of anomalous contributions to the Bianchi identity of $H$ and the two loop corrections to the
field equations, one can show using
\bea
\hat R_{A[B,CD]}=-{1\over3} \hat\nabla_AH_{BCD}-{1\over6}dH_{ABCD}~,
\eea
that (\ref{intdil}) is modified as
\bea
&&\hat R_{AB,CD} \Gamma^B \Gamma^{CD}\eps={1\over36} \hat\nabla_A \left(\Gamma^B\partial_A\Phi-{1\over12} H_{BCD} \Gamma^{BCD}\right)\eps
\cr
&&~~~~~
-{\alpha'\over 4} \left(\tilde R_{AB,EF} \tilde R_{CD,}{}^{EF}-F_{ABab} F_{CD}{}^{ab}\right) \Gamma^B \Gamma^{CD}\eps+{\cal O}(\alpha'^2)~.
\eea
If all the KSEs (\ref{kse})  are satisfied, and thus $\eps$ is a Killing spinor, then all the terms will vanish apart from that containing
$\tilde R$.  This is a restriction of the choice of $\tilde R$.  A solution is to choose the spacetime connection $\tilde R$ to satisfy
the same condition as that of the gaugino KSE on $F$.

In sigma model perturbation theory that will be described below, one can choose to this order $\tilde R=\check R$, where $\check R$ is the curvature
of the connection with torsion $-H$ evaluated at zeroth order in $\alpha'$.  This is because
\bea
\check R_{AB,CD}-\hat R_{CD,AB}={1\over 2} dH_{ABCD}
\eea
and $dH=0$ at zeroth order in $\alpha'$.  Then $\check R$ satisfies the gaugino KSE provided that all spinors that satisfy the gaugino KSE are $\hat\nabla$-parallel so that the holonomy of $\hat\nabla$ is contained in their isotropy group.

There are two points of view on how one should proceed from here.  In the sigma model perturbation approach, one begins at zeroth order
in $\alpha'$ with $H$ closed, $dH^{(0)}=0$.  Then this   gets corrected order by order in perturbation theory as has been indicated. From this point of view, the anomaly
correction is viewed as a first order correction and so on.
In such a case, the geometry of the supersymmetric heterotic backgrounds
at zeroth order is that we have already described in the previous sections.  This of course will be corrected order by order in perturbation theory but as all such corrections are not known, it is not clear what the final outcome will be.  However we do know that
it is corrected to all orders in $\alpha'$ \cite{sen-het, pshgp-het}.

An alternative point of view is to consider the anomalous Bianchi identity, the KSEs and field equations as exact at the order indicated in (\ref{kse}) and (\ref{hetfeqn}), respectively.
The KSEs of the theory for $dH\not=0$ have been solved  in \cite{sghet2} and the geometry of the backgrounds has been identified. However the description of the descendants is rather more involved.
Examples of K\"ahler   and hyper-K\"ahler   geometries with torsion  for which $dH\not=0$  have been first considered in \cite{aogp-het}.
 An existence theorem for solutions of the differential system given by the anomalous Bianchi identity (\ref{anb}) and the KSEs (\ref{kse}) on Hermitian  manifolds with  $\mathrm{hol}(\hat\nabla)\subseteq SU$ and $\tilde R$ the Chern connection  has been demonstrated  \cite{jlsty-het}. Solutions of this system on manifolds with other geometric structures have also been given in \cite{mfsiluvd-het}

The addition of higher curvature corrections up to and including the two loop order in sigma model perturbation theory does not alter the relation between the KSEs (\ref{kse}) and field equations (\ref{hetfeqn}) we have established for backgrounds with  $dH=0$ in sections \ref{noncompfeqs} and \ref{ccompfeqs}.  In particular if $\tilde R$ is chosen to satisfy the gaugino KSE, then for compact holonomy groups the KSEs imply all the field equations
provided that the (anomalous) Bianchi identities are satisfied.  For non-compact holonomy groups, the field equations given in
(\ref{noncompfeqn}) must also be satisfied.  This follows from an investigation of the integrability conditions of the
KSEs in (\ref{intcon}) and after taking into account the anomalous Bianchi identity of $H$ and the
two loop correction to the field equations.

\subsection{A brief overview of solutions}

All supersymmetric solutions of the heterotic theory that have been found can be organized
according to the classification of the solutions to the KSEs we have presented. The purpose here
is not to describe all solutions but rather to explain in which class  some of the most well known solutions  belong.

\subsubsection{WZW models}

Apart from the Minkowski vacuum, group manifold solutions to heterotic theory belong to descendants of backgrounds for which $\mathrm{hol}(\hat\nabla)=\{1\}$.
As we have not describe the geometry of descendants, we shall give a brief description of the descendants of $K=\{1\}$.  The backgrounds
are group manifolds which are parallelizable with a connection with skew-symmetric torsion.  Therefore they admit 16 parallel spinors.  If all parallel spinors also solve the gaugino
KSE, then $F=0$.

It remains to solve the dilatino KSE. For this there are two cases
to consider depending on whether or not the 1-form $d\Phi$ is null. Suppose that $d\Phi$ is not null
and $|d\Phi|^2\not=0$. In this case, one can show that the dilatino Killing spinor equation \cite{tksy-het}, \cite{foftksy-het} implies that
\bea
  \Pi = {1\over2} + \frac{\partial_\tM \Phi H_{\tN\tP\tQ} \Gamma^{\tM\tN\tP\tQ}}{24 |d\Phi|^2
 }   \,,
 \eea
is a projector, $\Pi^2=\Pi$. Since ${\rm tr}\, \Pi=8$, backgrounds with $dH=\hat R=0$ and $|d\Phi|\not=0$
preserve half of the supersymmetry. Moreover one can also show that
$d\Phi$ is $\nabla$-parallel, spacelike and $i_{d\Phi} H=0$, see e.g.~\cite{tksy-het, sghet1}.  Thus the spacetime
up to discrete identifications is a product $M=N\times \bR$, where $\bR$ is spanned by
$d\Phi$. Using this and the classification of Lorentzian Lie groups up to dimension 9, one finds  that up to local isometries the spacetime is one of the following groups
\bea
&&SL(2,\bR)\times SU(2)\times SU(2)\times \bR~,~~~SL(2,\bR)\times SU(2)\times\bR^4~,~~~
\cr
&&
\bR^{1,1}\times SU(3)~,~~~\bR^{3,1}\times SU(2)\times SU(2)~,~~~
\cr
&&\bR^{6,1}\times SU(2)~,~~~CW_4\times SU(2)\times \bR^3~,~~~CW_6\times SU(2)\times \bR~,
\eea
where  $CW_n$ are the Cahen-Wallach group manifolds described in appendix  \ref{cwmanifolds}.  Moreover $H$ is determined from the structure constants of the group manifolds and
all these are linear dilaton backgrounds.

On the other hand if $(d\Phi)^2=0$, i.e.~either $d\Phi$  is null or $d\Phi=0$, then $H$ is null. The condition
$i_{d\Phi} H=0$, implies that these backgrounds preserve at least eight  supersymmetries. Such solutions are locally
isometric to $CW_n\times \bR^{10-n}$ for $n=2,4,6$.
The dilaton is linear if $d\Phi\not=0$ otherwise it is constant, and $H$ is determined by the structure constants
of $CW$ spaces.
The group manifolds  $CW_{10}$ are also solutions  and for generic  structure constants  $\beta\in \Lambda^2(\bR^8)$ preserve
8 supersymmetries. Moreover for some special choice of $\bet$, these manifolds exhibit supersymmetry enhancement to 10, 12 and 14 supersymmetries.

\subsubsection{$N=8$ solutions with $SU(2)$ and $\bR^8$ holonomy}

The half-supersymmetric solutions of heterotic theory including $\alpha'$ corrections have been investigated in \cite{gp2-het}.
Here we shall take $dH=0$ and  first consider the $SU(2)$ holonomy backgrounds. From the results in section \ref{su2} the spacetime metric and $H$ can be written as
\bea
ds^2&=&\eta_{ab} \lambda^a \lambda^b+e^{2\Phi} \, ds_{\rm hk}^2~,~~
\cr
H&=&{1\over3} \eta_{ab} \lambda^a\wedge d\lambda^b+{2\over3} \eta_{ab} \lambda^a\wedge {\cal F}^b-\star_{\rm hk} \tilde de^{2\Phi}~,
\eea
where  $ds_{\rm hk}^2$ is a $d=4$ hyper-K\"ahler metric and $\lambda$ is an anti-self-dual principal bundle connection with
gauge group $\bR^{5,1}$, $SL(2,\bR)\times SU(2)$ or $CW_6$ with self-dual structure constants.

To find explicit examples, one has to specify a $d=4$ hyper-K\"ahler manifold, an anti-self dual instanton connection over it
and to determine the dilaton. The latter  is found by exploring the Bianchi identity (\ref{anb}) of $H$, i.e.~$dH=0$. This gives
\bea
-\nabla^2_{\rm hk} e^{2\Phi}-{1\over2} \eta_{ab}\,{\cal F}^a_{ij}\,\, {\cal F}^{bij}=0~.
\la{fineqn}
\eea
There are many solutions that can be constructed using the above data. These include the 5-brane solution of \cite{ccjhas-het} which
is given by setting $B^4=\bR^4$ with the Euclidean metric and $G=\bR^{5,1}$. In such a case, one has
\bea
&&ds^2=ds^2(\bR^{5,1})+e^{2\Phi} ds^2(\bR^4)~,~~~H=-\star dh~,~~~
\cr
&&e^{2\Phi}=1+{Q\over |x|^2}~,
\eea
where $Q$ is related to the charge of the brane.
More general classes of solutions have been constructed in \cite{gp2-het}.
The $SU(2)$ holonomy class of solutions also includes the Kaluza-Klein monopole for which $B^4$ is equipped with
the Gibbons-Hawking metric, the dilaton is constant and $H=0$.

Next consider the holonomy the $\bR^8$ solutions. In this case, there is a choice of coordinates $(u,v, x^i)$ such that

\bea
ds^2=2 \fe^- \fe^++ ds^2(\bR^8)~,~~~~H= d(\fe^-\wedge \fe^+)~,~~~
\cr
\fe^-=h^{-1} dv~,~~~\fe^+=du+V dv+ n_i dx^i~.
\eea
All components of the metric, $\Phi$ and $H$ depend on $v$ and $x$, and $X=\partial_u$ is the null parallel vector field.

The solutions of the Killing spinor equations are determined up to the functions $h$ and $V$, and the 1-form $n$.
These in turn can be found by solving the field equations (\ref{noncompfeqn}). In addition if  one assumes that
$h$, $V$ and $n$ are $v$ independent, then the field equations imply that
\bea
\partial_i^2 h=\partial_i^2 V=0~,~~~\partial^i dn_{ij}=0~,
\eea
and $e^{2\Phi}=h^{-1}$.
So $h$ and $V$ are harmonic functions of $\bR^8$ and $dn$ satisfies the Maxwell equations on $\bR^8$.
The solution is a superposition of fundamental strings \cite{adggjhrr-het}, pp-waves and  null rotations.
The NS5-brane solution, the fundamental string, the pp-wave and the Kaluza-Klein monopole are considered  the elementary
branes of heterotic theory. In perturbation theory, the  string solution  is considered as the back-reaction
of the elementary string  of theory while the  5-brane is its magnetic dual and it is solitonic.

\subsubsection{Compactification vacua }

Minkowski space compactification vacua are warped product solutions $\bR^{n-1,1}\times_w N^{d-n}$ with fields which are invariant under
the Poincar\'e symmetry of $\bR^{n-1,1}$ and  $N^{10-n}$  restricted to be a compact manifold without boundary.  In supergravity theories, it is known that there are no such smooth solutions, $n>2$,  with non-vanishing fluxes \cite{gwg-comp, jmcn-comp}, see also section \ref{sec:nonexfcomp}.  In  heterotic theory this non-existence theorem applies to the zeroth order in $\alpha'$ sector of the theory. To see this suppose that the warp factor is constant  and that  $H$ does not have  a non-vanishing component along $\bR^{n-1,1}$.  In such a case, the theorem follows from the dilaton field equation in (\ref{hetfeqn}) upon an application of   the Hopf maximum principle.  Indeed as $N^{10-n}$ is compact without boundary, $\Phi$ has an absolute maximum and an absolute  minimum which are critical points
 and so its hessian is negative or positive definite, respectively.  However the $H^2$ term in the dilaton field equation without the two-loop contribution is positive definite which is a contradiction unless $H=0$ and $\Phi$ is constant.  Thus no smooth solutions exist with non-trivial fluxes.

 The same conclusion holds whenever  $H$ has also non-vanishing components
 along $\bR^{n-1,1}$ and there is a non-trivial warp factor. The theorem again follows upon application of the Hopf maximum principle on both the warp
 factor and the dilaton field equations.  Again as can be seen from the dilaton field equation in (\ref{hetfeqn})
 higher order corrections can  modify this conclusion as the terms proportional to $\alpha'$ may not have a definite sign, see also
 \cite{sigp-het, sigp2-het}.

Suppose that $H=0$, the dilaton constant and let $N^{10-n}$ be a compact manifold without boundary. In such case, $N^{10-n}$ up to discrete identifications is
a product of the appropriate dimension  of the Berger type of manifolds  $N^8$ with holonomy $Spin(7)$, $N^7$ with holonomy $G_2$, Calabi-Yau $CY_8$ and $CY_6$ with holonomy
$SU(4)$ and $SU(3)$, respectively,  $K_3$ with holonomy $SU(2)$ and tori $T^k$ with holonomy $\{1\}$.

Amongst these, the compactifications on $N^8$ with holonomy $Spin(7)$, $SU(4)$ and $SU(2)\times SU(2)$ to two dimensions belong to the class
of heterotic solutions for which the holonomy of the connection with torsion, $\mathrm{hol}(\hat\nabla)$, is  in one of
the   non-compact  groups $Spin(7)\ltimes \bR^8$, $SU(4)\ltimes \bR^8$ and $(SU(2)\times SU(2))\ltimes\bR^8$,
respectively, see table \ref{tableholsigma}.  The two dimensional theories have chiral supersymmetry.  The compactification on $G_2$ manifolds to three dimensions
belongs to the heterotic solutions for which  $\mathrm{hol}(\hat\nabla)=G_2$ and has been investigated in \cite{Gunaydin:1995ku, gppkt2-11d}. Similarly the compactification of heterotic
theory on $CY_6$ belongs to the holonomy $\mathrm{hol}(\hat\nabla)=SU(3)$ class of solutions of the heterotic theory.  This leads to an ${\cal N}=1$ theory
in four dimensions and as a result it has been extensively investigated from the phenomenological point of view \cite{pcghasew-het}.
It is also instrumental in the understanding  of mirror symmetry, see e.g.~\cite{pcmlrs-het, bgmp-het, paclgr-het}.  The $K_3$ and $T^k$ compactifications belong to the
holonomy $\mathrm{hol}(\hat\nabla)=SU(2)$ and $\mathrm{hol}(\hat\nabla)=\{1\}$ solutions of the heterotic theory, respectively, and they have found applications in the understanding
of string dualities.

\subsubsection{Non-existence theorem for Euclidean signature solutions}

The KSEs (\ref{kse}) and field equations (\ref{hetfeqn}) of heterotic supergravity can also be considered on Euclidean signature manifolds
 in all  dimensions. Of course the representation of the supersymmetry parameter $\eps$ changes with dimension.  The parallel spinors have always compact isotropy groups  and therefore may not admit non-vanishing
1-form bilinears.  Nevertheless some of the results we have described carry over to the Euclidean case.  In particular,  the relation
between the KSEs and field equations remains unchanged.  A consequence of this is that the KSEs imply the field equations
provided the (anomalous) Bianchi identity (\ref{anb}) of $H$ is still valid.

As the KSEs imply the field equations,   and  in the absence of
$\alpha'$ corrections, one concludes that $H=0$ and $\Phi=\mathrm{const}$. This follows from   an application of the Hopf maximum principle on $\Phi$ and the dilaton field equation.  Therefore at zeroth order in $\alpha'$, there are no smooth Euclidean signature solutions which solve the KSEs of heterotic theory
with non-trivial fluxes.  For holonomy $SU(n)$ solutions this had already been shown by the authors  of   \cite{sigp-het, sigp2-het} using a complex geometry argument. In particular they  demonstrated the non-existence
of a certain holomorphic section in the canonical bundle for such manifolds and also extended the result under some conditions to include $\alpha'$ corrections.
Therefore the existence of smooth Euclidean signature compact solutions requires that $\alpha'$ corrections are included and
in particular $dH\not=0$.

\subsubsection{Brane superpositions, black holes and AdS/CFT}

Another class of solutions which has widespread applications is that of superpositions of the fundamental branes
of heterotic string theory.  Such a solution is the fundamental string within a 5-brane with a pp-wave propagating along the
string.   The solution depends on 3-harmonic functions on the hyper-K\"ahler manifold $B^4$, $h_1, h_5$ and $h_w$, which are related to the string, 5-brane and pp-wave solitons, respectively,  and reads
\bea
ds^2&=&h_1^{-1} \left(2 dv (du-{1\over2} (h_w-1) dv)\right)+ ds^2(\bR^4)+ h_5 ds_{\mathrm{hk}}^2(B^4)~,
\cr
H&=&dv\wedge du\wedge dh_1-\star_{\mathrm{hk}} dh_5~,~~~e^{2\Phi}= h_1^{-1} h_5~.
\eea
  Setting $B^4=\bR^4$, the solution can be written down explicitly \cite{at-het} and  gives upon reduction a black hole solution in 5-dimensions. This has been extensively investigated in \cite{ccjm-het}  as the entropy of this black hole
can be computed using a microscopic argument in IIB string theory as in \cite{ascv-het}.

Another widely investigated solution for microscopic  entropy computations  is that for which  $B^4$ is  the Gibbons-Hawking manifold  and  $h_1$, $h_5$ and $h_w$
are invariant under the tri-holomorphic isometry of $B^4$. This solution gives rise to a 4-dimensional black hole.  For more recent
investigations of this system that include higher order corrections see \cite{pcpmtopr-het}.  All these solutions are examples
of $N=4$ backgrounds with $\mathrm{hol}(\hat\nabla)\subseteq (SU(2)\times SU(2))\ltimes \bR^8$, see also \cite{gpat2-het} for more examples of solutions for which
$\mathrm{hol}(\hat\nabla)$ is a non-trivial group.

Amongst the heterotic solutions are also backgrounds which have been considered as gravitational duals of gauge theories
in the context of the AdS/CFT correspondence.  One such solution is that of \cite{acmv-het} which has been proposed as the gravitational
dual of minimal ${\cal N}=1$, $d=4$ supersymmetric gauge theory  in \cite{jmcn-het}.  This  is a solution of heterotic supergravity which preserves $N=4$ supersymmetries. It is an example  of a heterotic solution
with $\mathrm{hol}(\hat\nabla)\subseteq SU(3)$,  $G=\bR^{3,1}$ and $dH=0$.  This has been shown in \cite{gpat-het}, where the $\hat\nabla$-parallel  forms are also explicitly given. It evades the non-existence theorem of \cite{sigp-het} explained above because it is not compact.


\section{Geometry of $d=11$ supergravity backgrounds}

  The $d=11$ supergravity \cite{ecbjjs-11d} has  been proposed as the effective theory of M-theory \cite{pkt-11d}  which in turn  arises as a strong coupling limit of IIA strings \cite{wittendualities}.
 The supersymmetric solutions of $d=11$
supergravity   include the M2- and M5-branes \cite{mdks-11d, gueven-11d}, which are thought of as the ``elementary'' solitons of M-theory, and their superpositions and intersections \cite{gppkt-11d, at2-11d, jgdkjt-11d}.  These have extensively been used to give evidence in favour of the existence of M-theory and of string  dualities.   More recently they have found a key role in the AdS/CFT correspondence.    There are some extensive reviews on the supersymmetric solutions of $d=11$ supergravity theory, see e.g.~\cite{mdrkjl-11d, ks-11d}.
Here
we shall be mostly concerned with the systematics of solving the KSE of $d=11$ supergravity.    Later we shall explore some
applications in the theory of compactifications and the AdS/CFT correspondence. The bosonic fields of the theory are a metric and a 4-form field strength $F$, $dF=0$. The action, KSE and a summary of other properties of the theory
that are used in this review can be found in appendix \ref{11super}.

\subsection{Spinors and the KSE}

The theory has a single fermionic field, a gravitino, whose  supersymmetry
variation  gives the KSE
\bea
{\mathcal D}_\tM\eps=0~,
\label{11kse}
\eea
where
\bea
{\mathcal D}_\tM\defeq\nabla_\tM-{1\over288}\bigg(\Gamma_\tM{}^{\tA_1\tA_2\tA_3\tA_4} \elF_{\tA_1\tA_2\tA_3\tA_4}-8 \elF_{\tM\tA_1\tA_2\tA_3}
\Gamma^{\tA_1\tA_2\tA_3}\bigg)~,
\label{11supercor}
\eea
is the supercovariant derivative, $\nabla$ is the Levi-Civita connection of the spacetime and the supersymmetry parameter,  $\eps$, is in the 32-dimensional Majorana representation, $\Delta_{\bf 32}$,  of $Spin(10,1)$.

In anticipation of using the spinorial geometry method to solve (\ref{11kse}),  let us first realize $\Delta_{\bf 32}$ in terms
of forms. First, one can begin with the Dirac representation of $Spin(10)$.  This   is  identified with $\Lambda^*(\bC^5)$ as  has already been
described in appendix \ref{app:spinors}. This extends to a
real presentation of $Pin(10)$ and so to a real representation of $Spin(10,1)$. In particular, the gamma matrix along the time direction is
$\Gamma_0\defeq \Gamma_1\dots \Gamma_{\nat}$, where $\Gamma_{\nat}\defeq \Gamma_{10}$.
The reality condition is imposed using the anti-linear map $r_\tB=\Gamma_0\,b*$. Therefore a basis in $\Delta_{\bf 32}$
 is given by the forms
\bea
e_{a_1\cdots a_k}+(-1)^{[k/2]} \star e_{a_1\cdots a_k}~,~~
ie_{a_1\cdots a_k}-i(-1)^{[k/2]} \star e_{a_1\cdots a_k}~,~~
\label{d11timeb}
\eea
where   $a_1<\dots <a_k$,  $a_1,a_2,\dots ,a_5=1,\dots,5$, $k=0,1,2$,  and star is the Hodge duality operation in $\Lambda^*(\bC^5)$.

Alternatively, one can begin with the Dirac representation of $Spin(9,1)$ described in appendix \ref{app:spinors} and then construct
$\Delta_{\bf 32}$ by setting the gamma matrix along the 10-th direction to
$\Gamma_\nat=-\Gamma_{0123456789}$. Moreover, the reality condition is imposed using the anti-linear map
$r_\tB=-\Gamma_0 b*$. Therefore a basis in $\Delta_{\bf 32}$ can be chosen as
\bea
&&e_{a_1 \cdots a_k} +  (-1)^{[k/2]+k} \star e_{a_1 \cdots a_k}~,~~
ie_{a_1 \cdots a_k} -i  (-1)^{[k/2]+k} \star e_{a_1 \cdots a_k}~,~~
\cr
&&e_{a_1 \cdots a_k 5} + (-1)^{[k/2]+k} \star e_{a_1 \cdots a_k 5} \,,~~
i e_{a_1 \cdots a_k 5} - i (-1)^{[k/2]+k}  \star e_{a_1 \cdots a_k 5} \,,
\label{d11nullb}
\eea
where $a_1<\dots <a_k$, but now $a_1,\dots ,a_4=1,2,3,4$,  $k=0,1,2$, and star is the Hodge duality operation in $\Lambda^*(\bC^4)$.

There are two types of orbits of $Spin(10,1)$ on $\Delta_{\bf 32}$. One has isotropy group $SU(5)$ and the other has isotropy group $Spin(7)\ltimes \bR^9=(Spin(7)\ltimes \bR^8)\times \bR$,  \cite{bryant-11d, fof-br-11d}.
The former is an orbit of co-dimension 1. Representatives of the two orbits can be chosen as
\bea
1+e_{12345}~~\text{and}~~1+e_{1234}~,
\la{11tlksx}
\eea
written in the bases (\ref{d11timeb}) and (\ref{d11nullb}), respectively.

\subsection{$N=1$  $SU(5)$ backgrounds}

There are two  types of $N=1$ supersymmetric  $d=11$ supergravity backgrounds  depending on whether the Killing spinor
has isotropy group $SU(5)$ or $Spin(7)\ltimes \bR^9$ in $Spin(10,1)$.  The KSE has been solved using the bilinears method  in \cite{jgsp-11d, jgjgsp-11d}.  Here we shall present details of the solution of the KSE  for the $SU(5)$ invariant Killing spinor in the spinorial geometry method \cite{jguggp, uggpdr-11d} as the proof is shorter than when employing the bilinears method and  moreover  it can  be adapted  to classify the solutions that preserve a near maximal number of supersymmetries. We shall also outline the geometry
of backgrounds admitting a  $Spin(7)\ltimes \bR^9$ invariant Killing spinor.


\subsubsection{The solution of the linear system}

We begin by choosing the Killing spinor as
\bea
\eps=f (1+e_{12345})~,
\la{11tlks}
\eea
where $f$ is a (local) spacetime function.  This function appears because the orbits with isotopy group $SU(5)$ are of co-dimension 1
in $\Delta_{\bf 32}$ and they are not isolated.
Adapting on the spacetime the spinorial geometry coframe $\{\fe^0, \fe^i: i=1, \dots, \nat\}=\{\fe^0, \fe^\al, \fe^{\bar\al}:\al=1,\dots, 5\}$, we decompose the metric and fluxes as
\bea
&&ds^2=-(\fe^0)^2+\delta_{ij} \fe^i \fe^j~,
\cr
&&\elF={1\over 3!} \elE_{ijk}\, \fe^0\wedge \fe^i\wedge \fe^j\wedge \fe^k+{1\over 4!} \,\elM_{ijkl}\, \fe^i\wedge \fe^j\wedge \fe^k\wedge \fe^l~.
\eea
Note that $ds^2=-(\fe^0)^2+2 \delta_{\al\bar\beta} \fe^\al \fe^{\bar\beta}$ and  the   fluxes $\elE$ and $\elM$ can be decomposed in a similar way.
Substituting (\ref{11tlks}) into the Killing spinor equation (\ref{11kse}), one derives a linear system which reads
\bea
&&\partial_0 \log f+{1\over2} \Omega_{0,\al\bar\beta} g^{\al\bar\beta}
-{i\over24} \elM_{\al}{}^{\al}{}_{\beta}{}^{\beta}=0~,
\cr
&&i \Omega_{0,0 \bar\al}+{1\over3} G_{\bar\al\beta}{}^{\beta}{}+{i\over 72}
 \elM_{\beta_1\beta_2\beta_3\beta_4} \chi^{\beta_1\beta_2\beta_3\beta_4}{}_{\bar\al}=0~,
\cr
 &&\Omega_{0,\bar\al\bar\beta}-{i\over6} \elM_{\bar\al\bar\beta\gamma}{}^\gamma{}
 -{1\over 18} G_{\gamma_1\gamma_2\gamma_3}
 \chi^{\gamma_1\gamma_2\gamma_3}{}_{\bar\al\bar\beta}=0~,
 \eea
and
\bea
&&\partial_{\bar \al} \log f+{1\over2} \Omega_{\bar\al, \beta\bar\gamma} g^{\beta\bar\gamma}
+{i\over12} G_{\bar\al\gamma}{}^\gamma
-{1\over 72} \chi_{\bar\al}{}^{\beta_1\beta_2\beta_3\beta_4} \elM_{\beta_1\beta_2\beta_3\beta_4}=0~,
\cr
&&\partial_{\bar \al} \log f-{1\over2} \Omega_{\bar\al, \beta\bar\gamma} g^{\beta\bar\gamma}
+{i\over4} G_{\bar\al\gamma}{}^\gamma=0~,
\cr
&&i \Omega_{\bar\al,0\bar\beta}+{1\over6} \elM_{\bar\al\bar\beta\gamma}{}^\gamma
-{i\over 18} \chi_{\bar\al\bar\beta}{}^{\gamma_1\gamma_2\gamma_3}
G_{\gamma_1\gamma_2\gamma_3}=0~,
\cr
&&i\Omega_{\bar\al,0\beta}+{1\over12} g_{\bar\al\beta} \elM_\gamma{}^\gamma{}_\delta{}^\delta
+{1\over2} \elM_{\bar\al\beta\gamma}{}^\gamma=0~,
\cr
&&\Omega_{\bar\al, \bar\beta\bar\gamma}+{i\over 6} G_{\bar\al\bar\beta\bar\gamma}
-{1\over 12} \chi_{\bar\al\bar\beta\bar\gamma}{}^{\gamma_1\gamma_2}
\elM_{\gamma_1\gamma_2\delta}{}^\delta -{1\over 12} \elM_{\bar\al\gamma_1\gamma_2\gamma_3}
\chi^{\gamma_1\gamma_2\gamma_3}{}_{\bar\beta\bar\gamma}=0~,
\cr
&&\Omega_{\bar\al,\beta\gamma}-{i\over 2} G_{\bar\al\beta\gamma}-{i\over3} g_{\bar\al[\beta} G_{\gamma] \delta}{}^\delta
-{1\over 36} \elM_{\bar\al\bar\gamma_1\bar\gamma_2\bar\gamma_3}
\chi^{\bar\gamma_1\bar\gamma_2\bar\gamma_3}{}_{\beta\gamma}=0~.
\eea
where $\chi_{\al_1\al_2\al_3\al_4\al_5}=\sqrt{2}\, \epsilon_{\al_1\al_2\al_3\al_4\al_5}$~,

The solution of the linear system above expresses some of the fluxes in terms of geometry as
\bea
\elM_{\beta_1\dots\beta_4}&=&{1\over2} (-\Omega_{0,0\bar\al}+
2 \Omega_{\bar\al, \beta}{}^\beta) \chi^{\bar\al}{}_{\beta_1\dots\beta_4}~,~~~
G_{\bar\al \beta}{}^\beta=-2i \Omega_{\bar\al, \beta}{}^\beta-2i \Omega_{0,0\bar\al}~,
\cr
\elM_{\al}{}^\al{}_\beta{}^\beta&=&12 i \Omega_{\bar\al,0\beta} g^{\bar\al\beta}~,~~~
\elM_{\beta\bar\al\gamma}{}^\gamma=2i \Omega_{\bar\al,0\beta}+2i g_{\bar\al\beta} \Omega_{\bar\gamma, 0 \delta} g^{\bar\gamma\delta}~,
\cr
G_{\bar\al\beta\gamma}&=&-2i \Omega_{\bar\al,\beta\gamma}+2i g_{\bar\al[\beta} \Omega_{0,0\gamma]}~,~~~G_{\bar\al_1\bar\al_2\bar\al_3}=6i \Omega_{[\bar\al_1,\bar\al_2\bar\al_3]}
\cr
\elM_{\bar\al\beta_1\beta_2\beta_3}&=&{1\over2} [\Omega_{\bar\al,\bar\gamma_1\bar\gamma_2} \chi^{\bar\gamma_1\bar\gamma_2}{}_{\beta_1\beta_2\beta_3}
+ 3 \Omega_{\bar\gamma_1, \bar\gamma_2\bar\gamma_3} \chi^{\bar\gamma_1\bar\gamma_2\bar\gamma_3}{}_{[\beta_1\beta_2} g_{\beta_3]\bar\al}
\cr &&~~~
+12i \Omega_{[\beta_1,0\beta_2} g_{\beta_3]\bar\al}]~.
\label{11fluxgeom}
\eea
  In addition, one finds the conditions  on the geometry
\bea
&&\partial_0\log f=0~
,~~~
\Omega_{\bar\al,0\beta}+\Omega_{\beta,0\bar\al}=0~,~~~2 \partial_{\bar\alpha} \log f+\Omega_{0,0\bar\al}=0~,
\cr
&&\Omega_{0, \bar\al\bar\beta}=\Omega_{\bar\al, 0\bar\beta}~,~~~
\Omega_{0,\beta\bar\al} g^{\beta\bar\al}+\Omega_{\bar\al,0,\beta} g^{\bar\al\beta}=0~,
\cr
&&\Omega_{\bar\al, \beta\gamma} g^{\bar\al\beta}-\Omega_{\gamma,\beta}{}^\beta-\Omega_{0,0\gamma}=0~.
\label{11geomx}
\eea
It is clear  by construction that both the linear system and its solution are expressed in terms of the
representations of the isotropy group $SU(5)$ of the Killing spinor. Furthermore, observe that not all components of the
fluxes are expressed in terms of the geometry.  In particular the Hermitian-traceless component of $\elM_{\alpha\bar\beta\gamma\bar\delta}$
is not determined in terms of geometry.  This signals   that the implementation of field equations and Bianchi
identities will not be straightforward for solutions of the KSE with few supersymmetries.

\subsubsection{Geometry}\label{11geom}

The spacetime admits a 1-form, a 2-form and a 5-form bilinear of the Killing spinor $\eps=f(1+e_{12345})$.  These can be easily
computed to find
\bea
&&X=f^{2} \fe^0~,~~~~\omega=- f^2\, (\fe^1\wedge \fe^6+\dots+\fe^5\wedge \fe^\nat)~,~~~
\cr
&&\tau=f^2\,{\rm Im}[(e^1+i e^6)\wedge\ldots\wedge(e^5+i
e^\nat)]+\frac{1}{2} f^2\, e^{0}\wedge \omega\wedge \omega\,.
\label{11bilin}
\eea
We shall use these to interpret the conditions we have found on the geometry of spacetime in (\ref{11geomx}).

First it is straightforward to verify that the first three conditions together with the symmetric part of the fourth condition in (\ref{11geomx}) imply that $X$ is a Killing vector field.
Moreover upon using the Bianchi identity of $F$, one can also show that ${\cal L}_XF=0$ and therefore all the fields are invariant under the
action of $X$.  Furthermore, if one also uses in addition the antisymmetric part of the fourth condition and the fifth  condition in (\ref{11geomx}), one can establish that $\eps$ is  invariant
under the action on $X$.  Thus we have shown that
\bea
{\cal L}_X g={\cal L}_X F=0~,~~~{\cal L}_X\eps=0~,
\label{unigeom}
\eea
where ${\cal L}_X$ in the last condition above is the spinorial Lie derivative defined in (\ref{spinder}).  The last condition in (\ref{11geomx}) can be expressed  as
\be
 W_5+2df=0~,
 \label{11extra}
\ee
where $(W_5)_i={1\over 40} \chi^{j_1\dots j_5} \nabla_{[i}\chi_{j_1\dots j_5]}$.

As has already been mentioned, the geometric conditions  in (\ref{unigeom}) are universal and hold for the $N=1$ backgrounds of all  supergravity theories.  The last condition is required for the  Killing superalgebras to close, see section \ref{sec:killingsuper}.

 To locally  describe  the geometry of spacetime, notice that a consequence of the invariance of $\eps$ under the action of $X$ is that
all form bilinears of the Killing spinor are also invariant.  So  we have
\bea
{\cal L}_X\omega={\cal L}_X\tau=0~.
\eea
The spacetime can locally be described as a fibration whose fibres have a tangent space spanned by $X$ where the base space $B$ is a 10-dimensional space with an $SU(5)$ structure that satisfies (\ref{11extra}).  This follows because both $\omega$ and $\tau$ are invariant
under $X$ and that $i_X\omega=i_X\tau=0$.  Therefore both $\omega$ and $\tau$ ``descend'' as fundamental $SU(5)$ forms on the base space $B$. If one adapts a local coordinate $t$ along $X$, $X=\partial_t$, then $\fe^0=f^2 (dt+w_i\fe^i)$.

The conditions relating the fluxes to the geometry, (\ref{11fluxgeom}), can also be expressed in a real basis in terms of covariant and/or exterior derivatives  of form bilinears (\ref{11bilin}).  However, the final expressions are rather involved, see \cite{jguggp}, and we shall not describe them here.

\subsection{Geometry of $Spin(7)\ltimes \bR^9$ backgrounds}\label{11geom2}

To describe the geometry of $d=11$ backgrounds admitting  a Killing spinor which  has isotropy group
$Spin(7)\ltimes \bR^9$ in $Spin(10,1)$, one can choose $\eps=1+e_{1234}$, where the spinor is written in the basis (\ref{d11nullb}). The spacetime geometry is best described in a coframe with respect to which the  metric can be written
\be
ds^2 = 2 \fe^- \fe^+ + (\fe^9)^2 + \delta_{ij} \fe^i \fe^j~,
\ee
for $i,j=1, \dots, 8$. Note that this coframe is different from that used in \cite{jgjgsp-11d} to give the solution of the KSE. This coframe is determined up to $(Spin(7) \ltimes \bR^8)\times \bR$ gauge transformations which transform the coframe as that in (\ref{nullpatching}).

For this class of solutions, the form bilinears are
\be
X= \fe^-, \qquad \tau = \fe^- \wedge \phi~,
\ee
where
\be
\phi = {1 \over 4!} \phi_{ijk\ell}\, \fe^i \wedge \fe^j \wedge \fe^k \wedge \fe^\ell~,
\ee
is the fundamental $Spin(7)$ self-dual 4-form of $Spin(7)$.

In either the bilinears or spinorial geometry methods for solving the KSE, the solution yields the following conditions on the spacetime geometry
\begin{eqnarray}
\label{nullblock1}
\Omega_{(M,N)+} =0~,~~~
\Omega^{\bf{7}}_{[i,j]+} =0~,~~~
\Omega_{i,9+} =0~,~~~
\Omega_{+,9i} =0~,~~~
\Omega^{\bf{7}}_{+,ij} =0~,
\end{eqnarray}
and
\begin{eqnarray}
\label{nullblock2}
\Omega_{-,+9} = -{1 \over 4} \Omega^i{}_{,i9}~,~~~
\Omega^{\bf{7}}_{9,ij} = - \Omega^{\bf{7}}_{[i,j]9}~,
\end{eqnarray}
and
\begin{eqnarray}
\label{nullblock3}
\Omega_{9,9i} - 6 \Omega_{i,+-} &=& -\phi_i{}^{j_1 j_2 j_3} \Omega_{j_1,j_2 j_3}
-2 \Omega^j{}_{,ji}~,
\end{eqnarray}
where we have used the decomposition, $\Lambda^2(\bR^8) = \Lambda^2_{\bf{7}}\oplus\Lambda^2_{\bf{21}}$, of the space of 2-forms in $\bR^8$,  $\Lambda^2(\bR^8)$, in irreducible   $Spin(7)$ representations. Note that $\Lambda^2_{\bf{21}}=\mathfrak{spin}(7)$.
As the Killing  spinor is  $(Spin(7) \ltimes \bR^8)\times \bR$ invariant, it satisfies $\Gamma^-\epsilon=0$ and $\omega_{ij}\Gamma^{ij}\epsilon=0$ for all $\omega\in \Lambda^2_{\bf{21}}$.

As in the previous cases that the spacetime admits a null 1-form blinear, we define the space transverse to the lightcone ${\cal T}$.
${\cal T}$ is further decomposed as ${\cal T}=L\oplus {\cal Z}$, where $L$ is spanned by $\fe^9$ and ${\cal Z}$ is associated to the spinor representation of $Spin(7)$. The conditions ({\ref{nullblock1}}) can be rewritten as
\begin{eqnarray}
\label{nullblock1b}
{\cal{L}}_X g=0 \ , \qquad {\cal{L}}_X \epsilon =0 \ , \qquad i_{V}(X \wedge dX) \in \Lambda^2_{\bf{21}}~,
\end{eqnarray}
where $\Lambda^2_{\bf{21}}$ now denotes the space of sections of the vector bundle associated with the ${\bf 21}$ representation of $Spin(7)$ and  $V$ is the vector field dual to the 1-form $\fe^+$.  Similarly,
 ({\ref{nullblock2}}) is equivalent to
\be
\label{nullblock2b}
i_{V} ({\cal{L}}_{W} \tau ) \in \Lambda^4_{{\bf{35}}}~,
\ee
where   $W$ is the vector field dual to the 1-form $\fe^9$
and we have used the decomposition $\Lambda^4(\bR^8) = \Lambda^4_{\bf{35}}\oplus  \Lambda^4_{\bf{1}}\oplus \Lambda^4_{\bf{7}}\oplus \Lambda^4_{\bf{27}}$ in irreducible representations of $Spin(7)$.
Note that $\Lambda^4_{\bf{35}}$ is the space  of the anti-self-dual 4-forms.
The remaining condition ({\ref{nullblock3}}) can be rewritten in terms of the Lee form of the $Spin(7)$ 4-form $\phi$ as
\be
\label{nullblock3b}
\Omega_{9,9i} -6 \Omega_{i,+-} = -3 {\tilde{\theta}}_i~,
\ee
where ${\tilde{\theta}}=-{1 \over 6} \star_{{}_8} \big((\star_{{}_8} {\tilde{d}} \phi) \wedge \phi\big)$, $\star_{{}_8}$ denotes the Hodge dual operation in ${\cal Z}$  and ${\tilde{d}}$ denotes the restriction of the exterior derivative ${\cal Z}$.

\subsection{Geometry of IIA and IIB $N=1$ backgrounds } \label{n1iiaiib}

The investigation of the geometry of $N=1$ backgrounds in type IIB and IIA $d=10$ supergravities has been carried out in \cite{ugjggp-iib, ugjggp2-iib, ugjggpdr-iib} and
\cite{uggpcsI-iia, uggpcsII-iia, uggpcsIII-iia}, respectively,
where the explicit solution to the KSEs can be  found. Here we shall give a very brief description of the results.

Beginning with IIB supergravity, the gauge group of the theory in the formulation of \cite{jspw-iib, js-iib} is
$Spin(9,1)\cdot U(1)$.  The $U(1)$ arises because the supersymmetry parameter $\eps$, which is in the Weyl representation of $Spin(9,1)$,
  is additionally twisted with the pull-back of the canonical bundle of
the upper-half plane.  This is the scalar manifold of the axion and the dilaton, i.e.~the two IIB scalars.  More detailed description of the couplings of
IIB supergravity will be given in appendix  \ref{iibsuper}.  An investigation reveals that there are three types
of orbits of $Spin(9,1)$ in the space of Weyl spinors with  isotropy group  either
$Spin(7)\ltimes \bR^8$, or $SU(4)\ltimes \bR^8$, or $G_2$.

As in the more detailed analysis we have presented in section \ref{11geom} for $d=11$ supergravity,  the conditions on the geometry of the IIB $N=1$ backgrounds imposed by the KSEs include the existence of a Killing vector field $X$
which is constructed as a bilinear of the Killing spinor $\epsilon$.  In addition,  $X$ leaves  the other fields invariant as well as  the Killing spinor $\eps$.  One difference is that the spinorial Lie derivative in IIB is defined as in (\ref{spinder}) with $\nabla$ given in
(\ref{iibspin}), i.e.~it involves an additional connection term $-(i/2) Q$ associated with the $U(1)$ twist of $\eps$.  $X$ is timelike or null depending on whether the isotropy group is compact or non-compact, respectively. These
are the universal conditions expected from the IIB KSEs.  In all cases there are additional conditions on the spacetime which depend
on the type of orbit to which the Killing spinor $\eps$ belongs, for details see the references above.

In (massive) IIA supergravity \cite{mhmn-iia, fgmp-iia, icpw-iia, lr-iia} the supersymmetry parameter is in the Majorana representation of $Spin(9,1)$.  One can show that
there are four  types of  orbits  of  $Spin(9,1)$ in the space of Majorana  spinors with  isotropy groups
either $Spin(7)$, $Spin(7)\ltimes\bR^8$, $SU(4)$ or $G_2\ltimes \bR^8$. The solution of the KSEs of IIA
supergravity for one Killing spinor gives   restrictions
on the geometry of the spacetime.  These   again include the universal conditions.  Therefore there is
 a Killing vector $X$ constructed as  a bilinear of the Killing spinor $\epsilon$ which leaves all the fields invariant as well as $\epsilon$,
 where  the spinorial Lie derivative is given as in (\ref{spinder}).  There are additional conditions on the geometry
 of spacetime which depend on the type of orbit to which $\epsilon$ belongs and can be found in the original papers mentioned above.

\subsection{Global properties of the solutions}\label{gloprosol}

The description of the geometry   of $d=11$ and $d=10$ type II supergravity backgrounds with $N=1$ supersymmetry  we have given is local. It depends
on the assumption that at some open set of the spacetime the supercovariant connections preserve the type of orbit to which the Killing spinor belongs
under parallel transport.  This is not automatically the case as the holonomy of the supercovariant connections for generic backgrounds is in $SL(32,\bR)$, see also section \ref{gaugeholonomy}.  Therefore, under parallel transport the spinors are transformed with $SL(32,\bR)$ transformations  which  do not necessarily preserve their orbit type.  The properties of form bilinears also change under such parallel transport.  For $N=1$ backgrounds in $d=11$ supergravity an indication  that a Killing spinor with isotropy group   $SU(5)$  has changed  under parallel transport to another  one  with isotropy group $Spin(7)\ltimes \bR^9$ is that the Killing vector bilinear $X$ changes from timelike to null at some region of spacetime.  Such phenomenon is widespread in gravitational backgrounds and signals the existence of Killing horizons.

A priori the spinorial geometry method can be adapted  to  solve  this problem.  For example one can choose Killing spinor
representatives which include all orbit types.  However in such a case, the resulting linear system will be rather involved.

A related issue is the restriction of the $G$-structure of the spacetime as a consequence of the existence of a Killing spinor.
As a Killing spinor is a no-where vanishing section of an appropriate spin bundle, one expects that the $G$-structure of the spacetime may reduce to a subgroup of the isotropy group of the Killing spinor.  However in $d=11$ and $d=10$ type II theories  the relevant  spin bundle has a rank
much larger than the dimension of the spacetime.  As a consequence it always admits no-where vanishing sections.  So a priori the existence of a no-where vanishing section does not necessarily imply the reduction of the spacetime $G$-structure.  However, if one insists that  the orbit type of a Killing spinor is preserved under parallel transport everywhere on the spacetime, then the structure group reduces to a subgroup of the isotropy group of the Killing spinor.
These observations clarify the use of the $G$-structure language to describe the geometry of supersymmetric backgrounds in $d=11$, $d=10$  type II and other
supergravities.

\subsection{Killing superalgebras }\label{sec:killingsuper}

The Killing spinors and associated Killing vector bilinears on a  supersymmetric background can be endowed with a superalgebra structure. Superalgebras are $\bZ_2$-graded associative algebras with a compatible  bracket structure which satisfies the super-Jacobi identities, see e.g.~\cite{vk-super}.  In particular,   superalgebras decompose as $\mathfrak{g}=\mathfrak{g}_0+\mathfrak{g}_1$, where $\mathfrak{g}_0$ and $\mathfrak{g}_1$ are
 the even and odd subspaces of the superalgebra with grading 0 and 1, respectively. Given elements $\alpha, \beta\in\mathfrak{g}$ with grading $|\alpha|$ and $|\beta|$,  the bracket  is defined as
  \bea
  [\alpha, \beta]_{\mathfrak{g}}\defeq \alpha \beta-(-1)^{|\alpha| |\beta|} \beta\alpha~.
   \eea
   This satisfies the super-Jacobi identities
   \bea
   [[\alpha, \beta]_{\mathfrak{g}}, \gamma]_{\mathfrak{g}}+(-1)^{|\gamma| (|\alpha|+ |\beta|)} [[\gamma, \alpha]_{\mathfrak{g}}, \beta]_{\mathfrak{g}}+ (-1)^{|\alpha| (|\beta|+|\gamma|)}[[\beta,\gamma]_{\mathfrak{g}},\alpha]_{\mathfrak{g}}=0~.
\nonumber \\
   \label{superjac}
   \eea
  Note that the bracket $[\cdot,\cdot]_{\mathfrak{g}}$ between two odd elements of the superalgebra is an anticommutator which we denote with $\{\cdot,\cdot\}$ while all the rest of the brackets are commutators  $[\cdot,\cdot]$.

Returning to the definition \cite{jgrmpt-super, fof-super} of  a Killing superalgebra $\mathfrak{g}$ for supersymmetric backgrounds    the odd subspace $\mathfrak{g}_1$ of $\mathfrak{g}$ is spanned by  $Q_{\epsilon_{\mathbf{n}}}$, where we have associated a generator $Q_{\epsilon_{\mathbf{n}}}$ to every linearly independent Killing spinor $\epsilon_{\mathbf{n}}$ on the spacetime, $\mathbf{n}=1,\dots, N$. Similarly, $\mathfrak{g}_0$ is spanned by $V_{X_{\mathbf{m}\mathbf{n}}}$,  where we have associated a generator $V_{X_{\mathbf{m}\mathbf{n}}}$ to every linearly independent Killing vector bilinear $X_{\mathbf{m}\mathbf{n}}$ on the spacetime. For $d=11$ and (massive) IIA supergravities, the latter are defined in terms of the Killing spinors as
 \bea
 X_{\mathbf{m}\mathbf{n}}\defeq \langle(\Gamma_+-\Gamma_-) \eps_\mathbf{m}, \Gamma_\tA \eps_\mathbf{n}\rangle\,\fe^\tA~.
 \label{allbilinears}
 \eea
 For IIB supergravity, one takes the real part of the above expression. Note that the inner product used in (\ref{allbilinears}) is proportional to the Dirac inner product, see appendix \ref{app:spinors}, where the proportionality factor has been introduced for convenience.
  The proof that all $X_{\mathbf{m}\mathbf{n}}$ are Killing  follows from the linearity of the KSEs and the Killing property
  of the vector bilinear of  a single Killing spinor  that we have already demonstrated, see e.g.~sections \ref{11geom} and \ref{11geom2}
  for $d=11$ supergravity.
  Similar definitions for $X$ exist
 in all supergravity theories.  Observe that $X_{\mathbf{m}\mathbf{n}}=X_{\mathbf{n}\mathbf{m}}$.

The (anti-)commutators of the Killing superalgebra are defined as follows
\bea
&&\{Q_{\epsilon_{\mathbf{m}}}, Q_{\epsilon_{\mathbf{n}}}\}=V_{X_{\mathbf{m}\mathbf{n}}}~,~~~~[V_{X_{\mathbf{m}\mathbf{n}}}, Q_{\epsilon_{\mathbf{p}}}]=Q_{{\cal L}_{X_{\mathbf{m}\mathbf{n}}}\epsilon_{\mathbf{p}}}~,~~~
\cr
&&
[V_{X_{\mathbf{m}\mathbf{n}}}, V_{X_{\mathbf{p}\mathbf{q}}}]=V_{[X_{\mathbf{m}\mathbf{n}}, X_{\mathbf{p}\mathbf{q}}]}~,
\eea
where  $[X_{\mathbf{m}\mathbf{n}}, X_{\mathbf{p}\mathbf{q}}]$ is the Lie commutator of two vector fields,  and  in $d=11$ and (massive) IIA supergravities ${\cal L}_{X_{\mathbf{m}\mathbf{n}}}$ is the spinorial Lie derivative (\ref{spinder}) with respect to $X_{\mathbf{m}\mathbf{n}}$, while
in IIB supergravity the spinorial Lie derivative involves an additional $U(1)$ twist as explained in section \ref{n1iiaiib}.  It has been demonstrated in \cite{fofehjgm-super} that the super-Jacobi identities (\ref{superjac}) are satisfied in $d=11$ and IIB supergravities.  This is expected to hold for all supergravity theories.
It is worth pointing out that the universal condition ${\cal L}_X \epsilon=0$ we have found for $N=1$ backgrounds in all supergravity theories
we have investigated is required for the super-Jacobi identity  $[ \{Q_\eps, Q_\eps\}, Q_\eps]={\cal L}_X\eps=0$ to be satisfied.

\section{Maximally supersymmetric solutions in $d=10$ and $d=11$}
\label{secmaxsusy}

Maximally supersymmetric backgrounds are those that preserve all supersymmetries of a supergravity theory. Typically, these
have a special status amongst the other solutions. For example
in $d=10$ type II and $d=11$ supergravities,  the maximally supersymmetric    backgrounds
 preserve  32 supersymmetries and have found extensive applications in  compactifications
 and in the AdS/CFT correspondence. They
have been classified up to a local isometry in \cite{jfofgp1, jfofgp2}.  Here, we shall summarize
the main steps of the proof of the classification  theorem.

\subsection{$d=11$ supergravity}

The maximally supersymmetric solutions of $d=11$ supergravity are
locally isometric to one of the following solutions


 \begin{enumerate}

 \item[-] $AdS_4\times S^7$ with metric and flux
 \bea
 ds^2= \ell^2\, d\mathring{s}^2(AdS_4)+ 4 \ell^2\, d\mathring{s}^2(S^7)~,~~~F= \pm 3 \ell^3\, \mathring{{\rm dvol}} (AdS_4)~,
 \eea

 \item[-] $AdS_7\times S^4$ with metric and flux
 \bea
 ds^2=\ell^2\, d\mathring{s}^2(AdS_7)+ {1\over4} \ell^2\, d\mathring{s}^2(S^4)~,~~~F=\pm {3\over8} \ell^3 \mathring{{\rm dvol}} (S^4)~,
 \eea

 \item[-] the plane wave with metric and flux
 \bea
&& ds^2=2 dv du+ A_{ij} x^i x^j dv^2+\delta_{ij} dx^i dx^j~,~~~
\cr
&&F = \mu  dv \wedge dx^1\wedge dx^2 \wedge dx^3~,
\label{cwas}
 \eea
 with $A=-\frac{\mu^2}{36}\, {\rm
      diag}(4,4,4,1,1,1,1,1,1)$ and $\mu\not=0$~,

 \item[-] Minkowski spacetime  $\bR^{10,1}$ for which $F=0$~,

\end{enumerate}

where $d\mathring{s}^2$ and  $\mathring{{\rm dvol}}$ denote the metrics and volume forms  of the indicated spaces  with radii  normalized to  one, respectively,
and $\ell\in \bR_{>0}$. The $AdS_4\times S^7$ and $AdS_7\times S^4$ solutions are of the Freund-Rubin form \cite{pfmr-ads} and  have  been  found
in \cite{mdcp-max} and \cite{kppvnpt-max}, respectively.  The plane wave solution has been given in \cite{jkg-max}. Observe that plane wave  parameter $\mu$ can be absorbed
in a coordinate redefinition.  The plane wave solution is a Penrose limit of both the maximally supersymmetric AdS backgrounds of the theory
\cite{mbjfofch-iib}.

To prove the above statement observe that
maximal supersymmetry implies that the supercurvature of the supercovariant
connection, ${\mathcal R}_{\tM\tN}=[{\mathcal D}_\tM, {\mathcal D}_\tN]$, must vanish.
Expanding this in skew-symmetric products of gamma matrices as
\bea
{\mathcal R}_{\tM\tN}=\sum_{k=1}^5 {1\over k!}\,T_{\tM\tN,\tA_1\dots \tA_k}\, \Gamma^{\tA_1\dots \tA_k}~,
\la{11rt}
\eea
all components of  ${\mathcal R}$ in the Clifford algebra basis must vanish, i.e.
\bea
T_{\tM\tN,\tA_1\dots \tA_k}=0~.
\eea
The explicit expression for these components is given in (\ref{supelcurv}).

The vanishing of the term of  ${\mathcal R}$  linear in gamma matrices implies that
$\elF\wedge \elF=0$. In turn this gives
\bea
i_X\elF\wedge \elF=0~,
\la{oneg}
\eea
for any spacetime vector field $X$.

Substituting (\ref{oneg}) into the vanishing condition of the term cubic in gamma matrices, one finds that
\bea
\nabla_{\tB}\elF_{\tC\tA_1\tA_2\tA_3}-\nabla_{\tC}\elF_{\tB\tA_1\tA_2\tA_3}=0~.
\eea
This together with the Bianchi identity for $\elF$, $d\elF=0$, gives that
\bea
\nabla_{\tB}\elF_{\tA_1\tA_2\tA_3\tA_4}=0~.
\la{par}
\eea
Substituting this into the quadratic component, one concludes that the Riemann tensor is also parallel,
$\nabla R=0$,
and therefore the spacetime of maximally supersymmetric backgrounds is a Lorentzian symmetric space.

After some computation, the terms in quartic gamma matrices imply that
\bea
\elF_{\tC\tD[\tA_1\tA_2} \elF_{\tA_3\tA_4]}{}^{\tC\tD}=0~.
\eea
Using the results that arise from the terms quartic in gamma matrices, the terms quintic in gamma matrices give
\bea
i_X \elF\wedge i_Y\elF=0~,
\la{fiveg}
\eea
for any  vector fields $X,Y$.

Using (\ref{oneg}) and (\ref{fiveg}), one can show that
\bea
i_Yi_Z\elF\wedge \elF=0~,
\la{ofg}
\eea
for any vector field $Y,Z$.
Taking the inner derivation with respect to another vector field of both (\ref{fiveg}) and (\ref{ofg}), one finds that
\bea
&&i_Xi_Y\elF\wedge i_Z\elF-i_Xi_Z\wedge i_Y\elF=0~,
\cr
&&
i_Zi_Xi_Y\elF\wedge \elF+i_Xi_Y\elF\wedge i_Z\elF=0~,
\la{ofg1}
\eea
after an appropriate lexicographic relabeling of the vector fields, respectively. The first equation implies
that $i_Xi_Y\elF\wedge i_Z\elF$ is symmetric in the vector field $Z$ and $Y$, while the second implies that it is skew
symmetric. As a result the two terms of the second equation in (\ref{ofg1}) vanish separately. So in particular, one has
\bea
i_Zi_Xi_Y\elF\wedge \elF=0~.
\la{plu}
\eea
This condition is known as a Pl\"ucker relation and it implies that $\elF$ is decomposable, i.e.~it can be written as the
wedge product of four one forms
\bea
\elF=\theta^1\wedge \theta^2\wedge \theta^3\wedge \theta^4~.
\eea
As a result $\elF$  determines a 4-plane at every point in spacetime.

As we have already mentioned the spacetime is a Lorentzian symmetric spaces, $M=G/H$. In particular, $\mathfrak{g}=\mathfrak{h}\oplus \mathfrak{m}$,
and $\mathfrak{m}$ is identified as the tangent space of $G/H$ at the
origin. Moreover, the Lorentzian symmetric space have been classified. It can be shown that they are products of one of the Lorentzian spaces
Minkowski $\bR^{n-1,1}$, $\dS_{n}$, $\AdS_{n}$ or  Cahen-Wallach $\CW_n(A)$  with a Euclidean symmetric space \cite{mcnw-max}.  In particular, the metric of Cahen-Wallach spaces is given in (\ref{cwman}), appendix \ref{cwmanifolds}, with  $\det A\not=0$.

To continue,  since $\elF$ is decomposable and parallel, it spans an $H$-invariant four-plane $\mathfrak {n}\subset \mathfrak{m}$. If $\elF$ is either
time-like or space-like, then the normal $\mathfrak{n}^\perp$ is also $H$-invariant and the symmetric space decomposes into a product
of a four-dimensional and a seven-dimensional symmetric space, $M=X_4\times Y_7$.
Using this, and solving the equation quadratic in gamma matrices (\ref{11rt}), one finds
the $AdS_4\times S^7$ and $AdS_7\times S^4$ solutions for $\elF$ time-like
and space-like, respectively, as stated in the beginning of the section.

It remains to investigate the case in which $\elF$ is null. The only symmetric spaces that admit parallel null forms are those that
locally are products $\CW\times N$, where  $N$ is a Euclidean symmetric space. The
equation quadratic in gamma matrices (\ref{11rt}) implies that the only option is  the plane wave solution. The Minkowski space  arises whenever $F=0$. This completes the proof.

\subsection{IIB supergravity}

The maximally supersymmetric solutions of IIB supergravity are locally isometric to one of the following:


  \begin{enumerate}

  \item[-] $AdS_5\times S^5$ with non-vanishing fields
  \bea
  &&ds^2=\ell^2 d\mathring{s}^2(AdS_5)+\ell^2 d\mathring{s}^2 (S^5)~,~~~
  \cr
  &&\iiF=\pm \ell^4 \left(\mathring{\dvol}(\AdS_5) - \mathring{\dvol}(S^5)\right)~,
  \eea
  \item[-] the plane wave solution with non-vanishing fields
  \bea
 && ds^2=2 dv du+ A_{ij} x^i x^j dv^2+\delta_{ij} dx^i dx^j~,~~~A=-\mu^2 {\bf 1}~,
 \cr
 &&\iiF =  \mu\, dv
   \wedge (dx^1\wedge dx^2 \wedge dx^3 \wedge dx^4 + dx^5\wedge dx^6
    \wedge dx^7 \wedge dx^8)~,~~~
    \eea
    \item[-] Minkowski space $\bR^{9,1}$~,

    \end{enumerate}
  where $\ell\in \bR_{>0}$ and $d\mathring{s}^2$ and   $\mathring{\dvol}$ denote the metrics and volume forms
  of the corresponding spaces with radii normalized to one, respectively. The existence of a IIB maximally supersymmetric  $AdS_5\times S^5$ solution has been mentioned
  in \cite{js-iib}, see also the comment added there.  The plane wave solution has been found in \cite{mbfofchgp-max} and has been demonstrated in \cite{mbjfofch-iib} to be the Penrose limit of the AdS solution.  The parameter
  $\mu\not=0$ of the plane wave solution can be absorbed via a coordinate redefinition.

The proof for this proceeds as  in $d=11$. The algebraic Killing spinor
equation of IIB supergravity implies that
\bea
P=G=0~,
\la{van}
\eea
i.e. the one-form and three-form field strengths vanish.   To investigate the gravitino KSE, we again
consider the supercovariant curvature, ${\mathcal R}$,  with only five-form flux.  Expanded in skew-symmetric products of gamma matrices, ${\mathcal R}$
 is written as
\bea
{\mathcal R}_{\tM\tN}=\sum_{k=0}^2 {1\over (2k)!} T_{\tM\tN,\tA_1\dots \tA_{2k}} \Gamma^{\tA_1\dots \tA_{2k}}~.
\label{iibrexp}
\eea
Again maximal supersymmetry requires that
\bea
T_{\tM\tN,\tA_1\dots \tA_{2k}}=0~.
\eea
The condition that arises for $k=0$, together with (\ref{van}), imply that
the dilaton and axion can be taken to be constant. The term quartic in gamma matrices  implies that
\bea
\nabla_{\tB} \iiF_{\tC\tA_1\dots \tA_4}-\nabla_{\tC} \iiF_{\tB\tA_1\dots \tA_4}=0~,
\cr
\iiF_{\tD\tB[\tA_1\tA_2 \tA_3} \iiF_{\tA_4\tA_5 \tA_6] \tC}{}^\tD=0~.
\eea
The first equation together with the Bianchi identity for $\iiF$ imply that $\iiF$ is parallel
\bea
\nabla\tB\iiF_{\tA_1\dots \tA_5}=0~.
\eea
The second equation can also be written as
\bea
i_X\iiF_\tL\wedge i_Y\iiF^\tL=0~.
\eea
Observe that this also implies that $i_X\iiF_\tL\wedge \iiF^\tL=0$. Then a similar argument to that presented for eleven-dimensional
supergravity reveals that
\bea
i_Z i_Y i_X \iiF_\tL\wedge \iiF^\tL=0~.
\label{gplucker}
\eea
This relation is not a Pl\"ucker relation but a generalization. It has been solved in \cite{jfofgp2} to reveal that there is a decomposable
five-form $K$ such that
$\iiF=K+\star K$, where $K$ is a simple form. Note that  (\ref{gplucker}) and its generalization to (k+1)-forms can also be thought as the Jacobi identity of metric k-Lie algebras \cite{vf-max}.

It remains to solve the condition quadratic in gamma matrices (\ref{iibrexp}). This together with $\nabla\iiF=0$ imply that the spacetime
 is a symmetric space, $G/H$. Moreover, if $K$ is either
time-like or space-like, then $\iiF$ defines an $H$-invariant five-dimensional subspace $\mathfrak{n}$ of $\mathfrak{m}$ which has
an $H$-invariant normal $\mathfrak{n}^\perp$,
$\mathfrak{g}=\mathfrak{h}\oplus \mathfrak{m}$. Again the spacetime decomposes and the only solution is $AdS_5\times S^5$.
The remaining case is when $K$ is null. This gives the plane wave solution. The Minkowski space  arises whenever all form field strengths vanish.  This completes the proof.

\subsection{Other $d=10$ supergravities}

A similar analysis to the one presented above for the $d=11$ and IIB supergravities reveals that
the maximally supersymmetric backgrounds of IIA supergravity are locally isometric to $\bR^{9,1}$ with constant dilaton
and with all  remaining  form field strengths vanishing. The same applies to the heterotic or type I supergravities.
The massive IIA supergravity does not have a maximally supersymmetric background provided that the cosmological constant
is non-zero.

\section{Nearly maximally supersymmetric supergravity backgrounds}

Spinorial geometry can be adapted to classify backgrounds that preserve a near maximal number of supersymmetries.  In particular,
we shall present a brief description of the proof that the $N=31$ backgrounds of IIB  and $d=11$ supergravities are locally
maximally supersymmetric \cite{ugjggpdr-nmax, ugjggpdr2-nmax}. A similar  result for IIA supergravity has been
demonstrated  in \cite{ibjaov-nmax}.

To investigate the geometry of backgrounds of $d=10$ and $d=11$ supergravities with a near maximal number of supersymmetries it is more
convenient to use the gauge symmetry  to choose a canonical  form for the normals to the Killing spinors. To see this,
let us specialize to the $N=31$ case, and write the Killing spinors as,
\bea
\eps^r=\sum^{32}_{i=1} f^r_i \eta^i~,~~~r=1,\dots,31,
\eea
where $\eta^i$ is a basis in the space of spinors and  $f^r_i$ is a matrix of real spacetime functions of rank 31.
The main difficulty in solving the KSEs or their integrability conditions is that $f^r_i$ is not a
square invertible matrix. To overcome this, one uses the gauge symmetry of the KSEs to choose
the hyperplane of Killing spinors. It turns out that
the most efficient way to do this is to use the gauge symmetry to orient  the normal $\nu$ to the Killing spinors into a particular direction.
Having chosen the normal spinor $\nu$, the 31 Killing spinors are then defined by the orthogonality condition
\bea
\langle\nu, \eps^r\rangle_s=0~,
\la{orth31}
\eea
where $\langle\cdot, \cdot\rangle_s$ is a suitable $Spin$-invariant inner product in the space of spinors. Typically, there are several cases
that one should investigate corresponding to the number of canonical forms for $\nu$ up to supergravity gauge transformations, i.e.~the number
of orbit types of the gauge group on the space of  spinors.  Although the methodology to find nearly maximally supersymmetric backgrounds here is described in the context of $d=10$ and $d=11$
supergravities, it also applies to all other theories.

\subsection{N=31, IIB}

To begin the proof of the main result in IIB supergravity, a convenient basis in the space of IIB spinors can be chosen as $(\eta^p, i \eta^p)$, where $\eta^p$ is a basis in the space
of Majorana-Weyl spinors. In such a case, the Killing spinors can be written as
\bea
\eps^r=\sum_{p=1}^{16} f^r_p \eta^p+ i \sum_{p=1}^{16} f^r_{16+p} \eta^p~,
\la{iib31ks}
\eea
where $(f^r_p, f^r_{16+p})$ is a matrix of real spacetime functions of rank 31.
A choice of a $Spin(9,1)$-invariant inner product is the real part of the Majorana inner product of IIB spinors
\bea
\langle\eps_1, \eps_2\rangle_s\defeq {\mathrm {Re}}\, \mathrm {B}\, (\eps_1, \eps_2)~,
\eea
see appendix \ref{app:spinors}.
It turns out that $\mathrm {B}$  is skew-symmetric and vanishes when restricted to either chiral or anti-chiral spinors.
As a result, two spinors have a non-trivial inner product iff one of the spinors is chiral and the other anti-chiral. Therefore, since the IIB Killing spinors
are chosen to be chiral, the normal $\nu$  lies in the anti-chiral representation of $Spin(9,1)$.

The gauge group $Spin(9,1)$ has three different orbits in the space of anti-chiral spinors with representatives
\bea
&&(n+im)(e_5+e_{12345})~,~~~(n-\ell+im) e_5+(n+\ell+im)e_{12345}~,
\cr
&&n (e_5+e_{12345})+im (e_1+e_{234})
\la{3normals}
\eea
and with isotropy groups $Spin(7)\ltimes \bR^8$, $SU(4)\ltimes\bR^8$ and $G_2$, respectively. Therefore there are three different choices for the normal $\nu$ to the 31 Killing spinors.

The analysis for the three different cases is similar. Because of this, we shall outline the proof
for the first normal spinor and the details for the other two cases can be found in \cite{ugjggpdr-nmax}.
Substituting the first spinor in (\ref{3normals}) as a normal and the expression for the Killing spinors (\ref{iib31ks}) into the
orthogonality condition (\ref{orth31}), one  finds
\bea
f_1^r n-f_{17}^r m=0~.
\eea
After assuming without loss of generality that $n\not=0$ and solving this equation for $f^1_r$,  one finds that
the Killing spinors (\ref{iib31ks}) can be written as
\bea
\eps^r={f_{17}^r \over n} (m+in) (1+e_{1234})+\sum_{k\not =1, 17} f^r_k \eta^k~.
\la{31sp}
\eea
Observe now that the transformation from the Killing spinors $(\eps^r)$ to the basis $((n+im)(1+e_{1234}), \eta^k)$ is invertible. Substituting this
into the algebraic KSE of IIB supergravity, see appendix \ref{iibsuper},  one finds that
\bea
P_\tM \Gamma^\tM\eta^p=0~,~~~p=2,3,\dots,16~.
\eea
This is due to the complex conjugation operation in the algebraic KSE and the choice
of the basis in (\ref{31sp}).  The above equation implies that $P$ must be null. But some of the $\eta^p$ spinors
are annihilated by $\Gamma^-$ and some others are annihilated by $\Gamma^+$.  As a result the only solution that satisfies both light-cone
projections is $P=0$.

Next, if $P=0$, the algebraic KSE is linear over the complex numbers, as a result it has an
even number of solutions. Since it is required to have 31, one concludes that it should have 32. The only Clifford algebra
element which annihilates all spinors is the zero element and thus the 3-form flux vanishes, $G=0$.

Therefore the algebraic KSE gives $P=G=0$. If this is the case, then the gravitino KSE
becomes linear over the complex numbers, and therefore admits an even number of solutions. So if it is required to admit 31 Killing spinors, then it will have 32.
The same analysis holds for the other two normal spinors in (\ref{3normals}), see \cite{ugjggpdr-nmax} and therefore it follows that all IIB backgrounds
with 31 supersymmetries are  maximally supersymmetric.

\subsection{N=31, D=11}

As in the IIB case  outlined in the previous section, $d=11$ backgrounds with 31 supersymmetries are also maximally supersymmetric \cite{ugjggpdr2-nmax}.
The proof in $d=11$ though is different from that described   for IIB. This is because one has to solve directly the
gravitino KSE. In particular, one has to show that for the  backgrounds with 31 supersymmetries the integrability
condition of the gravitino KSE, ${\mathcal R}\eps^r=0$, implies that the supercovariant curvature vanishes,
${\mathcal R}=0$ .

To continue, it is convenient to write   ${\mathcal R}$  in terms of two different bases. In one of the bases, ${\mathcal R}$
automatically satisfies ${\mathcal R}\eps^r=0$. While in the other, one can easily impose the field
equations and Bianchi identities of supergravity theories. Comparing the two expressions, one can show the vanishing of the supercovariant
curvature.

To proceed further,
let $(\eta^i)$ be a basis in the space of  spinors.   Then observe that the supercovariant curvature
 for a background with $31$ Killing spinors can be written as
\bea
{\mathcal R}_{\tM\tN}= \sum^{32}_{i=1} u_{\tM\tN,i}\,\eta^i\otimes \nu~,
\la{supcurv12}
\eea
where the $u$'s are spacetime forms, the spinor indices have been suppressed
and $\nu$ is the normal to the Killing spinors. The orthogonality condition has been taken with respect to  a $Spin(10,1)$-invariant Majorana inner product.  In particular, ${\mathcal R}_{\tM\tN}\eps^r=0$ as required. Therefore
in terms of the $u$'s  the supercovariant curvature satisfies all the supersymmetry conditions.

To constrain further the components $u$ of ${\mathcal R}$, one has to impose the field equations and Bianchi identities
of 11-dimensional supergravity. These are most easily expressed in terms of the $T$ components. In particular, observe
that  $\Gamma^\tN{\mathcal R}_{\tM\tN}$ is a  linear combination of field equations and Bianchi identities, and therefore
it necessarily vanishes identically. In turn this leads to
  \bea
    && (T^1_{\tM\tN})^\tN = 0~,~~~  (T^2_{\tM\tN})_\tP{}^\tN = 0~,~~~(T^1_{\tM\tP_1})_{\tP_2} + \tfrac{1}{2} (T^3_{\tM\tN})_{\tP_1 \tP_2}{}^\tN =  0~, \cr
    && (T^2_{\tM[\tP_1})_{\tP_2 \tP_3]} - \tfrac{1}{3} (T^4_{\tM\tN})_{\tP_1 \tP_2 \tP_3}{}^\tN =  0~,~
    (T^3_{\tM[\tP_1})_{\tP_2 \tP_3 \tP_4]} + \tfrac{1}{4} (T^5_{\tM\tN})_{\tP_1 \cdots \tP_4}{}^\tN =  0~, \cr
    && (T^4_{\tM[\tP_1})_{\tP_2 \cdots \tP_5]} - \frac{1}{5 \cdot 5!} \epsilon_{\tP_1 \cdots \tP_5}{}^{\tQ_1 \cdots \tQ_6}
      (T^5_{\tM \tQ_1})_{\tQ_2 \cdots\tQ_6}  =  0~.
\nonumber \\
 \label{field-eqs}
 \eea
The second and third  of these equations are consequences of the Einstein
and $F$ field equations, respectively.
We  also use the additional conditions
 \bea
  && (T^1_{\tM\tN})_\tP = (T^1_{[\tM\tN})_{\tP]} \,, \qquad
  (T^2_{\tM\tN})_{\tP\tQ} = (T^2_{\tP\tQ})_{\tM\tN} \,, \qquad
  (T^3_{[\tM\tN})_{\tP\tQ\tR]} = 0 \,,
 \label{T-constraints}
 \eea
which can easily be derived by inspecting the explicit expressions of $T$ in terms of the physical fields in (\ref{supelcurv})
and by using
the Bianchi identity of $F$. Observe that the first condition in (\ref{field-eqs}) is a consequence
of the first condition in (\ref{T-constraints}).

Next comparing (\ref{supcurv12}) with (\ref{11rt}), one concludes that
\bea
(T_{\tM\tN}^k)_{\tA_1\tA_2\dots \tA_k}=\frac{(-1)^{k+1}}{32} u_{\tM\tN,i}\,
{\mathrm B}(\eta^i, \Gamma_{\tA_1\tA_2\dots \tA_k}\nu)~,~~~k=0, \dots, 5~,
\la{supcurv2}
\eea
where the relation
\bea
\eta\otimes \theta=\frac{1}{32}\sum^5_{k=0} {(-1)^{k+1}\over k!} {\mathrm B}(\eta,
\Gamma_{\tA_1\tA_2\dots \tA_k}\theta)\,  \Gamma^{\tA_1\tA_2\dots
\tA_k}~,
\eea
of bi-spinors to spacetime forms has been used.
Since $T^0$ vanishes identically, consistency requires that the $u$'s must satisfy
\bea
\sum_{i=1}^{32}u_{\tM\tN,i}\, B(\eta^i,\nu)=0~. \la{supcon2}
\eea
This equation is easily solved by choosing an appropriate basis $(\eta^i)$ and setting one of the $u$'s to zero.

 It remains to impose the conditions
(\ref{field-eqs}) and (\ref{T-constraints}) on the $u$'s. For this one uses  the relation (\ref{supcurv2}) and  a representative
for the normal spinor $\nu$ up to  $Spin(10,1)$ transformations. As  the normal spinors are in the same representation as the Killing spinors, and as $Spin(10,1)$ has two different orbits in $\Delta_{32}$ with isotropy groups $SU(5)$ and $Spin(7)\ltimes \bR^9$,  there are two different cases of backgrounds with 31 supersymmetries to be investigated.
We shall not proceed further to carry out the analysis as it is rather technical and can be found in \cite{ugjggpdr2-nmax}.  The key point
to stress though is that the proof requires the use of the field equations and Bianchi identities  in addition of course  to the requirement that the backgrounds preserve 31 supersymmetries.

The possibility remains that backgrounds with 31 supersymmetries can be constructed as discrete quotients of
maximally supersymmetric backgrounds. This possibility has been excluded in \cite{fofpreons11}.
Therefore all $d=11$ backgrounds with 31 supersymmetries are maximally supersymmetric.

\subsection{$N>16$ supersymmetric backgrounds}

In IIB supergravity, one can show that all $N>28$  backgrounds  are maximally supersymmetric \cite{ugjggpdr3-nmax}.
Moreover there is a unique $N=28$ plane wave solution \cite{ugjggp-nmax}.   This is a superposition of the maximally supersymmetric
plane wave and a common sector solution which preserves 28 supersymmetries \cite{ibrr-nmax}.

In $d=11$ supergravity, the results are less stringent. It can be shown though that all backgrounds that preserve $N\geq 30$ supersymmetries
are maximally supersymmetric \cite{ugjggp2-nmax}.
As has been mentioned all $N=31$  IIA supergravity backgrounds  are maximally supersymmetric \cite{ibjaov-nmax}.
It is likely that one can obtain in IIA supergravity stronger results similar to those of IIB. This is because apart from the
gravitino KSE, the theory has an algebraic KSE and therefore the techniques
used for IIB can be applied  in IIA. However no such investigation has  taken place.  It should be noted that there are several solutions known,  all plane waves, that preserve $16<N<32$ supersymmetries in $d=10$ and $d=11$
supergravities but these have not been systematically constructed.

\subsection{The homogeneity theorem} \label{homtheorem}

 The homogeneity conjecture states the following.
 \begin{enumerate}
 \item[-] All supergravity backgrounds that preserve more than half of the supersymmetry of a theory are
 locally isometric to Lorentzian homogenous spaces.
 \end{enumerate}
   The conjecture has been confirmed \cite{fofjh1} for $d=11$ and type II $d=10$ supergravities.
 So all backgrounds of these theories that preserve $N>16$ supersymmetries are locally isometric to $d=11$ and $d=10$ Lorentzian homogeneous spaces, respectively.  The proof of this result is remarkably simple and can be demonstrated as follows.

 The aim of the proof is to show that if a background preserves more than half of the supersymmetry of a supergravity theory, then
 its tangent space at every point will be spanned by the Killing vectors  constructed as   bilinears of Killing spinors.  As a result, it admits
 a transitive group action of isometries.  The calculation can be done point-wise on the spacetime $M$.  For this consider two Killing spinors
 $\epsilon$ and $\eta$ and  the Killing vector bilinear
 \bea
 X\vert_p=\langle \epsilon, \Gamma^\tA\eta \rangle_s~ \partial_\tA\vert_p
 \label{hombi}
 \eea
 evaluated at a point $p\in M$,
where $\langle \cdot, \cdot \rangle_s$ is a suitable spin invariant inner product over $\bR$ such that the above bilinear is associated
to a Killing vector field on $M$.
Such a bilinear always exists in supergravity theories and the particular choice is not relevant for the argument that follows.

Let us identify the tangent and co-tangent bundles using the spacetime metric. If for all Killing spinors $\epsilon$ and $\eta$ the bilinears (\ref{hombi}) span $T_pM$ at every $p\in M$, there is nothing to show.  Suppose instead that
they do not.  In such a case, there is a vector field $Y$, $Y_p\not= 0$,  which is normal to the span of all bilinears and therefore
\bea
(Y^\tA X_\tA)\vert_p= \langle \epsilon, Y^A \Gamma_\tA \eta\rangle_s\vert_p=0~.
\eea
The last relation implies that the Clifford algebra operation $Y^A \Gamma_A\vert_p$ is a map from the bundle  of Killing spinors ${\cal K}$ to its normal ${\cal K}^\perp$,  $Y^A \Gamma_A\vert_p:~~{\cal K}_p\rightarrow {\cal K}\vert_p^\perp$.  ${\cal K}$ is a subbundle
of the spin bundle of the supergravity theory whose fibre  at every point $p$, ${\cal K}_p$,  is spanned by the Killing spinors at that point and the normal ${\cal K}^\perp$ is taken with respect to $\langle \cdot, \cdot \rangle_s$.

However $(Y^A \Gamma_A)^2= Y^2 {\bf 1}$.  This implies that if $Y\vert_p$ is either timelike or spacelike, $(Y^\tA \Gamma_\tA)\vert_p$ is an injection
as the kernel is $\{0\}$.  But if the solutions preserve more than half of the supersymmetry, this is in conflict with the assumption that $\mathrm{rank}\, {\cal K}>\mathrm{rank}\, {\cal K}^\perp$.  Therefore $Y\vert_p=0$ and the bilinears (\ref{hombi}) span $T_pM$.

It remains to investigate the possibility that $Y\vert_p$ is null with $Y\vert_p\not=0$.  Focusing on the $d=11$ and $d=10$  type II  supergravities at hand,  if a solution preserves more than half of the supersymmetry, then at least one of the bilinears (\ref{hombi}) will have to be timelike. The maximal number of linearly independent  Killing spinors that can give only null vector bilinears is 16.  The presence of an  additional Killing spinor, which is the case here as $N>16$,  will give rise to a timelike vector bilinear.  This becomes rather
apparent
after looking at the description of the relevant spinor representations in terms of forms.  As a result, the normal $Y$ to the span
of the bilinears (\ref{hombi}) cannot be null because then it cannot be orthogonal to the timelike Killing vector bilinears.
This is a contradiction of our assumption that $Y\vert_p$ is null and therefore we must set again $Y\vert_p=0$.  This proves that all backgrounds
of $d=11$ and $d=10$  type II supergravities theories that preserve strictly more  than 16 supersymmetries must  locally be
Lorentzian homogeneous spaces.  This result also applies to heterotic supergravity and is expected to hold to many other theories as well.

\section{Horizons}

\subsection{Symmetry enhancement near black hole and brane horizons}

A key phenomenon which has spearheaded many of the most well-known  examples of the AdS/CFT correspondence is that
 there is a (super)symmetry enhancement   near certain black-hole and brane horizons, see e.g.~\cite{bc-hor, ggghpt-hor}.   In particular the near horizon geometry of the extreme Reissner-Nordstr\"om black hole is $AdS_2\times S^2$.  So
 the $\bR\times SO(3)$ isometry group of the black hole solution enhances near the horizon to $SL(2, \bR)\times SO(3)$. The observation that the near horizon geometry of D3-branes is
 $AdS_5\times S^5$ has led to the most celebrated example of the $AdS_5/CFT_4$ correspondence which  states that string theory on $AdS_5\times S^5$  is dual to the (maximally supersymmetric)  ${\cal N}=4$ $d=4$ gauge theory. The isometry group $SO(4,2)\times SO(6)$
 of $AdS_5\times S^5$ is identified with the product of the conformal times the R-symmetry groups of the  gauge theory.

To illustrate how symmetry enhances near horizons, consider   the  Reissner-Nordstr\"om black hole with mass $M$ and charge $Q$.
The metric can be written as
\bea
ds^2=-{\Lambda \over \rho^2} dt^2+ \rho^2 \Lambda^{-1} d\rho^2+\rho^2 ds^2(S^2)~,
\eea
where $\Lambda= \rho^2-2M\rho+ Q^2= (\rho-\rho_+) (\rho-\rho_-)$ and $\rho_\pm= M\pm\sqrt{M^2-Q^2}$ are the radii of inner
and outer horizons.
Introduce Eddington-Finkelstein coordinates as
$d\rho^*= \rho^2 \Lambda^{-1} d\rho$ and $u=t+\rho^*$
to rewrite the metric as
\bea
ds^2=-{\Lambda\over \rho^2} du^2+2du d\rho+ \rho^2 ds^2 (S^2)~.
\eea
Next,  define the coordinate $r=\rho-\rho_+$ centred at the outer horizon
and observe that the metric is  analytic in $r$. Expanding around $r=0$, one has

\bea
ds^2&=&2du \big[dr- {1\over2} \Big(r\,{\rho_+-\rho_-\over \rho_+^2}+ r^2 \,{2\rho_--\rho_+\over \rho_+^3}+{\cal O}(r^3)\Big) du\big]
\cr
&+&(\rho_+^2+2 r\, \rho_++ r^2) ds^2(S^2)~.
\eea
The linear term in $r$ is the surface gravity of the horizon.  The 2-sphere $S^2$ is the ``spatial horizon section'' of the horizon. For an extreme  black hole, $\rho_-=\rho_+$, one can scale the coordinates as
$u\rightarrow \ell^{-1} u$ and $r\rightarrow \ell\, r$ and take the limit $\ell\rightarrow 0$ to find
\bea
ds^2=2 du \big [dr -{1\over2} r^2\, {1\over \rho_+^2} du]+  \rho_+^2 ds^2(S^2)~,
\eea
which is a metric on $AdS_2\times S^2$.  The geometry in the limit  is the  ``near horizon geometry'' of the extreme black hole.
For non-extreme black holes, the limit $\ell\rightarrow 0$ diverges and therefore the  notion of a near horizon geometry is not well defined.

As has already been mentioned,   the $\bR\times SO(3)$ isometry group  of the Reissner-Nordstr\"om black hole in the limit enhances
to $SL(2, \bR)\times SO(3)$.  In addition, viewing the extreme Reissner-Nordstr\"om black hole as a solution of the  ${\cal N}=2$ $d=4$ minimal
supergravity,  the $N=4$ supersymmetry of the solution also enhances to $N=8$ near the horizon.  The emergence of
the conformal group $SL(2,\bR)$ has been extensively utilized in the microscopic counting of black hole entropy, see e.g.~\cite{as-hor}.

\subsection{The horizon conjecture}

Before, we proceed to state the horizon conjecture in detail, let us describe a model of a spacetime with an extreme Killing horizon and  its near horizon
geometry. Killing horizons  are  spacetime hypersurfaces  where a timelike Killing vector field becomes null.  The event horizons of many of the black holes of interest are Killing horizons.  In fact, under some natural assumptions all the event  horizons of $d=4$ black holes are Killing \cite{sh-hor}.
In what follows the focus will be on the metric but the analysis can be extended to include other fields like the form fluxes of
supergravity theories.  It has been shown in \cite{vmji-hor, hfirrw-hor} that near a smooth extreme Killing horizon,  one can adapt a coordinate system
such that the metric takes the form
\bea
ds^2=2 du \left(dr+ r\, h_\tI(r, y) dy^\tI-{1\over2} r^2\, \Delta(r, y) du\right)+\gamma_{\tI\tJ}(y,r) dy^\tI dy^\tJ~.
\eea
For $\Delta>0$, $\partial_u$ is a timelike Killing vector field which  becomes null at the hypersurface $r=0$.
The near horizon geometry of the spacetime is defined after scaling the coordinates $u, r$ as  $u\rightarrow \ell^{-1} u, r\rightarrow \ell r$, and then taking the limit $\ell\rightarrow 0$.  The resulting metric is
\bea
ds^2=2 du \left(dr+ r\, h_\tI(y) dy^\tI-{1\over2} r^2\, \Delta(y) du\right)+ \gamma_{\tI\tJ}(y) dy^\tI dy^\tJ~,
\label{gnhm}
\eea
 where $h$, $\Delta$ and $\gamma$ have been evaluated at $r=0$ and therefore they depend only on the $y$ coordinates. This is a metric on an open neighborhood $M$ of the horizon hypersurface.
  The co-dimension two subspace ${\cal S}$ defined by  $u=r=0$ is the ``spatial horizon section''
of the Killing horizon and it is equipped with the metric $d\tilde s^2({\cal S})=\gamma_{\tI\tJ} dy^\tI dy^\tJ$.  For black hole horizons,   ${\cal S}$ is expected to be compact without boundary.

 If the original spacetime with the Killing horizon is a solution of the Einstein equations, then this will also be the case for $M$ with  the near horizon metric (\ref{gnhm}).  Because of this,
 one can consider $M$ independently of the ``parent'' spacetime as a solution of the theory. This is the approach that will be adopted from now on in the analysis that follows.

  Let  $M$ be a spacetime with metric  (\ref{gnhm}), and possibly non-trivial fluxes, that solves the field equations of a supergravity theory and preserves at least one supersymmetry.  In addition, assume   that the fields are smooth and the spatial horizon section ${\cal S}$ is compact without boundary.     Then the horizon conjecture \cite{jggp-hor, ugjggp-hor} states  the following.

 \begin{enumerate}

 \item[-]
 The number of Killing  spinors $N$   of   $M$ are
 \bea
 N=2 N_-+\mathrm{Index}(D_E)~,
 \label{nindex}
  \eea
  where $N_-\in \bZ_{>0}$ and   $D_E$ is a Dirac operator, defined on the horizon sections ${\cal S}$, which is possibly twisted with vector bundle  $E$. The choice of $E$ depends on the gauge symmetries
of supergravity theory.

\item[-]  If $M$ has  non-trivial fluxes and $N_-\not=0$, then $M$ will admit an $\mathfrak{sl}(2,\bR)$ isometry subalgebra.
 \end{enumerate}
The conjecture has been proven for various theories which include $d=11$ \cite{jggp-hor},  (massive) IIA \cite{ugjgukgp-hor, ugjgukgp2-hor} , IIB \cite{ugjggp-hor} and heterotic supergravities \cite{afjggp-hor}. It has also been
demonstrated for the minimal gauged ${\cal N}=1$ $d=5$ supergravity \cite{jgjggpws-hor}, the ${\cal N}=2$ $d=4$ gauged supergravity coupled to any number
of vector fields \cite{jgtmgp-hor} and the ${\cal N}=1$ $d=5$ supergravity coupled to any number of vector fields \cite{uk-hor}.

We shall demonstrate the proof  of the horizon conjecture in $d=11$ supergravity but before we do this let us first explain some of its consequences.
First,  if the index vanishes, $\mathrm{Index}(D_E)=0$,  which is the case for non-chiral theories, then $N$ is  even. In particular, all
odd dimensional near horizon geometries  preserve an even number of supersymmetries.    Therefore if a near  horizon geometry  preserves
one supersymmetry, possibly inherited from the parent spacetime, it will necessarily preserve another one, and therefore
it will exhibit supersymmetry enhancement.

 The near horizon geometries  with non-trivial fluxes of all non-chiral supergravity theories admit  an $\mathfrak{sl}(2, \bR)$ isometry subalgebra.  Observe that the near horizon geometry (\ref{gnhm}) admits two Killing vector fields $\partial_u$ and $u\partial_u-r\partial_r$.  Their Lie bracket algebra is solvable. The conjecture states that there must be an additional
 isometry such that all three together generate $\mathfrak{sl}(2, \bR)$.  Therefore, all such near horizon geometries exhibit
 symmetry enhancement.

 On the other hand if $N_-=0$, then $N=\mathrm{index}(D_E)$.  The number of Killing spinors is determined by the topology of ${\cal S}$.  It turns out that such near horizon geometries are rather restricted. Typically all the form fields strengths vanish and the scalars are constant.
 Such near horizon geometries, up to discrete identifications,  are products of the form $\bR^{1,1}\times {\cal S}$, where ${\cal S}$ is a product of Berger
 manifolds that admit parallel spinors.  The formula $N=\mathrm{index}(D_E)$ becomes a well-known relation between the index of the Dirac
 operator and the number of parallel spinors on certain Berger type of manifolds.

 \subsection{Proof of the conjecture in $d=11$}

\subsubsection{Preliminaries}

Consider  a solution of $d=11$ supergravity with a Killing horizon that satisfies all the assumptions made for the validity
of the horizon conjecture.  The near horizon geometry apart from the metric also exhibits a non-trivial 4-form flux $F$.
The near horizon  fields  are
\bea
ds^2 &=& 2 {\bf{e}}^+ {\bf{e}}^- + ds^2({\cal S})~,~~~
F = {\bf{e}}^+ \wedge {\bf{e}}^- \wedge Y
+ r {\bf{e}}^+ \wedge d_h Y + Z~,~~~
\label{hd11fields}
\eea
where $Y$ and $Z$ are a 2-form and a 4-form on ${\cal S}$, respectively, which depend only of the coordinates $y$,  $d_hY=dY-h\wedge Y$, and
\bea
{\bf{e}}^+ = du~,~~~{\bf{e}}^- = dr + r h - {1 \over 2} r^2 \Delta du~,~~~
{\bf{e}}^i = e^i{}_\tJ dy^\tJ~,
\label{11hframe}
\eea
is a coframe with $d \tilde s^2({\cal S})=\delta_{ij} e^i{}_\tI e^j{}_\tJ dy^\tI dy^\tJ$. Clearly the metric is of the form in
(\ref{gnhm}).

The Bianchi identities (\ref{d11bids}) and  field equations (\ref{d11feqn})  of the flux $F$ are rewritten in terms of   the horizon fields as
\begin{eqnarray}
\label{f11feq}
dZ=0 \ ,~~~ d_h \star_{{}_9} Z-\star_{{}_9} d_hY = Y \wedge Z~,~~~-d \star_{{}_9} Y = {1 \over 2} Z \wedge Z~,
\end{eqnarray}
where $\star_{9}$ is the Hodge star operation on ${\cal S}$. The  spacetime volume form  has been decomposed as $\epsilon_{{}_{11}}= {\bf{e}}^+ \wedge {\bf{e}}^- \wedge \epsilon_{{}_9}$, where  $\epsilon_{{}_9}$
is the volume form of ${\cal S}$. Similarly, the independent  Einstein equations are

\begin{eqnarray}
\label{11ein1}
{\tilde {R}}_{ij} + {\tilde{\nabla}}_{(i} h_{j)} -{1 \over 2} h_i h_j = -{1 \over 2} Y^2_{ij}
+{1 \over 12} Z^2_{ij}
+ \delta_{ij} \bigg( {1 \over 12} Y^2
-{1 \over 144} Z^2 \bigg)~,
\end{eqnarray}
and
\begin{eqnarray}
\label{11ein2}
{\tilde{\nabla}}^i h_i = 2 \Delta + h^2 -{1 \over 3} Y^2
-{1 \over 72} Z^2~,
\end{eqnarray}
where $\tilde \nabla$ and  ${\tilde{R}}_{ij}$ are the Levi-Civita connection and the Ricci tensor of ${\cal{S}}$, respectively.
For the rest of the notation see appendix \ref{formnote}.

An outline of the proof  of the first part of the horizon conjecture is as follows.  First one integrates  the $d=11$ supergravity  KSE along the coordinates  $r,u$ which appear explicitly in the expressions
for the fields in (\ref{hd11fields}) and determines
the remaining independent  KSEs.  Typically these are    parallel transport equations acting on spinors  that depend on the fluxes and can be thought of as a restriction of the gravitino KSE of the theory on ${\cal S}$.
 From these, one can define certain Dirac like operators, the  horizon Dirac operators. A key next step is the proof of
two Lichnerowicz type theorems which  relate the zero modes of the horizon Dirac operators to the Killing spinors on  ${\cal S}$.
Then the index theorem is used to count the number of Killing spinors and establish the formula (\ref{nindex}).

For   the second part of the conjecture, one shows that   for horizons with non-trivial fluxes and $N_-\not=0$  there is always a pair of Killing spinors whose three vector bilinears are Killing  and their Lie bracket algebra is
$\mathfrak{sl}(2,\bR)$. This  establishes  the horizon conjecture  for $d=11$ supergravity.

\subsubsection{KSEs on the spatial horizon section} \label{horsolkse}

The KSE of $d=11$ supergravity (\ref{11kse}) can be integrated along the $u,r$ coordinates of the near horizon geometry
to yield
\bea
\epsilon= \phi_+ + u \Gamma_+ \Theta_- \phi_-+\phi_-+r \Gamma_-
\Theta_+\left(\phi_+ + u \Gamma_+ \Theta_- \phi_-\right)~,
\label{h11ks}
\eea
where the spinors $\phi_\pm$ satisfy $\Gamma_\pm \phi_\pm =0$, depend only of the coordinates
of ${\cal S}$, $\phi_\pm=\phi_\pm(y)$,  and
\bea
\Theta_\pm={1 \over 4} \slashed {h} +{1 \over 288} \slashed {Z} \pm {1 \over 12} \slashed {Y}~.
\eea
See appendix \ref{formnote} for the notation.
Substituting (\ref{h11ks}) back into  of the KSE (\ref{11kse}) leads
to a plethora of additional  equations on $\phi_\pm$. These include the conditions
\begin{eqnarray}
\label{cc1}
\bigg( {1 \over 2} \Delta
+2 \big({1 \over 4} {\slashed{h}} -{1 \over 288} {\slashed{Z}}+{1 \over 12} {\slashed{Y}} \big)
\Theta_+ \bigg) \phi_+ =0~,
\end{eqnarray}
and
\begin{eqnarray}
\label{cc3}
\bigg( -{1 \over 2} \Delta
+ 2 \big(-{1 \over 4} {\slashed{h}} +{1 \over 288} {\slashed{Z}} +{1 \over 12} {\slashed{Y}}\big)
\Theta_- \bigg)
\phi_- =0~,
 \end{eqnarray}
which will be used later in the investigation of warped AdS backgrounds.
However after some involved analysis described in \cite{jggp-hor}, which
makes an essential use of the field equations and Bianchi identities in (\ref{f11feq}), (\ref{11ein1}) and (\ref{11ein2}), one finds that  the remaining independent KSEs  are
\bea
\nabla_i^{(\pm)}\phi_\pm \equiv {\tilde{\nabla}}_i \phi_\pm + \Psi^{(\pm)}_i \phi_\pm =0~,
\label{s11kses}
\eea
where
\bea
\Psi^{(\pm)}_i &=& \mp{1 \over 4} h_i -{1 \over 288} {\slashed \Gamma\mkern-4.0mu Z}_i +{1 \over 36} \slashed{Z}_i
\pm{1 \over 24} {\slashed \Gamma\mkern-4.0mu Y}_i \mp{1 \over 6} \slashed{Y}_i~.
\eea
These can be thought of as suitable restrictions of the gravitino KSE (\ref{11kse}) on the spatial horizon section ${\cal S}$.
Because of this we also refer to $\phi_\pm$ as Killing spinors on ${\cal S}$.
In addition, it turns out that if $\phi_-$ is a Killing spinor, $\nabla_i^{(-)}\phi_-=0$, then $\phi'_+=\Gamma_+\Theta_-\phi_-$
will also be a Killing spinor, i.e.
\bea
\nabla_i^{(+)}\phi'_+ =0~.
\label{phi+phi-}
\eea
This is the first indication that there may be a doubling in the number of Killing spinors for near horizon spacetimes.

\subsubsection{Lichnerowicz type theorems}

To continue with the proof of the formula (\ref{nindex}),   the Killing spinors $\phi_\pm$  are related to  the zero modes of Dirac like operators on ${\cal S}$.
This is done via a Lichnerowicz type theorem.  As a reminder, the classic  Lichnerowicz theorem is as follows.
Suppose that ${\cal D}$ is the Dirac operator on a Riemannian manifold $W$ which is compact without boundary.  It can be established
that ${\cal D}^2=\nabla^2-(1/4) R$, where $\nabla$ is the Levi-Civita connection of $W$ and  $R$ its  Ricci scalar. After a partial integration, one has that
\bea
\int_W \parallel  {\cal D} \eta\parallel^2= \int_W \parallel  \nabla \eta\parallel^2+ {1\over4} \int_W  R  \parallel\eta\parallel^2~,
\eea
where all the inner-products are positive definite and $\eta$ is a spinor.
Clearly, if $R=0$, then all the zero modes of the Dirac operator are parallel and vice-versa.

Returning to the near horizon geometries, define the  ``horizon Dirac operators'' as
\bea
{\cal{D}}^{(\pm)}\phi_\pm\defeq \Gamma^i \nabla^{(\pm)}_i\phi_\pm=\Gamma^i {\tilde{\nabla}}_i \phi_\pm + \Psi^{(\pm)} \phi_\pm  \ ,
\eea
where
\bea
\Psi^{(\pm)} = \Gamma^i \Psi^{(\pm)}_i = \mp{1 \over 4} \slashed{h} +{1 \over 96} \slashed {Z} \pm{1 \over 8} \slashed {Y}~.
\eea

Clearly, if $\nabla_i^{(\pm)}\phi_\pm =0$ , then  ${\cal{D}}^{(\pm)}\phi_\pm=0$. The converse is also true, i.e.
\bea
\nabla_i^{(\pm)}\phi_\pm=0 \Longleftrightarrow {\cal{D}}^{(\pm)}\phi_\pm=0~.
\label{lichntype}
\eea
The proof of this for  the  ${\cal{D}}^{(+)}$ operator
relies on the use of the Hopf maximum principle. Using the field equations and Bianchi identities (\ref{f11feq}), (\ref{11ein1}), (\ref{11ein2}) and  assuming that  ${\cal{D}}^{(+)} \phi_+ =0$, one can establish that
\bea
{\tilde{\nabla}}^i {\tilde{\nabla}}_i \parallel \phi_+\parallel^2 -h^i {\tilde{\nabla}}_i \parallel \phi_+\parallel^2
= 2 \langle {{\nabla}}^{(+)}{}^i \phi_+ , {{\nabla}}^{(+)}_i \phi_+ \rangle~,
\label{plusmax}
\eea
where $\langle\cdot, \cdot\rangle$ is the $Spin(9)$ invariant Hermitian inner product, see appendix \ref{app:spinors}.
As $\langle\cdot, \cdot\rangle$ is positive definite, the right-hand-side of the equation above is positive semi-definite.
On the other hand, $\parallel \phi_+\parallel^2$ as a function on the compact manifold  ${\cal S}$  has a global maximum and a global minimum.  These are critical points and  the hessian is either negative  or positive definite, respectively. Therefore, the left-hand-side of (\ref{plusmax}) changes sign while the right-hand-side is definite. So consistency requires that both sides must vanish
establishing (\ref{lichntype}) for the ${\cal{D}}^{(+)}$ operator and
\bea
\parallel \phi_+\parallel^2=\mathrm{const}~.
\label{phi+const}
\eea
The constancy of the length of $\phi_+$ will later be used  in the investigation of the $\mathfrak{sl}(2, \bR)$ symmetry.

The proof of (\ref{lichntype}) for the ${\cal{D}}^{(-)}$ operator uses a partial integration argument as that
of the standard Lichnerowicz theorem stated in the beginning of the section. In particular after imposing the field equations and Bianchi identities (\ref{f11feq}), (\ref{11ein1})and  (\ref{11ein2}), one can establish that
\bea
\int_{\cal{S}} \parallel{\cal{D}}^{(-)} \phi_- \parallel^2&=&\int_{\cal{S}}  \parallel {{\nabla}}^{(-)}\phi_-  \parallel^2 + \int_{\cal{S}}\langle {\cal C}\phi_-, {\cal{D}}^{(-)}\phi_-\rangle~,
\eea
where ${\cal C}$ is a Clifford algebra element that depends on the fluxes. As $\langle\cdot, \cdot\rangle$ is positive definite, if $\phi_-$ is a zero mode of ${\cal{D}}^{(-)}$, then it will  satisfy ${{\nabla}}^{(-)}\phi_-=0$ which proves the statement.

\subsubsection{Counting the Killing spinors}

After the proof of  the Lichnerowicz type theorems in the previous section, the apparatus  to prove the first part of the horizon conjecture is in place. As ${\cal S}$ is an odd-dimensional manifold, the index of the Dirac operator vanishes  and therefore (\ref{nindex}) gives $N=2 N_-$.

To demonstrate this, the spacetime spin bundle $S$ restricted on ${\cal S}$  splits as  $S=S_+\oplus S_-$, where the sections of $S_\pm$ are $\phi_\pm$,   $\phi_\pm\in \Gamma(S_\pm)$. Note that $S_+$ and $S_-$ are isomorphic as $Spin(9)$ bundles. Observe that the horizon Dirac operator acts as  ${\cal D}^{(+)}: ~\Gamma(S_+)\rightarrow \Gamma(S_+)$ and similarly its adjoint  $({\cal D}^{(+)})^\dagger: ~\Gamma(S_+)\rightarrow \Gamma(S_+)$, where the adjoint has been taken  with respect to the $Spin(9)$ invariant inner product $\langle\cdot, \cdot\rangle$, see appendix \ref{app:spinors}.

The horizon Dirac operator ${\cal D}^{(+)}$ has the same principal symbol as the standard Dirac operator ${\cal D}$ and so $\mathrm{Index}({\cal D}^{(+)})=\mathrm{Index}({\cal D})=0$ as the index of ${\cal D}$ vanishes. Thus
 \bea
{\rm dim}\, {\rm ker}\, {\cal D}^{(+)}  = {\rm dim}\, {\rm ker}\, ({\cal D}^{(+)})^\dagger ~.
\label{index}
\eea
On the other hand,   $({\cal D}^{(+)})^\dagger\Gamma_+=\Gamma_+{\cal D}^{(-)}$,  and so
\bea
  {\rm dim}\, {\rm ker}\, ({\cal D}^{(+)})^\dagger={\rm dim}\, {\rm ker}\, {\cal D}^{(-)}~.
\eea
    Therefore, one establishes that
\bea
{\rm dim} \, {\rm ker}\, {\cal{D}}^{(+)} = {\rm dim}\, {\rm ker}\, {\cal{D}}^{(-)}~.
\eea
The number of supersymmetries of a near horizon geometry is the number of $\nabla^{(\pm)}$ parallel spinors  and so from the Lichnerowicz type theorems and the index argument above
\bea
N={\rm dim} \, {\rm ker}\, {\cal{D}}^{(+)} + {\rm dim}\, {\rm ker}\, {\cal{D}}^{(-)}=2\, {\rm dim}\, {\rm ker}\, {\cal{D}}^{(-)}=2 N_-.
\eea
 This proves that the  number of supersymmetries preserved by  M-horizon geometries is  even confirming the first part of the  horizon conjecture
 for $d=11$ supergravity.

\subsubsection{Emergence of conformal symmetry}

The main task is to select two spacetime Killing spinors and demonstrate that the associated vector bilinears satisfy an
$\mathfrak{sl}(2,\bR)$ Lie bracket algebra.  As the near horizon geometries under investigation preserve some supersymmetry, $N=2N_-\not=0$, there is a $\phi_-$ Killing spinor. As a consequence of (\ref{phi+phi-}),
$\phi_+= \Gamma_+\Theta_-\phi_-$ will also be a Killing spinor.  The existence of both $\phi_+$ and $\phi_-$ suffices to construct two spacetime
Killing spinors provided that $\phi_+\not=0$, i.e.~$\phi_-$ is not in the kernel of $\Theta_-$. Indeed one can show that
\bea
\mathrm{Ker}\, \Theta_-\not=\{0\}\Longleftrightarrow F=0,\, h=\Delta=0~.
\label{kertheta}
\eea
So if $\mathrm{Ker}\, \Theta_-\not=\{0\}$, the fluxes will vanish and therefore the near horizon geometries will be products $\bR^{1,1}\times S^1\times N^8$, where $N^8$ has holonomy contained in $Spin(7)$.

To sketch of proof of  (\ref{kertheta}) assume that the kernel of $\Theta_-$ is non-trivial and hence there is a $\phi_-\not=0$ such that
$\Theta_-\phi_-=0$. Taking the inner product of $\Theta_-\phi_-=0$ with $\phi_-$, one finds that $\Delta=0$ as $\parallel\phi_-\parallel$
is no-where vanishing.  Using the maximum principle and a partial integration argument,  one can similarly proceed to prove (\ref{kertheta}). The details of the proof can be found in \cite{jggp-hor}.

Therefore for horizons with non-trivial fluxes, for every $\phi_-$ Killing spinor there is an associated  non-trivial Killing spinor $\phi_+= \Gamma_+\Theta_-\phi_-$.  In turn, the near horizon spacetime admits  two Killing spinors given by
\bea
\epsilon_1&=&\epsilon(\phi_-, 0)=\phi_-+u \phi_++ru \Gamma_-\Theta_+\phi_+~,
\cr
\epsilon_2&=&\epsilon(\phi_-, \phi_+)=\phi_++r \Gamma_-\Theta_+\phi_+~,~~~~\phi_+=\Gamma_+\Theta_-\phi_-~.
\eea
These give rise to 3 Killing vector bi-linears, see also (\ref{allbilinears}), given by
\bea
X_1\defeq \langle(\Gamma_+-\Gamma_-) \eps_1, \Gamma^\tA \eps_2\rangle \partial_\tA&=&-2u \parallel\phi_+\parallel^2 \partial_u+ 2r \parallel\phi_+\parallel^2 \partial_r+ V^i \tilde \partial_i~,
\cr
X_2\defeq \langle(\Gamma_+-\Gamma_-) \eps_2, \Gamma^\tA \eps_2\rangle \partial_\tA&=&-2 \parallel\phi_+\parallel^2 \partial_u~,
\cr
X_3\defeq \langle(\Gamma_+-\Gamma_-) \eps_1, \Gamma^\tA \eps_1\rangle \partial_\tA&=&-2u^2 \parallel\phi_+\parallel^2 \partial_u +(2 \parallel\phi_-\parallel^2
\cr
&&+ 4ru \parallel\phi_+\parallel^2)\partial_r+ 2u V^i \tilde \partial_i~,
\eea
where $V=\langle \Gamma_+\phi_-, \Gamma^i \phi_+\rangle\, \tilde \partial_i$ is a Killing vector on ${\cal S}$ which leaves all the data invariant and $\tilde \partial_i=e_i^\tI \partial_\tI$.  To simplify somewhat the expressions for the Killing vector fields above, we have used that
\bea
-\Delta \parallel\phi_+\parallel^2+4 \parallel\Theta_+\phi_+\parallel^2=0~,~~~\langle\phi_+, \Gamma_i \Theta_+\phi_+\rangle=0~.
\eea
These follow either from the KSEs or equivalently from the Killing condition on $X_1$, $X_2$ and $X_3$.

A straightforward computation reveals that the Lie bracket algebra of $X_1$,  $X_2$ and $X_3$  is  $\mathfrak{sl}(2,\bR)$ ,
\bea
&&[X_1,X_2]=2 \parallel\phi_+\parallel^2 X_2~,~~~[X_2, X_3]=-4 \parallel\phi_+\parallel^2 X_1  ~,~~~
\cr
&&[X_3,X_1]=2 \parallel\phi_+\parallel^2 X_3~,
\eea
where $\parallel\phi_+\parallel$ is constant, see (\ref{phi+const}).  Note that the emergence of the $\mathfrak{sl}(2,\bR)$ symmetry is dynamical as the proof of its existence  requires the use of the field equations.  This completes the proof of the second part of the horizon conjecture.

A special case arises whenever $V=0$.  This  together with the Killing condition of $X_1$, $X_2$ and $X_3$ imply that $h=\Delta^{-1} d\Delta$, see \cite{jggp-hor} for the proof.  The spacetime is a warped product of $AdS_2$ with the horizon section ${\cal S}$,
$ AdS_2\times_w {\cal S}$.  Therefore, the warped $AdS_2$ solutions of supergravity theories are included in the near horizon geometries
and therefore all the properties proven for the latter also hold for the former.

As has already been mentioned, the horizon conjecture demonstrated for $d=11$ supergravity also holds for other theories including (massive) IIA,  IIB, $d=5$ and $d=4$ supergravities.  In particular, one can show after integrating over the $r,u$ coordinates  that the remaining independent KSEs are those naively expected from
restricting the gravitino and algebraic KSEs on the spatial horizon section ${\cal S}$.  The form of the Killing spinors is exactly as in
(\ref{h11ks}) though of course the field content of $\Theta_\pm$ is different. One of the
additional complications that arises in the proof of
Lichnerowicz type  theorems  is the presence  the  algebraic KSEs, like for example a  dilatino and/or a gaugino KSE.  Nevertheless after an appropriate
choice of Dirac horizon operators, it is possible to prove with the use of   maximum principle and  partial integration arguments that the zero
modes of the horizon Dirac operators are Killing, i.e.~they solve both the gravitino and algebraic KSEs. The $\mathfrak{sl}(2, \bR)$ conformal symmetry emerges in the same way as described for $d=11$ supergravity.

\section{AdS and Minkowski flux compactifications}

Amongst the $d=10$ and $d=11$ supersymmetric solutions which have found widespread applications in supergravity, string, and M-theory compactifications  and in the AdS/CFT correspondence  are warped products of Minkowski and AdS spaces with some internal space, for reviews see e.g.~\cite{mdbncp-ads, mg-ads} and \cite{oasgjmhoyo-ads}. Such backgrounds are characterized by the requirement that they are invariant under the isometry
 group of either the AdS or Minkowski subspaces. Many of the properties of these backgrounds can be investigated in a unified way irrespective on whether they are solutions of $d=11$,  (massive) IIA or IIB supergravities.  These properties include the counting of the number of preserved supersymmetries, as well as the Killing superalgebras. However to be concrete, we  shall mostly present the analysis for the $d=11$ backgrounds and only comment on the results  for  other theories.

\subsection{Warped AdS and Minkowski backgrounds from horizons}

Warped AdS and Minkowski backgrounds are  examples of near horizon geometries possibly allowing   for  non-compact spatial horizon sections \cite{ugjggp-ads}.  To see this, consider the metric
\bea
ds^2= A^2 ds^2(AdS_n)+ ds^2(N^{d-n})~,~~~
\eea
on a warped product of $AdS_n$ with an internal space $N^{d-n}$, $AdS_n\times_w N^{d-n}$, where $A$ is the warp factor which depends only on the coordinates of  $N^{d-n}$ and
\bea
ds^2(AdS_n)&=&e^{{2z\over\ell}} (2dudv+ \sum_{a=1}^{n-3} (dx^a)^2)+dz^2~,~~~n>2~,
\cr
ds^2(AdS_2)&=&2 du(dv-\ell^{-2} v^2 du)~,~~~n=2~,
\eea
is the metric on the Poincar\'e patch of $AdS_n$ space. The parameter  $\ell$ is the radius of $AdS_n$.

To see that the metrics above can be written in a near horizon form (\ref{gnhm}),  for $n>2$ perform the coordinate transformation
\bea
v= A^{-2} e^{-{2z\over\ell}} r~,~~~
\eea
with the rest of the coordinates unchanged, to find that the metric transforms to
\bea
ds^2&=&2 du \left(dr-r({2\over\ell} dz+d\log A^2)\right) +A^2 \left(dz^2+ e^{{2z\over\ell}} \sum_{a=1}^{n-3} (dx^a)^2\right)
\cr &&+ds^2(N^{d-n})~.
\label{adsnmetrl}
\eea
This is a near horizon metric (\ref{gnhm})  with $\Delta=0$,  $h=-({2\over\ell} dz+d\log A^2)$ and metric on the spatial horizon section
\bea
ds^2({\cal S})=A^2 \left(dz^2+ e^{{2z\over\ell}} \sum_{a=1}^{n-3} (dx^a)^2\right)+ds^2(N^{d-n})~.
\eea
The spatial horizon section is the warped product of the  hyperbolic space $H^{n-2}$ with the internal space $N^{d-n}$,
${\cal S}= H^{n-2}\times_w N^{d-n}$.

Similarly for the warped $AdS_2$ backgrounds, $AdS_2\times_w N^{d-n}$,  perform the coordinate transformation $r=vA^2$ to find that the metric can be put into the near horizon form (\ref{gnhm}) with $\Delta=\ell^{-2} A^{-2}$ and $h=-d\log A^2$.  The near horizon section ${\cal S}$ is identified with the internal space $N^{d-2}$, ${\cal S}=N^{d-2}$.

The rest of the form fluxes of the warped $AdS_n$ backgrounds of supergravity theories can also be put into a near horizon form.  In particular a typical $k$-form flux field strength $F$, $k\geq n$, can be written as
\bea
F= d\mathrm{vol}(AdS_n)\wedge W +Z= \fe^+\wedge \fe^-\wedge Y+Z~,
\eea
where $W$ and $Z$ are $(k-n)$- and $k$-forms on $N^{d-n}$ which depend only on the coordinates of $N^{d-n}$. If $k<n$, then $F$ will be written as above with $W=0$.  Note also that  the terms  $d_h Y$ in the fluxes, see (\ref{hd11fields}),  must vanish as they are not invariant under the isometries of $AdS_n$.

The warped Minkowski backgrounds can be viewed as a special case of AdS backgrounds which arise in the limit that the AdS radius $\ell$
goes to infinity, $\ell\rightarrow \infty$.  In particular  notice that in this limit the metric (\ref{adsnmetrl}) on $AdS_n\times_w N^{d-n}$  becomes a metric on
$\bR^{n-1,1}\times_w N^{d-n}$.

\subsection{Solution of KSEs for AdS backgrounds}

In the literature, the KSEs of supergravity theories  have been solved for warped AdS backgrounds in many different ways.  Some approaches involve a  factorization of the spacetime Killing spinors into Killing spinors on the AdS subspace and those on the internal space.  There is an extensive literature on AdS solutions and an incomplete list of works is \cite{pfmr-ads, lclrnw-ads, Klebanov:1998hh, Acharya:1998db, Cvetic:2000cj, Gauntlett:2004zh, Gauntlett:2005ww, Gauntlett:2006ux,  Gaiotto:2009gz, Lust:2009mb, Pilch:2000ej, Kim:2006qu, Itsios:2012zv, Macpherson:2014eza, DHoker:2016ysh, Apruzzi:2015wna, Apruzzi:2014qva, Beck:2015gqa}.
Here we shall adopt
the approach developed in \cite{jggp-ads, sbjggpiib-ads, sbjggpiia-ads} in which the spacetime KSEs are solved directly for these backgrounds without any assumptions. This  utilizes all the technology that has been developed to solve the KSEs
for near horizon geometries described in section \ref{horsolkse}. Apart from generality, this methods
 allows to treat some of the properties of the backgrounds simultaneously without reference to a particular  AdS background or the theory
that it is a solution of.  A comparison  of the different approaches can be found in \cite{ugjggp2-ads}.

To begin let us write the   $d=11$ warped AdS backgrounds  as
\bea
ds^2&=&2\fe^-\fe^++ (\fe^z)^2+ \sum_{a=1}^{n-3} (\fe^a)^2+ ds^2(N^{11-n})~,
\cr
F&=& d\mathrm{vol}(AdS_n)\wedge W+Z=\fe^+\wedge \fe^-\wedge Y+Z~,
\label{adsn11dfields}
\eea
where $W$ and $X$ are $(4-n)$- and 4-forms on $N^{11-n}$ and $Y$ has been introduced to facilitate progress in the analysis that follows.   The spacetime coframe is chosen as
\bea
&&\fe^-=du~,~~~\fe^+=dr -({2\over\ell} dz+d\log A^2)~,~~~\fe^z=A dz~,~~~
\cr
&&\fe^a= A e^{{z\over\ell}} dx^a~,~~~\fe^i=\fe^i{}_\tI dy^\tI~,
\label{adsframe}
\eea
the 2-form $Y$ is
\bea
&&Y=\fe^z\wedge W~,~~(n=3)~;~Y=W \, \fe^z\wedge\fe^1~,~~(n=4)~;~
\cr
&&Y=0~,~~n>4~,
 \eea
and  $ds^2(N^{11-n})=\delta_{ij} \fe^i\fe^j$.

The field equations of $d=11$ supergravity can be rewritten in terms of the component fields of the $AdS_n$ backgrounds as given in
(\ref{adsn11dfields}).  In particular, the field equation for the warp factor, which arises from the Einstein equations along the $AdS_n$ subspace, is
\bea
D^2\log \,A=-{n-1\over \ell^2 A^2}-n \partial^i\log\, A \partial_i\log\, A+{1\over 6} Y^2+{1\over144} Z^2~,
\label{warpfeq}
\eea
where $D$ is the Levi-Civita connection of the internal space $N^{11-n}$.  It can be argued that for smooth solutions $A$ is no-where zero. The rest of the field equations can be found  in \cite{jggp-ads} and they will not be repeated here.

The integration of the $d=11$ gravitino KSE (\ref{11kse}) over the $AdS_n$ subspace of $AdS_n\times N^{d-n}$ leads to an expression for the Killing spinor which is explicit in AdS coordinates. The remaining KSEs include  a restriction of  (\ref{11kse}) to the internal space $N^{11-n}$. In addition to these, there are also new  algebraic conditions on the spinors which arise as integrability conditions of the integration of (\ref{11kse}) over $AdS_n$.     In particular, a spacetime Killing spinor, $\eps$,
can be written as
\bea
\epsilon=\epsilon_1+\epsilon_2+\epsilon_3+\epsilon_4~,
\label{ksp1}
\eea
where
\bea
\epsilon_1&=&\sigma_+~,~~~\epsilon_2=\sigma_--\ell^{-1} e^{{z\over\ell}} x^a \Gamma_{az} \sigma_--\ell^{-1} A^{-1} u \Gamma_{+z} \sigma_-~,
\cr
\epsilon_3&=&e^{-{z\over\ell}}\tau_+-\ell^{-1} A^{-1} r e^{-{z\over\ell}} \Gamma_{-z} \tau_+-\ell^{-1} x^a \Gamma_{az} \tau_+~,~~~\epsilon_4=e^{{z\over \ell}} \tau_-~,
\label{ksp2}
\eea
and all the gamma matrices are in the coframe basis (\ref{adsframe}).
The $\sigma_\pm$ and $\tau_\pm$  spinors satisfy the lightcone projections $\Gamma_\pm\sigma_\pm=\Gamma_\pm\tau_\pm=0$, and depend only on the coordinates of $N^{11-n}$.
Furthermore, $\sigma_\pm$ and $\tau_\pm$ are parallel along the internal space $N^{11-n}$
\bea
{D}_i^{(\pm)} \sigma_\pm=0~,~~~{D}_i^{(\pm)}\tau_\pm=0~,
\label{kseads1}
\eea
and satisfy the algebraic conditions
\bea
\Xi^{(\pm)}\sigma_\pm=0~,~~~\Xi^{(\pm)}\tau_\pm=\mp{1\over \ell A}\tau_\pm~,
\label{kseads2}
\eea
where
\begin{eqnarray}
{D}_i^{(\pm)}&\defeq& D_i\pm {1\over 2} \partial_i \log A-{1\over 288} {\slashed \Gamma\mkern-4.0mu Z}_i
+{1\over 36} {\slashed{Z}}_i\pm {1\over24} {\slashed \Gamma\mkern-4.0mu Y}_i\mp {1\over6} \slashed Y_i~,
\cr
\Xi^{(\pm)}&\defeq& -{1 \over 2}
\Gamma_z {\slashed{\partial}}\log A
\mp{1 \over 2 \ell A} +{1 \over 288} \Gamma_z
{\slashed{Z}}
\pm {1 \over 6}{\slashed{Q}}~,
\label{d11der}
\end{eqnarray}
and where $Q$ is defined by the relation $Y=\fe^z\wedge Q$.
 Note that the last term in the first equation vanishes for $n=4$ as $i_XY=0$ for any vector field $X$ on the  internal space.

The conditions (\ref{kseads1}) on  $\sigma_\pm$ and $\tau_\pm$  are those that are thought of as   restrictions of the spacetime gravitino KSE to the internal space.  While those in  (\ref{kseads2}) are the new ones that arise as integrability conditions. We shall refer to both as the remaining KSEs.

 An outline of the  proof of the formulae  (\ref{ksp1}) and (\ref{ksp2}) for the Killing spinor $\eps$ is as follows. As the warped AdS backgrounds  can be put in a near horizon form, one can integrate along the $u,r$ coordinates to get
the expression of the Killing spinor as for near horizon geometries in (\ref{h11ks}).  Then the independent KSEs (\ref{s11kses}) on $\phi_\pm$  evaluated along the  $z$ coordinate can be written as
\bea
\partial_z\phi_\pm=A\,\Xi^{(\pm)} \phi_\pm~,
\label{zinter}
\eea
where $\Xi^{(\pm)}$ is as in (\ref{kseads2}). Taking another $z$ derivative
of the above equation and using the conditions (\ref{cc1}) and (\ref{cc3}), one finds that
\bea
\partial_z^2\phi_\pm\pm {1\over\ell} \partial_z\phi_\pm=0~.
\eea
Therefore, the solutions are $\phi_\pm=\kappa_\pm+e^{\mp{z\over\ell}} \lambda_\pm$, where $\kappa_\pm$ and $\lambda_\pm$
are independent of the $u, r,z$ coordinates.  Substituting $\phi_\pm$ back into (\ref{zinter}), one finds the algebraic conditions
in (\ref{kseads2}) on $\kappa_\pm$ and $\lambda_\pm$, respectively.    The integration over the remaining $x^a$ AdS coordinates  does not produce  additional
integrability conditions. Performing the integration over the $x^a$ coordinates and using the algebraic KSEs (\ref{kseads2}), one finds the expression for the Killing spinors as in (\ref{ksp1}) and (\ref{ksp2}).

The solution of the KSEs of $d=10$ supergravities for warped AdS   backgrounds proceeds as for the $d=11$ ones described above. The expression for the
Killing spinors is the same as (\ref{ksp1}) and (\ref{ksp2}).  The spinors $\sigma_\pm$ and $\tau_\pm$ satisfy  some remaining KSEs. These include a restriction of the original KSEs of these theories on the internal spaces $N^{10-n}$, like (\ref{kseads1}),  and some additional ones that arise as integrability conditions of the integration of the gravitino KSE over $AdS_n$, like (\ref{kseads2}). Of course the former, apart from parallel transport equations, like those of (\ref{kseads1}) associated with the gravitino KSE, include also  algebraic KSEs  which are restrictions of the algebraic KSE of these theories to $N^{10-n}$. For example in IIA supergravity $\tau_\pm$ and $\sigma_\pm$ satisfy  a condition which is a restriction of  dilatino KSE to the internal space.

\subsection{Counting supersymmetries for warped AdS backgrounds}

The number of supersymmetries preserved by warped AdS backgrounds come with multiplicities.  This is because Clifford algebra operations act on the solutions of the remaining  KSEs (\ref{kseads1}) and (\ref{kseads2}) generating new ones. In particular,  if $\sigma_\pm$ are solutions of (\ref{kseads1}) and (\ref{kseads2}), then
\begin{eqnarray}
\tau_\pm\defeq \Gamma_{za}  \sigma_\pm~,~~\forall\, a~,
\end{eqnarray}
will also be solutions.  As the Clifford algebra operations $\Gamma_{za}$ are invertible, the converse is also true.
\medskip

Similarly, if  $\sigma_+$ and $\tau_+$ are solutions, then
\begin{eqnarray}
\sigma_-\defeq A \Gamma_- \Gamma_z \sigma_+~,~~~\tau_-\defeq A \Gamma_- \Gamma_z \tau_+~,
\end{eqnarray}
will also be  solutions, and vice versa.

Furthermore, if $\sigma_+$ is Killing spinor of (\ref{kseads1}) and (\ref{kseads2}), then
\begin{eqnarray}
\sigma_+'\defeq\Gamma_{ab} \sigma_+~,~~~\forall\,a, b~,~\text{with}~a<b~,
\label{degsigma}
\end{eqnarray}
will also be  Killing spinors. Therefore, one can start from a solution and act with the Clifford algebra operations above to construct a whole multiplet.

The counting proceeds with the identification of the linearly independent solutions  in each multiplet.  The number of Killing spinors of an $AdS_n$ background is the number of Killing spinors in each multiplet times the number of multiplets that can occur. First, we have seen that warped $AdS_2$ backgrounds preserve an even number of supersymmetries as a special case of near horizon geometries. Since for $n\geq 3$,  the $\sigma_-$ and $\tau_-$ solutions are generated from those of $\sigma_+$ and $\tau_+$, it suffices to count the latter.  $AdS_3$ backgrounds can admit either $\sigma_+$ or $\tau_+$ or both $\sigma_+$ and $\tau_+$ Killing spinors. Therefore the multiplet contains always the pair $\sigma_\pm$ or $\tau_\pm$ or both,   and so these backgrounds preserve $2k$ supersymmetries.

 For $AdS_n$, $n>3$, the $\tau_+$ solutions are generated from those of $\sigma_+$.   Therefore the number of linearly independent Killing spinors in a multiplet is four times the linearly independent $\sigma_+$ spinors that arise from the application of (\ref{degsigma}). For warped $AdS_4$ backgrounds, (\ref{degsigma}) does not produce any degeneracy and so the number of supersymmetries preserved are 4k. For warped $AdS_5$ backgrounds, (\ref{degsigma}) yields two linearly independent $\sigma_+$ spinors for each multiplet.  Therefore these backgrounds preserve $8k$ supersymmetries. A similar counting leads to the conclusion that warped $AdS_6$ backgrounds preserve $16k$  supersymmetries.  Note though that to correctly count the number of Killing spinors for warped  $AdS_7$ backgrounds,  the $\sigma_+$ Killing spinor which is used to construct a  multiplet can be chosen to satisfy the condition $\Gamma_{1234}\sigma_+=\pm \sigma_+$.  Such a choice leads to the possibility of
 warped $AdS_7$ backgrounds that can  preserve $16$ supersymmetries.  These results are tabulated in table \ref{tab2c}.  The counting of supersymmetries of warped AdS backgrounds in $d=10$ type II supergravities can be done in the same way leading to the same results as for $d=11$ supergravity.

\begin{table}[h!]
\centering
\begin{tabular}{|c|c|c|}\hline
$$& $N~\text{for}~ AdS_n$ & $N ~\text{for}~\bbR^{n-1,1}$
 \\
\hline\hline
$n=2$&$2k$&$-$
\\
\hline
$n=3$&$2k$&$2k$
\\
\hline
$n=4$&$4k$&$4k$
\\
\hline
$n=5$&$8k$&$8k$
\\
\hline
$n=6$&${ 16k}$&$8k$
\\
\hline
$n=7$&${ 16k}$&$16k$
\\
\hline
\end{tabular}
\caption{The proof that warped $AdS_2$ backgrounds preserve $2k$ supersymmetries requires that the fields are smooth and the internal space is compact without boundary. For the rest of $AdS_n$ and $\bR^{n-1,1}$ backgrounds, no such assumptions are necessary.  In all cases $N\leq 32$. This couting of supersymmetries applies to all $d=11$ and $d=10$ type II  supergravities.}
 \label{tab2c}
\end{table}

\subsection{KSEs and counting supersymmetries for warped Minkowski backgrounds}

To find the Killing spinors of warped Minkowski backgrounds, one follows the same steps as in the AdS case.
It turns out that the Killing spinors can be written as
\begin{eqnarray}
\epsilon&=&\sigma_++ u \Gamma_+\Gamma_z\Xi^{(-)}\sigma_-+\sum_m x^m\Gamma_m A \Gamma_z\Xi^{(+)}\sigma_+
\cr
&&~~+\sigma_-+r \Gamma_-\Gamma_z\Xi^{(+)} \sigma_++\sum_m x^m\Gamma_m A \Gamma_z\Xi^{(-)}\sigma_-~,
\label{minksp1}
\end{eqnarray}
where the coordinates of the Minkowski space are $(u,r, x^m)=(u,r, z, x^a)$ and all the gamma matrices are in a coframe basis.  The remaining KSEs are
\bea
D^{(\pm)}_i\sigma_\pm=0~,~~~(\Xi^{(\pm)})^2\sigma_\pm=0~,
\label{minkremkses}
\eea
where $D^{(\pm)}_i$ is given in (\ref{d11der}) and
\bea
\Xi^{(\pm)}&\defeq& -{1 \over 2}
\Gamma_z {\slashed{\partial}}\log A
 +{1 \over 288} \Gamma_z
{\slashed{Z}}
\pm {1 \over 6}{\slashed{Q}}~.
\label{minkxi}
\eea
Before we describe the proof of this observe that the Killing spinors may depend on the coordinates of the Minkowski space provided that
$\sigma_\pm\notin \mathrm{ker}\,\Xi^{(\pm)}$.  This may seem a bit puzzling but it should be allowed. This is because
$AdS_n$ spaces in the Poincar\'e patch can be viewed as warped Minkowski $\bR^{n-2,1}$ backgrounds, $AdS_n=\bR^{n-2,1}\times_w \bR$, and we have demonstrated that
Killing spinors of AdS spaces (\ref{ksp1}) and (\ref{ksp2}) depend on all AdS coordinates including those of the $\bR^{n-2,1}$ subspace.  For more discussion
on this see \cite{ugjggp2-ads}.

 Returning to the proof of (\ref{minksp1}), one  first solves the gravitino KSE along the coordinates $(u,r)$, as for near horizon geometries,
 to yield the expression of the Killing spinor $\eps$ in terms of the $(u,r)$ coordinates and in terms of the $\phi_\pm$ spinors  as for near horizon geometries (\ref{h11ks}).  Then the gravitino KSE (\ref{s11kses}) on $\phi_\pm$ along the $z$ coordinate  reads
\bea
\partial_z\phi_\pm=A\, \Xi^{(\pm)} \phi_\pm~,
\eea
where $\Xi^{(\pm)}$ is as in (\ref{minkxi}).  Clearly a solution of this is $\phi_\pm=\kappa_\pm+z A\, \Xi^{(\pm)}\kappa_\pm$
provided that $(\Xi^{(\pm)})^2\kappa_\pm=0$, where $\kappa_\pm$ does not depend on $z$.  Proceeding in a similar way and solving  the
gravitino KSE along the remaining coordinates of the Minkowski subspace, one finds  (\ref{minksp1}) and (\ref{minkremkses}).

To count the multiplicities of Killing spinors observe that  if
$\sigma_-$ is a Killing spinor, then   $\sigma_+= A^{-1} \Gamma_+\Gamma_m \sigma_-$ will also be  Killing spinors for every $m$, and vice versa. Furthermore, if $\sigma_+$ is a Killing spinor, then $\sigma_+'=\Gamma_{mn} \sigma_+$ will also be Killing spinors for every $m<n$.  Counting the independent Killing spinors in a way similar to that presented for warped AdS backgrounds, one establishes the results of table \ref{tab2c}. Notice that the counting of supersymmetries of warped $\bR^{1,1}$ backgrounds   has been excluded from the results in table \ref{tab2c}.  This is because warped $\bR^{1,1}$ backgrounds with fluxes may  either be singular or the internal space may not compact.  Therefore, the  counting  of supersymmetries presented for the warped $AdS_2$ backgrounds cannot straightforwardly be adapted to this case.

\subsection{A non-existence theorem for smooth warped de-Sitter and Minkowski compactifications}\label{sec:nonexfcomp}

There are restrictions on the existence of smooth warped de-Sitter and Minkowski supergravity compactifications \cite{gwg-comp, jmcn-comp}. To see this consider a $d=10$ or a $d=11$  supergravity theory and seek   warped flux compactification solutions with metric
\bea
ds^2= e^{2\phi} ds^2(M^n)+ ds^2(N^{D-n})~,
\eea
where $e^{2\phi}$ is the warp factor,  $M^n$ is either Minkowski, $\bR^{n-1,1}$, de-Sitter, $dS_n$, or anti-de-Sitter, $AdS_n$, space, and $N^{d-n}$ is an internal space. The rest of the fields are non-vanishing but they are suppressed in the statements that follow.

The Einstein field equations along $M^n$, or equivalently the field equation of the warp factor,, can be written as
\bea
D^2 e^{n\phi}= q\, e^{(n-2) \phi} R(M^n)+e^{n\phi} S(F)~,
\eea
where $q\in \bR_{>0}$,  $R(M^n)$ is  the scalar curvature of $M^n$, $D$ is the Levi-Civita connection of the internal space $N^{d-n}$  and $S(F)$ a function that depends on the other fields
of the theory.  See also the warp factor field equation for $AdS_n$ backgrounds in (\ref{warpfeq}).
Clearly, $S(F)$ depends on both the choice of background and the theory under investigation.  But the key observation is that for $d=11$, (massive) IIA and IIB supergravities $S(F)\geq 0$ and vanishes whenever the fields are zero.

As for de-Sitter backgrounds,  $R(dS_n)>0$, an application of the Hopf maximum principle, or equivalently a partial integration argument, reveals that there are no smooth warped   compactifications with compact, without boundary, internal space. Moreover there are no smooth warped Minkowski compactifications, $R(\bR^{n-1,1})=0$,   with non-trivial fluxes and compact, without boundary, internal space. The only smooth such compactifictions are those with trivial fluxes
and with constant warp factor, e.g.~Calabi-Yau type of compactifications.
 It is essential to stress that this argument does not depend on whether or not the backgrounds preserve some  supersymmetry.  It is solely based on the Einstein field equation and in particular the  field equation of the warp factor. This non-existence theorems have
 consequences for the applications of supergravity and string theory compactifications  in particle physics and  cosmology, see e.g.~\cite{udvt-ads} for a review and references therein.
The former require flux compactifications for moduli stabilization while the latter rely on the existence of  de-Sitter vacua.

\subsection{Killing superalgebras for warped AdS backgrounds}

To make further progress towards the investigation of the geometry of warped AdS backgrounds, one may proceed  to find their Killing superalgebras, see section \ref{sec:killingsuper}. As
we shall demonstrate later, these are sufficient to determine all AdS backgrounds which preserve $N>16$ supersymmetries in $d=11$ and type II $d=10$ supergravities.

Before we present the key steps  of the proof  identifying all Killing superalgebras of AdS backgrounds  \cite{sbugjggp-ads}, let us state our  assumptions, see also \cite{rdapf-ads} for an early superalgebra computation. Take $\mathfrak{g}$ to be the Killing superalgebra of a warped  $AdS_n$ background. The even part of the superalgebra $\mathfrak{g}_0$ contains the isometries $\mathfrak{so}(n-1,2)$ of the $AdS_n $ subspace. This can be verified after an explicit computation of the vector bilinears of the Killing spinors (\ref{ksp1}).  One may also expect that  $\mathfrak{g}_0=\mathfrak{so}(n-1,2)\oplus\mathfrak{t}_0$, $n\not=3$,  where $\mathfrak{t}_0$ is the Lie algebra of isometries
of the internal space $N^{d-n}$.  However this is not always  the case. One way to see this is to observe that  $AdS_k$ backgrounds can be written as warped products, $AdS_k=AdS_n\times_w \bR^{k-n}$,  of $AdS_n$ spaces for $n<k$.  The internal space $N^{d-n}$ of $AdS_n$ is
$N^{d-n}=\bR^{k-n}\times_w N^{d-k}$, where $N^{d-k}$ is the internal space of the $AdS_k$ background.    From the perspective of $AdS_n$, there exist Killing vector fields  with components
on the both $AdS_n$ and its internal space $N^{d-n}$ which cannot be separated into isometries of $AdS_n$ and isometries of $N^{d-n}$. This is because $\mathfrak{so}(k-1,2)$ cannot be decomposed as $\mathfrak{so}(k-1,2)=\mathfrak{so}(n-1,2)\oplus \mathfrak{m}$, where $\mathfrak{m}$ is also a Lie algebra.  To avoid such a phenomenon developing, we shall assume that either $\mathfrak{g}_0=\mathfrak{so}(n-1,2)\oplus\mathfrak{t}_0$, where  $\mathfrak{t}_0$ is the algebra of isometries of the internal space, or that the internal space of $AdS_n$ backgrounds is compact without boundary. In either case, one finds that the following conditions on the bilinears
\bea
\langle \sigma_+, \Gamma_i \Gamma_a \sigma'_+\rangle~=0~,~~~\langle \tau_+, \Gamma_i \Gamma_z \sigma_+\rangle~=0~,
\label{billI}
\eea
for every Killing spinor $\sigma_+$, $\sigma_+'$ and $\tau_+$.  Of course for $AdS_n$, $n>3$, backgrounds the two  conditions
are equivalent while for $AdS_3$ backgrounds  only the latter applies. A consequence of the algebraic KSEs (\ref{kseads2}) is then that
\bea
\langle \tau_+, \sigma_+\rangle=0~.
\eea

Another consequence of the requirement that $\mathfrak{g}_0=\mathfrak{so}(n-1,2)\oplus\mathfrak{t}_0$,  and so  $[\mathfrak{so}(n-1,2), \mathfrak{t}_0]=0$, and the Killing condition on the spacetime vector bilinears (\ref{allbilinears}) of the Killing spinors (\ref{ksp1}) is that
\bea
\tilde X^i\partial_i A=0~,~~~\langle\sigma_+,\sigma'_+\rangle=\mathrm{const}~,
\label{superconk}
\eea
where $\tilde X=\langle \sigma_+, \Gamma^i \Gamma_z \sigma'_+\rangle \tilde\partial_i$ are the Killing vector bilinears of the internal space.
Therefore, the warp factor is invariant under the action of $\mathfrak{g}_0$.  As the inner product of Killing spinors is constant from now on  without loss of generality we shall set
\bea
\langle\sigma_+^r,\sigma_+^s\rangle={1\over2} \delta^{rs}~,
\label{spinnorm}
\eea
where for $AdS_n$, $n>3$, $r, s=1,\dots, N/4$,  and for $AdS_2$ and  $AdS_3$, $r, s=1,\dots, N/2$.
The identification of Killing superalgebras  is somewhat different for $AdS_n$, $n>3$ and $AdS_n$, $n\leq3$ backgrounds and therefore they will treated differently.

\subsubsection{Killing superalgebras for $AdS_n$, $n>3$, backgrounds}

Supposing   that the conditions on the bilinears (\ref{billI}) hold,  one can demonstrate that the Killing superalgebras  $\mathfrak{g}$ of warped  $AdS_n$, $n>3$, backgrounds are those tabulated in table \ref{tablesupern}.  Moreover
 the isometry algebras $\mathfrak{t}_0$ of their internal space are presented in table \ref{tableison}.

\begin{table}[h]
\begin{center}
\vskip 0.3cm
\underline { Killing superalgebras of $d=10, 11$ warped  AdS$_n$, $n>3$,  solutions}
 \vskip 0.3cm
 \begin{tabular}{|c|c|c|c|c|c|}
  \hline
  $N$ & AdS$_4$ & AdS$_5$ & AdS$_6$ & AdS$_7$
  \\ \hline
  4  &$\mathfrak{ osp}(1\vert 4)$ & -       & -    & -
  \\ \hline
  8  & $\mathfrak{osp}(2\vert 4)$ & $\mathfrak{sl}(1\vert 4) $& -    & -
  \\ \hline
  12 &$\mathfrak{ osp}(3\vert 4)$ & -       & -    & -
  \\ \hline
  16 &$ \mathfrak{osp}(4\vert 4)$ & $\mathfrak{ sl}(2\vert 4)$ &$\mathfrak{ f}^*(4) $& $\mathfrak{osp}(6,2\vert 2)$
  \\ \hline
  20 &$\mathfrak{ osp}(5\vert 4)$ & -       & -    & -
  \\ \hline
  24 &$\mathfrak{ osp}(6\vert 4)$ &$ \mathfrak{sl}(3\vert 4)$ & -    & -
  \\ \hline
  28 &$\mathfrak{ osp}(7\vert 4)$ & -       & -    & -
  \\ \hline
  32 & $\mathfrak{osp}(8\vert 4)$ & $\mathfrak{ sl}(4\vert 4)/1_{8\times 8}$ & -    & $\mathfrak{osp}(6,2\vert 4)$
  \\ \hline
 \end{tabular}
 \end{center}
 \caption{ $\mathfrak{ f}^*(4) $ is a different
 real form of $\mathfrak{f}(4)$,  which appears in the AdS$_3$ case. }
 \label{tablesupern}
\end{table}

\begin{table}[h]
\begin{center}
\vskip 0.3cm
\underline {Isometry algebras of internal spaces}
 \vskip 0.3cm
 \begin{tabular}{|c|c|c|c|c|c|}
  \hline
  $N$ & AdS$_4$ & AdS$_5$ & AdS$_6$ & AdS$_7$
  \\ \hline
  4  & \{0\} & - & - & -
  \\ \hline
  8  & $\mathfrak{so}(2)$ & $\mathfrak{u}(1)$ & - & -
  \\ \hline
  12 & $\mathfrak{so}(3)$ & - & - & -
  \\ \hline
  16 & $\mathfrak{so}(4)$ & $\mathfrak{u}(2)$ & $\mathfrak{so}(3)$ & $\mathfrak{so}(3)$
  \\ \hline
  20 & $\mathfrak{so}(5)$ & - & - & -
  \\ \hline
  24 & $\mathfrak{so}(6)$ & $\mathfrak{u}(3) $ & - & -
  \\ \hline
  28 & $\mathfrak{so}(7)$ & - & - & -
  \\ \hline
  32 & $\mathfrak{so}(8)$ & $\mathfrak{su}(4)$ & - & $\mathfrak{so}(5)$
  \\ \hline
 \end{tabular}
 \end{center}
 \caption{These algebras must act effectively on the internal spaces of $AdS_n$ backgrounds}
 \label{tableison}
\end{table}

The proof of these results relies on the fact that the dependence of the Killing spinors,  (\ref{ksp1}) and (\ref{ksp2}),  on the coordinates of the $AdS_n$ subspace
of the warped backgrounds is  known.  As a result, one can explicitly compute the (anti-) commutators
\bea
&&\{\mathfrak{g}_1, \mathfrak{g}_1\}= \mathfrak{so}(n-1,2)+{\mathfrak{t}_0}~,
\cr
&&[\mathfrak{so}(n-1,2), \mathfrak{g}_1]\subseteq \mathfrak{g}_1~.
\eea
The key commutator that needs to be found is $[\mathfrak{t}_0, \mathfrak{g}_1]$.  This typically requires some information on the underlying geometry of the internal space $N^{d-n}$. However for $AdS_n$, $n>3$,  backgrounds this is not necessary and the result follows as a consequence of the super-Jacobi identities (\ref{superjac})  of the Killing superalgebra. The remaining commutator $[\mathfrak{t}_0, \mathfrak{t}_0]$
also follows from the super-Jacobi identities.  The details of this computation can be found in \cite{sbugjggp-ads}.  It is remarkable that
for each $AdS_n$ background the Killing superalgebra is specified uniquely by the number $N$ of supersymmetries that are preserved.
Another important  consequence of the computation of the  super-Jacobi identities    is that the Lie algebra $\mathfrak{t}_0$ acts (almost) effectively on the internal space,
i.e. all elements of $\mathfrak{t}_0$ generate a non-trivial vector field on $N^{d-n}$.  If this is not the case, the super-Jacobi identities cannot be satisfied.

\subsubsection{Killing superalgebras of $AdS_2$ and $AdS_3$ backgrounds}

Let us now turn to investigate the Killing superalgebras of warped $AdS_3$ backgrounds.
$AdS_3$ is locally a group manifold and the Killing superalgebra $\mathfrak{g}$ decomposes as
$\mathfrak{g}=\mathfrak{g}_L\oplus \mathfrak{g}_R$ into left and right superalgebras, $[\mathfrak{g}_L, \mathfrak{g}_R]=0$.  The left superalgebra  $\mathfrak{g}_L$ is associated with the $\sigma_+$ Killing spinors while
$\mathfrak{g}_L$ is associated with the $\tau_+$ Killing spinors. It suffices to identify only $\mathfrak{g}_L$ as the same techniques can be used to identify  the
$\mathfrak{g}_R$ Killing superalgebras. The list of $\mathfrak{g}_L$  Killing superalgebras is the same as that of the $\mathfrak{g}_R$ Killing superalgebras.  Though  a given background may exhibit a different $\mathfrak{g}_L$ from  a $\mathfrak{g}_R$ Killing superalgebra. Assuming the conditions (\ref{billI}) on the bilinears, the $\mathfrak{g}_L$ Killing superalgebras of
$AdS_3$ backgrounds are given in table \ref{tablesuper3}.  Furthermore, the isometry algebras of the internal space are given in
table \ref{tableiso3}.  The Killing superalgebras of $AdS_2$ backgrounds with compact, without boundary, internal space can be
identified with the left copies $\mathfrak{g}_L$ of the Killing superalgebras of warped $AdS_3$ backgrounds.

\begin{table}[h!]
\centering
\underline { Killing superalgebras of $d=10, 11$ warped AdS$_2$ and $AdS_3$ solutions}
 \vskip 0.3cm
 \begin{tabular}{|c|c|}
  \hline
  $N_L$ & $\mathfrak{g}_L/\mathfrak{c}_L$
  \\ \hline
  $2k$  &$\mathfrak{osp}( k\vert 2)$
  \\ \hline
 $ 4k, k>1 $ & $\mathfrak{sl}(k\vert 2)$
  \\ \hline
 $8k, k>1$  &  $ \mathfrak{osp}^*(4\vert 2k)$
 \\ \hline
$16$  &  $ \mathfrak{f}(4)$
  \\ \hline
  $14$  &  $\mathfrak{g}(3) $
  \\ \hline
  $8$  &  $ \mathfrak{D}(2,1,\alpha)$
  \\ \hline
  $8$  &  $ \mathfrak{sl}(2\vert 2)/1_{4\times 4}$
  \\ \hline
 \end{tabular}
  \caption{ $(\mathfrak{g}_L/\mathfrak{c}_L)_0=\mathfrak{so}(1,2)\oplus \mathfrak{t}_0/\mathfrak{c}_L$, where
  there may be a central term $\mathfrak{c}_L$.  The superalgebras $ \mathfrak{osp}^*(4\vert 2k)$ are different real forms of the $ \mathfrak{osp}(4\vert 2k)$ superalgebras, see  table \ref{tablesupern}.}
  \label{tablesuper3}
\end{table}

 \begin{table}
 \centering
 \vskip 0.3cm
	\underline{Isometries of internal spaces}
\vskip 0.3cm
	
	\begin{tabular}{|c|c|c|c|}
		\hline
		$N_L$ & $\mathfrak{g}_L/\mathfrak{c}_L$ & $(\mathfrak{t}_L)_0/\mathfrak{c}_L$&$\mathrm{dim}\, \mathfrak{c}_L $ \\
		\hline
		$2k$& $\mathfrak{osp}(k|2)$ & $\mathfrak{so}(k)$ & 0\\
		$4k,~k>2$ & $\mathfrak{sl}(k|2)$ & $\mathfrak{u}(k)$& 0 \\
		$8k, k>1$ & $\mathfrak{osp}(4|2k)$ & $\mathfrak{sp}^*(k) \oplus \mathfrak{sp}^*(1)$ &0\\
		16 & $\mathfrak{f}(4)$ & $\mathfrak{spin}(7)$ &0\\
		14 & $\mathfrak{g}(3)$ & $\mathfrak{g}_2$&0 \\
		8 & $\mathfrak{D}(2,1,\alpha)$ & $\mathfrak{so}(3) \oplus \mathfrak{so}(3)$&0 \\
		8 & $\mathfrak{sl}(2|2)/1_{4\times 4}$ & $\mathfrak{su}(2)$& $\leq 3 $\\ [1ex]
		\hline
	\end{tabular}
\caption{$ \mathfrak{sp}^*( k)$ is the compact symplectic algebra with (real) dimension $k (2k+1)$ which  is a real form of $\mathfrak{sp}(2 k)$}
	\label{tableiso3}
\end{table}

The identification of Killing superalgebras in table  \ref{tablesuper3}  for warped $AdS_2$ and  $AdS_3$ backgrounds is more involved than that presented in the previous section   for warped $AdS_n$, $n>3$ backgrounds.  To outline the main steps of  the proof,
consider first the Killing superalgebra $\mathfrak{g}_L$ of  $N=2$ warped $AdS_3$ backgrounds. A direct computation reveals that
\bea
\{Q_\tA, Q_\tB\}=V_{\tA\tB}~,~~~[V_{\tA\tB}, Q_\tC]=-\ell^{-1} (\epsilon_{\tC\tA} Q_\tB+\epsilon_{\tC\tB} Q_\tA)~,
\eea
where $A,B,C=1,2$,  $V_{\tA\tB}$ are the generators of $\mathfrak{g}_0=\mathfrak{so}(1,2)=\mathfrak{sp}(2)$ and $Q_\tA$ are  odd generators
of the superalgebra associated to the two Killing spinors (\ref{ksp1}) constructed from $\sigma_\pm$. There are no internal space isometries generated from vector bilinears and so $\mathfrak{t}_0=\{0\}$. This superalgebra is isomorphic to $\mathfrak{osp}(1|2)$. Note that in this section $A,B,C$ are not frame coframe indices.

Suppose now that we have $N=2k$ supersymmetries. As the dependence of the Killing spinors (\ref{ksp1})  on the $AdS_3$ coordinates is known, one finds after a direct computation that
\bea
&&\{Q_{Ar}, Q_{Bs}\}=V_{AB}\delta_{rs}+\epsilon_{AB} \tilde V_{rs}~,~~~
\cr
&&[V_{AB}, Q_{Cr}]=-\ell^{-1} (\epsilon_{CA} Q_{Br}+\epsilon_{CB} Q_{Ar})~,
\eea
where $r, s=1,\dots, k$ and $\tilde V_{rs}\in\mathfrak{t}_0$.  It remains to compute $[\tilde V_{rs}, Q_{At}]$.  As the spinorial Lie derivative
along isometries $\tilde X$ of the internal space does not change the functional dependence of the Killing spinors $\eps$ on the $AdS_3$ coordinates, one concludes
that
 \bea
 [\tilde V_{rs}, Q_{At}]&=&-\ell^{-1}(\delta_{tr} Q_{As}- \delta_{ts} Q_{Ar})+\ell^{-1}\alpha_{rst}{}^p Q_{Ap}~,
 \eea
 for some structure constants $\alpha$ which remain to be determined.  The super-Jacobi identities (\ref{superjac}) together with
 the identity
 \bea
 \langle {\cal L}_{\tilde X}\sigma_+^r, \sigma_+^s\rangle +\langle \sigma_+^r, {\cal L}_{\tilde X}\sigma_+^s\rangle=0~,
 \label{invnormcon}
 \eea
 imply that  $\alpha$ is a 4-form.  To prove (\ref{invnormcon}), take the Lie derivative of the normalization condition
 of the $\sigma_+$ Killing spinors in  (\ref{spinnorm}) with respect to internal space isometries $\tilde X$.

 Furthermore $\alpha$ is invariant under the representation $D$ of $\mathfrak{t}_0$ on $\mathfrak{g}_1$, where
 \bea
 D(\tilde V_{rs}) Q_{\tA t}\defeq [\tilde V_{rs}, Q_{\tA t}]~.
 \eea
Note that $\mathfrak{g}$ may have a centre $\mathfrak{c}_L\defeq \{\tilde V\in \mathfrak{t}_0\vert D(\tilde V)=0\}$. It turns out that
$\mathfrak{c}_L=\{0\}$ apart from one case where it can have a dimension of at most 3, see table \ref{tableiso3}.

The key observation which identifies the   representations
  $D$  that can occur is that $D$ acts transitively (and effectively) on spheres in $\bR^{{N\over2}}$.
 For this it suffices to show that given  two linearly independent vectors  $u,w\in \bR^{{N\over2}}$, there is an element $R(u,w)\in \mathfrak{t}_0$ such that $R(u,w)$ generates $SO(2)$ rotations
on the 2-plane spanned by $u$ and $w$ in $\bR^{{N\over2}}$. The statement then follows as such  $SO(2)$ rotations act transitively on all directions in the 2-plane spanned by $u$ and $w$.

For this set $R(u,w)=u^r w^s\tilde V_{rs}$  and $p\cdot Q_A= p^r Q_{\tA r}$, and   observe that
\bea
D(R(u,w))(p\cdot Q_A)&=& [u^r w^s\tilde V_{rs}, p^t  Q_{\tA t}]
\cr
&=&-\ell^{-1} ( p\cdot u~ w^r- p\cdot w~ u^r )  Q_{\tA r}~,
\eea
for any $p$ that lies in the 2-plane spanned by $u$ and $w$ as $\alpha(u,w,p,\cdot)=0$.
So indeed  $R(u,w)$ acts as an infinitesimal orthogonal rotation on the 2-plane spanned by $u$ and $w$.
As this can be done for any
$u,w\in \bR^{{N\over2}}$, it follows that $\mathfrak{t}_0$ acts transitively on $S^{{N\over2}-1}\subset \bR^{{N\over2}}$.
The groups that act effectively and transitively on spheres have been classified in \cite{dmhs-ads} and they have been used in
the Berger classification of irreducible simply connected Riemannian manifolds \cite{js-ads}.  They are given in table \ref{tabletransx}.   Some further analysis
which can be found in \cite{sbugjggp-ads} reveals that only some of these groups occur in the investigation of superalgebras and the final
result is described in table \ref{tablesuper3}.

\begin{table}
\centering
\vskip 0.3cm
\underline {Lie algebras of groups acting transitively on spheres}
 \vskip 0.3cm
 \begin{tabular}{|c|c|c|}
  \hline
 $\text{ Algebra}$ & $\text{Sphere}$ & $N/2$
  \\ \hline
  $\mathfrak{so}(k)$  &$S^{k-1}$ & $k$
  \\ \hline
 $ \mathfrak{u}(k) $ & $S^{2k-1}$ & $2k$
  \\ \hline
 $ \mathfrak{su}(k)$  & $S^{2k-1}$ & $2k$
  \\ \hline
 $ \mathfrak{sp}^*(k)\oplus \mathfrak{sp}^*(1) $ & $S^{4k-1}$ & $4k$
  \\ \hline
  $ \mathfrak{sp}^*(k)\oplus \mathfrak{u}(1)$  & $S^{4k-1}$ &  $4k$
  \\ \hline
  $ \mathfrak{sp}^*(k)$  & $S^{4k-1}$ &  $4k$
  \\ \hline
  $\mathfrak{g}_2 $ & $S^6$ & $7$
  \\ \hline
 $ \mathfrak{spin}(7)$  & $S^7$ & $8$
  \\ \hline
 $ \mathfrak{spin}(9) $ & $S^{15}$ & $16$
  \\ \hline
 \end{tabular}
 \vskip 0.2cm
  \caption{  $ \mathfrak{spin}(9)$  cannot be realized as a symmetry of the internal space of warped $AdS_2$ and $AdS_3$ backgrounds as there are no such maximally supersymmetric backgrounds.}
 \label{tabletransx}
\end{table}

\subsection{ $N>16$ AdS backgrounds}\label{s:n16adsb}

As an application of the technology developed so far, we shall provide a classification of smooth
 warped AdS backgrounds with compact, without boundary, internal space that preserve $N>16$ supersymmetries in $d=11$ and $d=10$ type II supergravities \cite{jfofgp1, sbjggp-cads, ahslgp1-cads, ahslgp2-cads}.  In particular,
one can show the  following.

\begin{enumerate}

\item[-] There are no warped $AdS_n$, $n=2,3,6$, backgrounds that preserve $N>16$ supersymmetries.

\item[-] The only warped $AdS_4$ backgrounds that preserve $N>16$  supersymmetries are locally isometric to the $N=24$,  $AdS_4\times \bC P^3$, solution of IIA supergravity
of \cite{bncp-cads} and the maximally supersymmetric  solution,  $AdS_4\times S^7$, of   $d=11$ supergravity.

\item[-] The only warped $AdS_7$ backgrounds that preserve $N>16$ supersymmetries are locally isometric the the maximally
supersymmetric solution, $AdS_7\times S^4$,  of $d=11$ supergravity.

\item[-] The only warped $AdS_5$ backgrounds that preserve $N>16$ supersymmetries are locally isometric to the maximally
supersymmetric solution, $AdS_5\times S^5$,  of IIB supergravity.

\end{enumerate}

The result above follows immediately for warped $AdS_6$ and $AdS_7$ backgrounds as these preserve either 16 or 32 supersymmetries.
Therefore if they exist, they must be maximally supersymmetric.  The maximally supersymmetric backgrounds have been classified in
\cite{jfofgp1} and this has already been reviewed in section \ref{secmaxsusy} yielding the result stated above.

The next new case that  arises is that of warped $AdS_5$ backgrounds that may preserve $N=24$ supersymmetries.  The maximally
supersymmetric $AdS_5$ backgrounds have already been dealt with in section  \ref{secmaxsusy} as part of the classification of maximally
supersymmetric backgrounds in $d=10$ and $d=11$ supergravities.

\subsubsection{ A non-existence theorem for warped $N=24$ $AdS_5$ backgrounds}

Here we shall present the main points of the proof in the context of $d=11$ supergravity, see also \cite{sbjggp-cads} and for the rest of the theories.
The fields of  warped  $AdS_5$ backgrounds in $d=11$ supergravity are
\bea
&&ds^2= 2 du (dr+ rh)+A^2 (ds^2+ e^{{2z\over\ell}} (dx^a)^2)+ds^2(M^6)~,~~
\cr
&&F=Z~,~~~h=-{2\over\ell} dz-2 A^{-1} dA~,
\eea
and the supercovariant connection on the internal space is
\bea
D_i^{(\pm)}=D_i\pm {1\over2} \partial_i\log A-{1\over288} {\slashed \Gamma\mkern-4.0mu Z}_i+{1\over 36} \slashed{Z}_i~.
\eea
Using that $\parallel\sigma_+\parallel^2$ is constant, see (\ref{superconk}),  and  $D_i^{(+)}\sigma_+=0$, one finds that
\bea
-\parallel\sigma_+\parallel^2 \partial_i\log A+{1\over144} \langle\sigma_+, {\slashed \Gamma\mkern-4.0mu Z}_i \sigma_+\rangle=0~.
\label{ads5g16con1}
\eea
The Killing vectors along the internal space are
 \bea
 \tilde X_i=A\langle\sigma_+, \Gamma_{z12i} \sigma_+\rangle~.
 \eea
 Using this, (\ref{ads5g16con1}) can be written as
 \bea
 i_{\tilde X}\star_{{}_6}Z=6 \parallel\sigma_+\parallel^2  dA~.
 \eea
 Taking the inner derivation of the above equation with $\tilde X$, one also finds that
 \bea
i_{\tilde X} dA=0~.~~~
\eea
An adaptation of the homogeneity theorem argument \cite{fofjh1}, reviewed in section \ref{homtheorem},  leads to the conclusion that $\tilde X$ span the tangent space of the
internal space.  As a result $Z=0$ and $A$ is constant.  However in such a case,
 the warp factor field equation
\bea
D^2\log A=-{4\over \ell^2 A^2} -5 (d\log A)^2+{1\over144} Z^2~,
\eea
cannot be satisfied.   This excludes the existence of warped $AdS_5$ backgrounds in $d=11$ supergravity that preserve
$N>16$ supersymmetries.

\begin{table}\renewcommand{\arraystretch}{1.3}
	\caption{7-dimensional compact, simply connected,  homogeneous spaces}
	\centering
	\begin{tabular}{c l}
		\hline
		& $M^7=G/H$  \\  
		\hline
		(1)& { $\frac{\mathrm{Spin}(8)}{\mathrm{Spin}(7)}= S^7$}, symmetric space\\
		(2)&{ $\frac{\mathrm{Spin}(7)}{G_2}=S^7$} \\
		(3)& { $\frac{SU(4)}{SU(3)}$ }diffeomorphic to $S^7$ \\
        (4) & { $\frac{Sp(2)}{Sp(1)}$ }diffeomorphic to $S^7$ \\
		(5) & { $\frac{Sp(2)}{Sp(1)_{max}}$}, Berger space \\
		(6) & $ \frac{Sp(2)}{\Delta(Sp(1))}=V_2(\bR^5)$ , not spin\\
        (7) & $\frac{SU(3)}{\Delta_{k,l}(U(1))}=W^{k,l}$~~ $k, l$ coprime, Aloff-Wallach space\\
		(8)&$\frac{SU(2) \times SU(3) }{\Delta_{k,l}(U(1))\cdot (1\times SU(2))}=N^{k,l}$ ~$k,l$ coprime\\
		(9) & $\frac{SU(2)^3}{\Delta_{p,q,r}(U(1)^2)}=Q^{p,q,r}$ $p, q, r$ coprime\\
(10)&$M^4\times M^3$,~~$M^4=\frac{\mathrm{Spin}(5)}{\mathrm{Spin}(4)}, ~\frac{ SU(3)}{S(U(1)\times U(2))}, ~\frac{SU(2)}{U(1)}\times \frac{SU(2)}{U(1)}$\\
&~~~~~~~~~~~~~~~~$M^3= SU(2)~,~\frac{SU(2)\times SU(2)}{\Delta(SU(2))}$\\
(11)&$M^5\times \frac{SU(2)}{U(1)}$,~~$M^5=\frac{\mathrm{Spin}(6)}{\mathrm{Spin}(5)}, ~\frac{ SU(3)}{SU(2)}, ~\frac{SU(2)\times SU(2)}{\Delta_{k,l}(U(1))},~ \frac{ SU(3)}{SO(3)} $\\
[1ex]
		\hline
	\end{tabular}
	\label{table:nonlin7}
\end{table}

\subsubsection{ Existence and uniqueness  theorems for warped $N>16$ $AdS_4$ backgrounds}

The proof presented in the previous section to find the $N>16$ warped $AdS_5$ backgrounds cannot be adapted
to investigate the warped $AdS_4$ backgrounds that preserve $N>16$ supersymmetries.  Instead, a more detailed investigation is required  of the homogeneous structure of spacetime which is implied by the homogeneity theorem.

First, one establishes that the warp factor $A$ is constant.  To prove this one first uses  the algebraic KSE, $\Xi^{(+)}\sigma_+=0$, to find  $i_{\tilde X} dA=0$, where ${\tilde X}$ is a Killing vector field of the internal space. As $N>16$,
the homogeneity theorem implies that ${\tilde X}$ span the tangent of the internal space  which gives that $A$ is constant.   Thus the spacetime is a product $AdS_4\times N^{d-4}$ and $N^{d-4}$ is a homogeneous space $G/H$.

Therefore, one has to identify the homogeneous spaces $G/H$ that can occur as internal spaces of $N>16$ $AdS_4$ backgrounds.
To do this, one  uses the classification of all Killing superalgebras of AdS backgrounds and in particular that of the Lie algebra of isometries
of the internal spaces tabulated in table  \ref{tableison}. This together with the homogeneity theorem imply that $\mathfrak{Lie}\,G=\mathfrak{t}_0=\mathfrak{so}(N/4)$  for $AdS_4$ backgrounds with $N>16$.

Further progress is made utilizing the classification of homogeneous spaces, see \cite{lclrnw-ads, sk-cads, yn-cads}. In particular for $d=11$ supergravity $AdS_4$ backgrounds have a 7-dimensional homogeneous internal space.  All simply connected 7-dimensional  homogeneous spaces have been tabulated in table \ref{table:nonlin7}.  It is straightforward to observe that the requirement that $\mathfrak{Lie}\,G=\mathfrak{so}(N/4)$ restricts the number of homogeneous spaces that can occur as internal spaces of $AdS_4$  backgrounds that preserve $N>16$ supersymmetries to the first
five in table \ref{table:nonlin7}.  The rest of the proof proceeds with the analysis of each case separately and  gives that the only warped $AdS_4$ backgrounds of $d=11$ supergravity are locally isometric to the maximally supersymmetric $AdS_4\times S^7$ solution.

One uses   a similar methodology to establish the classification statement for $AdS_4$ backgrounds  in section \ref{s:n16adsb} for (massive) IIA and IIB supergravities.   This method can also be extended to prove  a non-existence result for  $AdS_2$ and $AdS_3$ backgrounds that preserve
$N>16$ supersymmetries \cite{ugjggp-cads, ahslgp2-cads}.

\section{Conclusions}

Significant progress has been made the last 15 years to classify the supersymmetric backgrounds in all supergravity theories and explore their applications in the context of string theory, M-theory, gauge theory,  black holes and the AdS/CFT correspondence.  The task has been completed for a substantial class of theories which include all those with a small number of supercharges in each spacetime dimension. This  has given  an insight into the structure of all supersymmetric solutions and has led to a plethora of existence and uniqueness theorems for backgrounds, including those of black holes and warped AdS spaces, which otherwise would have been out of reach. We presented the classification of supersymmetric solutions in terms of only a few examples and there are many other significant theories that could have been included.  However, we endeavored to be concise and give a  taste of how such  proofs and calculations  can be carried out.

The emphasis in this review has been to describe  the bilinears and spinorial geometry methods that have been used  to  solve the KSEs of supergravity theories.  These methods cover all the theories that have been treated in the literature.  The solution of the KSEs is the first key step towards the classification of  supersymmetric backgrounds.
After obtaining the solution of the KSEs, we proceeded to explain how to use it to identify the geometry of the spacetime.

 For the description of the geometry of supersymmetric backgrounds, we have used partly local techniques and partly borrowed the language
 of bundles and  G-structures. These are sufficient for all practical purposes.  We also
presented a taste of how powerful global techniques like index theory, Lichnerowicz type theorems and the Hopf maximum principle  can be used to prove general properties of supersymmetric backgrounds.

 We have  included in the review some other key properties of the
supersymmetric solutions like their Killing superalgebras and the homogeneity theorem, and also the non-existence theorem for de-Sitter and Minkowski flux compactifications in supergravity. As applications, we demonstrated that the emergence of conformal symmetry near supersymmetric Killing horizons is a generic phenomenon in supergravity, which does not depend on the details of the black hole solutions. We also  classified
the warped AdS backgrounds that preserve $N>16$ supersymmetries in $d=11$ and $d=10$ supergravities.

One of the last remaining challenges in this field is to solve the KSEs in $d=11$ and $d=10$ type II supergravities for
backgrounds preserving any number of supersymmetries.  As has already been described, the geometry of  $N=1$  backgrounds
 is known and there is a classification of the maximally and nearly maximally supersymmetric backgrounds.
However very little is known about the geometry of solutions that preserve an intermediate number of supersymmetries. The final objective
is to give a description of the geometries similar to those of heterotic backgrounds as presented in section \ref{sec:heterotic}.
It is encouraging that there are strong constraints on the existence of special backgrounds that preserve $N>16$ supersymmetries like those
for the AdS backgrounds we have described.  This  indicates that the bulk of the task will be to understand the geometry  of backgrounds that preserve $N\leq 16$ supersymmetries.

Of course in many applications the interest is focused on special types of solutions, e.g.~black holes or warped Minkowski and  AdS flux compactifications.  For those there are many simplifications and a complete identification of all such backgrounds may be possible.  It is encouraging that there is increasing detail in the understanding of AdS backgrounds which have applications in the context of the AdS/CFT correspondence.  It is very likely that in the next few years there will be a complete understanding of the structure of all such solutions.

\section*{Acknowledgments}

UG is supported by the Swedish Research Council. GP would like to thank the Yau Mathematical Sciences Center, Tsinghua University and Theoretical Physics Department at CERN for hospitality and support. Part of this review has been presented by one of us, GP, at the `` School on Theoretical Approaches to Black Hole Physics’’ Mesoamerican Centre for Theoretical Physics, Universidad Autonoma de Chiapas, Sept 2017,  and as a short course at Tsinghua University. JG is supported by the STFC Consolidated Grant ST/L000490/1. GP is partially supported from the  STFC consolidated grant  ST/P000258/1.



\appendix

\section{Notation for forms} \label{formnote}


Let $M$ be a manifold with a (local) coframe $\fe^i$ and coordinates $y^\tI$. The exterior derivative  on a k-form,
\begin{eqnarray}
\omega={1\over k!}\, \omega_{\tI_1\dots \tI_k}\, dy^{\tI_1}\wedge\dots \wedge dy^{\tI_k}={1\over k!}\, \omega_{i_1\dots i_k}\, \fe^{i_1}\wedge\dots \wedge \fe^{i_k}~,
\end{eqnarray}
is
\begin{eqnarray}
d\omega\defeq {1\over k!}\, \partial_{\tI_1} \omega_{\tI_2\dots \tI_{k+1}} dy^{\tI_1}\wedge\dots \wedge dy^{\tI_{k+1}}~.
\end{eqnarray}
Therefore, one has
$(d\omega)_{\tI_1\dots \tI_{k+1}}= (k+1) \partial_{[\tI_1} \omega_{\tI_2\dots \tI_{k+1}]}~.$
The inner derivation $i_X$ of a k-form $\omega$ with respect to a vector field $X$ is
\bea
i_X\omega\defeq {1\over (k-1)!} X^j \omega_{ji_1\dots i_{k-1}}  \fe^{i_1}\wedge\dots \wedge \fe^{i_{k-1}}~.
\eea
Furthermore, it is convenient to set
\begin{eqnarray}
\omega^2\defeq \omega_{i_1\dots i_k} \omega^{i_1\dots i_k}~,~~~\omega^2_{i_1 i_{2}}\defeq \omega_{i_1j_1\dots j_{k-1}} \omega_{i_2}{}^{j_1\dots j_{k-1}} \ ,
\end{eqnarray}
where the indices are raised with respect to a metric, $ds^2=g_{ij}\, \fe^i \fe^j$,  on $M$.
The inner product of two k-forms $\chi$ and $\omega$ is
\begin{eqnarray}
(\chi, \omega)\defeq {1\over k!}\, \chi_{i_1\dots i_k} \omega^{i_1\dots i_k} \ .
\end{eqnarray}

Given a volume form $d\mathrm{vol}={1\over n!} \epsilon_{i_1\dots i_n} dx^{i_1}\wedge \dots \wedge dx^{i_n}$, the Hodge dual of the k-form $\omega$ is defined as
\begin{eqnarray}
\chi\wedge *\omega= (\chi, \omega) d\mathrm{vol}~,
\end{eqnarray}
for every k-form $\chi$.

It is well-known that for every form $\omega$, one can define a Clifford algebra element ${\slashed \omega}$ given by
\begin{eqnarray}
{\slashed\omega}\defeq\omega_{i_1\dots i_k} \Gamma^{i_1\dots i_k}~,
\end{eqnarray}
where $\Gamma^i$, $i=1,\dots n$, are the Dirac gamma matrices. In addition we have introduced the notation
\begin{eqnarray}
{\slashed\omega}_{i_1}\defeq \omega_{i_1 i_2 \dots i_k} \Gamma^{i_2\dots i_k}~,~~~{\slashed \Gamma\mkern-4.0mu \omega}_{i_1}\defeq \Gamma_{i_1}{}^{
i_2\dots i_{k+1}} \omega_{i_2\dots i_{k+1}}~.
\end{eqnarray}
This significantly shorten some of expressions for the KSEs.

\section{Spinors and forms} \label{app:spinors}

There is an extensive literature on the representations of Clifford algebras and $Spin$ groups, see e.g.
\cite{lawson, harvey}. Here the emphasis is on an explicit realization  of the spinor representations of the $Spin$ groups
in terms of forms which is used in the spinorial geometry approach to solving KSEs.
We follow the construction of \cite{wang} and  \cite{jguggp} for the Euclidean and Lorentzian cases, respectively.

\subsection{Euclidean}

To realize  the Dirac representation, ${}^c\Delta$,  of $Spin(2n)$ in terms of forms, consider the space of all (complex) forms on $\bC^n$,
$\Lambda^*(\bC^n)$, equipped with a Hermitian inner product $\langle \cdot, \cdot\rangle$ and set ${}^c\Delta=\Lambda^*(\bC^n)$. Then  gamma matrices act as
\bea
\Ga_i\zeta= \se_i\wedge \zeta+i_{\se_i} \zeta~,~~~\Ga_{i+n}\zeta=i(\se_i\wedge \zeta-i_{\se_i} \zeta)~,~~~i=1,\dots, n~,
\la{cbasis}
\eea
where $\se_i$ is a Hermitian  basis in $\Lambda^1(\bC^n)$, $\langle \se_i,\se_j\rangle=\delta_{ij}$,  and $\zeta$ is a multi-degree
form in $\Lambda^*(\bC^n)$. The operation $i_{\se_i}$ is the inner derivation with respect to the vector constructed from $e_i$ using
$\langle \cdot, \cdot\rangle$.
The gamma matrices, $\Ga_\tA$,  $A=1,\dots, 2n$,  defined above are Hermitian with respect to
 $\langle \cdot, \cdot\rangle$ and satisfy the Clifford algebra relations $\Gamma_\tA\Gamma_\tB+\Gamma_\tB\Gamma_\tA=2\, {\bf 1}\,\delta_{\tA\tB}$.  It is usual to label bases in the space of forms with upper indices. Here  the basis $\{\se_i\}$ has been
 labelled with lower indices in order to distinguish the notation of spinors in terms of forms from that used for  forms on the spacetime.

In the Euclidean case, the Dirac inner
product, $D$, is identified with  the Hermitian inner product
 $\langle\cdot, \cdot\rangle$  on $\Lambda^*(\bC^n)$, $D(\cdot, \cdot)=\langle\cdot, \cdot\rangle$. As the gamma matrices are Hermitian with respect to $\langle\cdot, \cdot\rangle$,
 \bea
 \langle\Gamma_{\tA\tB} \eta, \zeta\rangle+\langle \eta, \Gamma_{\tA\tB}\zeta\rangle=0~,
 \eea
 and so $D$ is invariant under the Lie algebra $\mathfrak{spin}(2n)$ of $Spin(2n)$.  In fact $D$ is invariant under the action of
 $Spin(2n)$ which is the double cover of $SO(2n)$.

 The Dirac representation of $Spin(2n)$ is reducible and decomposes
as ${}^c\Delta={}^c\Delta^+\oplus {}^c\Delta^-$ into chiral and anti-chiral Weyl representations according to the decomposition
of $\Lambda^*(\bC^n)$ into forms of even and odd degree, ${}^c\Delta^+=\Lambda^{\rm ev}(\bC^n)$ and ${}^c\Delta^-=\Lambda^{\rm odd}(\bC^n)$,
respectively.

Next consider the linear maps $a\defeq\prod_{i=1}^n \Ga_i$ and $b\defeq \prod_{i=1}^n \Ga_{i+n}$.
There are two $Spin(2n)$-invariant bi-linears, the Majorana inner products, which can be constructed on ${}^c\Delta$ as
\bea
{\mathrm A}(\eta, \zeta)\defeq \langle a \eta^*, \zeta\rangle~,~~~~{\mathrm B}(\eta, \zeta)\defeq \langle b\eta^*, \zeta\rangle~,
\eea
where $\eta, \zeta\in {}^c\Delta$ and $\eta^*$ is the complex conjugate of $\eta$. The bi-linearity of $A$ and $B$ is assured because $\langle\cdot, \cdot\rangle$ is anti-linear in the left entry.
A straightforward computation following the definitions reveals that
\bea
&&{\mathrm A}(\eta, \Ga_\tA\zeta)=(-1)^{n-1}{\mathrm A}(\Ga_\tA \eta, \zeta)~,~~{\mathrm A}(\eta, \Ga_\tA\zeta)=(-1)^{{(n+2) (n-1)\over2}}{\mathrm A}(\zeta, \Ga_\tA\eta)~,
\cr
&&{\mathrm B}(\eta, \Ga_\tA\zeta)=(-1)^n{\mathrm B}(\Ga_\tA\eta, \zeta)~,~~{\mathrm B}(\eta, \Ga_\tA\zeta)=(-1)^{{n (n-1)\over2}}{\mathrm B}(\zeta, \Ga_\tA\eta)~.
\eea
This confirms that both ${\mathrm A}$ and ${\mathrm B}$ are invariant under $Spin(2n)$ and in addition that one
of them is also invariant under $Pin(2n)$. Note that the Lie algebra of the $Pin$ group is spanned by $\Gamma_{\tA\tB}$ and $\Gamma_\tA$.
Furthermore, after lowering the spinor indices of the  gamma matrices with respect to $\mathrm A$ or  $\mathrm B$, these become     either symmetric or skew-symmetric.

So far, we have dealt with complex representations of $Spin$ groups. Real representations exist whenever a reality condition can be imposed
on the complex representations, i.e.~there is an antilinear map ${\mathcal R}$ which commutes with the action of
$Spin$ on ${}^c\Delta$ and ${\mathcal R}^2={\bf 1}$. Such maps are not unique since if ${\mathcal R}$ is a reality condition, then
$e^{i\theta} {\mathcal R}$ is also a reality condition for any angle $\theta$. To proceed consider the anti-linear maps
 maps  $r_{{\mathrm A}}=a *$ and $r_{{\mathrm B}}=b *$. One can verify that
\bea
r_{{\mathrm A}}^2&=&(-1)^{{n (n-1)\over2}} {\bf 1}~,~~~r_{{\mathrm A}}\, \Ga_\tA=(-1)^{n-1} \Ga_\tA\, r_{{\mathrm A}}~,~~~
\cr
r_{{\mathrm B}}^2&=&(-1)^{{n (n+1)\over2}} {\bf 1}~,~~~r_{{\mathrm B}}\, \Ga_\tA=(-1)^{n} \Ga_\tA\, r_{{\mathrm B}}~.
\eea
Therefore there are real representations with reality conditions either ${\mathcal R}_{{\mathrm A}}= e^{i\theta}r_{{\mathrm A}}$
or ${\mathcal R}_{{\mathrm B}}= e^{i\theta}r_{{\mathrm B}}$ provided that $[{n\over2}]\in 2\bZ$. If in addition
${\mathcal R} {}^c\Delta^\pm\subset {}^c\Delta^\pm$, then there are Majorana-Weyl representations. So there are Majorana-Weyl
representations iff $n\in 4\bZ$.

There is an oscillator basis in the space of Dirac spinors which we use to solve the KSEs.
To see this, write the gamma matrices   (\ref{cbasis}) in a Hermitian basis as
\bea
\hga_\al={1\over \sqrt{2}} (\Ga_\al-i \Ga_{\al+n})=\sqrt{2}\, \se_\al\wedge,~~\hga_{\bar\al}={1\over \sqrt{2}} (\Ga_\al+i \Ga_{\al+n})=
\sqrt{2}\, i_{\se_\al}~,
\label{hermgammax}
\eea
and set $\hga^{\bar\al}=\delta^{\bar\al\beta} \hga_\beta$ and $\hga^{\al}=\delta^{\al\bar\beta} \hga_{\bar\beta}$. One can verify that
$\gamma_\alpha \gamma_{\beta}+\gamma_\beta\gamma_\alpha=0$ and $\gamma_\alpha \gamma_{\bar\beta}+\gamma_{\bar\beta}\gamma_\alpha=2 {\bf 1}\, \delta_{\alpha\bar\beta}$.  It is clear that the whole Dirac
representation can be constructed by acting with the ``creation operators'' $\hga^{\bar\al}$ on the Clifford vacuum represented
by the 0-degree form $1$. In curved spaces, a choice of a spacetime coframe which is compatible with the realization of gamma matrices as in
(\ref{hermgammax}) is referred to as either an (almost) Hermitian coframe or a ``spinorial geometry coframe''.

One way to realize the spinor representation of $Spin(2n+1)$ in terms of forms is to add an  additional gamma
matrix $\Ga_{2n+1}$ to those of $Spin(2n)$ proportional to $\prod_{A=1}^{2n} \Ga_A$. The Dirac representation of $Spin(2n+1)$
will coincide with the complex representation of $Pin(2n)$. Moreover for $n$ even, $Spin(2n+1)$ will admit
a Majorana representation provided that $Pin(2n)$ admits one. Similarly for $n$ odd, $Spin(2n+ 1)$ will admit
a Majorana representation provided that $Pin(2n)$ admits one.

\subsection{Lorentzian}

The realization  of spinor representations of $Spin(2n-1, 1)$ in terms of forms proceeds as in the Euclidean case described in the appendix  above. Let  $\bC^n$ be equipped with a Hermitian inner product $\langle \cdot, \cdot\rangle$ and a Hermitian basis $\{e_1, \dots, e_n\}$.  The Dirac representation
is identified with ${}^c\Delta=\Lambda^*(\bC^n)$ and the gamma matrices act as
\bea
\Gamma_0\zeta&=&-\se_n\wedge \zeta+i_{\se_n} \zeta~,~~~\Gamma_n\zeta=\se_n\wedge \zeta+i_{\se_n} \zeta~,
\cr
\Gamma_i\zeta&=&\se_i\wedge \zeta+ i_{\se_i} \zeta~,~~~\Gamma_{i+n}\zeta=i (\se_i\wedge \zeta- i_{\se_i} \zeta)~,~~~i=1,\dots, n-1~,
\eea
where $\zeta\in \Lambda^*(\bC^n)$,  and $i_{\se_n}$ and   $i_{\se_i}$ are  the inner derivations with respect to the vectors constructed from $e_n$ and $e_i$.
  A straightforward computation reveals that the gamma matrices satisfy the Clifford
algebra relation $\Ga_\tA \Ga_\tB+\Ga_\tB \Ga_\tA=2 {\bf 1} \eta_{\tA\tB}$,  where $\eta$ is the Lorentzian metric with mostly plus signature.

The Dirac inner product $D$ is defined as
\bea
\label{dprdx}
D(\eta, \zeta)\defeq \langle\Gamma_0\eta, \zeta\rangle~,
\eea
and it can be shown to be $Spin(2n-1, 1)$ invariant, where $Spin(2n-1, 1)$ is the component of the spin group connected to identity element. Observe that the standard Hermitian inner product $\langle \cdot, \cdot\rangle$ is not invariant. This inner product is also written as $D(\eta, \zeta)=\bar\eta\zeta$, where $\bar\eta$
is called the Dirac conjugate of $\eta$.
Again the Dirac representation of $Spin(2n-1,1)$ is reducible and decomposes
as ${}^c\Delta={}^c\Delta^+\oplus {}^c\Delta^-$ into Weyl chiral and anti-chiral representations according to the decomposition
of $\Lambda^*(\bC^n)$ into forms of even and odd degree,
respectively.

Next consider the linear maps $a=\prod_{i=1}^n \Ga_i$ and $b= \Ga_0\prod_{i=1}^{n-1} \Ga_{i+n}$.
There are two $Spin(2n-1,1)$-invariant bi-linears which can be constructed on ${}^c\Delta$ given by
\bea
{\mathrm A}(\eta, \zeta)=\langle a \eta^*, \zeta\rangle~,~~~~{\mathrm B}(\eta, \zeta)=\langle b\eta^*, \zeta\rangle~,
\eea
where $\eta, \zeta\in {}^c\Delta$.
A straightforward computation following the definitions reveals that
\bea
&&{\mathrm A}(\eta, \Ga_\tA\zeta)=(-1)^{n-1}{\mathrm A}(\Ga_\tA\eta, \zeta)~,~~{\mathrm A}(\eta, \Ga_\tA\zeta)=(-1)^{{(n+2) (n-1)\over2}}{\mathrm A}(\zeta, \Ga_\tA\eta)~,
\cr
&&{\mathrm B}(\eta, \Ga_\tA\zeta)=(-1)^n{\mathrm B}(\Ga_\tA\eta, \zeta)~,~~{\mathrm B}(\eta, \Ga_\tA\zeta)=(-1)^{{n (n-1)\over2}}{\mathrm B}(\zeta, \Ga_\tA\eta)~.
\eea
It is clear that both ${\mathrm A}$ and ${\mathrm B}$ are invariant under $Spin(2n-1,1)$ and in addition that one
of them is also invariant under $Pin(2n-1,1)$. Also the gamma matrices are either symmetric or skew-symmetric with respect to
these bi-linears.

In the Lorentzian case, real representations are constructed by relating the Dirac and Majorana conjugates. So one considers the
anti-linear maps
$r_{{\mathrm A}}\defeq\Gamma_0a *$ and $r_{{\mathrm B}}\defeq\Gamma_0b *$ and after some straightforward calculation finds that
\bea
&&r_{{\mathrm A}}^2=(-1)^{{(n+2) (n-1)\over2}} {\bf 1}~,~~~r_{{\mathrm A}} \Ga_\tA=(-1)^{n} \Ga_\tA r_{{\mathrm A}}~,
\cr
&&r_{{\mathrm B}}^2=(-1)^{{n (n-1)\over2}} {\bf 1}~,~~~r_{{\mathrm B}} \Ga_\tA=(-1)^{n-1} \Ga_\tA r_{{\mathrm B}}~.
\eea
Imposing a reality condition ${\mathcal R}$ which is proportional to $r_{{\mathrm A}}$ and $r_{{\mathrm B}}$ up to a phase, one finds that $Spin(2n-1,1)$ has real representations provided that $[{n-1\over2}]\in 2\bZ$. Moreover, there are Majorana-Weyl
representations provided $n\in 4\bZ+1$. The linear maps $C_{{\mathrm A}}=\Gamma_0 a$ and $C_{{\mathrm B}}=\Gamma_0 b$
are called charge conjugation matrices.

The Dirac representation admits an oscillator basis as in the  Euclidean case.  In particular, one has that
\bea
&&\hga_-={1\over \sqrt{2}} (\Ga_n-  \Ga_{0})=\sqrt{2}\, \se_n\wedge~,~~\hga_\al={1\over \sqrt{2}} (\Ga_\al-i \Ga_{\al+n})=\sqrt{2}\, \se_\al\wedge ~,
\cr
&&\hga_+={1\over \sqrt{2}} (\Ga_n+  \Ga_{0})=\sqrt{2}\, i_{\se_n}~,~~~\hga_{\bar\al}={1\over \sqrt{2}} (\Ga_\al+i \Ga_{\al+n})=
\sqrt{2}\, i_{\se_\al}
\label{Lor-Her}
\eea
satisfy the Clifford algebra relation $\gamma_\tA \gamma_\tB+\gamma_\tB\gamma_\tA=2 {\bf 1} \eta_{\tA\tB}$, where now
the non-vanishing component of the Lorentzian metric $\eta$ in this basis are $\eta_{+-}=1$ and $\eta_{\alpha\bar\beta}=\delta_{\alpha\bar\beta}$.   It is clear that the whole Dirac
representation can be constructed by acting  with the ``creation operators'' $(\hga^+, \hga^{\bar\al})$ on the Clifford vacuum represented
by the 0-degree form $1$, where $\gamma^\tA=\eta^{\tA\tB} \gamma_\tB$.  This basis is analogous to the Hermitian basis in the Euclidean
case described in the previous appendix.  The difference is the two light-cone directions it contains. On curved spaces,
there is a (local) spacetime coframe for which the gamma matrices take the form (\ref{Lor-Her}).  In the spinorial geometry approach, the solutions to the KSEs are expressed in such a coframe.  Because of this,  we shall refer to such coframe as a ``spinorial geometry coframe''.

The construction of spinor representations of $Spin(2n,1)$ can be done in a way similar to the one we have explained
for $Spin(2n+1)$. Though here for the realization of the Majorana representations of $Spin(2n,1)$ in terms of forms, one can begin from
Majorana representations of either $Spin(2n)$ or $Spin(2n-1,1)$.  This has been utilized in the solution of the KSE
of $d=11$ supergravity for $N=1$ backgrounds.

\section{Group manifolds, symmetric and homogeneous spaces}

\subsection{Homogeneous spaces}\label{homospaces}

A detailed exposition of the geometry of group manifolds,  symmetric and homogeneous spaces can be found in \cite{skkn-append}. Here we summarize some basic properties of the latter which have been used throughout the review.
Consider the left coset space $G/H$, where $G$ is a  Lie group which acts  effectively from the left on $G/H$ and $H$ is a closed Lie subgroup of $G$. Let us denote the Lie algebras of $G$ and $H$ with  $\mathfrak{g}$ and $\mathfrak{h}$, respectively, and  assume that
there is a decomposition $\mathfrak{g}=\mathfrak{h} \oplus \mathfrak{m}$ such that

\bea\label{commutation}
&&[h_\alpha, h_\beta] = f_{\alpha\beta}{}^\gamma \, h_\gamma~,~~~
[h_\alpha, m_\tA] = f_{\alpha \tA}{}^\tB \, m_\tB~,
\cr
&&[m_\tA,m_\tB] = f_{\tA\tB}{}^\tC \, m_\tC + f_{\tA\tB}{}^\alpha \, h_\alpha~,
\eea
where $h_\alpha$, $\alpha=1,2,..., \dim{\mathfrak{h}}$ and $m_\tA$, $A=1,..., \dim{\mathfrak{g}}-\dim{\mathfrak{h}}$ are  bases in $\mathfrak{h}$ and $\mathfrak{m}$, respectively.
If $f_{\tA\tB}{}^\tC=0$, that is $[\mathfrak{m},\mathfrak{m}] \subseteq \mathfrak{h}$, $G/H$ will be a symmetric space.

Let $s: U\subset G/H\rightarrow G$  be a local section of the coset. The decomposition of the Maurer-Cartan form in components along  $\mathfrak{h}$ and  $\mathfrak{m}$ is
\begin{align}
s^{-1} ds = \bbl^\tA \, m_\tA + \Psi^\alpha \, h_\alpha~,
\label{MC}
\end{align}
which  defines a local left-invariant coframe $\bbl^\tA$  and a canonical left-invariant connection  $\Psi^\alpha$ on  $ G/H$. The curvature and torsion of the canonical connection are
\bea\label{dei}
&&R^\alpha \defeq d\Psi^\alpha+\frac12 f_{\beta\gamma}{}^\alpha \Psi^\beta\wedge \Psi^\gamma=-\frac12 f_{\tB\tC}{}^\alpha \bbl^\tB\wedge \bbl^\tC~,
\cr
&&T^\tA\defeq d\bbl^\tA+f_{\beta \tC}{}^\tA \Psi^\beta\wedge \bbl^\tC=-\frac12 f_{\tB\tC}{}^\tA \bbl^\tB\wedge \bbl^\tC~,
\eea
respectively, where the equalities follow after  taking the exterior derivative of (\ref{MC}) and  using (\ref{commutation}).  If $G/H$ is symmetric, then
the torsion vanishes.

Left-invariant metrics $ds^2$ and    p-forms $\omega$ on $ G/H$ can be written as
\bea
ds^2=g_{\tA\tB} \bbl^\tA \bbl^\tB~,~~~
\omega= \frac{1}{p!} \, \omega_{\tA_1 ... \tA_p} \, \bbl^{\tA_1} \wedge ... \wedge \bbl^{\tA_p}~,
\eea
respectively, where the components $g_{\tA\tB}$ and   $\omega_{A_1...A_p}$ are constant and satisfy
\begin{align}\label{hinvariance}
f_{\alpha (\tA}{}^\tC g_{\tB)\tC}~,~~~ f_{\alpha[\tA_1}{}^\tB \, \omega_{\tA_2...\tA_p]\tB} =0~.
\end{align}
The latter condition is required for invariance under the right action of $H$ on $G$. All left-invariant forms are  parallel with respect to the canonical connection.

For symmetric spaces, the canonical connection coincides with the Levi-Civita connection of invariant metrics. However for the rest of the homogeneous spaces this is not the case as the canonical connection has non-vanishing torsion.
Let $\Omega$ be the Levi-Civita connection of an invariant metric $ds^2$ in the left-invariant coframe. As the difference of two connections is a tensor, we set
\bea
\Omega{}^\tA{}_\tB= \Psi^\alpha f_{\alpha \tB}{}^\tA+\bbl^\tC Q_{\tC,}{}^\tA{}_\tB~.
\eea
Requiring that $\Omega$ is metric and torsion free,
\begin{align}\label{levi-civ}
\Omega_{\tA\tB} + \Omega_{\tB\tA} = 0~,~~~
d\bbl^\tA + \Omega^\tA{}_\tB \, \wedge \, \bbl^\tB =0~,
\end{align}
respectively, one finds that
\begin{align}
\Omega^\tA{}_\tB =  \Psi^\alpha\, f_{\alpha \tB}{}^\tA + \frac{1}{2} \left( g^{\tA\tD} \, f_{\tD\tB}{}^\tE \, g_{\tC\tE} +  g^{\tA\tD} \, f_{\tD\tC}{}^\tE \, g_{\tB\tE} + f_{\tC\tB}{}^\tA\right) \, \bbl^\tC~.
\end{align}
In turn, the Riemann  curvature 2-form $R^\tA{}_\tB$ is
\begin{align}
R^\tA{}_\tB = \frac{1}{2} \left( Q_{\tC,}{}^\tA{}_{\tE} \, Q_{\tD,}{}^\tE{}_{\tB} - Q_{\tD,}{}^\tA{}_{\tE} \, Q_{\tC,}{}^\tE{}_{\tB} - Q_{\tE,}{}^\tA{}_{\tB} \, f_{\tC\tD}{}^\tE - f_{\tC\tD}{}^\alpha \, f_{\alpha \tB}{}^\tA \right) \, \bbl^\tC \wedge \bbl^\tD~.
\end{align}
  Note that the expression for $\Omega^\tA{}_\tB$ is considerably  simplified whenever the coset space is naturally reductive because the structure constants $f_{\tA\tB\tC}=f_{\tA\tB}{}^\tE \, g_{\tC\tE}$ are then skew-symmetric.

\subsection{Cahen-Wallach spaces} \label{cwmanifolds}

Cahen-Wallach spaces $CW_n$ are plane-wave spacetimes which are also symmetric spaces. In Brinkmann coordinates, their metric
can be written as
\bea
ds^2= 2dv (du+A_{ij} y^i y^j dv)+\delta_{ij} dy^i dy^j~,
\label{cwman}
\eea
where $A$ is a constant matrix.

A  subclass of  $CW_n$ spaces   are also group manifolds.  To identify these, consider the non-vanishing Lie bracket commutators
\bea
[t_i, t_j]=-\beta_{ij}\, t_+~,~~~[t_-, t_i]= \beta^j{}_i\, t_j~,
\label{cwncomrel}
\eea
where $\{t_+, t_-, t_i: i=1, \dots, n-2\}$ are some generators and  $(\beta_{ij})$ is a non-degenerate skew-symmetric matrix, $\beta_{ij}=\delta_{ik} \beta^k{}_j$.  These give rise to the Maurer-Cartan relations
\bea
d\bbl^-=0~,~~~d\bbl^i=-\beta^i{}_j \bbl^-\wedge \bbl^j~,~~~d\bbl^+={1\over2} \beta_{ij} \bbl^i\wedge \bbl^j~,
\eea
where $\bbl=\bbl^- t_-+\bbl^+ t_++ \bbl^i t_i$ is a left invariant coframe.
These can be solved as
\bea
&&\bbl^-=dv~,~~~\bbl^i= dx^i+ \beta^i{}_j x^j dv~,~~~
\cr
&&\bbl^+=du+\beta_{ij} x^i dx^j-{1\over2} \beta_{ki} \beta^k{}_jx^i x^j dv~,
\label{cwcoord}
\eea
for some coordinates $(u,v, x^i)$.
The most general bi-invariant metric up to an overall scale and a redefinition of $\bbl^+$ as $\bbl^+\rightarrow \bbl^++\lambda \bbl^-$ is
\bea
ds^2=2 \bbl^-\bbl^+ + \delta_{ij} \bbl^i \bbl^j~.
\eea
Substituting (\ref{cwcoord}) into the metric, and after a further coordinate transformation $y=e^{{1\over2}v\beta } x$, one finds (\ref{cwman}) for $A_{ij}=-{1\over 8} \beta_{ki} \beta^k{}_j$. Observe that without loss of generality $\beta$ can be chosen to be in
block-diagonal form in which case $A$ becomes  a  diagonal negative definite  matrix.
Amongst the $CW_n$ spaces that are group manifolds, $CW_6$ with self-dual structure constants $\beta$ appears in the description
of supersymmetric backgrounds in $d=5$ and $d=6$ supergravities as well as in the heterotic theory.

\section{Fierz identities for $d=5$ supergravity} \label{fierzfive}

Let $\epsilon_1, \epsilon_2, \epsilon_3, \epsilon_4$ be Dirac spinors of $Spin(4,1)$.  These  satisfy the following Fierz identity
\begin{eqnarray}
\label{fierz5a}
D(\epsilon_1, \epsilon_2) D(\epsilon_3, \epsilon_4)
&=& {1 \over 4}D(\epsilon_1, \epsilon_4) D(\epsilon_3, \epsilon_2)
+{1 \over 4}D(\epsilon_1, \Gamma_\tA \epsilon_4) D(\epsilon_3, \Gamma^\tA \epsilon_2)
\nonumber \\
&-&{1\over 8}D(\epsilon_1, \Gamma_{\tA \tB} \epsilon_4) D(\epsilon_3, \Gamma^{\tA \tB} \epsilon_2)~,
\end{eqnarray}
where $D$ is the Dirac inner product given in ({\ref{dprdx}}). This Fierz identity
is equivalent to
\begin{eqnarray}
\label{fierz5b}
D(\epsilon_3, \epsilon_4) \epsilon_2 &=& {1 \over 4} D(\epsilon_3, \epsilon_2) \epsilon_4
+{1 \over 4} D(\epsilon_3, \Gamma_\tA \epsilon_2) \Gamma^\tA \epsilon_4
\nonumber \\
&-&{1 \over 8} D(\epsilon_3, \Gamma_{\tA \tB} \epsilon_2) \Gamma^{\tA \tB} \epsilon_4 \ .
\end{eqnarray}

These Fierz identities
differ from those used in \cite{gghpr}, because in that work a mostly minus signature spacetime metric was used, whereas here we use a mostly plus metric. Given a spinor $\epsilon$, we use the Fierz identities
to obtain algebraic conditions on the   bilinears
defined in ({\ref{sp5bilin1}}) and ({\ref{sp5bilin2}}).

\begin{itemize}
\item[(i)] Setting $\epsilon_1=\epsilon_2=\epsilon_3=\epsilon_4=\epsilon$ gives
\begin{equation}
-{3 \over 4} f^4 = {1 \over 4}X^2 -{1 \over 8}(\omega_1)^2 \ .
\end{equation}
\item[(ii)] Setting $\epsilon_1=\epsilon_4= r_{\mathrm A} \epsilon $, $\epsilon_2=\epsilon_3=\epsilon$
gives
\begin{equation}
{1 \over 4} f^4 = {1 \over 4} X^2 +{1 \over 8} (\omega_1)^2 \ .
\end{equation}
\item[(iii)] Setting $\epsilon_1= \epsilon_2= r_{\mathrm A} \epsilon $, $\epsilon_3 = \epsilon_4=\epsilon$ gives
\begin{equation}
\xi_{\tA \tB} {\bar{\xi}}^{\tA \tB}= 8f^4 \ .
\end{equation}

\item[(iv)] Setting $\epsilon_1=\epsilon_2=\epsilon_3=\epsilon$, $\epsilon_4=r_{\mathrm A} \epsilon $
gives
\begin{equation}
\label{aux5a}
(\omega_1)_{\tA \tB} \xi^{\tA \tB}=0 \ .
\end{equation}
\item[(v)] Setting $\epsilon_1=\epsilon_3=\epsilon$, $\epsilon_2=\epsilon_4=r_{\mathrm A} \epsilon $
gives
\begin{equation}
\label{aux5d}
\xi^2 =0 \ .
\end{equation}

\end{itemize}

In particular, these conditions also imply that
\begin{equation}
X^2 = -f^4~,
\end{equation}
so the vector  bilinear is either null or timelike depending on whether
$f=0$ or $f \neq 0$ respectively. Furthermore, we also have
\begin{equation}
(\omega_1)^2 = 4f^4 \ .
\end{equation}

In obtaining these expressions we have made use of the identities
\begin{equation}
D(r_{\mathrm A} \epsilon_1, r_{\mathrm A} \epsilon_2) = D(\epsilon_2,\epsilon_1)~,
\end{equation}
and
\begin{equation}
D(\Gamma_\tA \epsilon_1, \epsilon_2)=-D(\epsilon_1, \Gamma_\tA \epsilon_2),
\qquad D(\Gamma_{\tA \tB} \epsilon_1, \epsilon_2) =- D(\epsilon_1, \Gamma_{\tA \tB} \epsilon_2)~,
\end{equation}
which imply
\begin{equation}
D(r_{\mathrm A} \epsilon, \Gamma_\tA r_{\mathrm A} \epsilon)=D(\epsilon,\Gamma_\tA \epsilon),
\qquad D(r_{\mathrm A} \epsilon, \Gamma_{\tA \tB} r_{\mathrm A} \epsilon)=-D(\epsilon,\Gamma_{\tA \tB} \epsilon) \ .
\end{equation}

Further algebraic conditions, which are also useful in determining the various types of projection conditions which the spinors must satisfy are obtained by considering the
Fierz identity (\ref{fierz5b}). In particular, on setting $\epsilon_2=\epsilon_3=\epsilon_4=\epsilon$, and also $\epsilon_2=\epsilon_3=r_{\mathrm A} \epsilon,
\epsilon_4 =\epsilon$ and comparing the expressions gives
\begin{equation}
\label{prj5a}
X_\tA \Gamma^\tA \epsilon = if^2 \epsilon \ ,
\end{equation}
and
\begin{equation}
\label{prj5b}
(\omega_1)_{\tA \tB} \Gamma^{\tA \tB}\epsilon = -4if^2 \epsilon \ .
\end{equation}
Also, setting $\epsilon_3=\epsilon_4=\epsilon$, $\epsilon_2=r_{\mathrm A} \epsilon$ gives
\begin{equation}
\label{prj5c}
\xi_{\tA \tB} \Gamma^{\tA \tB} \epsilon = -8if^2 \epsilon~,
\end{equation}
and setting $\epsilon_3 = \epsilon$, $\epsilon_2=\epsilon_4 = r_{\mathrm A} \epsilon$
leads to
\begin{equation}
\label{prj5d}
{\bar{\xi}}_{\tA \tB} \Gamma^{\tA \tB} \epsilon=0 \ .
\end{equation}

The condition ({\ref{prj5b}}) implies that
\begin{equation}
D(\epsilon, \Gamma_L (\omega_1)_{\tA \tB} \Gamma^{\tA \tB} \epsilon) = D(\epsilon, -4if^2 \Gamma_L \epsilon)~,
\end{equation}
which on taking real and imaginary parts gives
\begin{equation}
i_X  \omega_1=0~,
\end{equation}
and
\begin{equation}
(\omega_1)_{\tA \tB} {}^* (\omega_1)_\tC{}^{\tA \tB} = 4f^2 X_\tC \ .
\end{equation}
Similarly, ({\ref{prj5c}}) and ({\ref{prj5d}}) lead to
\begin{equation}
i_X \xi=0, \qquad \xi_{\tA \tB} {}^* (\omega_1)_\tC{}^{\tA \tB} = 0 , \qquad \xi_{\tA \tB} {}^* {\bar{\xi}}_\tC{}^{\tA \tB} = 8f^2 X_\tC,
\qquad \xi_{\tA \tB} {}^* \xi_\tC{}^{\tA \tB}=0~.
\end{equation}

Next, note that ({\ref{prj5a}}) implies that
\begin{equation}
D(\epsilon, \Gamma_{\tA \tB} X_\tC \Gamma^\tC \epsilon) = if^2 (\omega_1)_{\tA \tB}, \qquad
D(\epsilon, \Gamma_{\tA \tB} X_\tC \Gamma^\tC r_{\mathrm A} \epsilon) = if^2 \xi_{\tA \tB}~,
\end{equation}
which in turn gives that
\begin{equation}
i_X {}^* (\omega_1)_{\tA \tB} = -f^2 (\omega_1)_{\tA \tB}, \qquad i_X {}^* \xi_{\tA \tB} = -f^2 \xi_{\tA \tB} \ .
\end{equation}

Additional  bilinear identities are then obtained from ({\ref{fierz5a}}) on setting
\begin{itemize}
\item[(a)] Setting $\epsilon_1=\epsilon_3 = \epsilon$, $\epsilon_2=\Gamma_\tB \epsilon$, $\epsilon_4=\Gamma_\tA \epsilon$ gives
\begin{equation}
(\omega_1)_{\tC \tA} (\omega_1)^\tC{}_\tB = X_\tA X_\tB + f^4 g_{\tA \tB} \ .
\end{equation}
\item[(b)] Setting $\epsilon_1 =\epsilon_3 =\epsilon$, $\epsilon_2 = \Gamma_\tB r_{\mathrm A} \epsilon$,
$\epsilon_4 = \Gamma_\tA \epsilon$ gives
\begin{equation}
(\omega_1)_{\tC(\tA} \xi^\tC{}_{\tB)} =0 \ .
\end{equation}
Furthermore, the identity ({\ref{prj5b}}) implies that
\begin{equation}
D(\epsilon, \Gamma_{\tA \tB} (\omega_1)_{\tC \tD} \Gamma^{\tC \tD} r_{\mathrm A} \epsilon ) = 4if^2 \Gamma_{\tA \tB} r_{\mathrm A} \epsilon~,
\end{equation}
which implies that
\begin{equation}
(\omega_1)_{\tB \tC} \xi_\tA{}^\tC - (\omega_1)_{\tA \tC} \xi_\tB{}^\tC = 2if^2 \xi_{\tA \tB} \ .
\end{equation}
It follows that
\begin{equation}
(\omega_1)_{\tB \tC} \xi_\tA{}^\tC = if^2 \xi_{\tA \tB} \ .
\end{equation}
\item[(c)] Setting $\epsilon_1 = \epsilon_3=\epsilon$, $\epsilon_2 = \Gamma_\tB r_{\mathrm A} \epsilon$,
$\epsilon_4= \Gamma_\tA r_{\mathrm A} \epsilon$ gives
\begin{equation}
\xi_{\tC \tA} \xi^\tC{}_\tB =0~,
\end{equation}
on using ({\ref{aux5d}}).
\item[(d)] Setting $\epsilon=r_{\mathrm A} \epsilon$, $\epsilon_2 = \Gamma_\tB r_{\mathrm A} \epsilon$,
$\epsilon_3 = \epsilon$, $\epsilon_4 = \Gamma_\tA \epsilon$ gives
\begin{equation}
{\bar{\xi}}_{\tC \tA} \xi^\tC{}_\tB + {\bar{\xi}}_{\tC \tB} \xi^\tC{}_\tA = 4 X_\tA X_\tB + 4f^4 g_{\tA \tB} \ .
\end{equation}
Also, the condition ({\ref{prj5c}}) implies that
\begin{equation}
D(r_{\mathrm A} \epsilon, \Gamma_{\tA \tB} \xi_{\tC \tD} \Gamma^{\tC \tD} \epsilon)
= -8if^2 D(r_{\mathrm A} \epsilon, \Gamma_{\tA \tB} r_{\mathrm A} \epsilon)~,
\end{equation}
and hence
\begin{equation}
{\bar{\xi}}_\tA{}^\tC \xi_{\tB \tC} - {\bar{\xi}}_\tB{}^\tC \xi_{\tA \tC} = 4if^2 (\omega_1)_{\tA \tB} \ .
\end{equation}
It follows that
\begin{equation}
{\bar{\xi}}_{\tC \tA} \xi^\tC{}_\tB = 2if^2 (\omega_1)_{\tA \tB} +2 X_\tA X_\tB +2f^4 g_{\tA \tB} \ .
\end{equation}

\end{itemize}
This completes the Fierz identities needed to solve the KSEs of ${\cal N}=1$ $d=5$ supergravity.

\section{$d=11$ and type II $d=10$ supergravities}

Here we  summarize key properties  of $d=11$ and  IIB $d=10$ supergravities
that we are using throughout the review. Some additional formulae which include the integrability conditions of the KSEs
 are also given.

\subsection{$d=11$ supergravity} \label{11super}

The action of the bosonic fields of  $d=11$ supergravity \cite{ecbjjs-11d} is
\bea
 I= \int_M\, ( \tfrac{1}{2}\, R\, \dvol + \tfrac14 F\wedge\star F + \tfrac1{12}
    F \wedge F \wedge A )~,
\eea
where $\elF=dA$, $A$  is the 3-form gauge potential, $R$ is the scalar
curvature of the metric $g$ and $\dvol$ is the spacetime volume form. For a superspace formulation see \cite{Brink:1980az}.

The KSE of $d=11$ supergravity has already been given in (\ref{11kse}) and the supercovariant derivative ${\mathcal D}$ has been
presented in (\ref{11supercor}).  The supersymmetry parameter $\eps$  is
in the 32-dimensional Majorana representation  $\Delta_{32}$ of $Spin(10,1)$.

The integrability condition of the gravitino KSE  is
\bea
[{\mathcal D}_\tM, {\mathcal D}_\tN]\eps\equiv{\mathcal R}_{\tM\tN}\eps=0~,
\eea
where ${\mathcal R}$ is the curvature of the supercovariant connection \cite{jfofgp1}
\bea
  \label{supelcurv}
  {\mathcal R}_{\tM\tN}&=& \tfrac14 R_{\tM\tN,\tA\tB}  \Gamma^{\tA\tB} + \frac{2}{(288)^2}
  \elF_{\tA_1\dots \tA_4} \elF_{\tB_1\dots \tB_4}
  \epsilon_{\tM\tN}{}^{\tA_1\dots \tA_4 \tB_1\dots \tB_4}{}_\tC \Gamma^\tC
  \cr
  &&+ \tfrac{48}{(288)^2}\big[4 \elF_{\tM\tA_1\tA_2\tA_3 } \elF^{\tA_1\tA_2\tA_3}{}_{\tB} \Gamma^{\tB}{}_\tN-
  4 \elF_{\tN\tA_1\tA_2\tA_3} \elF^{\tA_1\tA_2\tA_3}{}_{\tB}\Gamma^{\tB}{}_\tM
  \cr
&&-36 \elF_{\tA\tB\tM\tC} \elF^{\tA\tB}{}_{\tN\tD} \Gamma^{\tC\tD}
  +\elF_{\tA_1\dots \tA_4} \elF^{\tA_1\dots \tA_4} \Gamma_{\tM\tN}\big]
  \cr
 && +\tfrac{1}{36} \big[\nabla_\tM \elF_{\tN\tA_1\tA_2\tA_3 }-\nabla_\tN \elF_{\tM\tA_1\tA_2\tA_3 }\big]
  \Gamma^{\tA_1\tA_2\tA_3 }
  \cr
  &&-\tfrac{8}{(288)^2 3}\big[\elF_{\tB_1\dots \tB_4} \elF_{\tC_1\tC_2 \tC_3\tN}
  \epsilon_\tM{}^{\tB_1\dots \tB_4\tC_1\tC_2 \tC_3}{}_{\tA_1\tA_2\tA_3}
  - (\tN\leftrightarrow \tM)\big] \Gamma^{\tA_1\tA_2\tA_3}
  \cr
  &&-\tfrac{1}{ 432}\big[4 \elF_{\tC\tA_1\tA_2 \tA_3} \elF^\tC{}_{\tM\tN \tA_4}
  \Gamma^{\tA_1\dots \tA_4}
  \cr
  &&+3 \elF_{\tB\tC\tA_1\tA_2} \elF^{\tB\tC\tA_3}{}_\tN \Gamma^{\tA_1\tA_2}{}_{\tM \tA_3}-3 \elF_{\tB\tC\tA_1\tA_2}
  \elF^{\tB\tC\tA_3}{}_\tM\Gamma^{\tA_1\tA_2}{}_{\tN\tA_3}\big]
  \cr
  &&
  -\tfrac{1}{288}\big[\nabla_\tM\elF_{\tA_1\dots \tA_4} \Gamma^{\tA_1\dots \tA_4}{}_\tN-
  (\tN \leftrightarrow \tM)\big]
  \cr
  &&
  -\tfrac{1}{(72)^2 5!} \big[-6 \elF_{\tM \tB_1\tB_2 \tB_3} \elF_{\tN \tC_1\tC_2 \tC_3}
  \epsilon^{\tB_1\tB_2 \tB_3\tC_1\tC_2 \tC_3}{}_{\tA_1\dots \tA_5}
  \cr &&
  - 6 F_{\tM \tP \tB_1\tB_2} F^\tP{}_{\tC_1\tC_2 \tC_3}
  \epsilon_\tN{}^{\tB_1\tB_2\tC_1\tC_2 \tC_3}{}_{\tA_1\dots \tA_5}
  \cr
  &&
  +6 \elF_{\tN\tP\tB_1\tB_2} \elF^\tP{}_{\tC_1\tC_2 \tC_3}
  \epsilon_\tM{}^{\tB_1\tB_2\tC_1\tC_2 \tC_3}{}_{\tA_1\dots \tA_5}
  \cr
  &&
  +9\elF_{\tL\tP\tB_1\tB_2} \elF^{\tL\tP}{}_{\tC_1\tC_2 }
  \epsilon_{\tM\tN}{}^{\tB_1\tB_2\tC_1\tC_2 }{}_{\tA_1\dots \tA_5}\big]
  \Gamma^{\tA_1\dots \tA_5}~,
\eea
and where we have used that
\begin{equation}
  \Gamma^{\tA_1\dots \tA_{2k}}=-\tfrac{(-1)^k}{ (11-2k)!} \epsilon^{\tA_1\dots
  \tA_{2k}}{}_{\tB_1\dots \tB_{11-2k}} \Gamma^{\tB_1\dots \tB_{11-2k}}~,
\end{equation}
with $\epsilon_{01\dots 9\nat}=-1$.

One expects that $\Gamma^\tB {\mathcal R}_{\tA\tB}$ can be expressed in terms of the field equations and Bianchi identities
as it arises from the supersymmetry variation of the gravitino field equation. A direct computation reveals \cite{uggpdr-11d} that
\begin{eqnarray}
\Gamma^\tB {\mathcal R}_{\tA\tB} & = & E_{\tA\tB} \Gamma^{\tB}
  -\tfrac{1}{36} L\elF_{\tC_1\tC_2\tC_3} (\Gamma_\tA{}^{\tC_1\tC_2\tC_3}
  - 6 \delta_\tA^{\tC_1} \Gamma^{\tC_2\tC_3}) + \nonumber \\
&& \;\; +\tfrac{1}{6!} B\elF_{\tC_1\dots \tC_5} (\Gamma_\tA{}^{\tC_1 \cdots \tC_5}
 - 10 \delta_\tA^{\tC_1} \Gamma^{\tC_2 \cdots \tC_5}) \,,
 \label{intcond}
 \end{eqnarray}
 where
 \bea
 E_{\tA\tB}&\defeq& R_{\tA\tB} -\tfrac{1}{12} \elF_{\tA\tC_1\tC_2\tC_3}\elF_\tB{}^{\tC_1\tC_2\tC_3}
 +\tfrac{1}{144} g_{\tA\tB}\elF_{\tC_1\cdots \tC_4} \elF^{\tC_1\cdots
 \tC_4}\,, \nonumber \\
 L\elF_{\tA\tB\tC}&\defeq& *(d*\elF+\tfrac{1}{2} \elF\wedge \elF)_{\tA\tB\tC}\,, \nonumber \\
 \label{d11feqn}
 \eea
 are the Einstein and 4-form flux field equations, respectively, and
 \bea
 B\elF_{\tA_1\dots \tA_5}\defeq   (d\elF)_{\tA_1\dots \tA_5}~,
 \label{d11bids}
 \eea
is the Bianchi identity of $\elF$. Clearly, this vanishes  provided that $\elF$ is closed which is the case
in the context of $d=11$ supergravity.

\subsection{IIB supergravity} \label{iibsuper}

The bosonic fields of IIB supergravity \cite{jspw-iib, js-iib}  are the spacetime metric $g$, two real scalars,
 the axion $\sigma$
and the dilaton $\phi$, which are combined into a complex 1-form field strength $P$,
two 3-form field strengths $G^1$ and $G^2$ which are combined to a complex 3-form
field strength $G$, and a self-dual
5-form field strength $F$. To describe these, we introduce
a $SU(1,1)$ matrix $U=(V_+^a, V_-^a)$, $a=1,2$ such that
\bea
V_-^a V_+^b-V_-^b V_+^a=\epsilon^{ab}~,
\eea
where $\epsilon^{12}=1=\epsilon_{12}$,  $(V_-^1)^*=V_+^2$ and $(V_-^2)^*=V_+^1$.  The signs denote
$U(1)\subset SU(1,1)$ charge. Then set
\bea
P_\tM\defeq -\epsilon_{ab} V_+^a \partial_\tM  V_+^b~,~~~
Q_\tM\defeq -i \epsilon_{ab} V_-^a\partial_\tM V_+^b~.
\eea
The 3-form  field strengths $G_{\tM\tN\tR}^a=3\partial_{[\tM} A^a_{\tN\tR]}$, with $(A^1_{\tM\tN})^*= A^2_{\tM\tN}$ combine into the complex
field strength
\bea
G_{\tM\tN\tR}\defeq -\epsilon_{ab} V^a_+ G_{\tM\tN\tR}^b~.
\eea
The five-form self-dual field strength is
\bea
F_{\tM_1\tM_2\tM_3\tM_4\tM_5}\defeq 5\partial_{[\tM_1} A_{\tM_2\tM_3\tM_4\tM_5]}+{5i\over8} \epsilon_{ab} A^a_{[\tM_1\tM_2} G^b_{\tM_3\tM_4\tM_5]}~,
\eea
where
$F_{\tM_1\dots \tM_5}= {1\over 5!}
\epsilon_{\tM_1\dots \tM_5}{}^{\tN_1\dots \tN_5}
F_{\tN_1\dots \tN_5}$ and $\epsilon_{01\dots9}=-1$.
The axion $\sigma$ and the dilaton $\phi$ fields  can be combined into a complex scalar, $\tau\defeq \sigma+i e^{-\phi}$.  In turn this is related to $V$ as
\bea
{V_-^2\over V_-^1}={1+i\tau\over 1-i\tau}~.
\eea
This completes the description of the bosonic fields of the theory.

The  KSEs of IIB supergravity are the gravitino KSE which is the parallel transport equation
\bea
{\cal D}\eps=0~,
\la{kseqna}
\eea
 of the
supercovariant connection
\be
{\mathcal D}_\tM\defeq \tilde\nabla_\tM +{i\over 48} \Gamma^{\tN_1\dots \tN_4 }
 F_{\tN_1\dots \tN_4 \tM}
 -{1\over 96} (\Gamma_{\tM}{}^{\tN_1\tN_2\tN_3}
G_{\tN_1\tN_2\tN_3}-9 \Gamma^{\tN_1\tN_2} G_{\tM\tN_1\tN_2}) C*~,
\ee
and the algebraic KSE
\be
{\mathcal A}\epsilon\defeq P_\tM \Gamma^\tM C\epsilon^*+ {1\over 24} G_{\tN_1\tN_2\tN_3} \Gamma^{\tN_1\tN_2\tN_3} \epsilon=0~,
\la{kseqnb}
\ee
where
$$
\tilde\nabla_\tM=D_\tM+{1\over4} \Omega_{\tM,\tA\tB} \Gamma^{\tA\tB}~,~~~~~~D_\tM=\partial_\tM-{i\over2}Q_\tM~,
\label{iibspin}
$$
is the
spin connection, $\nabla_\tM=\partial_\tM+{1\over4} \Omega_{\tM,\tA\tB} \Gamma^{\tA\tB}$,
twisted with the $U(1)$ connection $Q_\tM$, $Q_\tM^*=Q_\tM$.  The supersymmetry parameter, $\epsilon$,
is a complex Weyl spinor,
$\Gamma_{0\dots 9}\epsilon=\epsilon$, and  $C$ is a charge conjugation
 matrix.

The integrability conditions of the KSEs are
\be
[{\mathcal D}_\tM,{\mathcal
D}_\tN]\eps={\mathcal R}_{\tM\tN}\eps=0~,
\la{inta}
\ee
and
 \be
 [{\mathcal
D}_\tM,{\mathcal A}]\eps=0~,
\la{intb}
\ee
where ${\mathcal R}$ is the supercovariant curvature  given in \cite{gptsimpisiib}.
The components  ${\mathcal I}_\tA= \tfrac{1}{2} \Gamma_\tA{}^{\tB\tC}{\mathcal R}_{\tB\tC}$ and
${\mathcal I}=\Gamma^\tM[{\mathcal D}_\tM,{\mathcal A}]$ of the
 integrability conditions can be expressed in terms of field equations and Bianchi identities \cite{ugjggpdr-iib} as
\bea
{\mathcal I}_\tA\eps&=&\big[\tfrac{1}{2}\Gamma^\tB
E_{\tA\tB}- i \Ga^{\tB_1\tB_2 \tB_3}\LF_{\tA\tB_1\tB_2 \tB_3}\big]\eps
\cr
&&~~~~
-\big[\Ga^\tB LG_{\tA\tB} -\Ga_\tA{}^{\tB_1\ldots \tB_4}BG_{\tB_1\ldots \tB_4}
\big]C\eps^*~,
\la{IAint}
 \eea
and similarly
as
\bea {\mathcal
I}\epsilon=\big[\tfrac{1}{2}\Ga^{\tA\tB}LG_{\tA\tB}+\Ga^{\tA_1\ldots
\tA_4}BG_{\tA_1\ldots
\tA_4}\big]\eps+\big[LP+\Ga^{\tA\tB}BP_{\tA\tB}\big]C\eps^*~,
\la{Iint}
\eea
where
\bea
 E_{\tA\tB}&:=& R_{\tA\tB}-\tfrac{1}{2}g_{\tA\tB}R -\tfrac{1}{6}F_{\tA\tC_1\ldots \tC_4}F_{\tB}{}^{\tC_1\ldots \tC_4}
 -\tfrac{1}{4}G_{(\tA}{}^{\tC_1\tC_2}G^*_{\tB)\tC_1\tC_2}\nn\\
 &&+\tfrac{1}{24}g_{\tA\tB}G^{\tC_1\tC_2\tC_3}G^*_{\tC_1\tC_2\tC_3}  -2P_{(\tA}P^*_{\tB)}+g_{\tA\tB}P^\tC P^*_\tC\,,\nn\\
 \LG_{\tA\tB}&:=& \tfrac{1}{4}(\tilde\nabla^\tC G_{\tA\tB\tC}-P^\tC G^*_{\tA\tB\tC}+\tfrac{2i}{3}F_{\tA\tB\tC_1\tC_2\tC_3}G^{\tC_1\tC_2\tC_3})
 \,,\nn\\
 \LP&:=& \tilde\nabla^\tA P_\tA+\tfrac{1}{24}G_{\tA_1\tA_2\tA_3}G^{\tA_1\tA_2\tA_3}\,,\nn\\
 \LF_{\tA_1\ldots \tA_4}&:=& \tfrac{1}{3!}(\nabla^\tB F_{\tA_1\ldots
 \tA_4 \tB}-\tfrac{i}{288} \epsilon_{\tA_1 \ldots \tA_4}{}^{\tB_1 \ldots \tB_6} G_{\tB_1\tB_2\tB_3}G^*_{\tB_4\tB_5\tB_6})\,,\nn\\
 \BF_{\tA_1\ldots \tA_6}&:=& \tfrac{1}{5!}(\partial_{[\tA_1}F_{\tA_2\ldots
 \tA_6]}-\tfrac{5i}{12}G_{[\tA_1\tA_2\tA_3}G^*_{\tA_4\tA_5\tA_6]})\,,\nn\\
 BG_{\tA_1\ldots
 \tA_4}&:=&\tfrac{1}{4!}(D_{[\tA_1}G_{\tA_2\tA_3\tA_4]}+P_{[\tA_1}G^*_{\tA_2\tA_3\tA_4]})\,,\nn\\
 BP_{\tA\tB}&:=& D_{[\tA}P_{\tB]}\,.
\eea
One can show that $\LF$ and $\BF$ are not independent but are related by the self-duality condition on $F$.
The field strengths $P$ and $G$ have different $U(1)\subset SU(1,1)$ charges. In particular, one has
$D_\tM P_\tN\defeq \partial_\tM P_\tN-2i Q_\tM P_\tN$ and $
D_\tM G_{\tN_1\tN_2\tN_3}\defeq \partial_\tM G_{\tN_1\tN_2\tN_3}-i Q_\tM G_{\tN_1\tN_2\tN_3}$.
This concludes the description of the KSEs, field equations and Bianchi identities of the theory.

\bibliography{review_v2}

\bibliographystyle{utphysmodb}







\end{document}